\documentclass[fleqn,usenatbib]{mnras}

\usepackage{newtxtext,newtxmath}

\usepackage[T1]{fontenc}
\usepackage{ae,aecompl}

\usepackage{graphicx}	
\usepackage{amsmath}	

\usepackage[dvipsnames]{xcolor}

\usepackage{tikz,xcolor,hyperref}

\def\arcsec{^{\prime\prime}}

\def\deg{{$^\circ$}} 
\def\fdg{\hbox{$.\!\!^\circ$}}  
\def\apjl{ApJL}
\def\apj{ApJ}
\def\apjs{ApJS}

\def\aj{AJ}
\def\aap{A\&A}

\definecolor{lime}{HTML}{A6CE39}
\DeclareRobustCommand{\orcidicon}{%
    \begin{tikzpicture}
    \draw[lime, fill=lime] (0,0) 
    circle [radius=0.16] 
    node[white] {{\fontfamily{qag}\selectfont \tiny ID}};
    \draw[white, fill=white] (-0.0625,0.095) 
    circle [radius=0.007];
    \end{tikzpicture}
    \hspace{-2mm}
}

\newcommand{\orcidChrisO}{\href{https://orcid.org/0000-0003-0017-349X}{\orcidicon}}
\newcommand{\orcidChrisW}{\href{https://orcid.org/0000-0002-4569-016X}{\orcidicon}}
\newcommand{\orcidSamuel}{\href{https://orcid.org/0000-0001-9372-4611}{\orcidicon}}
\newcommand{\orcidFuyan}{\href{https://orcid.org/0000-0002-1620-0897}{\orcidicon}}
\newcommand{\orcidXiaohui}{\href{https://orcid.org/0000-0003-3310-0131}{\orcidicon}}
\newcommand{\orcidPatrick}{\href{https://orcid.org/0000-0003-4237-0520}{\orcidicon}}

\def\alphaK{$-1.11$}
\def\alphaN{$-2.00$}

\def\betaKlo{$-3.53\pm 0.24$}
\def\MstarKlo{$-26.02\pm 0.18$}
\def\logPhistarKlo{$-7.17$}
\def\betaNlo{$-3.92\pm 0.32$}
\def\MstarNlo{$-27.32\pm 0.13$}
\def\logPhistarNlo{$-8.24$}
\def\Nlow{54}

\def\betaKhi{$-3.38\pm 0.32$}
\def\MstarKhi{$-25.87\pm 0.33$}
\def\logPhistarKhi{$-7.33$}
\def\betaNhi{$-3.60\pm 0.37$}
\def\MstarNhi{$-27.09\pm 0.30$}
\def\logPhistarNhi{$-8.32$}
\def\Nhi{28}

\def\betaThreelo{$-3.18\pm 0.22$}
\def\logPhiThreelo{$-9.18\pm 0.05$}
\def\NThreelo{49}
\def\betaThreehi{$-3.43\pm 0.32$}
\def\logPhiThreehi{$-9.56\pm 0.06$}
\def\NThreehi{30}

\title[Ultra-luminous quasars from SkyMapper. II.]{Ultra-luminous high-redshift quasars from SkyMapper -- \\ II. New quasars and the bright end of the luminosity function}

\author[C.~A.\ Onken et al.
]{Christopher A.\ Onken$^{1,2}$\orcidChrisO, Christian Wolf$^{1,2}$\orcidChrisW, Fuyan Bian$^{3}$\orcidFuyan, Xiaohui Fan$^{4}$\orcidXiaohui, 
\newauthor Wei Jeat Hon$^{5}$, David Raithel$^{1}$, Patrick Tisserand$^{6}$\orcidPatrick, Samuel Lai$^{1}$\orcidSamuel\\
$^1$Research School of Astronomy and Astrophysics, Australian National University, Canberra ACT 2611, Australia\\
$^2$Centre for Gravitational Astrophysics, Australian National University, Canberra ACT 2600, Australia \\
$^3$European Southern Observatory, Alonso de C\'{o}rdova 3107, Casilla 19001, Vitacura, Santiago 19, Chile\\
$^4$Steward Observatory, University of Arizona, 933 North Cherry Avenue, Tucson, AZ 85721, USA\\
$^5$School of Physics, University of Melbourne, Parkville, Victoria 3010, Australia \\
$^6$Sorbonne Universit\'{e}s, UPMC Univ Paris 6 et CNRS, Institut d'Astrophysique de Paris, 98 bis bd Arago, F-75014 Paris, France
}

\date{Accepted XXX. Received YYY; in original form ZZZ}

\pubyear{2021}

\begin{document}
\label{firstpage}
\pagerange{\pageref{firstpage}--\pageref{lastpage}}
\maketitle

\begin{abstract}
We search for ultra-luminous Quasi-Stellar Objects (QSOs) at high redshift using photometry from the SkyMapper Southern Survey Data Release 3 (DR3), in combination with 2MASS, VHS DR6, VIKING DR5, AllWISE, and CatWISE2020, as well as parallaxes and proper motions from {\it Gaia} DR2 and eDR3. 
We report 142 newly discovered Southern QSOs at $3.8<z<5.5$, of which 126 have $M_{145} <-27$~ABmag and are found in a search area of 14\,486~deg$^2$.
This Southern sample, utilising the {\it Gaia} astrometry to offset wider photometric colour criteria, achieves unprecedented completeness for an ultra-luminous QSO search at high redshift. 
In combination with already known QSOs, we construct a sample that is $>80$ per cent complete for $M_{145}<-27.33$~ABmag at $z=4.7$ and for $M_{145}<-27.73$~ABmag at $z=5.4$. 
We derive the bright end of the QSO luminosity function at restframe 145~nm for $z=4.7-5.4$ and measure its slope to be $\beta =$\betaNhi\ and $\beta =$\betaKhi\ for two different estimates of the faint-end QSO density adopted from the literature.
We also present the first $z\sim 5$ QSO luminosity function at restframe 300~nm.
\end{abstract}

\section{Introduction}\label{intro}

Supermassive black holes can be observed across vast distances, provided they accrete matter at a sufficient rate. Radiation released in the accretion process makes these objects the most luminous in the entire Universe. As long as they are not obscured by local dust, they will be seen as Quasi-Stellar Objects (QSOs) with characteristic spectral signatures. 
Ultra-luminous QSOs point us to the most massive and fastest growing black holes in the Universe. The demographics of this type of object are of particular interest at high redshift, an era in the early universe in which black holes undergo their most dramatic and least-explained growth.

As these most extreme objects are intrinsically rare, any search for them has always been for the proverbial needles in a haystack. Over the last two decades, useful samples of high-redshift ultra-luminous QSOs have been detected, supported by a range of massive data sets including the iconic Sloan Digital Sky Survey \citep[SDSS;][]{SDSS}, and later supported by all-sky data from the Widefield Infrared Survey Explorer \citep[{\it WISE};][]{Wright10}. Two more useful steps have been the recent addition of a Southern analogue to SDSS, the SkyMapper Southern Survey \citep[SMSS; ][]{Wolf18a,Onken19} as well as infrared sky surveys such as the VISTA Hemisphere Survey \citep[VHS;][]{VHS}. At all steps in this journey, candidate lists for the rare high-redshift QSOs were swamped with the tails of the distribution from cool, red stars in our own Milky Way Galaxy. A major simplification of these searches was delivered by the {\it Gaia} satellite mission \citep{GaiaDR2} of the European Space Agency (ESA); {\it Gaia} measured proper motions and parallaxes for a billion objects. This data set revealed the cool-star nature for a large fraction of the candidates, allowing effective cleaning of the candidate lists. Within two days of the release of {\it Gaia} DR2, \citet{Wolf18b} identified the most luminous known QSO, SMSS\,J215728.21-360215.1 (a.k.a. SMSS\,J2157-3602) at $z=4.692$, which has since been shown to be powered by a supermassive black hole with 34 billion solar masses \citep{Onken20}. 

As ultra-luminous QSOs at high redshift are extremely rare, the size and completeness of a sample matters for any inference on the bright end of their luminosity function, their evolution with cosmic time in the early universe, their contribution to reionisation in the early universe, and the evolution of their host galaxies and early galaxies in general. These objects also lend a helping hand to studies of the intergalactic medium and the build-up of chemical elements traced with absorption lines in foreground galaxies \citep{Ryan-Weber09,Simcoe11}. For all these reasons, enlarging the existing samples is a worthwhile undertaking. 

Currently, the best reference for the bright end of the QSO luminosity function (LF) at redshift $\sim $5 is the work by \citet[hereafter Y16]{Yang16}, which is based on a QSO sample from \citet[hereafter W16]{Wang16}. Although various other works in the literature have updated the $z\sim 5$ QSO LF \citep[e.g.][]{McGreer18,Kim20,Niida20}, progress in this area is currently focused on the persistent uncertainties at the faint end. At the bright end, the exploitation of new data sources such as Pan-STARRS \citep{2016arXiv161205560C} led to the discovery of many new QSOs \citep{Yang19,Schindler19}, but no updates to the LF bright-end parameters. The work by Y16 inherited from its data source, the SDSS, a focus on the Northern hemisphere, and searching in the South offers an opportunity to at least double the sample and refine the sparsely populated bright end of the LF.

Hence, we had set out to discover ultra-luminous high-redshift QSOs in the Southern sky, armed with a combination of the SMSS, the AllWISE data set\footnote{Explanatory Supplement to AllWISE Data Release, \url{http://wise2.ipac.caltech.edu/docs/release/allwise/expsup}}
and {\it Gaia} DR2. The new {\it Gaia} data allowed \citet[hereafter Paper~I]{Wolf20} to push QSO colour selection criteria closer to the main stellar locus, relative to the work by W16 and Y16, given that most stars could be identified and removed from the list by proper motions. The extended colour selections then revealed the rare QSOs with higher completeness than before. Similarly, \citet{Calderone19} found that searches for QSO at all redshifts based on machine learning benefit from the {\it Gaia} data.

Paper~I published a sample of 21 bright $z>4$ QSOs, selected from SMSS DR2 with $i_{\rm PSF}<18.2$~ABmag. In this paper, we extend the search with SMSS DR3, reaching deeper in magnitude and lower in Galactic latitude, and purify the selection by adding JHK photometry from VHS and the VIKING Survey \citep{Edge13}, filling in data from the shallower 2 Micron All-Sky Survey \citep[2MASS;][]{2MASS} where needed. We present the results of our spectroscopic follow-up of candidates down to $z_{\rm PSF}\approx 19$~ABmag. We examine our evolving selection criteria in light of known high-redshift QSOs in the Southern sky and suggest selection rules going forward.
We compare the results of our search to previous campaigns in the Northern sky, and provide an update to the bright end of the high-redshift QSO luminosity function in the often quoted restframe 145~nm band. Finally, we present, for the first 
time, direct measurements for the bright end of the luminosity function at restframe 300~nm.

Sect.~\ref{search_area} describes the data sources from which we construct our set of QSO candidates. In Sect.~\ref{sec:candidate_selection}, we discuss properties of the known QSOs and the selection rules we adopt. In Sect.~\ref{sec:spectroscopy}, we describe our spectroscopic follow-up of candidates, present the list of high-redshift QSOs we found, and discuss the completeness of the current sample. In Sect.~\ref{sec:LFs}, we construct luminosity functions and examine their evolution. Throughout the paper, we use Vega magnitudes for {\it Gaia} and IR data, and AB magnitudes for the SkyMapper passbands: $griz$. 
We adopt a flat $\Lambda$CDM cosmology with $\Omega_{\rm m}=0.3$ and a Hubble-Lema\^itre constant of $H_0=70$~km~sec$^{-1}$~Mpc$^{-1}$.

\section{Search area and data sources}\label{search_area}

We start from the {\tt master} catalogue\footnote{See \url{https://skymapper.anu.edu.au} for catalogue details.} of SMSS DR3, which covers nearly all the sky at declination $\delta < +2$\deg; missing parts are primarily found very close to the Galactic plane and the Galactic Centre. In this work, we avoid areas with high object density, where photometry may be challenging, and high reddening, where high-redshift QSOs may be dimmed beyond our search depth. Hence, we focus on Galactic latitudes $|b|>15$\deg. These two simple geometric rules define an area that covers 38.35 per cent of the full sky, or 15\,821~deg$^2$. 

We further exclude specific areas: by default, bright stars in DR3 are surrounded by exclusion zones \citep[for details see][]{Onken19}, where detected sources are flagged and prevented from inclusion in the {\tt master} table. For this work, the exclusion zones in $z$-band are relevant, which add up to 134~deg$^2$ within the search area. We also exclude areas around nearby galaxies, where, e.g., red supergiants and long-period variable stars can contaminate the QSO candidate list. We use the Updated Nearby Galaxy Catalog \citep{Karachentsev13} of 869 galaxies from the Local Volume, and mask the sky within 1.3$\times$ the major angular diameter $a_{26}$ (which corresponds to the Holmberg isophote of $\sim $26.5~mag(Vega)~arcsec$^{-2}$ in the $B$ band) of all the galaxies. The largest resulting exclusion zones are around the LMC and SMC with 6.99 and 4.12~deg radius, respectively, as well as the region around the core of Sgr dSph that reaches to $|b|>15$\deg. All further galaxies have exclusion zones smaller than 0.5~deg, with the Sculptor dwarf galaxy, NGC~55, NGC~253 and Centaurus~A being the next largest objects. The total excluded area is 276.5~deg$^2$.  

We then select objects in the SkyMapper $z$-band, which has an effective mean wavelength of 916~nm and a FWHM of 84~nm. The bandpass efficiency curve is asymmetric as it rises steeply at the blue edge but rolls off to the red with the declining sensitivity of the CCD detectors \cite[see][]{2011PASP..123..789B}. We limit the candidate selection to $z_{\rm PSF} < 19.5$~ABmag, although source incompleteness sets in at $z_{\rm PSF} \ga 18.5$~ABmag, with a dependence on sky location originating from the SMSS progress as of DR3. Magnitude dependence of the incompleteness is discussed in detail below (Sect.~\ref{sec:mag_incomp}).

From this SMSS DR3 list, we then use position-based cross-matches to other large-area surveys, such as those from {\it Gaia}, {\it WISE}, and VISTA. As new versions of the {\it WISE} and {\it Gaia} data sets were released in December 2020, we replaced the AllWISE photometry for the $W1$ and $W2$ bands with the CatWISE2020 catalogue \citep[hereafter CatWISE]{Marocco21}, which is more precise for non-variable objects; with the release of {\it Gaia} eDR3 \citep{Gaia_eDR3}, we updated our original selection from {\it Gaia} DR2 in order to further reduce our candidate lists with improved parallax and proper motion (PPM) information.

Using cross-matches with {\it Gaia} eDR3 contained in the DR3 {\tt master} table, we consider objects where the nearest {\it Gaia} source is within 0.5~arcsec, and the second-nearest is at least 5~arcsec away. Close pairs of sources in {\it Gaia} often appear as one single SMSS source, causing the primary match to be more than 0.5~arcsec offset and showing a secondary match within 1 to 2~arcsec. Neighbours within 5~arcsec may affect the {\it WISE} photometry, given a {\it WISE} $W1$/$W2$ PSF with $\sim 6\arcsec$ FWHM, which would be detrimental to separating QSOs from cool stars. We require that {\it Gaia} has PPM data for the source and that the SMSS DR3 $z$-band photometry flags are $<$ 4 (indicating reliable measurements). We estimate the fractional loss of objects due to these requirements by considering all known QSOs from the Milliquas v7.1 catalogue \citep{Flesch15} in our search area that have $z>3$ and $z_{\rm PSF} = 16-19$~ABmag. We find that we lose 6 per cent of objects, mostly due to the requirement of no neighbour being present within 5~arcsec. At this point, we end up with an effective search area of 35.1 per cent of the full sky, or 14\,486~deg$^2$.

Then we take advantage of cross-matches between the SMSS {\tt master} table and each of AllWISE and CatWISE, using only matches within 2~arcsec. Not all DR3 objects have a counterpart, but all known QSOs at $z>4$ with $z_{\rm PSF}<19$~ABmag do so. \citet{Marocco21} quote the 90 per cent completeness depth for CatWISE as $W1=17.7$~mag and $W2=17.5$~mag. To examine the impact on our QSO search, we estimate the mean expected $z$-band magnitude of $z\sim 5$ QSOs at the $W2$ completeness limit from the average colour of bright ($W2<15$~mag) known $z\sim 5$ QSOs, which is $z_{\rm PSF}-W2\approx 3.75$. Hence, at $W2=17.5$~mag we expect $z_{\rm PSF}\approx 21.25$~ABmag, which exceeds the depth of our planned search by $\sim 2$~mag. Brighter DR3 objects without {\it WISE} matches are most likely stars and not QSOs, because cool stars have bluer optical-minus-MIR colours. We also use data from the $W3$ band where it is available, but we do not require it for selection, because the 5-sigma sensitivity of $W3=11.3$~mag is too shallow for our purposes. Overall, we assume that the {\it WISE} data introduces no incompleteness into the high-redshift QSOs selection. In contrast, Y16 did find ALLWISE-imposed incompleteness to be important; however, such constraints only appeared at fainter magnitudes than are relevant for our search ($W1>17$~mag, $W2>16$~mag).

We note that about 6 per cent of $z>4$ QSOs from Milliquas v7.1 lack a CatWISE entry, although all of them are present in AllWISE. Thus, we use CatWISE data where available and fill in AllWISE data where needed. As the $W1-W2$ colour has been found useful for separating QSOs from cool stars and is also used in this work, we compared this colour for QSOs between AllWISE and CatWISE. We found the quoted uncertainties in $W1-W2$ to shrink by a factor of 2.3 and the colour changes per object from AllWISE to CatWISE to be statistically consistent with the errors quoted in AllWISE. We also find that the colour range of high-redshift QSOs shrinks slightly, in line with the quoted error properties, although this test has only moderate significance given the small numbers of objects.

\begin{figure*}
\begin{center}
\includegraphics[angle=270,width=0.98\textwidth,clip=true]{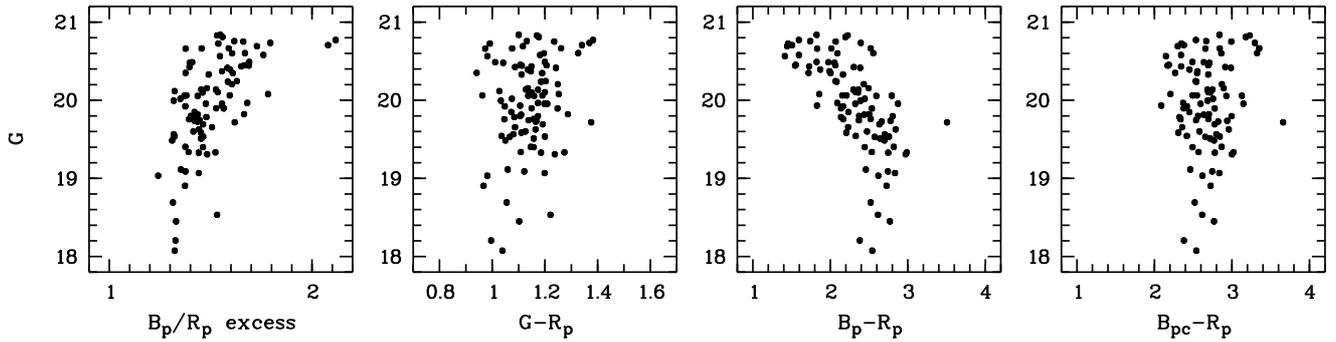} 
\caption{{\it Gaia} photometry of redshift $4.7<z<5.3$ QSOs near the faint end, where little flux is expected in the $B_p$ channel, while the $G$ band and the $R_p$ channel should still see significant flux: the Bp/Rp Excess Factor increases from a value of $\sim 1.3$, which is typical for point sources of this colour, to values around 2 near the detection limit (left panel); the $G-R_p$ colour shows no trend with magnitude (centre-left), but the $B_p-R_p$ colour drifts bluewards toward the detection limit (centre-right), suggesting extra flux being recorded. We use the ad-hoc quadratic correction from Eq.~\ref{bp_corr} to rectify the $B_p$ bias (right panel).
\label{bp_bias}}
\end{center}
\end{figure*}

\section{Candidate Selection}\label{sec:candidate_selection}

In this section, we discuss the properties of known QSOs brighter than $z_{\rm PSF}=19$~ABmag at redshift $4-6$. We include all known QSOs in our search area, as listed in Milliquas v7.1, which includes QSOs published until 14 Feb 2021\footnote{While this paper was under review, new releases of Milliquas (up to v7.3c on 28 Nov 2021) added 24 new QSOs discovered by other groups with confirmed redshifts between 4 and 6, mostly from \citet{2021AJ....162...72W}. However, only three are both in our search area and have $z_{\rm PSF}<19$~ABmag; two were unconfirmed at the time of observation and are reported as "discoveries" in our sample, as indicated in Table~\ref{qso4755}, and the third has $z_{\rm PSF}=18.9$~ABmag, at which depth our spectroscopic follow-up remains incomplete.}. 
Based on the distribution of these known objects, we propose selection criteria for high-redshift QSOs that we assume to be complete except for rare cases with unusual colours. The sample of candidates resulting from these criteria has not been observed completely with spectroscopy but can inform future observations and the completeness of the verified samples.

\subsection{PPM information from {\it Gaia}}

{\it Gaia} eDR3 provides measurements of the parallax $\pi$, the proper motion $\mu_\alpha,\mu_\delta$, and their errors $\sigma_\pi$, $\sigma_\alpha$ and $\sigma_\delta$ for most objects. For QSOs we demand that these are all consistent with zero within the errors. We calculate a $\chi^2$-style measure of consistency using 

\begin{equation}\label{PPM_SN}
    SN_{\rm PPM} = \sqrt{ \frac{1}{3} \left( \left(\frac{\max(0,\pi)}{\sigma_\pi}\right)^2 + \left(\frac{\mu_\alpha}{\sigma_\alpha}\right)^2 + \left(\frac{\mu_\delta}{\sigma_\delta}\right)^2 \right) }
\end{equation}

Our survey footprint contains 137 QSOs known before this work at magnitude $z_{\rm PSF}<19$~ABmag and redshift $z\ge 4$. Of these, 132 objects (96 per cent) have $SN_{\rm PPM}<2$ and 88 (64 per cent) have $SN_{\rm PPM}<1$, in excellent agreement with the statistics expected for a population without parallax and proper motion. We also find no noticeable trend in the PPM properties with magnitude or redshift, which all lends high credibility to the quoted {\it Gaia} PPM measurement uncertainties. 

\begin{figure*}
\begin{center}
\includegraphics[angle=270,width=\textwidth,clip=true]{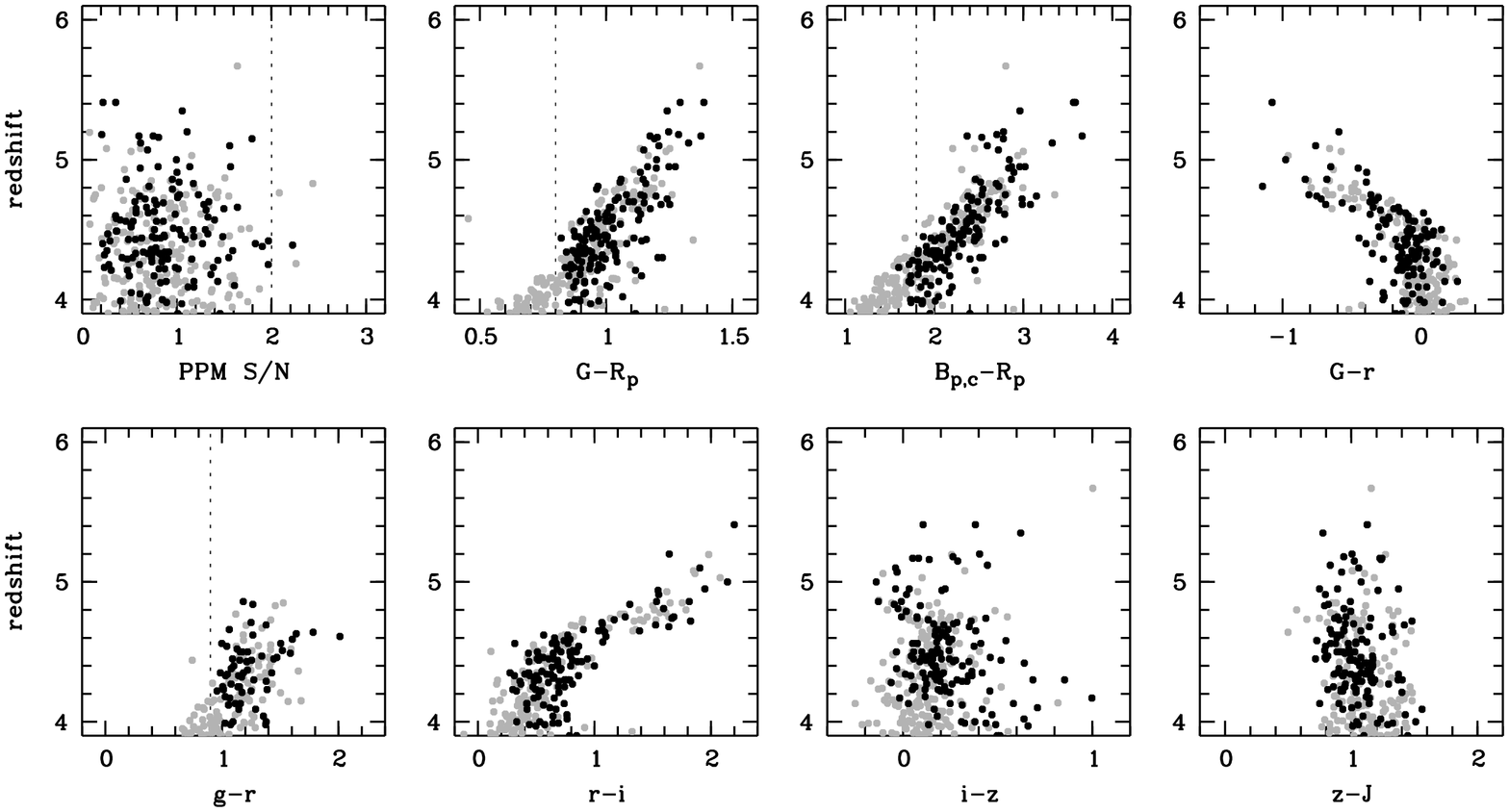} \\
\vspace{4mm}
\includegraphics[angle=270,width=\textwidth,clip=true]{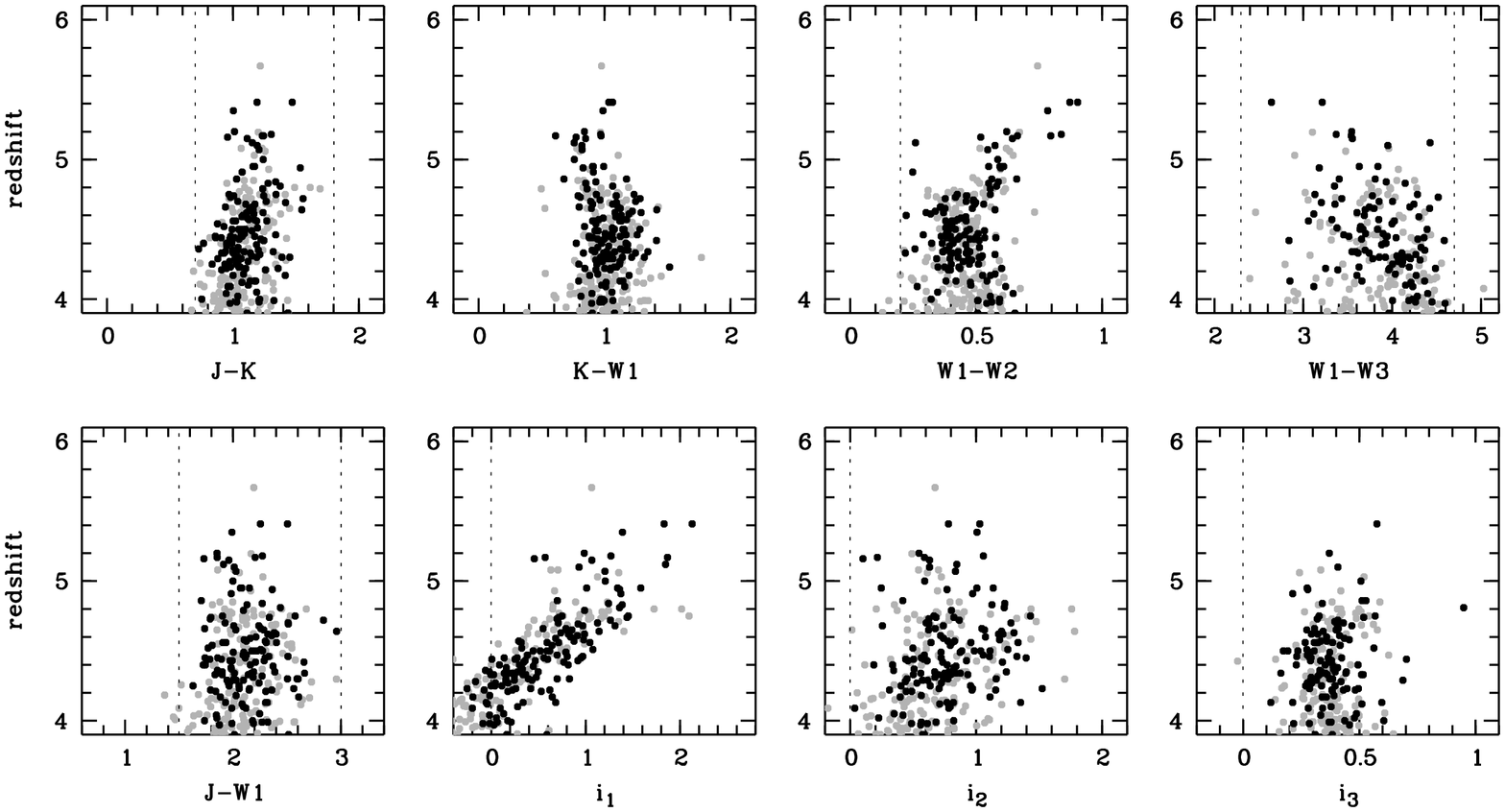} 
\caption{Characteristics of known high-redshift QSOs with $z_{\rm PSF}<19$~ABmag at redshift $z>3.9$: significance of parallax and proper motion signal (top left) and colour indices (all other panels). Grey points are previously known QSOs and black points are new objects presented in this work. Dashed lines indicate the adopted selection criteria, which eliminate only four of the QSOs at $z\ge 4.4$. The colour indices $i_1$, $i_2$, and $i_3$ are defined in Eq.~\ref{indices}. A few of the $z<4.4$ QSOs discovered in this work were inside the selection criteria before we corrected the faint {\it Gaia} $B_p$ magnitudes as explained in Sect.~\ref{photinfo}.
\label{col_redshift}}
\end{center}
\end{figure*}

\subsection{Photometric information}\label{photinfo}

Altogether, we consider the following photometric data: from SMSS DR3, the PSF magnitudes in the passbands $griz$ \citep[for full filter information, see][]{Wolf18a} as well as photometric flag information; from {\it Gaia} eDR3, the passbands $G$, $B_p$, and $R_p$, as well as the Bp/Rp Excess Factor; from VHS DR6 and VIKING DR5, the near-infrared passbands $JHK$; from CatWISE, the mid-infrared passbands $W1$ and $W2$; and from AllWISE, the passband $W3$. We correct the colours for interstellar foreground reddening using the extinction map from \citet[hereafter SFD]{SFD98} with extinction coefficients for {\it Gaia} from \citet{CasaVDB18} for use in $A_{\rm band} = R_{\rm band} \times 0.86 \times E(B-V)_{\rm SFD}$ and for SkyMapper from \citet{Wolf18a} for use in $A_{\rm band} = R_{\rm band} \times E(B-V)_{\rm SFD}$, as the \citet{SF11} correction factor 0.86 is already absorbed into the extinction coefficients. 
In the NIR, we had chosen $(A_J,A_H,A_K) = (0.8,0.5,0.3) \times 0.86 E(B-V)$, which is a little different from the values $(0.75,0.41,0.24)$ suggested by \citet{WangChen19}, but the effects are mostly on the order of 0.01~mag in the low-reddening areas we are searching. We assume no extinction in the {\it WISE} passbands. 

First we address an issue with {\it Gaia} photometry of faint red sources: in Fig.~\ref{bp_bias} we show the Bp/Rp Excess Factor and the colours $G-R_p$ and $B_p-R_p$ vs. $G$ magnitude of all QSOs known in the search area within a narrow redshift range of $4.7<z<5.3$. While there is no trend of $G-R_p$ colour with brightness in $G$, there is a strong trend of $B_p-R_p$ getting bluer and the Excess Factor increasing as the magnitude gets fainter. \citet{Riello20} discuss the origin of this bias towards higher $B_p$ fluxes in red objects that are near the detection limit in the $B_p$ channel. Here, we attempt to rectify the $B_p-R_p$ colour with an ad-hoc quadratic correction, only for objects with $G>19$~mag, of the form 
\begin{equation}\label{bp_corr} 
    B_{p,\rm c}-R_p = B_p-R_p + 0.3 \times (\max(0,G-19))^2 ~.
\end{equation}
The right panel in the figure shows the result of this correction, and in the following we adopt such corrected $B_{p,\rm c}-R_p$ colours.

Figure~\ref{col_redshift} shows the redshift trends of several colour indices formed from the passband measurements and corrected for foreground reddening. We see well-known features:
\begin{enumerate}
    \item Strong trends of colour with redshift arise as the spectral step from the unabsorbed QSO continuum to the Ly\,$\alpha$ forest near $\lambda_{\rm rest}\simeq 120$~nm is redshifted through the spectral range, see e.g. $G-R_p$, $B_{p,\rm c}-R_p$, $G-r_{\rm PSF}$, $g_{\rm PSF}-r_{\rm PSF}$, and $r_{\rm PSF}-i_{\rm PSF}$.
    \item Where both passbands in a colour index probe unabsorbed continuum, it reflects the spectral slope of the continuum, which has only moderate scatter in the bulk of the QSO population, see e.g. $z_{\rm PSF}-J$, $J-K$ and $K-W1$.
    \item Emission lines vary colours slightly as they are redshifted through the passbands, most notably the strong line H$\alpha$ in the colours $K-W1$ and $W1-W2$, where variations with redshift trace the structured passband efficiency curve of $W1$.
\end{enumerate}

\subsection{QSO selection criteria}\label{sec:sel_criteria}

Based on the distribution of colours observed for the existing QSO sample, we propose the following set of selection criteria to define the volume of QSO candidates at $z\ge 4.4$:

\begin{equation}
\begin{array}{cccclc}
 0.8 & < & G-R_p   & < & 1.8  \\
 1.8 & < & B_{p,\rm c}-R_p   \\
 0.9 & < & g_{\rm PSF}-r_{\rm PSF} &   &   & \quad {\rm if \,\, measured} \\
 0.7 & < & J-K & < & 1.8  \\
 1.5 & < & J-W1 & < & 3  \\
 0.2 & < & W1-W2 & < & 1.1  \\
 2.3 & < & W1-W3 & < & 4.7 & \quad {\rm if \,\, measured} 
\end{array}
\end{equation}

We further define three indices, $i_1$, $i_2$ and $i_3$, 
to reduce the contamination from broad absorption line QSOs (BALQSOs) at $z<4$ and from distant cool stars:

\begin{equation}
\begin{array}{ccccc}\label{indices}
   0 & < & (J-K) + (B_{p,\rm c}-R_p) - (z_{\rm PSF}-J) - 1.8 & = i_1 \\
   0 & < & (J-W1) - 1.4 (z_{\rm PSF}-J) + 0.1  & = i_2 \\
   0 & < & 0.6 - 0.5 (r_{\rm PSF}-i_{\rm PSF}) - (G-r_{\rm PSF}) & = i_3
\end{array}
\end{equation}

We include objects that are undetected in $r$-band. Considering the colour trends with redshift, it would seem reasonable to make stronger use of the $r_{\rm PSF}-i_{\rm PSF}$ colour. Unfortunately, this requires $r$-band data at the depth of the SMSS Main Survey, but in DR3 the hemisphere is only complete with deep data in $i$-band and $z$-band, while deep $r$-band covers only two thirds of the hemisphere. 
A non-detection in $r$-band may thus be either a sign of high redshift or a sign of shallow data, depending on the sky position. Hence, we do not attempt to fully exploit the $r$-band information at this stage. Finally, we apply a cut in PPM significance (see Eq.~\ref{PPM_SN}): 
\begin{equation}
    SN_{\rm PPM}<2 ~,
\end{equation}
which should make us lose only around 4 per cent of the QSO sample, while keeping contamination by Milky Way stars low.

These simple geometric selection rules may seem suboptimal, as several authors have convincingly argued that other techniques -- a general Bayesian selection \citep[e.g.][]{WMR01,Richards09,Mortlock12,Reed17}, or machine learning approaches \citep[e.g.,][]{Calderone19, 2021AJ....162...72W, 2021MNRAS.506.2471G} -- ought to be superior in modelling the amorphous locus of QSOs in high-dimensional SED space. However, previous QSO samples at high redshift were incomplete, with a highly structured selection function, which we do not want to propagate into our study. After all, this work aims at improving the completeness of observed samples. Hence, we chose not to apply elaborate statistical methods to inferior training samples when informing our follow-up observations. 

The final candidate list within our selection volume includes 197 objects brighter than $z_{\rm PSF}=18.7$~ABmag. Of these, 64 were known QSOs as of Milliquas v7.1, with 17 reported by \citep{Wolf18b} and Paper~I, and a further 47 reported by a variety of authors including W16, \citet{Yang17}, \citet{Schindler19}, \citet{Yang19}, and \citet{Lyke20}. 
This left 133 new candidates with $z_{\rm PSF}<18.7$~ABmag to be followed up.
The list includes a further 497 fainter candidates with $z_{\rm PSF}\ga 18.7$~ABmag, of which 21 were known QSOs as of Milliquas v7.1, leaving 476 candidates to follow up in this and future work.

\begin{figure*}
\begin{center}
\includegraphics[angle=270,width=0.78\textwidth,clip=true]{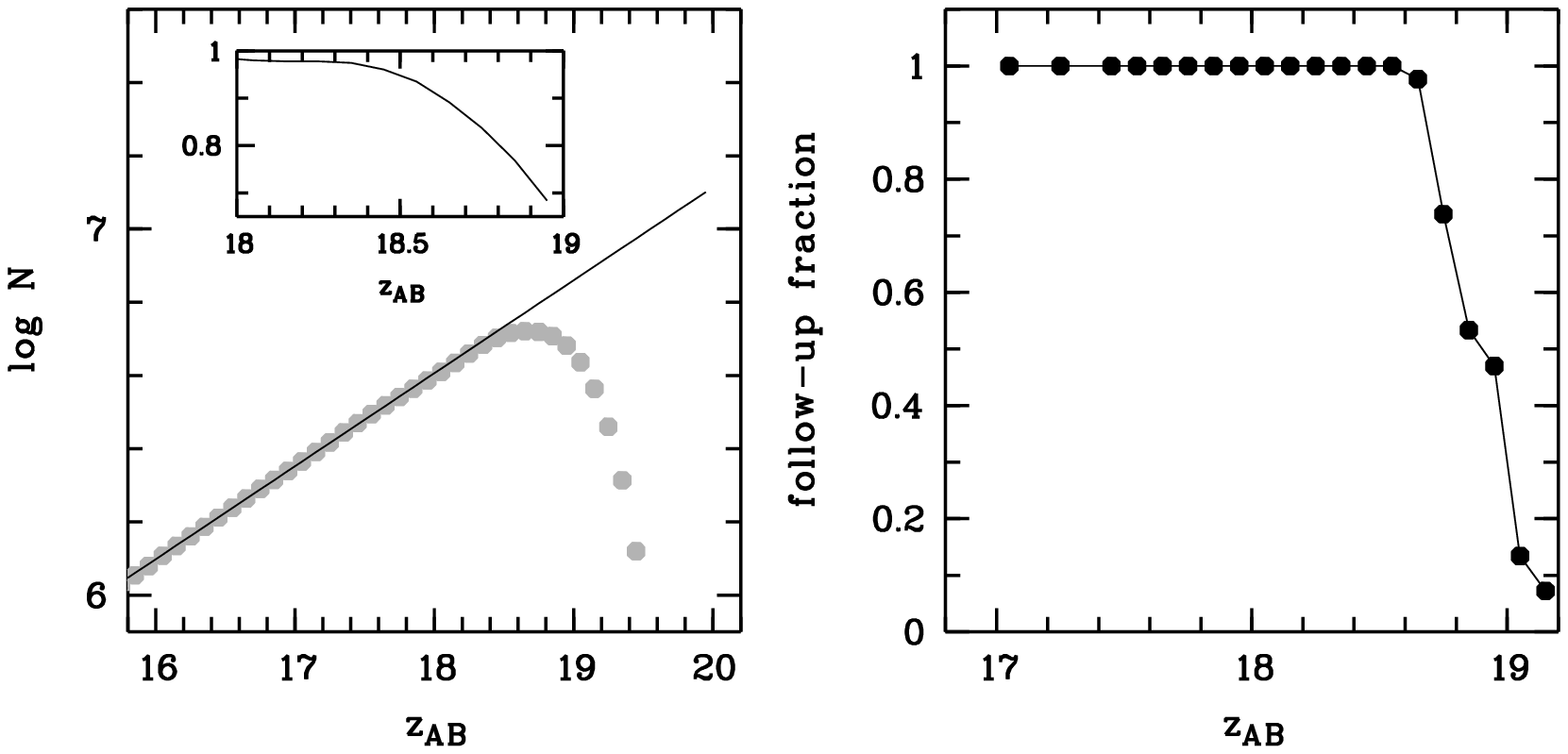} 
\caption{Completeness diagnostics: the SMSS $z$-band source list is incomplete at $z_{\rm PSF}>18.5$~ABmag (left panel), as diagnosed by the source number counts in the search area; the derived source completeness as shown in the inset declines from 95 per cent at $z_{\rm PSF}=18.5$~ABmag to 63 per cent at $z_{\rm PSF}=19$~ABmag. Starting from the source list, the spectroscopic follow-up is complete at $z_{\rm PSF} < 18.7$~ABmag, while completeness drops to $\sim $10 per cent at $z_{\rm PSF}=19$~ABmag. 
\label{nc_cplt}}
\end{center}
\end{figure*}

\section{Spectroscopy at the ANU 2.3m telescope: 2017 to 2021}\label{sec:spectroscopy}

This work builds on the study of Paper~I and extends it to fainter magnitudes from SMSS DR3. We have followed up 132 of the 133 unstudied objects from our $z_{\rm PSF}<18.7$~ABmag candidate list. Also, we followed up 154 fainter objects that extend to $z_{\rm PSF}\simeq 19$~ABmag, about 1 mag deeper than Paper~I. 

We note that during the early phase of this work, we had not yet applied the faint correction to the {\it Gaia} $B_p$ magnitudes mentioned in Sect.~\ref{photinfo}. At that stage, we had additional candidates with $B_p-R_p>1.8$~mag but $B_{p,c}-R_p<1.8$~mag, a small number of which we have followed up, as shown by the black points outside the selection cut in the $B_{p,c}-R_p$ panel of Fig.~\ref{indices}. These objects, however, did not end up in our complete sample discussed later, as they were all in the incomplete redshift range at $z<4.4$.

In addition, we have pursued a complementary search by occasionally sampling objects outside our selection criteria, in order to explore whether we are missing an unknown QSO population at high redshift. However, the only high-redshift QSO discovered this way was one at $z=4.39$, whose colours were indeed all within the selection criteria, while its PPM signal was slightly larger than our cutoff.

We have used the Wide Field Spectrograph \citep[WiFeS;][]{Dopita10} on the ANU 2.3m telescope at Siding Spring Observatory for a number of nights between December 2017 and October 2021. In conjunction with the standard RT560 beam-splitter, we primarily used the WiFeS B3000 and R3000 gratings in the blue and red arm, respectively, which together cover the wavelength range from 360~nm to 980~nm \ at a resolution of $R=3000$. This setup allows us to see QSO spectra from the Ly\,$\alpha$ forest to the \ion{C}{iv} line for all objects at redshift $z<5$. At redshift $z> 5$, the \ion{C}{iv} line starts to get lost in the sky noise and for the highest-redshift objects it is even outside the data range. However, in these cases the Ly\,$\alpha$ forest, in combination with the Ly\,$\alpha$ and \ion{Si}{iv} line, are sufficient to confirm objects as high-redshift QSOs. Because of mechanical issues with the instrument, two nights used alternative gratings in the red arm: I7000 in December 2020 and R7000 in April 2021. Despite the restricted wavelength range in each case, numerous candidates were successfully classified. Exposure times ranged from 600~sec to 2\,400~sec, and observing conditions varied in terms of cloud cover and seeing. For the faintest objects, we obtained and co-added two spectra of up to 2\,400~sec each. 

The data were reduced using the Python-based pipeline PyWiFeS \citep{Childress14}. PyWiFeS calibrates the raw data with bias, arc, wire, internal-flat and sky-flat frames, and performs flux calibration and telluric correction with standard star spectra. Flux densities were calibrated using a number of standard stars, which are usually observed on the same night, although not necessarily under the same cloud conditions.\footnote{This lack of robust flux calibration precludes an independent estimation of the absolute magnitudes from the spectroscopy.} We then extracted spectra from the calibrated 3D data cube using QFitsView\footnote{\url{https://www.mpe.mpg.de/~ott/QFitsView/}}.
Reduced spectra were visualised with the MARZ software \citep{Hinton16} and template spectra were overplotted to aid the classification and redshift determination. We estimate redshifts from broad \ion{Si}{iv}, \ion{N}{v}, and \ion{C}{iv} lines where available, and consider the blue edge of the Ly\,$\alpha$ line when necessary. Because of frequent absorption within the emission line profiles and the possibility of \ion{C}{iv} blueshifts, we determined the redshifts manually, and thus we estimate that the redshift uncertainties range from 0.01 to $\sim $0.05 for weak-lined objects. 

Since December 2017, spectra were taken for 739 objects, of which 577 provided sufficient signal for a confident classification. Most of the targets are not candidates anymore after using the refined PPM data of {\it Gaia} eDR3. 252 targets were observed to be QSOs, 67 per cent of which are at $z>3.8$. (Among those 170 QSOs were seven discovered by other groups and mistakenly included in our observing lists.) QSOs at lower redshift include objects with red continua as well as extreme BALQSOs and overlapping iron trough low-ionisation BALQSOs (OFeLoBALQSOs), which mimic the optical colours of high-redshift QSOs, but have MIR colours typical of their redshift, hence we have learned how they can largely be avoided from the start. Several stars contaminated the candidate lists, especially before the {\it Gaia} eDR3 release. Altogether 13 stars turned out to be likely red supergiants in nearby galaxies including NGC~300, the Sculptor and Fornax dwarf galaxies as well as the SMC and LMC and later we chose to exclude the sky areas covered by these.

\begin{figure*}
\begin{center}
\includegraphics[angle=270,width=0.98\textwidth,clip=true]{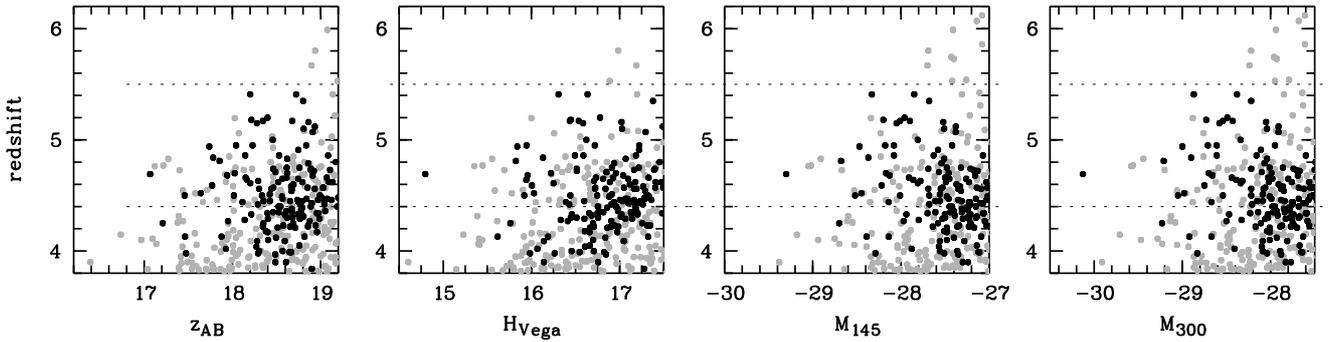} 
\caption{The current Southern bright high-redshift QSO sample: apparent optical and IR magnitude (left and centre-left panels) and luminosity at 145~nm and 300~nm (centre-right and right). Dark points are QSOs found with SkyMapper and reported by \citet{Wolf18b}, Paper~I, and this work. Light points are QSOs from the literature in the search area targeted in this work. The relative lack of objects at redshift $z>5.5$ suggests the sample is incomplete there. 
\label{sample_mz}}
\end{center}
\end{figure*}

\subsection{Newly identified high-redshift QSOs}

Among the 197 candidates defined by the final selection criteria in the nearly complete magnitude range at $z_{\rm PSF}<18.7$~ABmag, there are 116 $z\ge 4$ QSOs, 29 (mostly BAL) QSOs at $z<4$ and 19 stars. The remaining objects are not high-redshift QSOs, although we cannot type them confidently, either due to low signal in the spectra or due to unusual spectral features. As described in Sect.~\ref{sec:sel_effects}, only 4 known high-redshift QSOs were excluded by these criteria.

In Tables~\ref{qso3844}-\ref{qso4755}, we report our 142 newly identified high-redshift QSOs, split into three redshift ranges. The highest-redshift objects are at $z=5.41$, and together with the objects from Paper~I, the total number of $z>3.8$ QSOs discovered with this project is 163, of which 90 per cent have $M_{145}<-27$~ABmag.

The tables list $z$-band AB and $H$-band Vega magnitudes, which are useful for the LF construction and for planning more detailed follow-up observations.\footnote{One faint source in Table~\ref{qso4447} has no $z$-band photometry in SMSS DR3, and here we substituted the photometry from the NOIRLab Source Catalog (NSC) DR2 \citep{2021AJ....161..192N}. We have been unable to locate any NIR photometry for the source, but the lack of an $M_{300}$ estimate does not affect our analysis, because it falls well below the brightness regime in which we are complete and from which we construct the LFs.} In many cases, the $H$-band magnitude is not observed but inferred from $J$ and $K$ magnitudes, or if necessary, from the $z_{\rm PSF}$ and $W1$ photometry. We use the mean colour relations that we find in the sample, which are \mbox{$H-K = 0.41(J-K)$} and \mbox{$H-W1=-0.007+0.4677 (z_{\rm PSF}-W1)$}. The relations are roughly linear across colour, with a root mean square (RMS) scatter of 0.1~mag and 0.15~mag, respectively,
and have no trend with redshift in our range of concern.

\begin{table*}
\centering
\caption{List of newly discovered QSOs. Redshift $3.8\le z<4.4$; these are not part of a complete sample and are hence not used for the luminosity functions. Epoch is the date of the first spectrum taken by us. Comments refer to non-standard observing setups or unusual properties. See Appendix~\ref{app:A} for the spectra.}
\label{qso3844} 
\begin{tabular}{ccccccccccl}
\hline \noalign{\smallskip}  
SMSS ID  & Redshift  	& $R_p$  & $z_{\rm PSF}$  & $H$  & $W1$ & $W1-W2$ & $M_{145}$ & $M_{300}$ & Epoch & Comments \\
        &               & (Vega) & (AB)           & (Vega)   & (Vega) & (Vega)  & (AB)      & (AB) & (spectrum) \\
\noalign{\smallskip} \hline \noalign{\smallskip}
 J000458.11-213044.7 &  4.09 &  18.365 &  18.28 & $ 16.14 \pm 0.04$ &  14.58 &  0.27 & $-27.56$ & $-28.50$ &  20191128 &   \\
 J000948.16-371708.1 &  4.27 &  17.679 &  17.95 & $ 16.41 \pm 0.04$ &  15.02 &  0.45 & $-27.97$ & $-28.31$ &  20201208 &   \\
 J003222.03-035807.5 &  4.12 &  18.318 &  18.49 & $ 16.84 \pm 0.04$ &  15.30 &  0.48 & $-27.40$ & $-27.83$ &  20201209 & I7000 \\
 J005938.27-422704.1 &  4.25 &  18.463 &  18.69 & $ 17.26 \pm 0.03$ &  15.77 &  0.51 & $-27.21$ & $-27.45$ &  20201208 &   \\
 J010201.12-211545.9 &  4.29 &  18.281 &  18.40 & $ 16.60 \pm 0.04$ &  15.12 &  0.40 & $-27.53$ & $-28.12$ &  20191023 &   \\
 J014615.01-635639.2 &  4.28 &  17.695 &  18.14 & $ 16.33 \pm 0.04$ &  14.96 &  0.48 & $-27.80$ & $-28.39$ &  20200926 &   \\
 J021218.22-394548.8 &  4.13 &  17.222 &  17.46 & $ 15.62 \pm 0.03$ &  13.99 &  0.38 & $-28.39$ & $-29.04$ &  20201209 & I7000 \\
 J021638.50-033737.8 &  4.31 &  18.546 &  18.81 & $ 17.01 \pm 0.04$ &  15.44 &  0.52 & $-27.14$ & $-27.72$ &  20201001 &   \\
 J022654.20-322924.0 &  4.30 &  18.973 &  18.78 & $ 16.51 \pm 0.01$ &  14.80 &  0.52 & $-27.15$ & $-28.21$ &  20210118 &   \\
 J024535.48-445433.0 &  4.02 &  18.027 &  17.93 & $ 16.07 \pm 0.03$ &  14.82 &  0.56 & $-27.88$ & $-28.55$ &  20190703 &   \\
 J030427.83-780559.2 &  4.00 &  18.400 &  18.57 & $ 16.71 \pm 0.05$ &  15.18 &  0.59 & $-27.36$ & $-27.95$ &  20210118 &   \\
 J034132.56-411920.9 &  4.30 &  18.940 &  18.68 & $ 16.45 \pm 0.04$ &  14.65 &  0.51 & $-27.25$ & $-28.28$ &  20200927 &   \\
 J042529.68-395805.8 &  4.36 &  18.890 &  19.05 & $ 17.54 \pm 0.08$ &  16.17 &  0.54 & $-26.93$ & $-27.46$ &  20200810 &   \\
 J043802.34-530028.9 &  4.21 &  18.563 &  18.71 & $ 16.67 \pm 0.02$ &  15.00 &  0.39 & $-27.17$ & $-28.02$ &  20210203 &   \\
 J045459.29-420742.6 &  4.17 &  19.398 &  18.96 & $ 16.97 \pm 0.04$ &  15.19 &  0.56 & $-26.91$ & $-27.70$ &  20211022 &   \\
 J045623.58-085248.5 &  3.90 &  18.676 &  18.49 & $ 16.14 \pm 0.04$ &  14.44 &  0.65 & $-27.30$ & $-28.44$ &  20191130 &   \\
 J045639.34-402753.6 &  4.18 &  18.087 &  18.38 & $ 16.61 \pm 0.04$ &  15.21 &  0.57 & $-27.51$ & $-28.08$ &  20201209 & I7000 \\
 J052439.12-304800.4 &  4.17 &  18.052 &  18.40 & $ 16.66 \pm 0.05$ &  15.36 &  0.53 & $-27.48$ & $-28.02$ &  20201209 & I7000 \\
 J055901.33-855138.3 &  3.90 &  18.986 &  18.61 & $ 16.76 \pm 0.06$ &  15.35 &  0.51 & $-27.24$ & $-27.85$ &  20200927 &   \\
 J061751.60-240610.8 &  4.09 &  18.413 &  18.59 & $ 17.05 \pm 0.10$ &  15.61 &  0.62 & $-27.28$ & $-27.61$ &  20210118 &   \\
 J083905.36-222215.9 &  4.30 &  18.423 &  18.51 & $ 16.63 \pm 0.05$ &  15.20 &  0.43 & $-27.52$ & $-28.13$ &  20200227 & $b_{\rm gal} = 11\fdg 5$ \\
 J091926.16-094920.5 &  4.38 &  18.338 &  18.67 & $ 17.18 \pm 0.07$ &  15.74 &  0.39 & $-27.31$ & $-27.57$ &  20210411 & R7000 \\
 J092623.99-261620.9 &  4.35 &  18.453 &  18.46 & $ 16.77 \pm 0.06$ &  15.04 &  0.47 & $-27.61$ & $-28.01$ &  20210306 &   \\
 J095759.35-230941.8 &  4.32 &  18.451 &  18.60 & $ 16.68 \pm 0.06$ &  15.11 &  0.57 & $-27.39$ & $-28.06$ &  20210405 &   \\
 J104046.17-294635.5 &  4.33 &  18.486 &  18.78 & $ 17.14 \pm 0.09$ &  15.64 &  0.40 & $-27.21$ & $-27.61$ &  20210407 &   \\
 J110144.63-260043.5 &  4.37 &  18.682 &  18.78 & $ 16.98 \pm 0.06$ &  15.58 &  0.51 & $-27.24$ & $-27.78$ &  20210405 &   \\
 J113649.79-161939.6 &  4.25 &  18.192 &  18.35 & $ 16.70 \pm 0.05$ &  15.32 &  0.40 & $-27.58$ & $-28.02$ &  20210307 &   \\
 J120828.90-044419.9 &  4.33 &  18.642 &  18.74 & $ 17.20 \pm 0.06$ &  15.47 &  0.22 & $-27.22$ & $-27.54$ &  20210408 &   \\
 J131656.26-301244.3 &  4.29 &  18.761 &  18.76 & $ 17.14 \pm 0.09$ &  15.69 &  0.34 & $-27.23$ & $-27.60$ &  20210408 &   \\
 J132343.04-052951.6 &  4.04 &  18.174 &  18.27 & $ 16.61 \pm 0.04$ &  15.29 &  0.65 & $-27.57$ & $-28.02$ &  20210404 &   \\
 J132626.48-341716.4 &  4.36 &  18.562 &  18.61 & $ 16.98 \pm 0.07$ &  15.53 &  0.26 & $-27.40$ & $-27.78$ &  20210410 &   \\
 J134633.39-275654.3 &  3.97 &  18.678 &  18.49 & $ 16.85 \pm 0.07$ &  15.52 &  0.54 & $-27.35$ & $-27.77$ &  20210404 &   \\
 J135729.80-234839.2 &  4.34 &  18.823 &  18.96 & $ 17.48 \pm 0.08$ &  16.14 &  0.37 & $-27.06$ & $-27.28$ &  20200627 &   \\
 J151045.68-045007.4 &  4.05 &  18.903 &  18.76 & $ 16.66 \pm 0.04$ &  14.98 &  0.35 & $-27.14$ & $-27.99$ &  20210408 &   \\
 J151904.00-095230.8 &  4.22 &  18.432 &  18.65 & $ 16.90 \pm 0.05$ &  15.60 &  0.60 & $-27.36$ & $-27.84$ &  20210409 &   \\
 J155141.82-153448.9 &  4.33 &  18.230 &  18.21 & $ 16.46 \pm 0.04$ &  15.13 &  0.46 & $-27.94$ & $-28.35$ &  20200221 & weak-lined \\
 J160013.90-112800.2 &  4.29 &  18.863 &  18.99 & $ 17.07 \pm 0.07$ &  15.50 &  0.47 & $-27.30$ & $-27.78$ &  20210704 &   \\
 J164147.78-775029.8 &  4.13 &  18.090 &  17.88 & $ 16.19 \pm 0.04$ &  14.86 &  0.54 & $-28.14$ & $-28.53$ &  20200819 & See Appendix~\ref{sec:J1641} \\
 J172847.61-643404.6 &  4.22 &  18.261 &  18.34 & $ 16.71 \pm 0.05$ &  15.17 &  0.44 & $-27.64$ & $-28.02$ &  20200926 &   \\
 J181348.59-742305.6 &  4.33 &  18.547 &  18.59 & $ 17.02 \pm 0.08$ &  15.60 &  0.40 & $-27.50$ & $-27.76$ &  20200927 &   \\
 J182616.06-474625.4 &  4.23 &  18.629 &  18.94 & $ 17.35 \pm 0.11$ &  15.78 &  0.43 & $-27.03$ & $-27.38$ &  20210703 &   \\
 J185854.75-414702.1 &  4.34 &  18.762 &  18.97 & $ 16.75 \pm 0.05$ &  14.82 &  0.48 & $-27.07$ & $-28.02$ &  20210704 &   \\
 J193603.72-702734.4 &  4.39 &  18.580 &  18.79 & $ 16.86 \pm 0.06$ &  15.15 &  0.46 & $-27.26$ & $-27.92$ &  20201001 &   \\
 J194707.16-443732.3 &  4.13 &  18.876 &  18.82 & $ 17.32 \pm 0.11$ &  15.69 &  0.42 & $-27.09$ & $-27.36$ &  20210704 &   \\
 J195455.50-222254.5 &  4.30 &  18.611 &  18.62 & $ 16.76 \pm 0.05$ &  14.88 &  0.52 & $-27.50$ & $-28.04$ &  20210411 &   \\
 J201808.60-390820.6 &  4.10 &  18.701 &  18.40 & $ 16.70 \pm 0.06$ &  15.24 &  0.48 & $-27.48$ & $-27.96$ &  20201001 &   \\
 J202712.66-245732.8 &  4.34 &  18.959 &  18.96 & $ 17.20 \pm 0.07$ &  15.75 &  0.37 & $-27.06$ & $-27.55$ &  20210711 &   \\
 J202914.99-015504.0 &  4.23 &  18.630 &  18.86 & $ 17.31 \pm 0.08$ &  15.30 &  0.38 & $-27.20$ & $-27.45$ &  20210711 & extended? \\
 J205659.53-513055.1 &  3.84 &  19.262 &  18.90 & $ 17.28 \pm 0.06$ &  15.73 &  0.53 & $-26.80$ & $-27.26$ &  20210711 &   \\
 J214938.09-805805.1 &  4.22 &  18.987 &  18.86 & $ 16.99 \pm 0.07$ &  15.35 &  0.39 & $-27.23$ & $-27.78$ &  20201002 &   \\
 J222820.62-422626.3 &  3.99 &  18.718 &  18.67 & $ 16.55 \pm 0.04$ &  15.00 &  0.51 & $-27.11$ & $-28.05$ &  20210711 &   \\
 J230318.14-603123.8 &  4.35 &  18.136 &  18.52 & $ 16.82 \pm 0.05$ &  15.23 &  0.38 & $-27.44$ & $-27.92$ &  20201210 &   \\
 J230513.15-031958.9 &  4.35 &  18.494 &  18.68 & $ 17.13 \pm 0.05$ &  15.77 &  0.39 & $-27.31$ & $-27.62$ &  20210711 &   \\
 J230824.53-453905.6 &  3.99 &  18.640 &  18.65 & $ 16.83 \pm 0.04$ &  15.16 &  0.51 & $-27.12$ & $-27.77$ &  20210711 &   \\
 J230857.62-500914.9 &  3.98 &  17.372 &  17.47 & $ 15.98 \pm 0.01$ &  14.41 &  0.59 & $-28.29$ & $-28.61$ &  20190904 &   \\
 J231141.53-142215.7 &  3.98 &  18.571 &  18.29 & $ 16.26 \pm 0.04$ &  14.77 &  0.56 & $-27.51$ & $-28.34$ &  20210702 &   \\
 J231500.34-364049.9 &  4.27 &  18.842 &  18.88 & $ 16.91 \pm 0.05$ &  15.32 &  0.47 & $-27.05$ & $-27.81$ &  20200627 &   \\
 J232438.44-201956.9 &  4.23 &  18.285 &  18.58 & $ 16.94 \pm 0.06$ &  15.45 &  0.35 & $-27.34$ & $-27.77$ &  20200926 &   \\
\noalign{\smallskip} \hline
\end{tabular}

\end{table*}

\begin{table*}
\centering
\caption{List of newly discovered QSOs. Redshift $4.4\le z<4.7$ QSOs. One faint object has a $z$-band magnitude from the NOIRLab Source Catalog (NSC) DR2. See also caption of Table~\ref{qso3844}.}\label{qso4447}
\begin{tabular}{ccccccccccl}
\hline \noalign{\smallskip}  
SMSS ID  & Redshift  	& $R_p$   & $z_{\rm PSF}$  & $H$            & $W1$  & $W1-W2$ & $M_{145}$ & $M_{300}$ & Epoch & Comments \\
        &               & (Vega) & (AB)           & (Vega)           & (Vega) & (Vega)  & (AB)      & (AB) & (spectrum) \\
\noalign{\smallskip} \hline \noalign{\smallskip}
 J003438.57-762140.5 &  4.47 &  18.423 &  18.57 & $ 17.07 \pm 0.07$ &  15.59 &  0.38 & $-27.46$ & $-27.73$ &  20200627 &   \\
 J003656.23-241844.1 &  4.62 &  18.414 &  18.52 & $ 16.89 \pm 0.04$ &  15.29 &  0.37 & $-27.49$ & $-27.95$ &  20191022 &   \\
 J004022.28-333316.7 &  4.42 &  18.375 &  18.29 & $ 16.43 \pm 0.01$ &  14.94 &  0.51 & $-27.68$ & $-28.32$ &  20201208 &   \\
 J004541.14-553511.9 &  4.45 &  18.149 &  18.35 & $ 17.04 \pm 0.05$ &  15.80 &  0.36 & $-27.61$ & $-27.73$ &  20201208 &   \\
 J004917.31-081723.4 &  4.56 &  18.588 &  18.68 & $ 17.16 \pm 0.07$ &  15.81 &  0.40 & $-27.35$ & $-27.66$ &  20200630 &   \\
 J010255.67-574246.3 &  4.56 &  18.252 &  18.44 & $ 16.84 \pm 0.05$ &  15.18 &  0.42 & $-27.55$ & $-27.97$ &  20200926 &   \\
 J014423.70-114948.1 &  4.44 &  18.536 &  19.04 & $ 16.93 \pm 0.11$ &  15.10 &  0.44 & $-26.93$ & $-27.46$ &  20210711 &  $z$-band from NSC \\
 J020348.61-231453.8 &  4.50 &  18.898 &  18.99 & $ 17.32 \pm 0.06$ &  15.72 &  0.39 & $-26.99$ & $-27.47$ &  20200701 &   \\
 J024422.99-515931.6 &  4.46 &  18.779 &  19.12 & $ 17.36 \pm 0.06$ &  15.79 &  0.42 & $-26.86$ & $-27.39$ &  20200810 &   \\
 J024622.42-603753.7 &  4.57 &  18.341 &  18.47 & $ 16.72 \pm 0.04$ &  15.28 &  0.43 & $-27.53$ & $-28.09$ &  20191023 &   \\
 J025445.42-275042.5 &  4.54 &  18.628 &  18.67 & $ 17.29 \pm 0.02$ &  16.04 &  0.40 & $-27.32$ & $-27.51$ &  20211022 &   \\
 J025629.33-402041.8 &  4.51 &  18.393 &  18.42 & $ 16.91 \pm 0.04$ &  15.35 &  0.38 & $-27.56$ & $-27.88$ &  20191023 &   \\
 J030333.78-225121.4 &  4.57 &  18.662 &  18.85 & $ 17.05 \pm 0.05$ &  15.40 &  0.43 & $-27.14$ & $-27.77$ &  20200701 &   \\
 J031139.48-404428.8 &  4.43 &  18.997 &  18.90 & $ 17.10 \pm 0.06$ &  15.72 &  0.47 & $-27.07$ & $-27.66$ &  20200701 &   \\
 J032233.76-594328.1 &  4.42 &  18.546 &  18.77 & $ 16.91 \pm 0.06$ &  14.97 &  0.40 & $-27.21$ & $-27.85$ &  20191024 &   \\
 J033951.43-473959.9 &  4.45 &  18.507 &  18.65 & $ 17.24 \pm 0.04$ &  15.63 &  0.40 & $-27.31$ & $-27.52$ &  20191024 &   \\
 J034029.49-322353.9 &  4.66 &  18.925 &  19.07 & $ 17.33 \pm 0.06$ &  15.73 &  0.34 & $-26.95$ & $-27.48$ &  20200802 &   \\
 J034342.87-155923.6 &  4.57 &  18.995 &  19.03 & $ 17.40 \pm 0.07$ &  15.98 &  0.30 & $-27.02$ & $-27.55$ &  20200810 &   \\
 J034831.18-764305.5 &  4.66 &  18.617 &  18.62 & $ 16.93 \pm 0.06$ &  15.31 &  0.36 & $-27.50$ & $-27.97$ &  20200226 &   \\
 J035647.24-122512.1 &  4.66 &  18.162 &  18.15 & $ 16.73 \pm 0.06$ &  15.53 &  0.36 & $-27.91$ & $-28.14$ &  20200630 &   \\
 J044514.74-422805.6 &  4.48 &  18.949 &  19.01 & $ 17.28 \pm 0.07$ &  15.95 &  0.55 & $-26.96$ & $-27.49$ &  20200820 &   \\
 J045926.71-443855.2 &  4.65 &  18.684 &  18.57 & $ 16.81 \pm 0.04$ &  15.45 &  0.49 & $-27.45$ & $-28.04$ &  20200120 &   \\
 J051047.28-505722.5 &  4.44 &  18.890 &  18.75 & $ 17.22 \pm 0.06$ &  15.96 &  0.53 & $-27.22$ & $-27.54$ &  20210118 &   \\
 J093037.26-321931.1 &  4.43 &  18.072 &  18.32 & $ 16.47 \pm 0.04$ &  14.85 &  0.43 & $-27.76$ & $-28.33$ &  20201209 & $b_{\rm gal} = 13\fdg 7$, I7000 \\
 J095735.42-263039.4 &  4.49 &  18.381 &  18.56 & $ 16.88 \pm 0.07$ &  15.43 &  0.46 & $-27.48$ & $-27.92$ &  20200621 &   \\
 J104556.31-073145.9 &  4.47 &  18.684 &  18.84 & $ 17.09 \pm 0.04$ &  15.51 &  0.51 & $-27.15$ & $-27.69$ &  20200226 &   \\
 J105124.18-114529.8 &  4.43 &  18.769 &  18.90 & $ 17.21 \pm 0.03$ &  15.83 &  0.49 & $-27.10$ & $-27.56$ &  20200227 &   \\
 J105702.65-231013.7 &  4.50 &  18.325 &  18.33 & $ 16.75 \pm 0.05$ &  15.15 &  0.47 & $-27.68$ & $-28.05$ &  20200130 &   \\
 J110948.79-222757.7 &  4.49 &  18.922 &  18.99 & $ 17.33 \pm 0.08$ &  15.95 &  0.53 & $-27.02$ & $-27.46$ &  20200621 &   \\
 J114230.62-082330.6 &  4.46 &  19.145 &  18.97 & $ 17.00 \pm 0.05$ &  15.33 &  0.45 & $-27.03$ & $-27.78$ &  20200228 &   \\
 J122749.22-300600.1 &  4.63 &  18.120 &  18.29 & $ 16.71 \pm 0.06$ &  14.98 &  0.34 & $-27.78$ & $-28.16$ &  20210403 &   \\
 J125559.18-110812.7 &  4.56 &  18.671 &  18.93 & $ 17.19 \pm 0.05$ &  15.54 &  0.46 & $-27.09$ & $-27.63$ &  20200228 &   \\
 J133821.00-245845.8 &  4.60 &  18.537 &  18.69 & $ 17.12 \pm 0.07$ &  15.51 &  0.22 & $-27.37$ & $-27.74$ &  20200228 &   \\
 J134930.11-382208.4 &  4.40 &  18.815 &  18.89 & $ 17.18 \pm 0.09$ &  15.57 &  0.41 & $-27.15$ & $-27.60$ &  20200626 & redshift uncertain \\
 J145904.70-160326.6 &  4.46 &  18.617 &  18.74 & $ 17.16 \pm 0.03$ &  15.48 &  0.29 & $-27.35$ & $-27.64$ &  20210408 & extended/multiple \\ 
 J152219.16-271543.7 &  4.44 &  17.942 &  18.11 & $ 16.39 \pm 0.05$ &  15.02 &  0.38 & $-28.14$ & $-28.47$ &  20210404 &   \\
 J181812.74-564925.1 &  4.62 &  19.106 &  19.09 & $ 17.47 \pm 0.10$ &  15.98 &  0.46 & $-27.03$ & $-27.56$ &  20200813 &   \\
 J192427.57-582750.5 &  4.45 &  18.823 &  18.79 & $ 17.12 \pm 0.08$ &  15.78 &  0.44 & $-27.32$ & $-27.70$ &  20200616 &   \\
 J210743.01-494803.3 &  4.48 &  19.027 &  19.20 & $ 17.61 \pm 0.06$ &  15.93 &  0.39 & $-26.80$ & $-27.33$ &  20200812 &   \\
 J212109.58-404824.4 &  4.68 &  18.934 &  18.91 & $ 17.04 \pm 0.05$ &  15.48 &  0.53 & $-27.16$ & $-27.84$ &  20210711 &   \\
 J212132.33-651825.6 &  4.40 &  19.057 &  19.19 & $ 17.71 \pm 0.11$ &  16.19 &  0.45 & $-26.80$ & $-27.33$ &  20200813 &   \\
 J212358.26-391345.4 &  4.57 &  19.014 &  19.03 & $ 17.49 \pm 0.10$ &  16.00 &  0.23 & $-26.99$ & $-27.52$ &  20200812 &   \\ 
 J220004.31-291438.6 &  4.62 &  18.617 &  18.79 & $ 17.18 \pm 0.02$ &  15.44 &  0.30 & $-27.23$ & $-27.67$ &  20200616 &   \\
 J221150.10-525343.5 &  4.40 &  18.323 &  18.52 & $ 16.96 \pm 0.04$ &  15.67 &  0.32 & $-27.45$ & $-27.79$ &  20210702 &   \\
 J221423.26-142456.0 &  4.62 &  19.082 &  19.14 & $ 17.41 \pm 0.09$ &  15.78 &  0.44 & $-26.91$ & $-27.44$ &  20200812 &   \\
 J225551.95-122635.7 &  4.44 &  18.633 &  18.77 & $ 17.14 \pm 0.06$ &  15.82 &  0.58 & $-27.23$ & $-27.63$ &  20200630 &   \\
 J231425.23-665057.9 &  4.47 &  19.064 &  19.08 & $ 17.29 \pm 0.09$ &  15.86 &  0.45 & $-26.91$ & $-27.44$ &  20200813 &   \\
 J231614.93-461027.6 &  4.65 &  18.704 &  18.75 & $ 17.10 \pm 0.04$ &  15.51 &  0.38 & $-27.26$ & $-27.76$ &  20200627 &   \\
 J232009.94-605703.5 &  4.58 &  18.773 &  18.67 & $ 17.27 \pm 0.06$ &  15.81 &  0.44 & $-27.33$ & $-27.56$ &  20210712 &   \\
 J232518.43-584301.2 &  4.40 &  18.476 &  18.55 & $ 16.75 \pm 0.02$ &  15.63 &  0.36 & $-27.42$ & $-27.99$ &  20210119 &   \\
 J234842.90-214749.0 &  4.45 &  18.256 &  18.43 & $ 16.60 \pm 0.05$ &  15.11 &  0.50 & $-27.56$ & $-28.17$ &  20200615 &   \\
 J235834.56-493325.3 &  4.42 &  18.933 &  19.04 & $ 17.26 \pm 0.03$ &  15.64 &  0.35 & $-26.92$ & $-27.45$ &  20200627 &   \\
\noalign{\smallskip} \hline
\end{tabular}

\end{table*}

\begin{table*}
\centering
\caption{List of newly discovered QSOs. Redshift $4.7\le z<5.5$ QSOs. "W21" indicates two QSOs recently reported by \citet{2021AJ....162...72W}. See also caption of Table~\ref{qso3844}.}\label{qso4755}
\begin{tabular}{ccccccccccl}
\hline \noalign{\smallskip}  
SMSS ID  & Redshift  	& $R_p$  & $z_{\rm PSF}$  & $H$  & $W1$ & $W1-W2$ & $M_{145}$ & $M_{300}$ & Epoch & Comments \\
        &               & (Vega) & (AB)           & (Vega)           & (Vega) & (Vega)  & (AB)      & (AB) & (spectrum) \\
\noalign{\smallskip} \hline \noalign{\smallskip}
 J010924.31-503749.7 &  5.12 &  19.271 &  18.93 & $ 17.49 \pm 0.05$ &  16.18 &  0.26 & $-27.38$ & $-27.55$ &  20210929 &   \\
 J012736.71-300649.6 &  4.73 &  18.909 &  19.11 & $ 17.17 \pm 0.02$ &  15.78 &  0.52 & $-26.97$ & $-27.50$ &  20200810 &   \\
 J012938.99-582942.7 &  4.75 &  18.594 &  18.76 & $ 17.22 \pm 0.05$ &  15.88 &  0.56 & $-27.32$ & $-27.68$ &  20200120 &   \\
 J014534.68-163950.3 &  4.71 &  18.497 &  18.49 & $ 16.85 \pm 0.11$ &  15.43 &  0.49 & $-27.57$ & $-28.03$ &  20191128 &   \\
 J020240.11-294314.4 &  4.86 &  18.894 &  19.08 & $ 17.19 \pm 0.02$ &  15.80 &  0.55 & $-27.07$ & $-27.75$ &  20200818 &   \\
 J020436.66-252315.2 &  4.86 &  18.249 &  18.53 & $ 16.59 \pm 0.11$ &  14.90 &  0.66 & $-27.63$ & $-28.35$ &  20191020 &   \\
 J021739.31-125025.0 &  4.73 &  18.862 &  18.89 & $ 17.39 \pm 0.07$ &  16.17 &  0.42 & $-27.18$ & $-27.50$ &  20200630 &   \\
 J022009.01-352745.3 &  5.07 &  18.942 &  18.91 & $ 17.30 \pm 0.02$ &  15.94 &  0.55 & $-27.38$ & $-27.72$ &  20211020 &   \\
 J023648.56-114733.7 &  5.20 &  18.634 &  18.40 & $ 16.78 \pm 0.04$ &  15.51 &  0.62 & $-27.96$ & $-28.29$ &  20200926 &   \\
 J024133.93-543853.0 &  4.74 &  18.723 &  18.89 & $ 17.40 \pm 0.06$ &  16.18 &  0.46 & $-27.21$ & $-27.51$ &  20200630 &   \\
 J031431.12-573152.7 &  5.10 &  18.732 &  18.81 & $ 17.03 \pm 0.05$ &  15.71 &  0.57 & $-27.51$ & $-28.00$ &  20200801 &   \\
 J032933.94-410056.1 &  4.83 &  18.596 &  18.62 & $ 17.07 \pm 0.14$ &  15.69 &  0.55 & $-27.51$ & $-27.86$ &  20200626 &   \\
 J033703.05-254831.5 &  5.15 &  18.493 &  18.28 & $ 16.59 \pm 0.04$ &  15.27 &  0.64 & $-28.05$ & $-28.46$ &  20191128 & z=5.08? \\
 J035210.62-214544.9 &  5.16 &  19.012 &  18.86 & $ 17.02 \pm 0.06$ &  15.83 &  0.52 & $-27.52$ & $-28.04$ &  20211022 &   \\
 J040732.95-281031.3 &  4.75 &  18.467 &  18.62 & $ 16.91 \pm 0.04$ &  15.11 &  0.43 & $-27.48$ & $-27.99$ &  20200120 &   \\
 J050328.89-195623.0 &  5.00 &  18.316 &  18.46 & $ 16.62 \pm 0.05$ &  15.34 &  0.59 & $-27.81$ & $-28.37$ &  20210203 &   \\
 J050928.30-183435.1 &  4.70 &  18.579 &  18.62 & $ 16.96 \pm 0.05$ &  15.62 &  0.46 & $-27.46$ & $-27.93$ &  20200820 &   \\
 J082450.79-674241.5 &  5.41 &  18.775 &  18.72 & $ 16.64 \pm 0.05$ &  14.95 &  0.90 & $-27.85$ & $-28.38$ &  20210118 &   \\
 J084347.76-253155.6 &  4.75 &  18.629 &  18.62 & $ 16.92 \pm 0.06$ &  15.53 &  0.41 & $-27.56$ & $-28.01$ &  20200119 & $b_{\rm gal} = 10\fdg 5$, W21  \\
 J113522.01-354838.8 &  4.95 &  18.407 &  18.23 & $ 16.72 \pm 0.05$ &  15.29 &  0.59 & $-28.08$ & $-28.29$ &  20190705 &   \\
 J121921.12-360933.0 &  4.74 &  18.710 &  18.46 & $ 16.51 \pm 0.05$ &  14.88 &  0.46 & $-27.69$ & $-28.41$ &  20210203 &   \\
 J162551.54-043049.4 &  5.17 &  18.659 &  18.68 & $ 16.54 \pm 0.04$ &  15.32 &  0.80 & $-27.90$ & $-28.61$ &  20210409 &   \\
 J162758.93-083343.6 &  4.79 &  18.945 &  18.86 & $ 16.94 \pm 0.05$ &  15.46 &  0.56 & $-27.54$ & $-28.08$ &  20210704 &   \\
 J165333.86-761426.1 &  5.41 &  18.423 &  18.20 & $ 16.31 \pm 0.04$ &  14.72 &  0.87 & $-28.33$ & $-28.87$ &  20200227 &   \\
 J192600.92-314202.6 &  5.35 &  19.237 &  18.80 & $ 17.38 \pm 0.08$ &  15.95 &  0.78 & $-27.68$ & $-27.79$ &  20210703 &   \\
 J194124.59-450023.6 &  5.18 &  18.488 &  18.22 & $ 16.45 \pm 0.05$ &  14.91 &  0.84 & $-28.20$ & $-28.64$ &  20200615 &   \\
 J205559.20-601147.3 &  4.95 &  18.926 &  18.84 & $ 17.22 \pm 0.05$ &  15.73 &  0.55 & $-27.42$ & $-27.76$ &  20200616 &   \\
 J211002.60-454548.3 &  4.80 &  18.878 &  19.17 & $ 17.46 \pm 0.03$ &  16.16 &  0.57 & $-26.96$ & $-27.49$ &  20200813 &   \\
 J214608.20-485819.5 &  5.17 &  18.325 &  18.34 & $ 16.44 \pm 0.02$ &  14.88 &  0.66 & $-28.01$ & $-28.63$ &  20210429 &   \\
 J223419.12-804013.3 &  4.91 &  18.941 &  18.75 & $ 17.22 \pm 0.09$ &  15.81 &  0.25 & $-27.58$ & $-27.79$ &  20210710 &   \\
 J232952.75-200038.7 &  5.05 &  18.548 &  18.45 & $ 16.55 \pm 0.05$ &  14.95 &  0.65 & $-27.84$ & $-28.46$ &  20200625 & W21  \\
 J233435.28-365708.8 &  4.72 &  19.165 &  18.92 & $ 16.51 \pm 0.03$ &  14.59 &  0.38 & $-27.13$ & $-28.37$ &  20200813 &   \\
\noalign{\smallskip} \hline
\end{tabular}

\end{table*}

\subsection{Completeness considerations}\label{completeness}

In this section, we explore how to define a complete sample for use in the luminosity functions. Our selection criteria were adopted to provide a complete sample of QSOs at redshift $z\ge 4.4$, at least to a depth of $z_{\rm PSF}\approx 19$~ABmag. At fainter magnitudes, the uncertainties in colours broaden the QSO distribution and the significance of the PPM information degrades, making our approach less effective. 

\subsubsection{Colour and PPM selection effects}\label{sec:sel_effects}

Currently, there are 143 QSOs known in our search area with $z\ge 4.4$ and $z_{\rm PSF}<19$~ABmag. Here, we have assumed that all literature QSOs within these limits are contained in the compilation Milliquas v7.3 \citep{Flesch15}. Of the known QSOs, four objects are rejected by the selection criteria of Sect.~\ref{sec:sel_criteria}: two because of the PPM signal and two more because of their colours. One additional known $z=4.87$ QSO is nearly blended with a cool star and thus eliminated by our requirement for no close neighbours, which has reduced our effective search area. Hence, we conclude that our PPM and colour selection criteria may miss $\sim$3 per cent of the candidates across our search area of 14,486~deg$^2$.

While considering the known QSOs implies that we inherit traditional selection biases, we note that we were able to relax and extend the selection beyond previous rules, because the {\it Gaia} PPM data vastly reduced the stellar contamination. The majority of objects in the sample considered here are new QSOs identified using our extended cuts. By considering our selection boundaries relative to the shape of the QSO distributions against colour in Fig.~\ref{col_redshift}, we see that we capture not only the bulk of the QSO population but also its tails, and that the remaining incompleteness based on the photometric selection cuts must be small.

\subsubsection{Upper redshift limit?}

While we make no attempt to avoid the detection of QSOs at the highest possible redshifts, we assume that our search is only complete to $z\simeq 5.4$, where we find our two highest-redshift objects. Inspecting Fig.~\ref{sample_mz} suggests that the sample is conspicuously devoid of bright QSOs at $z>5.5$.

Here, it is worth highlighting our reliance on {\it Gaia} data and the redshift constraints this imposes. Beyond redshift $\sim 5.5$, very little flux remains in the {\it Gaia} $G$ passband from which the astrometric measurements are made. In fact, none of the known QSOs beyond z=5.7 have {\it Gaia} parallax measurements in eDR3. In the regime between $z=5.4$ and $z=5.7$, we expect the Ly\,$\alpha$ equivalent width to play an important role in determining whether {\it Gaia} can measure a parallax or proper motion.

\subsubsection{Incompleteness at the faint end}\label{sec:mag_incomp}

A significant consideration in this work is sample incompleteness at the faint end due to two different reasons: first, there is the source incompleteness of SMSS DR3, which we use as a parent sample. We use the $z$-band number counts of objects listed in our search area of DR3 and obtain a linear fit to the brightness range of $z_{\rm PSF}=16-19$~ABmag. The downturn from the linear fit (with slope $0.254\pm 0.001$) indicates the magnitude-dependent source incompleteness of the parent sample (see left panel in Fig.~\ref{nc_cplt}). We find that the completeness declines from 100 per cent at $z_{\rm PSF}=17$~ABmag to 95 per cent at $z_{\rm PSF}=18.5$~ABmag, to 86 per cent at $z_{\rm PSF} =18.7$~ABmag, and to 63 per cent at $z_{\rm PSF}=19$~ABmag. Secondly, the spectroscopic follow-up of the selected candidate list by the literature and by this work is only complete (bar one candidate) to $z_{\rm PSF}<18.7$~ABmag; fainter than that, the completeness drops quickly to $\sim$30 per cent at $z_{\rm PSF}\simeq19$~ABmag.

Secondly, we use NIR photometry from the three surveys 2MASS, VHS and VIKING, with a 2~arcsec matching radius. We find that 6 per cent of Milliquas QSOs with $z>3$ and $z_{\rm PSF} = 16-19$~ABmag have no $JK$ data, which we showed in Sect.~\ref{sec:sel_criteria} as being helpful in selecting high-redshift QSOs. This affects mostly fainter QSOs without 2MASS counterparts, in areas where neither the deeper VHS nor the VIKING survey is available at present. At $z_{\rm PSF}<18$~ABmag this incompleteness amounts to 2 per cent, and reaches 7 per cent by $z_{\rm PSF}=18.7$~ABmag and $\sim $10 per cent by $z_{\rm PSF}=19$~ABmag.

In conclusion, we have a highly complete sample of $z=4.4-5.4$ QSOs at $z_{\rm PSF}<18.7$~ABmag and a substantial further sample down to $z_{\rm PSF}\simeq19$~ABmag. The completeness at $z_{\rm PSF}>18.7$~ABmag is difficult to estimate reliably given the interplay of source incompleteness in the parent sample, incompleteness in spectroscopic follow-up, and the fraction of true QSOs among the candidates, which will likely decline as we go fainter and contamination by non-QSOs in the candidate sample increases.
We thus leave the faint-end incompleteness untreated in this work; further follow-up is progressing to find the missing QSOs at $z_{\rm PSF}>18.7$~ABmag.

Cumulative number counts as a function of apparent magnitude ($z$- and $H$-band), split into two redshift ranges ($4.4\le z<4.7$ and $4.7\le z<5.4$; the latter selected to match that of Y16), are shown in the left-hand panels of Fig.~\ref{NCs}.

\subsection{Sample comparisons between North and South}

\begin{figure}
\begin{center}
\includegraphics[angle=270,width=\columnwidth,clip=true]{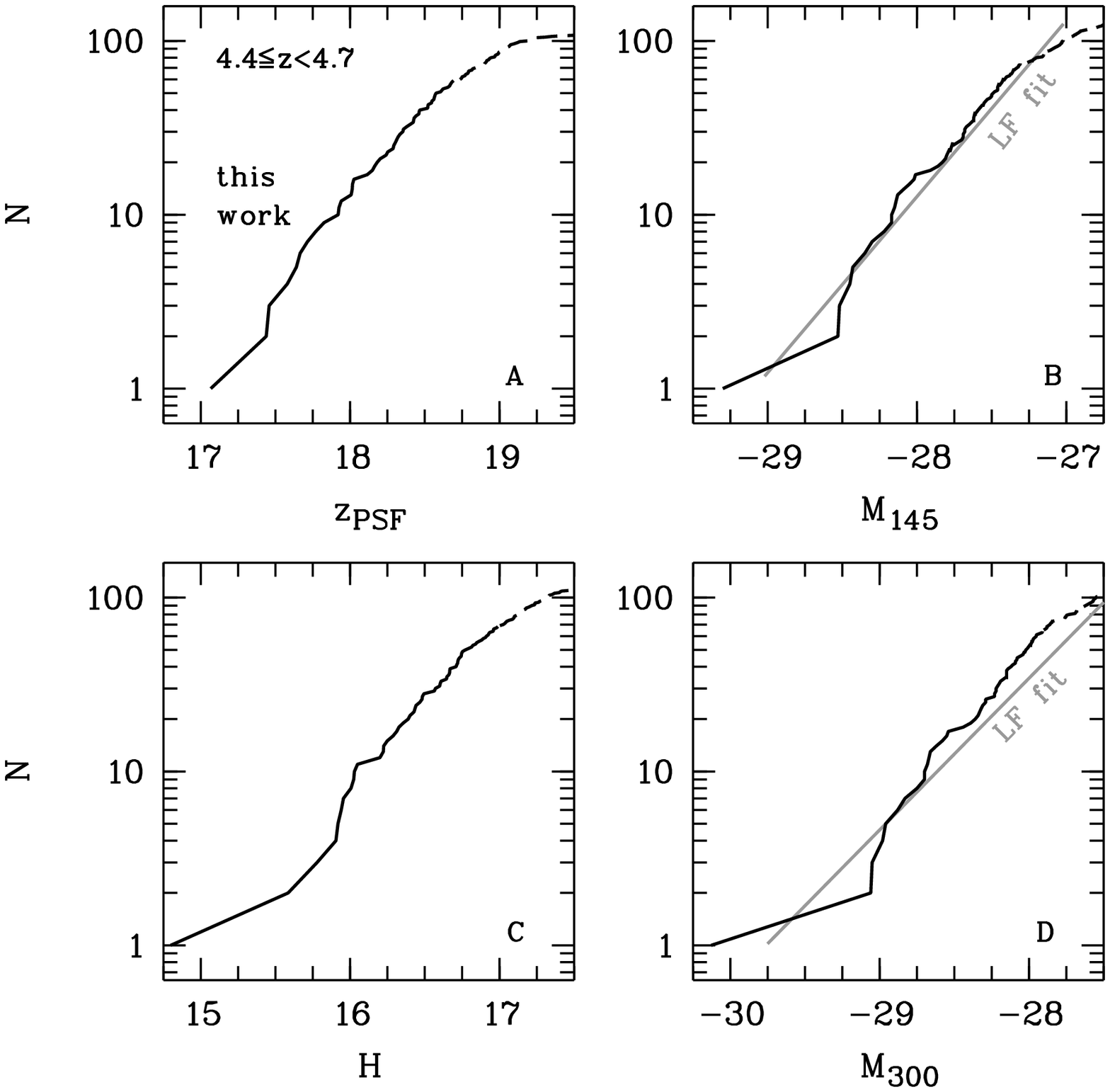}
\includegraphics[angle=270,width=\columnwidth,clip=true]{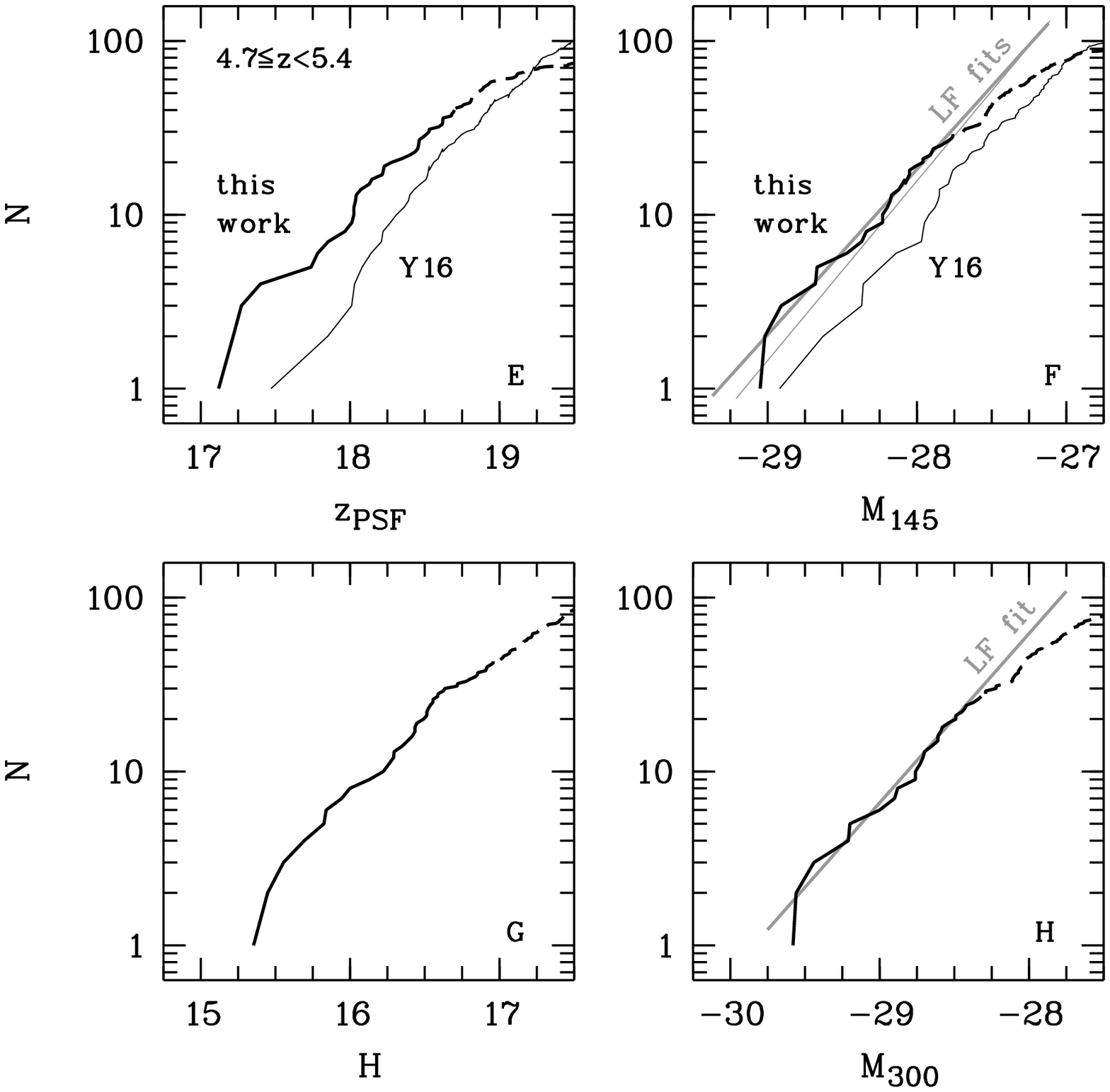} 
\caption{Cumulative QSO number counts vs. apparent magnitude in $z$- and $H$-band (panels ACEG) and vs. absolute magnitude $M_{145}$ and $M_{300}$ (panels BDFH). The top 4 panels (ABCD) show $4.4\le z < 4.7$, while the bottom 4 panels (EFGH) show $4.7\le z < 5.4$. The incomplete tail of our sample line is dashed. Panels E and F include a comparison with the Y16 sample. Thick grey lines are predictions from our luminosity function fits (the thin grey line in panel F is the Y16 fit, renormalised to our cosmology and search area).  
\label{NCs}}
\end{center}
\end{figure}

We now want to ask whether the new sample in the Southern hemisphere and the established literature sample in the Northern hemisphere are statistically consistent. Therefore, we look specifically at the redshift range $4.7<z<5.4$ in our sample and compare it with the W16/Y16 sample in the same redshift range that was searched in a Northern area of similar size (within 1 per cent). The comparison between the cumulative number counts for this redshift range is shown in Fig.~\ref{NCs}(E). Our sample is $>80$ per cent complete to $z_{\rm PSF}\simeq $18.7~ABmag, where it contains 38 objects, compared to 25 objects in the Y16 sample (they have another 21 objects down to their spectroscopic completeness limit of $z_{\rm SDSS}=19$~ABmag). This factor of $\sim 1.5$ is easily explained by the different completeness of candidate selection: calculations by Y16 suggest that their candidate selection is 50 to 60 per cent complete, while we assume $>80$ per cent completeness for our sample. In contrast, the slight difference in the $z$ bandpasses of SDSS and SkyMapper plays very little role: comparing the SDSS and SMSS $z$-band AB magnitudes for eight objects in common between Y16 and our Southern sample at $z_{\rm PSF}<18.7$~ABmag, we find a mean $z_{\rm SDSS}-z_{\rm SMSS}=-0.01$~mag with an RMS of 0.14~mag.

Next, we ask whether the completeness of Y16 appears well-estimated in hindsight: we try to apply the selection cuts of Y16 to our sample and count what fraction of objects we lose. The largest difference in the selection is that Y16 require \mbox{$W1-W2>0.5$~mag}, which was needed in the pre-{\it Gaia} era to combat the vast stellar contamination, while we relaxed the cut to \mbox{$W1-W2>0.2$~mag} given {\it Gaia}'s PPM data. At $z_{\rm PSF}<19$~ABmag, our inclusive selection rules revealed 55
QSOs, while using \mbox{$W1-W2>0.5$~mag} retains 34 
of these (62 per cent), in agreement with the original completeness correction used by Y16. 
Clearly, there are ultra-luminous $z\sim 5$ QSOs waiting to be discovered in the Northern hemisphere using updated {\it WISE} colour cuts and {\it Gaia} data. 

Among the QSOs selected by our criteria, there are 12 in the SDSS footprint, for which we can fully replicate the Y16 selection rules. We find only one object that the Y16 \mbox{$W1-W2$} colour cut would have retained, but which the other Y16 rules would exclude (SDSS~J013127.34-032100.1). Although based on only one object, this suggests a possible, modest $\sim10$ per cent correction to the Y16 completeness, in addition to the {\it WISE}-based factor above. 

\subsection{Estimation of $M_{145}$ and $M_{300}$}\label{sec:selsing}

Because of (a) limited S/N and significant sky emission line residuals in the WiFeS spectra, (b) the frequent presence of BAL features blueward of \ion{C}{iv}, and (c) the desire to compile a homogeneous dataset for the LF that includes QSOs from the literature, we estimate the QSO absolute magnitude at restframe 145~nm, $M_{145}$, utilising a composite spectrum of bright QSOs from $z=1-2$ \cite[][S16, hereafter]{2016A&A...585A..87S}. In comparison to other QSO composite spectra that incorporate lower-luminosity objects \cite[e.g.,][]{2002ApJ...565..773T, 2001AJ....122..549V}, the S16 composite\footnote{Retrieved from \url{https://github.com/jselsing/QuasarComposite}} provides a better match to the low-equivalent-width QSOs that one typically finds at high luminosity \cite[as expected from the Baldwin Effect;][]{1977ApJ...214..679B, 2020ApJS..249...17R}.

The S16 composite is convolved with the relevant filter curves to normalise the spectrum with the observed photometry. The $z$-band filter curves for SMSS and SDSS, and the $H$-band filter curve for VISTA were retrieved from the Spanish Virtual Observatory (SVO) Filter Profile Service\footnote{\url{http://svo2.cab.inta-csic.es/theory/fps/}} \citep{2012ivoa.rept.1015R,2020sea..confE.182R}.

We validated the application of the S16 composite by comparing the associated $M_{145}$ predictions against the subsample of W16/Y16 data that was drawn from SDSS spectroscopy (in contrast to their own follow-up data, calibrated from SDSS $i$-band photometry). From 52 QSOs, the median $M_{145}$ difference, after adding 0.07~mag to the values from Y16\footnote{The Y16 cosmological parameters ($\Omega_\Lambda=0.728$) predict 3.3 per cent larger luminosity distances at $z=5$ relative to ours.}, was only 0.06~mag (with the S16 composite slightly brighter), with an RMS of just 0.12~mag. We take this as evidence that the combination of the observed photometry and the S16 composite spectrum will give robust results for computing the $M_{145}$ LFs.\footnote{In Paper~I (and the initial arXiv version of this work), our $M_{145}$ estimates relied on simple interpolation between the SMSS $i_{\rm psf}$ and $z_{\rm psf}$ photometry. However, by ignoring the contributions of the UV emission lines, either photometric interpolation or the assumption of a fixed power-law continuum anchored to a single photometric band are likely to yield $M_{145}$ estimates that are systematically too bright. In particular, over the redshift interval $z=4.7-5.4$, the S16 composite gives $M_{145}$ values between 0.12 and 0.25~mag fainter than suggested by a ($\alpha_{\nu}=-0.3$) power-law continuum anchored to $z$-band. Recent work by \citet{2021arXiv211013736G} at $z\sim5$ have used similar power-law estimates (anchored to SMSS $i$-band), and our composite-based $M_{145}$ values are, in the mean, 0.18~mag fainter for the same 14 objects. Such use of $i$-band is particularly fraught as the redshift increases above $z=4.8$, where the flux through the filter is first boosted by Ly\,$\alpha$ emission, and then is soon diminished by Ly\,$\alpha$ absorption.}

For $M_{300}$, we apply a similar procedure, but based on the $H$-band photometry (measured or inferred). From the S16 composite, $M_{145}-M_{300}$ = 0.53~mag, but extrapolating the $M_{300}$ magnitudes from $z$-band would typically produce a slightly brighter $M_{300}$ value than our $H$-band-based estimates (up to 0.5~mag). This indicates that the spectral slopes of our QSOs are slightly bluer than the composite, as confirmed by the mean colour of the QSOs in our sample, $M_{145}-M_{300}=0.38$~mag. The bluer colours are consistent with the findings of \citet{Xie16}, in which more luminous QSOs had bluer UV spectral slopes. However, by anchoring our $M_{300}$ estimates to $H$-band, we minimise any associated error arising from variations in the spectral slope.

Cumulative number counts for $M_{145}$ and $M_{300}$ are shown in the right-hand panels of Fig.~\ref{NCs}, split into our two redshift ranges.

\section{Luminosity functions}\label{sec:LFs}

At the bright end, the number counts in the Southern sample exceed the number counts used previously for deriving the luminosity function of high-redshift QSOs. This allows us to be less dependent on completeness corrections. Hence, we re-derive the luminosity function for the bright end and update estimates
of the luminosity density and redshift evolution. 
Using the S16 composite spectrum, we can relate the apparent-magnitude completeness limit of $z_{\rm PSF}=18.7$~ABmag to a absolute-magnitude limit ranging from $M_{145}=-27.25$~ABmag at $z=4.4$ to $-27.73$~ABmag at $z=5.4$, and from $M_{300}\simeq -27.8$~ABmag at $z=4.4$ to $-28.3$~ABmag at $z=5.4$ (although we note that the $M_{300}$ magnitudes used in the LF calculation are estimated from $H$-band rather than extrapolating all the way from $z$-band). 
In the half-magnitude bin fainter than $z_{\rm PSF}=18.7$~ABmag, the completeness of our sample collapses from $>80$ per cent to $\sim 10$ per cent, which will show in the figures we present for the binned luminosity function. 

Prior to the application of any magnitude constraints, this final sample contains 171 QSOs in our search area of 14,486~deg$^2$, which is defined by $\delta<+2$\deg, $|b|>15$\deg\ and the exclusion zones detailed in Sect.~\ref{search_area}. This sample excludes our two highest-redshift QSOs as well as a large number of QSOs identified at $z<4.4$. One of the literature QSOs does not have $z$-band magnitude from SMSS DR3, and we use the value from NSC DR2 instead ($z=18.68$~ABmag). The sample is assumed to be overall $\sim 95$ per cent complete at $z_{\rm PSF}<18$~ABmag and $>80$ per cent at $z_{\rm PSF}<18.7$~ABmag, which we take into account for our statistical analysis.

For $4.4\le z < 4.7$, we have \Nlow\ QSOs brighter than $M_{145}=-27.33$~ABmag and which are used in the LF fitting below. That sample has a median redshift of 4.52. For the $4.7\le z < 5.4$ sample, which covers the same redshift interval as Y16 (a median redshift of 4.83 vs. 4.92 from Y16), we retain \Nhi\ QSOs brighter than $M_{145}=-27.73$~ABmag.

For the first time, NIR photometry yields direct measurements of $M_{300}$ for a wide-area sample of QSOs at the bright end. This is due to the simultaneously deep and wide coverage of the VHS, which has no equivalent in the Northern hemisphere, where the samples for previous determinations of the bright end of the luminosity functions were found. Similar to our treatment of $M_{145}$, we apply cuts in $M_{300}$ to define the samples for fitting the LFs. The thresholds of $M_{300}<-27.86$ and $-28.26$~ABmag for the low and high redshift bins, respectively, are the S16 equivalents for $z_{\rm PSF}=18.7$~ABmag. As discussed in Sect.~\ref{sec:selsing}, because the spectral slopes of our QSOs are, on average, bluer than the S16 composite spectrum, the $M_{300}$ thresholds we adopt from the $z$-band limit are conservative. Nonetheless, we end up with similar numbers of QSOs in the $M_{300}$ LF fits for our two redshift ranges: \NThreelo\ and \NThreehi.

\subsection{Parametric luminosity function fits}

We compare our data with the best-fit model of the LF from Y16, which is parametrised with a double power-law as
\begin{equation}
    \Phi(M,z) = \frac{\Phi^*(z)}{10^{0.4(\alpha+1)(M-M^*)}+10^{0.4(\beta+1)(M-M^*)}} ~,
\end{equation}\label{eq:lf}
having a turnover in the density centred at luminosity $M^*=-26.98 \pm 0.23$~ABmag, a faint-end slope of $\alpha=-2.03$ and a bright-end slope of $\beta=-3.58 \pm 0.24$. The normalisation, in units of number~mag$^{-1}$~Mpc$^{-3}$, is taken to evolve as \mbox{$\log_{10}\Phi^*(z)=-8.82-0.47(z-6)$} \citep{Fan2001b}, and we evaluate it at the median redshifts of our two bins. As noted in Sect.~\ref{sec:selsing}, we also add 0.07~mag to the $M^*$ value from Y16 and adjust their comoving volume, as their cosmological parameters predict larger luminosity distances at $z=5$. 
The incompleteness of the Southern sample at the faint end is evident as a slope change in the right-hand panels of Fig.~\ref{NCs}, particularly in the higher redshift bin.

To fit the LF, we adopt a Maximum Likelihood approach like that used by Y16 \cite[see][]{1983ApJ...269...35M}. 
We incorporate our full completeness model and consider two recent versions of the $z\sim5$ faint-end LF: those of \citet{Niida20} and \citet{Kim20}, hereafter N20 and K20, respectively. To probe such faint regimes, both surveys naturally covered much smaller areas, 340 and 85 deg$^2$, respectively, which yielded 72 and 32 QSOs across the $M_{145}$ range of -23 to -27~ABmag\footnote{As illustrated by \citet{2021arXiv210801090H}, at the faint end of the N20 and K20 absolute magnitude ranges, the population of UV sources is rapidly transitioning from being QSO-dominated to being galaxy-dominated, so contamination in the QSO samples becomes a serious concern.}. We take their respective faint-end slopes and normalisations, and fit our bright QSO sample (\Nhi\ QSOs) with the double power-law of Eq.~\ref{eq:lf}. For each of the two faint-end models, we allow $M^*$ to vary as part of the fit, but adjust $\Phi^*$ in a manner that maintains the existing fit to the faint-end data, given their respective $\alpha$ parameters. In this way, we fit $\beta$ and $M^*$, but also recompute $\Phi^*$.

First, we utilised the N20 faint-end parameters, namely $\alpha=-2.00$ and a faint-end normalisation that remains consistent with their combination of $\log_{10}\Phi^*=-8.26$~mag$^{-1}$~Mpc$^{-3}$ for $M^*=-27.15$~ABmag. (N20 fixed their bright-end slope based on a fit to Y16 and \citet{2013ApJ...768..105M}, but utilised an independent selection of bright QSOs from SDSS DR7 in their MLE fit.) We adopted the redshift evolution of $\Phi^*$ from \citet{Fan2001b} to scale the normalisation from their mean redshift of $z=4.75$ to our mean redshift of $4.91$. With our bright-end data, we then derive $\beta$=\betaNhi\ and $M^*$=\MstarNhi~ABmag, which gives $\log_{10}\Phi^*$=\logPhistarNhi~mag$^{-1}$~Mpc$^{-3}$. 

We then performed a similar fit with the K20 parameters, i.e., $\alpha=-1.11$ and $\log_{10}\Phi^*=-7.35$~mag$^{-1}$~Mpc$^{-3}$ for $M^*=-25.81$~ABmag at $<z>=4.95$. (In contrast to N20, K20 explicitly used the Y16 data to constrain the bright end in their MLE fit.) We find $\beta$=\betaKhi\ and $M^*$=\MstarKhi~ABmag, implying $\log_{10}\Phi^*$=\logPhistarKhi~mag$^{-1}$~Mpc$^{-3}$. 

The results from fitting with each faint-end constraint are provided in Table~\ref{DPL_pars}, and the bright-end LFs are overplotted as straight lines in panels (B) and (F) of Fig.~\ref{NCs} (using the parameters from the K20 faint end). We also plot the Y16 LF fit in Fig.~\ref{NCs}(F), where we have scaled their parameters to our cosmology and to our search area.

\begin{figure*}
\begin{center}
\includegraphics[width=0.92\textwidth,clip=true]{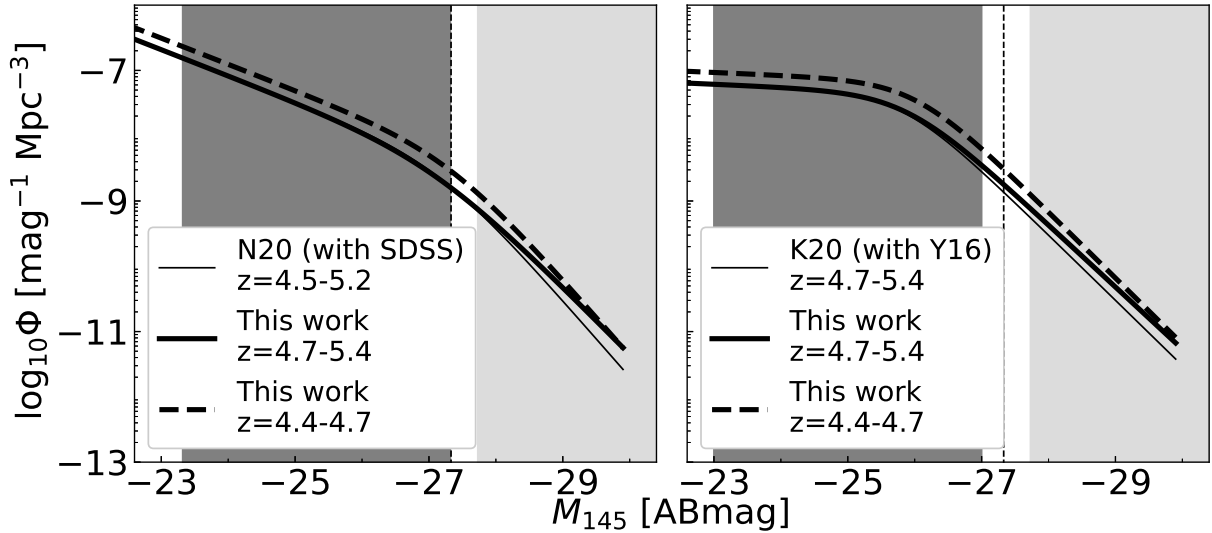} \\
\caption{
Parametrised QSO LF models at $z\sim5$, combining our bright-end data with the faint-end slopes and normalisations of ({\it left}) N20 and ({\it right}) K20. The solid lines show the fits at $z=4.7-5.4$: ours (thick), and the original N20/K20 models (thin). The bright-end fitting regions for our data are shown in the light grey shaded areas for $z=4.7-5.4$. Our fits to the $z=4.4-4.7$ sample are shown with thick dashed lines and the lower magnitude bounds for $z=4.4-4.7$ are shown as vertical dotted lines. The fits include sources up to a maximum of $M_{145}=-29.1$~ABmag. The faint-end samples cover the dark grey regions of $M_{145}\sim-23$ to $\sim-27$~ABmag.
\label{LF_2panel}}
\end{center}
\end{figure*}

The overall $z\sim5$ LF shape is illustrated in Figure~\ref{LF_2panel}, with the prior results shown in thin lines, and the new fits shown in thick lines for high redshift (solid) and low redshift (dashed). For the N20 parameters, there is some tension with the QSO densities found at the faint end of our highly complete survey region, where the close proximity between $M^*$ and our faint-end cutoff of $-27.73$~ABmag results in a steeper bright-end slope. In contrast, the shallower slope of K20 (in conjunction with its high normalisation) has no direct impact on the bright-end fit, and $M^*$ is more than 1~mag fainter, well below our sample's completeness limit.

While the LF in the low-redshift bin is generally higher at all magnitudes, the steeper bright-end slopes may suggest a change in behaviour at the brightest magnitudes (see Sect.~\ref{sec:downsizing}).

\begin{table}
\centering
\caption{Restframe 145~nm LF parameters for double power-law.}
\label{DPL_pars}
\begin{tabular}{ccccc}
\hline \noalign{\smallskip}  
median z & $\beta$  & $M^*$ & $\alpha$  & $\log_{10}\Phi^*$ \\
\noalign{\smallskip} \hline \noalign{\smallskip}
\multicolumn{5}{|c|}{Faint-end constraints from \citet{Kim20}}\\
\noalign{\smallskip} \hline
 $4.52$ & \betaKlo & \MstarKlo & \alphaK & \logPhistarKlo  \\
 $4.83$ & \betaKhi & \MstarKhi & \alphaK & \logPhistarKhi  \\
\noalign{\smallskip} \hline \noalign{\smallskip}
\multicolumn{5}{|c|}{Faint-end constraints from \citet{Niida20}}\\
\noalign{\smallskip} \hline
 $4.52$ & \betaNlo & \MstarNlo & \alphaN & \logPhistarNlo \\
 $4.83$ & \betaNhi & \MstarNhi & \alphaN & \logPhistarNhi \\
\noalign{\smallskip} \hline \noalign{\smallskip}
\end{tabular}
\end{table}

Previous work, principally based on the Y16 sample, found broadly similar slopes, with data-driven 'best' free fits ranging from $\beta=-3.94$ by N20 and $\beta=-3.80$ by Y16 to $\beta=-3.50$ by K20. 
At a slightly lower redshift of $z\sim 3.9$, \citet{Boutsia21} found a similar bright-end slope of $\beta = -4.0^{+0.6}_{-0.4}$. 
The Extremely Luminous QSO Survey \citep[ELQS;][]{Schindler19b} found bright-end slopes for the redshift range of $2.8 < z < 4.5$ in the range from $\beta =-4$ to $-4.5$ depending on the details of the fitting boundaries.

For the $M_{300}$ LF at $z\sim 5$, we lack the deep $H$-band photometric data to provide the counterparts to the N20 and K20 constraints, so we focus on fitting just a single power-law to the bright end of the LF. Applying the same MLE technique to that above, we fit the LF as a pure power-law. Retaining the redshift dependence of \citet{Fan2001b} for the LF normalisation, we fit the $\beta$ and $\Phi$ values at $M_{145}=-28.5$~ABmag at $z=4.4-4.7$, and $M_{145}=-28.75$~ABmag at $z=4.7-5.4$.

At $z=4.4-4.7$, from \NThreelo\ QSOs, we find the best fit to the bright end of the $M_{300}$ LF to be $\beta=$\betaThreelo\ and \mbox{$\log_{10}\Phi(M_{300}=-28.5)=$} \logPhiThreelo~mag$^{-1}$~Mpc$^{-3}$. In the $z=4.7-5.4$ bin, from \NThreehi\ QSOs, we find $\beta=$\betaThreehi\ and \mbox{$\log_{10}\Phi(M_{300}=-28.75)=$} \logPhiThreehi~mag$^{-1}$~Mpc$^{-3}$. The LF fits are shown with straight lines in panels (D) and (H) of Fig.~\ref{NCs}.

\subsection{Redshift evolution and downsizing}\label{sec:downsizing}

Decades of survey work in the evolution of galaxies and QSOs have established a paradigm of cosmic downsizing: in this paradigm, the galaxies reaching the highest stellar masses have formed the bulk of their stars at high redshift during short epochs of particularly intense star formation; in contrast, galaxies of progressively lower mass formed at later cosmic epochs during more extended and less intense activity \citep[e.g.][among many others]{Cowie96}. Given well-established relations between the mass of central supermassive black holes and the mass of their host galaxy bulges \citep{Ferrarese00,KH13}, we expect that the most massive black holes have thus grown most rapidly in the most luminous QSOs at high redshift. Surveys of QSOs have revealed a similar and matching trend of higher-luminosity objects peaking in activity at higher redshift, while lower-luminosity QSOs seem to have later activity peaks  \citep[e.g.][]{Hasinger05,Hopkins07}.

\begin{figure}
\begin{center}
\includegraphics[angle=270,width=0.8\columnwidth,clip=true]{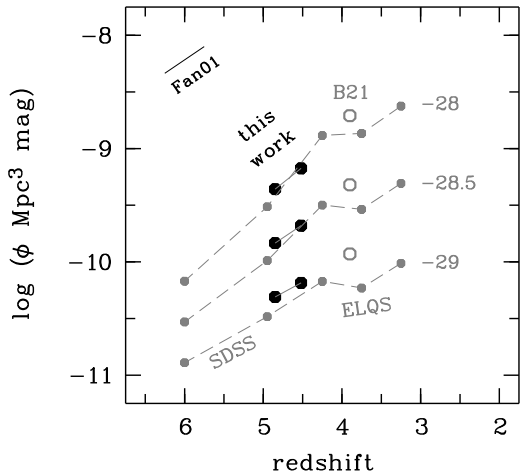} \\
\caption{
Evolution of $\Phi(M)$ with redshift at three $M_{145}$ points. Large symbols are from Southern samples in the SkyMapper survey area, with black points from this work and the grey open symbol from \citet{Boutsia21}. Small points are dominated by Northern samples including \citet{Jiang16} at $z\simeq 6$, Y16 at $z\simeq 5$ and ELQS points at $z<4.5$. The \citet{Fan2001b} evolution of $d\log \Phi^* /dz$ is shown in the top left corner.\label{QLFfits}}
\end{center}
\end{figure}

However, most of the past works could not draw on complete samples of QSOs at the extreme luminosities of $M_{145}<-28$, and hence we now compare the newest bright-end measurements of the QSO LF spanning the redshift range from $z\sim 4$ to $z\sim 6$ to reassess its evolution. In Fig.~\ref{QLFfits}, we complement our work and that of W16/Y16 with the LFs from ELQS and from \citet{Boutsia21} on the lower-z side and with the LF by \citet{Jiang16} on the higher-z end. Here, we visualise space densities from the actual best-fit parametrisations, because, due to the range of different bright-end slopes, it would be insufficient to consider only $M^*$ and $\Phi^*$ parameters. 

The evolution parameter of $\gamma = d\log \Phi(M)/dz=0.47$ estimated by \citet{Fan2001b} is shown in Fig.~\ref{QLFfits} as well and broadly consistent with the evolution at the highest redshifts. We see no strong evidence of a peak or plateau, although there is a hint of flattening around redshift 3 to 4.

\section{Summary}

We continued our program to search for QSOs at $z\ga 4$ in the Southern hemisphere using data from SkyMapper, {\it Gaia}, and several infrared surveys. 
We focus on the ultra-luminous end of the QSO distribution, the luminosity range of $M_{145}<-27$~ABmag, where true QSOs are outnumbered by cool stars to the most extreme degree possible. In this paper, we extend the search area to an effective 14\,486~deg$^2$. We have obtained full spectroscopic coverage of our candidate list at $z_{\rm PSF}<18.7$~ABmag, at which depth we are $>80$ per cent complete.

We discover 126 new QSOs with $M_{145} < -27$~ABmag, on top of 21 such objects presented by Paper~I, as well as a small number of less luminous objects. Previously, the most complete sample of bright high-redshift QSOs was that of W16 \& Y16, found within a similarly sized search area in the Northern sky, with a focus on the redshift range of $4.7<z<5.4$. 

We use our sample to redetermine the bright end of the QSO luminosity function at restframe 145~nm, updating that of Y16, and also present for the first time the bright end of the high-redshift QSO luminosity function at restframe 300~nm, capitalising on the depth of wide-field NIR photometry available in the Southern hemisphere. 

We measure the slope of the $M_{145}$ luminosity at $z=4.7-5.4$ as $\beta=$\betaKhi \ and \betaNhi, depending on the faint-end constraint that is used, both of which are in line with several earlier estimates. In addition, we measure $\beta=$\betaThreehi \ for $M_{300}$ from a fit with no faint-end constraint.

In the $z=4.4-4.7$ range, we measure slightly steeper bright-end slopes of $\beta=$\betaKlo \ and \betaNlo \ for the $M_{145}$ LF, and $\beta=$\betaThreelo\ for $M_{300}$.

A final determination of the QSO space densities at the luminous end and their evolution will require completing searches in the Northern hemisphere to create a complete all-sky sample of high-redshift QSOs. While this is one of our next steps, another is to push the completeness limit deeper to $z_{\rm PSF}\approx 19$~ABmag in order to obtain more robust statistics in the region where this sample connects to and overlaps with deeper literature samples. This, however, will require significant observational effort given that the contamination by stars increases with the fading discriminating power of {\it Gaia} parallaxes and proper motions at 19~ABmag and beyond.

\section*{Acknowledgements}

A number of observers contributed spectroscopy of candidates through time-swapping arrangements or extreme generosity, and we thank them for their efforts: Harrison Abbot, Katie Auchettl, and Mike Bessell.
We also thank the referee for suggestions that greatly improved the quality of the manuscript.
CAO was supported by the Australian Research Council (ARC) through Discovery Project DP190100252.
The national facility capability for SkyMapper has been funded through ARC LIEF grant LE130100104 from the Australian Research Council, awarded to the University of Sydney, the Australian National University, Swinburne University of Technology, the University of Queensland, the University of Western Australia, the University of Melbourne, Curtin University of Technology, Monash University and the Australian Astronomical Observatory. SkyMapper is owned and operated by The Australian National University's Research School of Astronomy and Astrophysics. The survey data were processed and provided by the SkyMapper Team at ANU. The SkyMapper node of the All-Sky Virtual Observatory (ASVO) is hosted at the National Computational Infrastructure (NCI). Development and support the SkyMapper node of the ASVO has been funded in part by Astronomy Australia Limited (AAL) and the Australian Government through the Commonwealth's Education Investment Fund (EIF) and National Collaborative Research Infrastructure Strategy (NCRIS), particularly the National eResearch Collaboration Tools and Resources (NeCTAR) and the Australian National Data Service Projects (ANDS).
This project has made use of data from the European Space Agency (ESA) mission {\it Gaia} (\url{https://www.cosmos.esa.int/gaia}), processed by the {\it Gaia} Data Processing and Analysis Consortium (DPAC, \url{https://www.cosmos.esa.int/web/gaia/dpac/consortium}). Funding for the DPAC has been provided by national institutions, in particular the institutions participating in the {\it Gaia} Multilateral Agreement.
This publication makes use of data products from the Wide-field Infrared Survey Explorer, which is a joint project of the University of California, Los Angeles, and the Jet Propulsion Laboratory/California Institute of Technology, and NEOWISE, which is a project of the Jet Propulsion Laboratory/California Institute of Technology. {\it WISE} and NEOWISE are funded by the National Aeronautics and Space Administration.
We have used data products from the Two Micron All Sky Survey, which is a joint project of the University of Massachusetts and the Infrared Processing and Analysis Center/California Institute of Technology, funded by the National Aeronautics and Space Administration and the National Science Foundation.
This paper uses data from the VISTA Hemisphere Survey ESO programme ID: 179.A-2010 (PI. McMahon).
Based on observations obtained as part of the VISTA Hemisphere Survey, ESO Progam, 179.A-2010 (PI: McMahon). The VISTA Data Flow System pipeline processing and science archive are described in \citet{2004SPIE.5493..411I}, \citet{2008MNRAS.384..637H} and \citet{2012A&A...548A.119C}.
This publication has made use of data from the VIKING survey from VISTA at the ESO Paranal Observatory, programme ID 179.A-2004. Data processing has been contributed by the VISTA Data Flow System at CASU, Cambridge and WFAU, Edinburgh.
This research has made use of the SVO Filter Profile Service (\url{http://svo2.cab.inta-csic.es/theory/fps/}) supported from the Spanish MINECO through grant AYA2017-84089. The results presented here have utilised modules from the {\sc SciPy} software package \citep{SciPy}.

\section*{DATA AVAILABILITY}
The SMSS data underlying this article are available at the SkyMapper node of the All-Sky Virtual Observatory (ASVO), hosted at the National Computational Infrastructure (NCI) at \url{https://skymapper.anu.edu.au}. The data from SMSS Data Release 3 are currently accessible only to Australia-based researchers and their collaborators.

\appendix

\section{Spectrum gallery}\label{app:A}
\counterwithin{figure}{section}

In the Figures below, we present the WiFeS spectra for the newly discovered QSOs from the redshift range $z=3.8-5.5$, ordered by RA.

\begin{figure*}
\begin{center}
\includegraphics[angle=270,width=0.32\textwidth,clip=true]{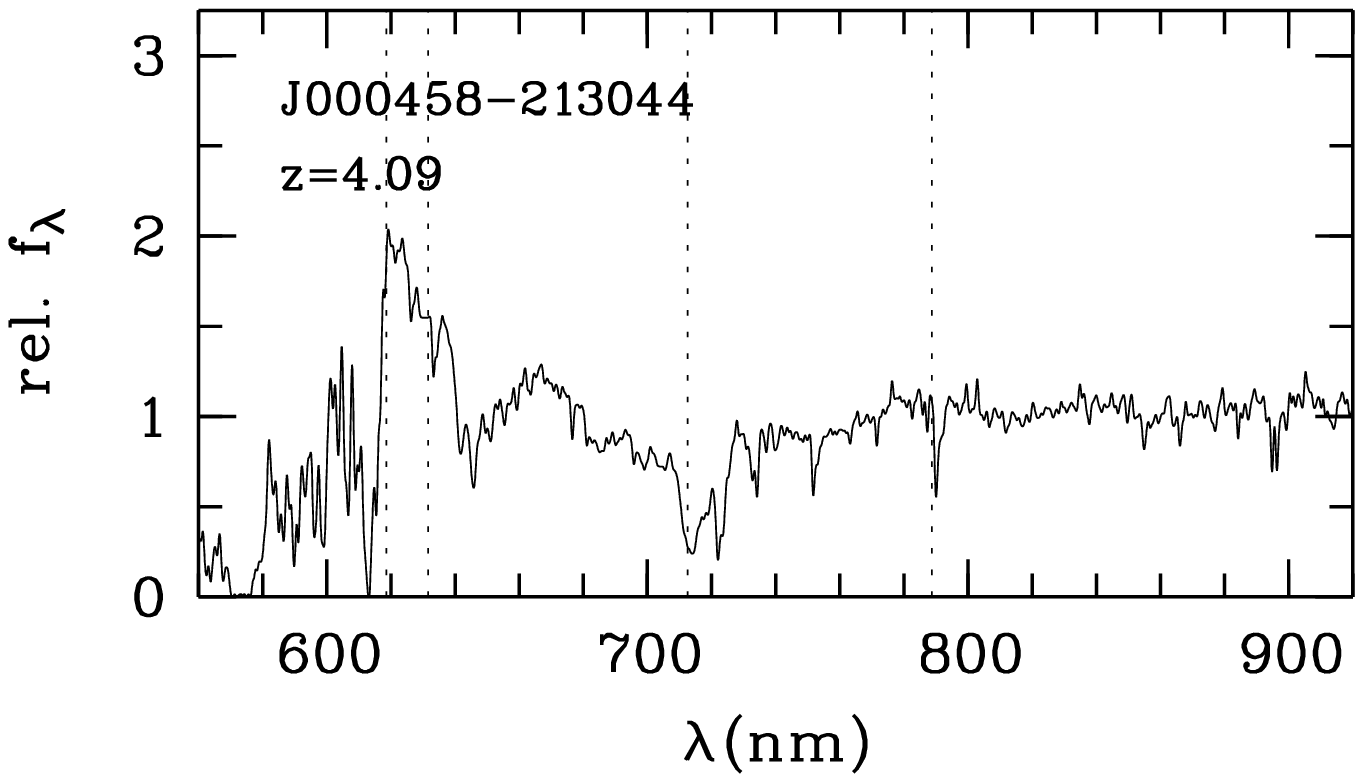}
\includegraphics[angle=270,width=0.32\textwidth,clip=true]{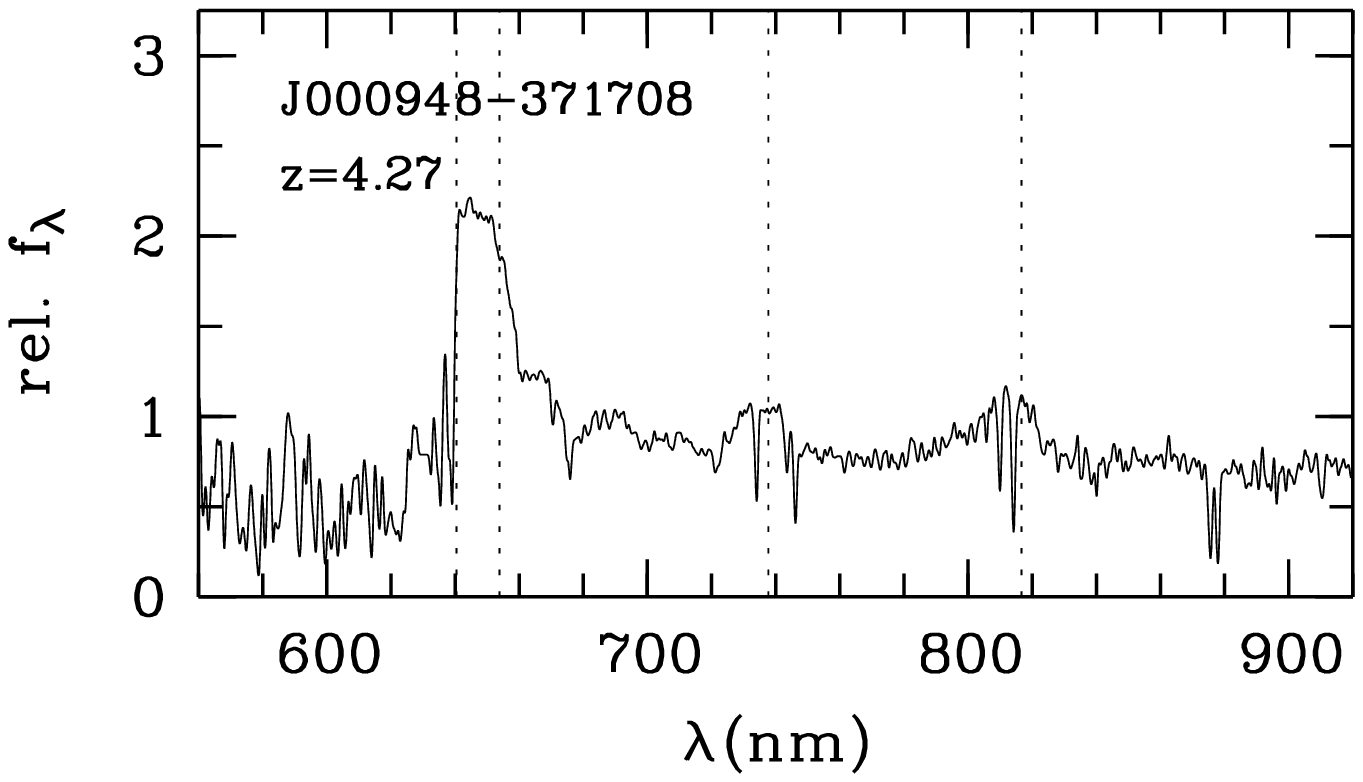}
\includegraphics[angle=270,width=0.32\textwidth,clip=true]{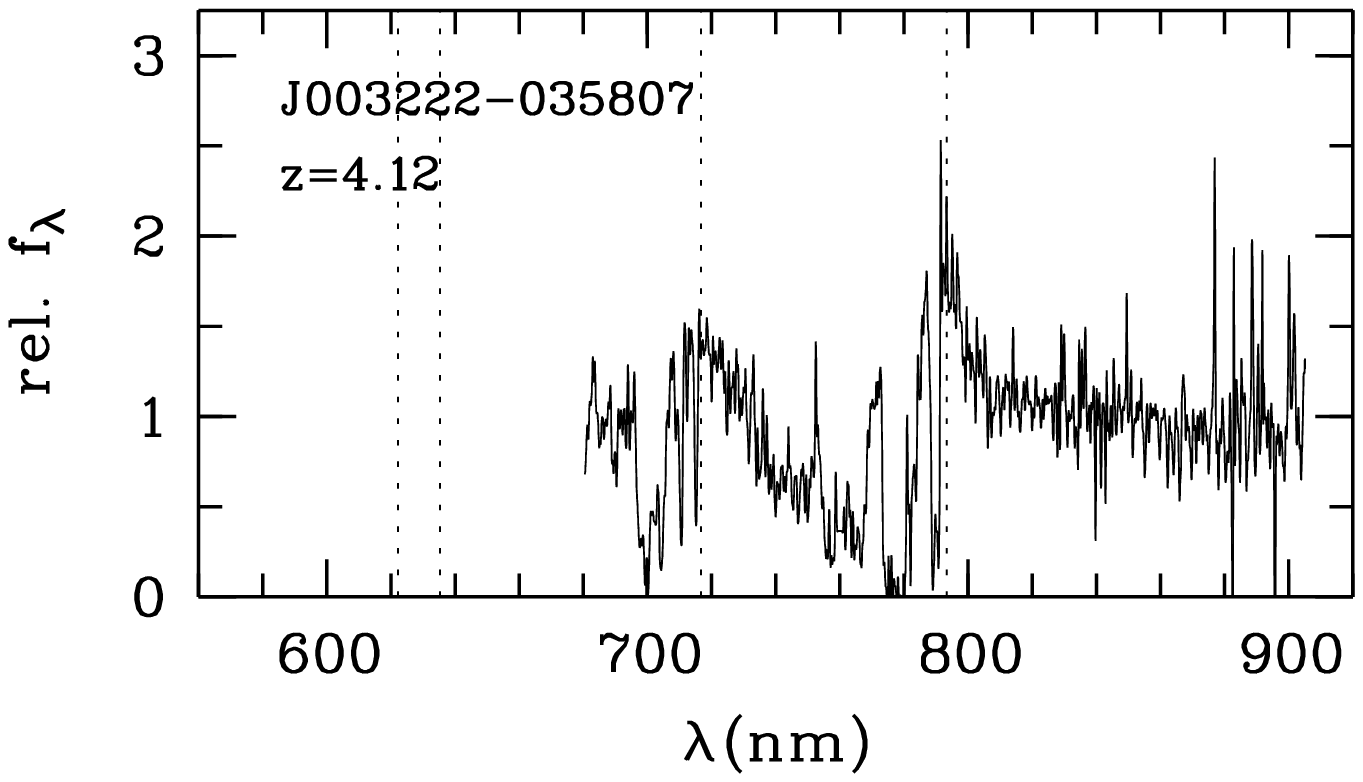}
\includegraphics[angle=270,width=0.32\textwidth,clip=true]{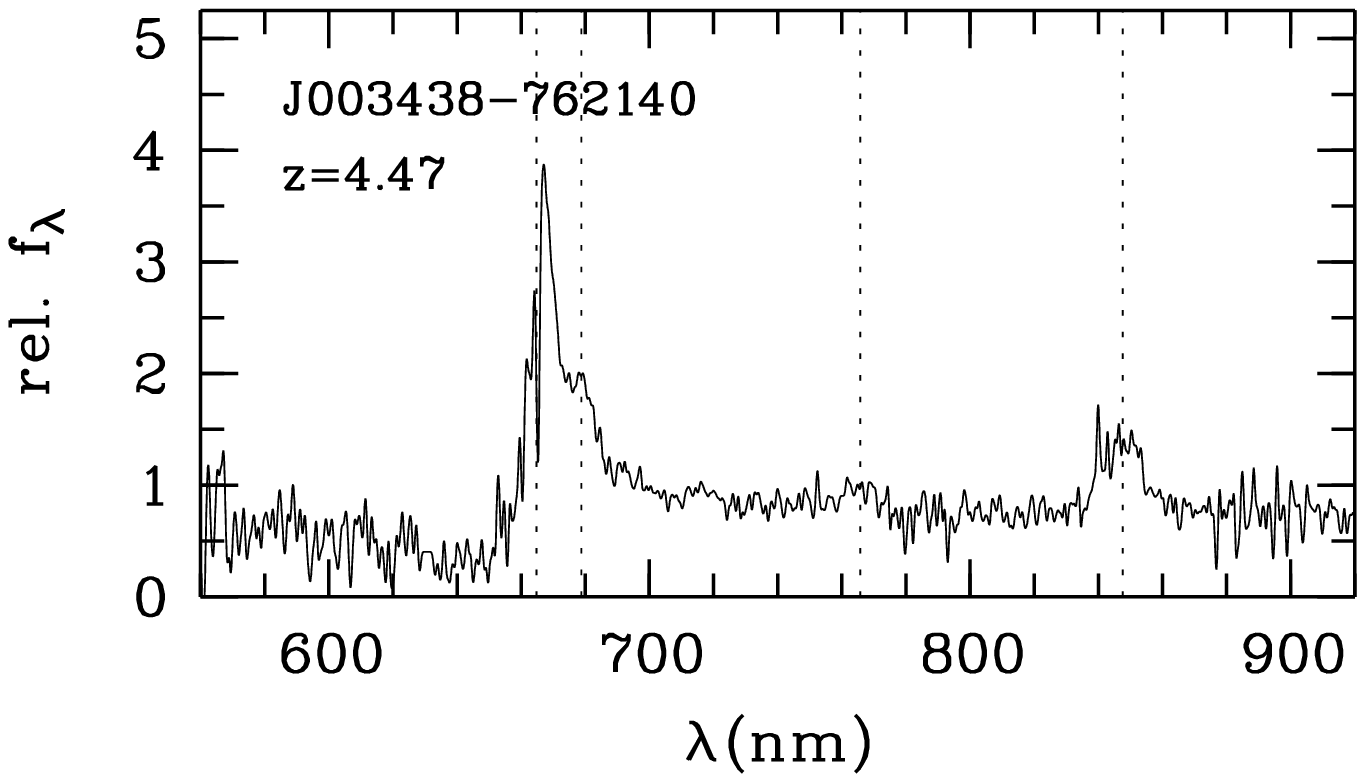}
\includegraphics[angle=270,width=0.32\textwidth,clip=true]{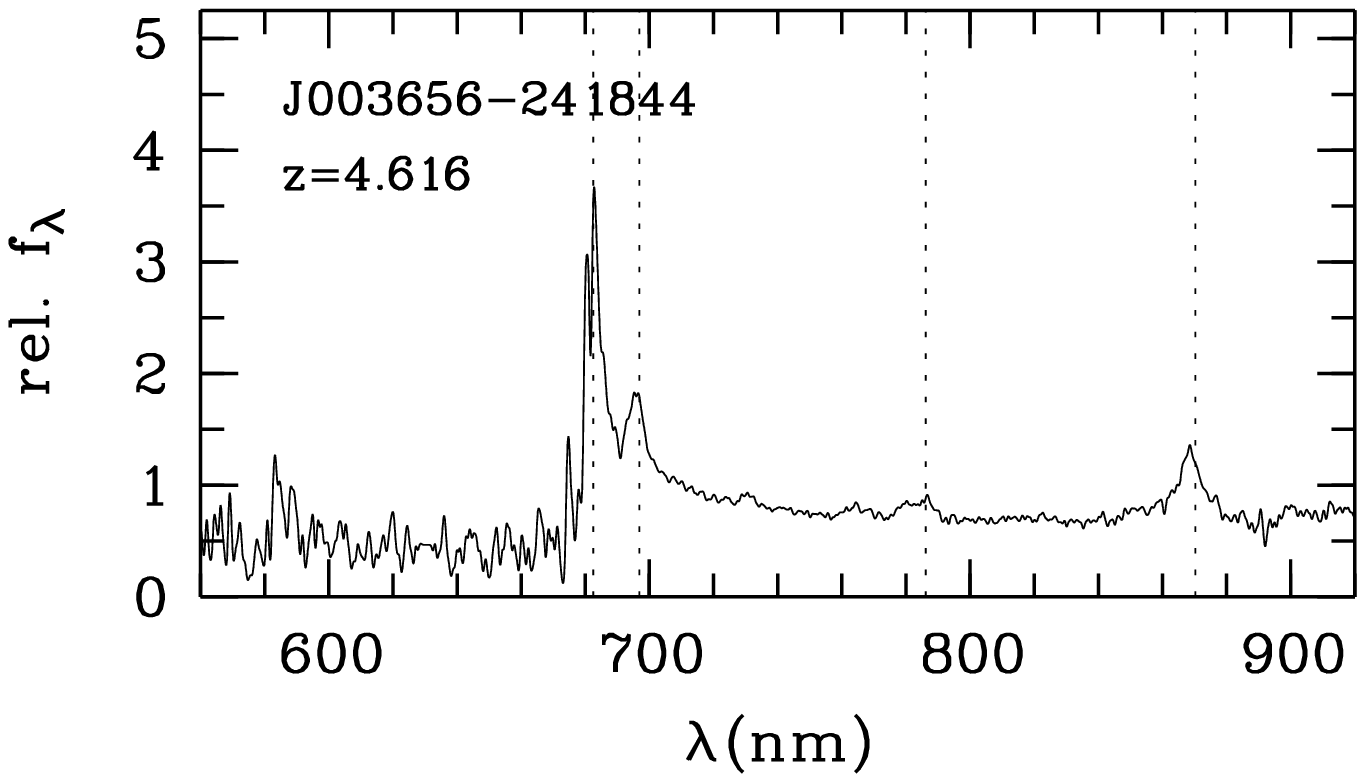}
\includegraphics[angle=270,width=0.32\textwidth,clip=true]{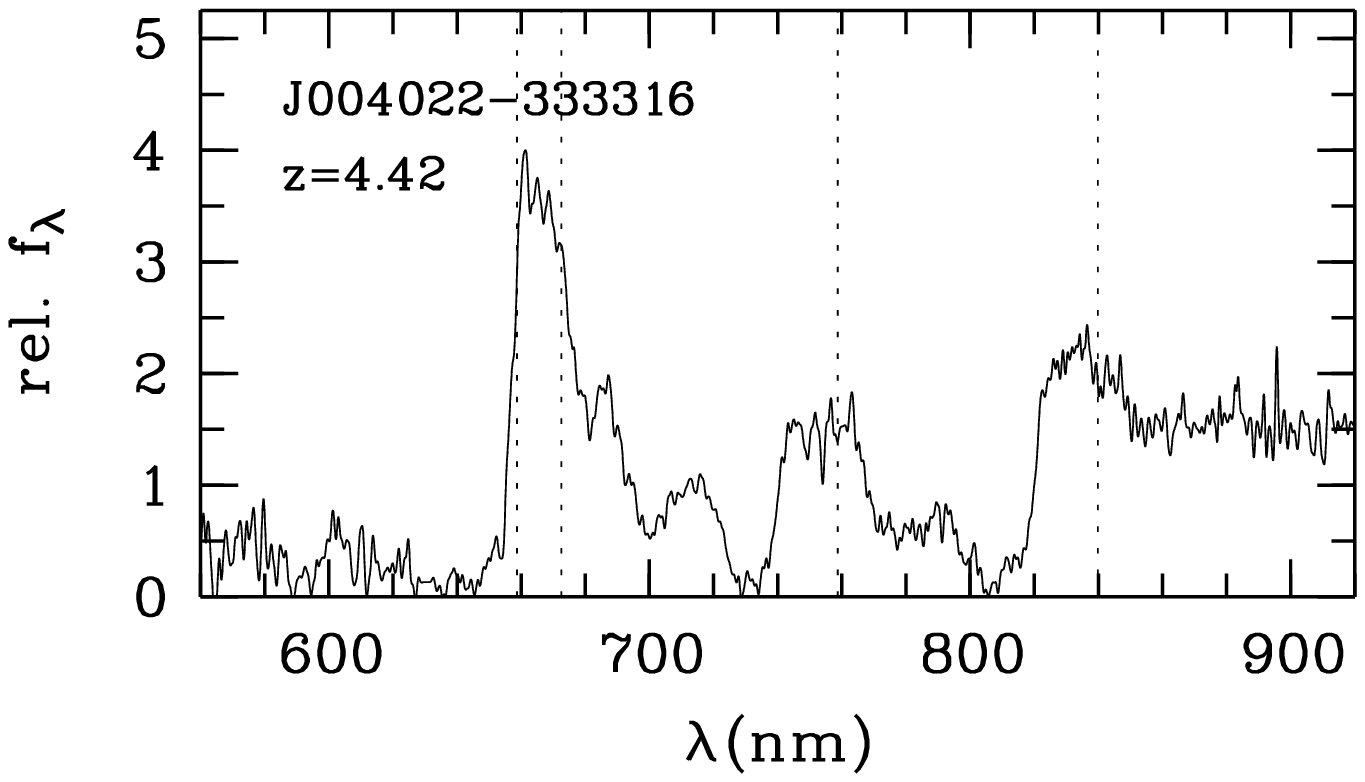}
\includegraphics[angle=270,width=0.32\textwidth,clip=true]{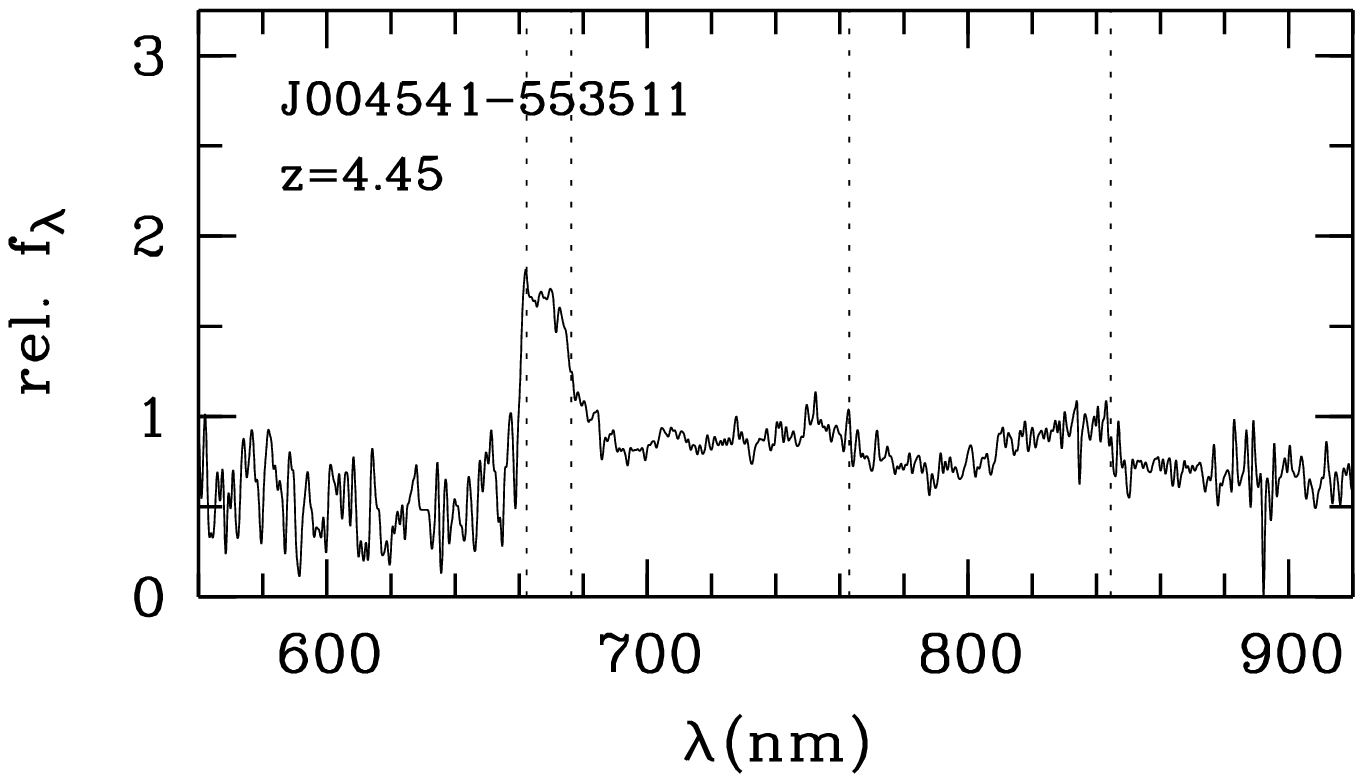}
\includegraphics[angle=270,width=0.32\textwidth,clip=true]{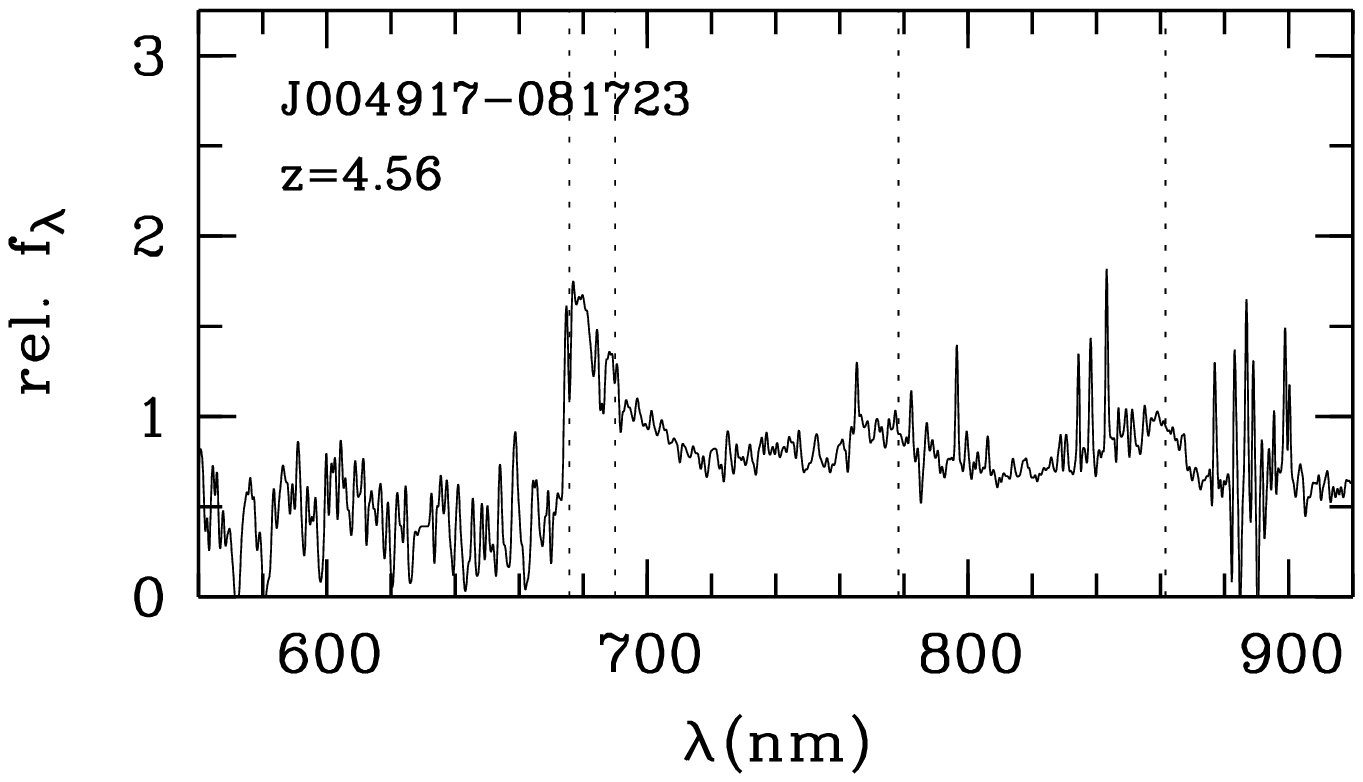}
\includegraphics[angle=270,width=0.32\textwidth,clip=true]{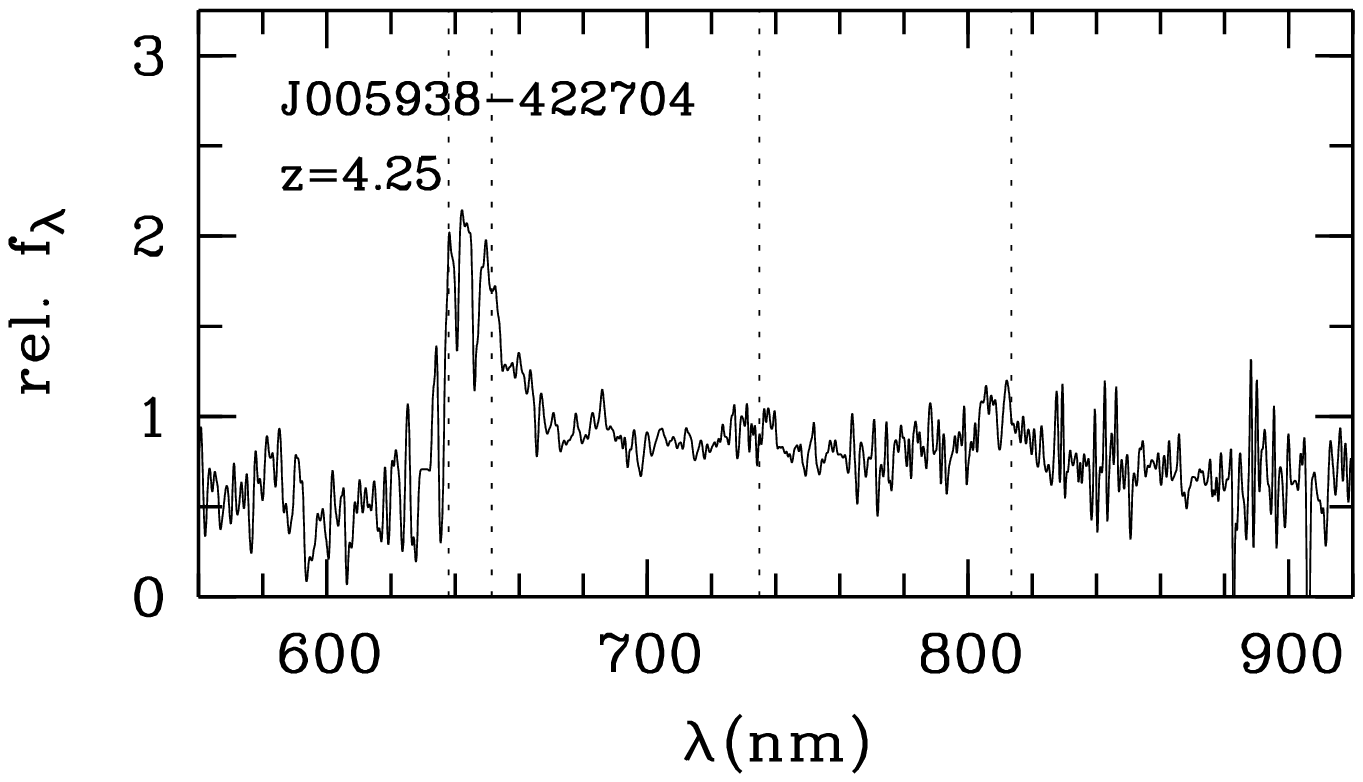}
\includegraphics[angle=270,width=0.32\textwidth,clip=true]{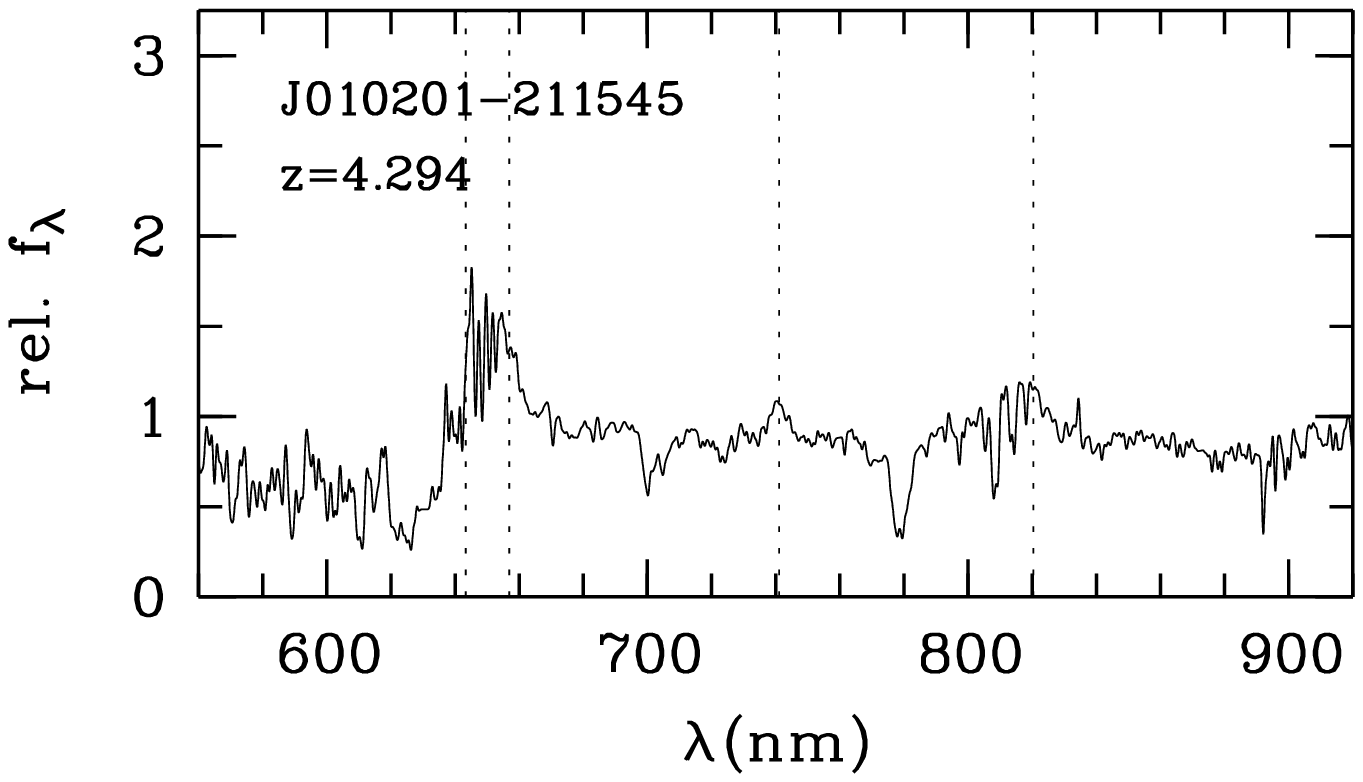}
\includegraphics[angle=270,width=0.32\textwidth,clip=true]{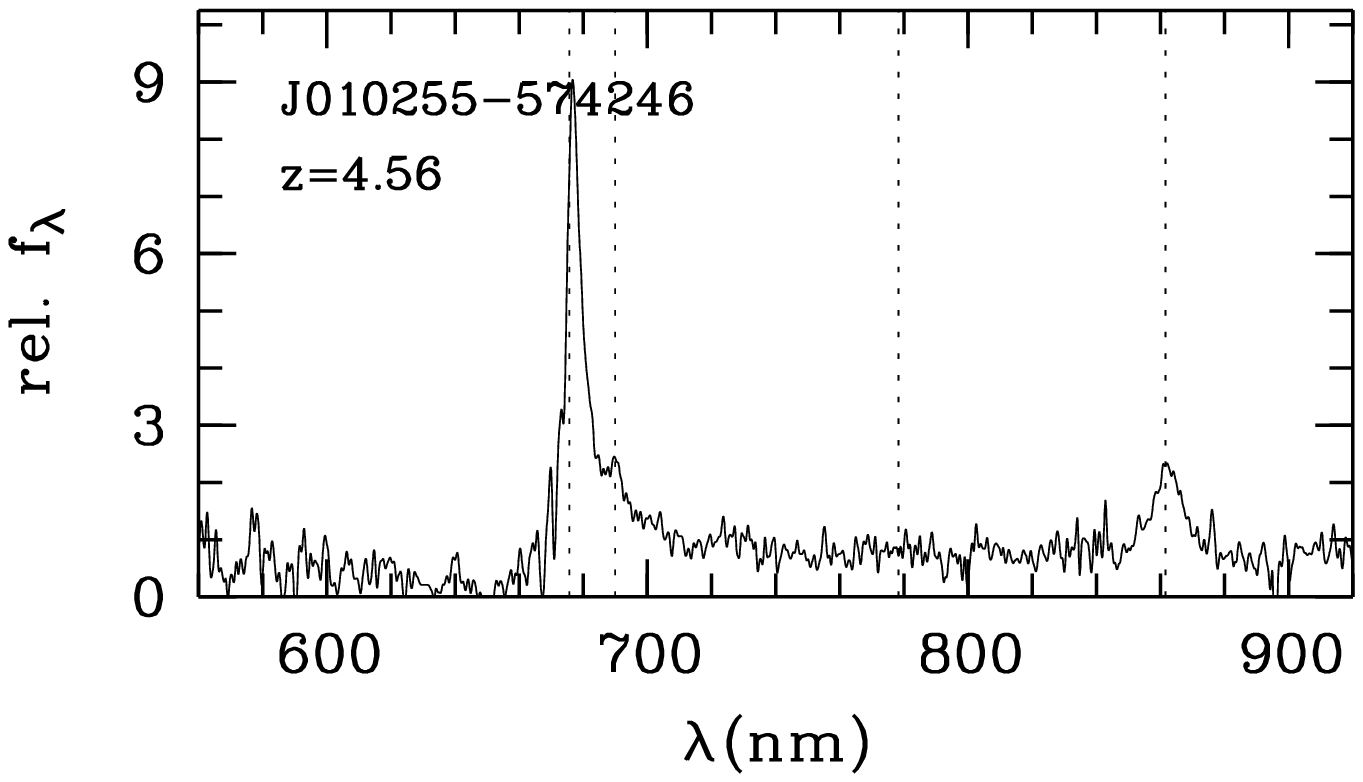}
\includegraphics[angle=270,width=0.32\textwidth,clip=true]{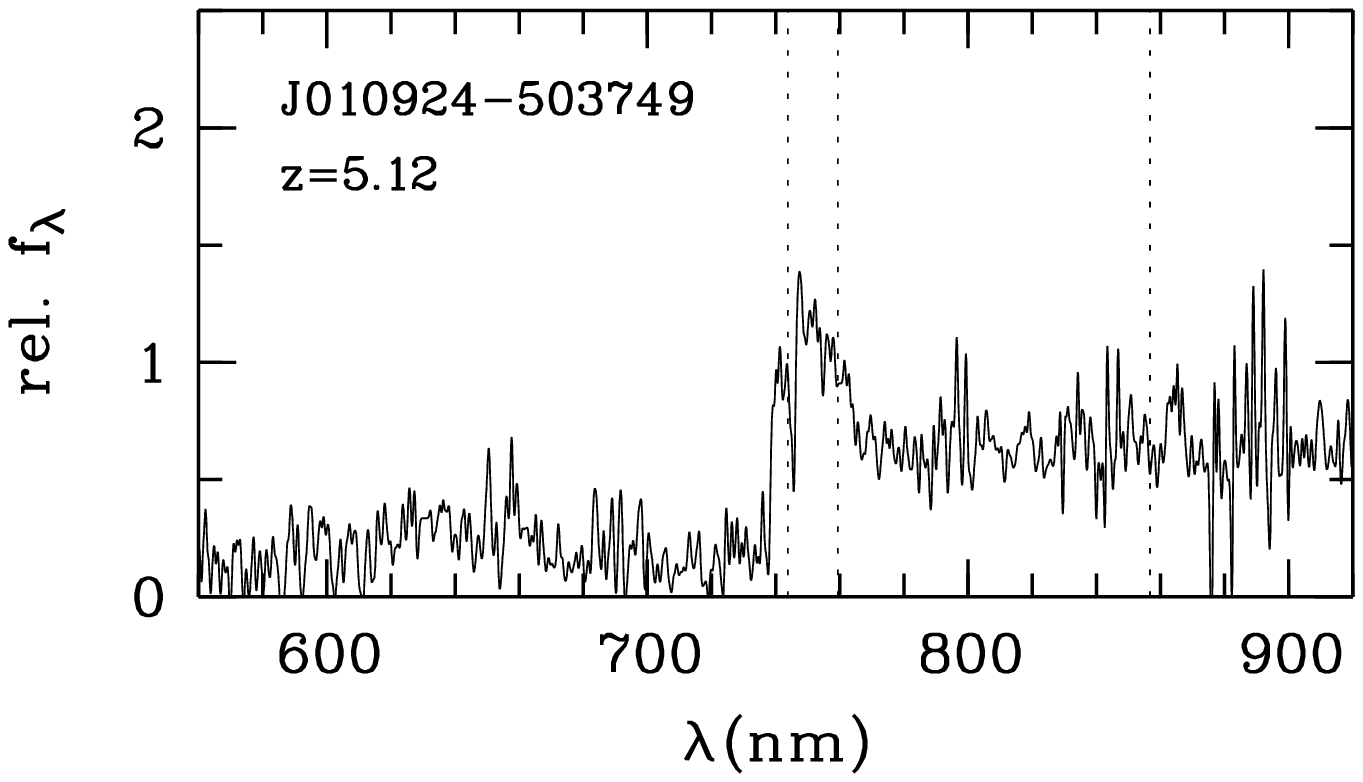}
\includegraphics[angle=270,width=0.32\textwidth,clip=true]{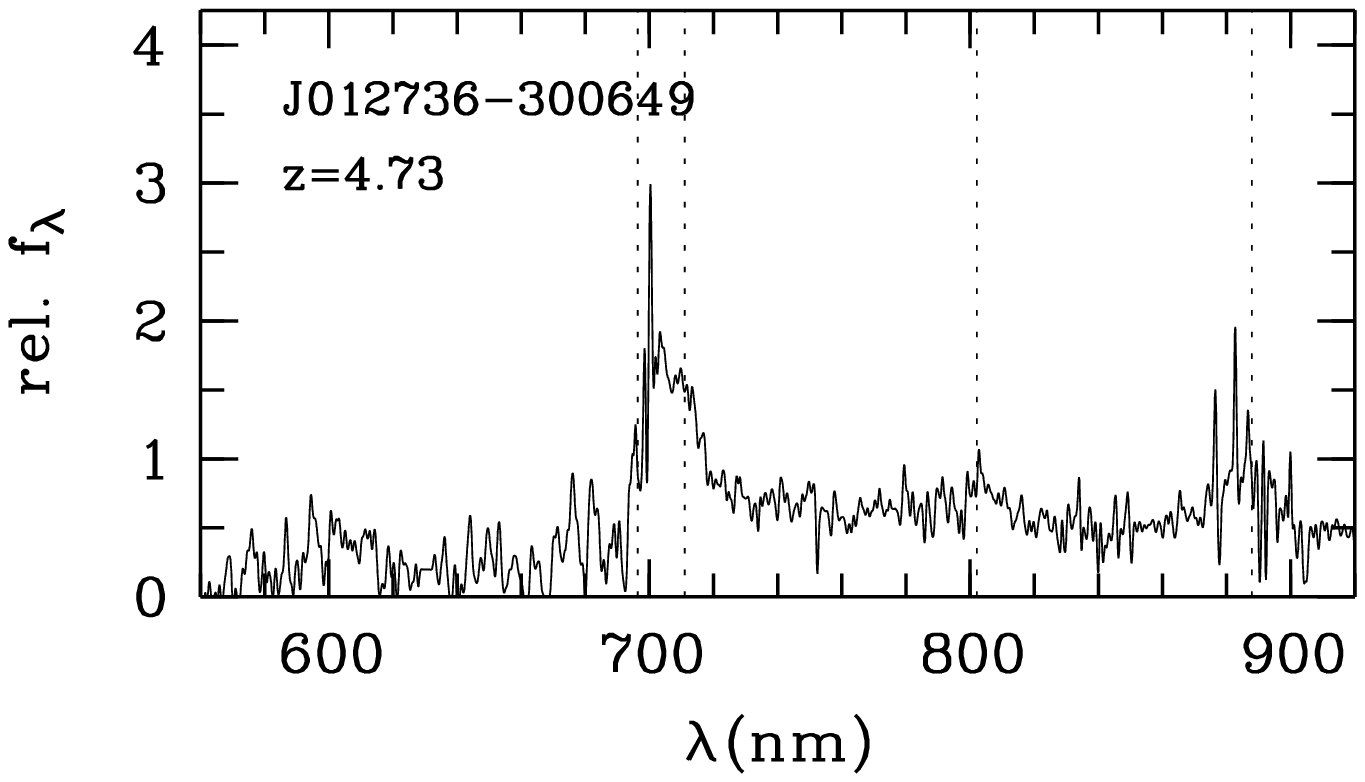}
\includegraphics[angle=270,width=0.32\textwidth,clip=true]{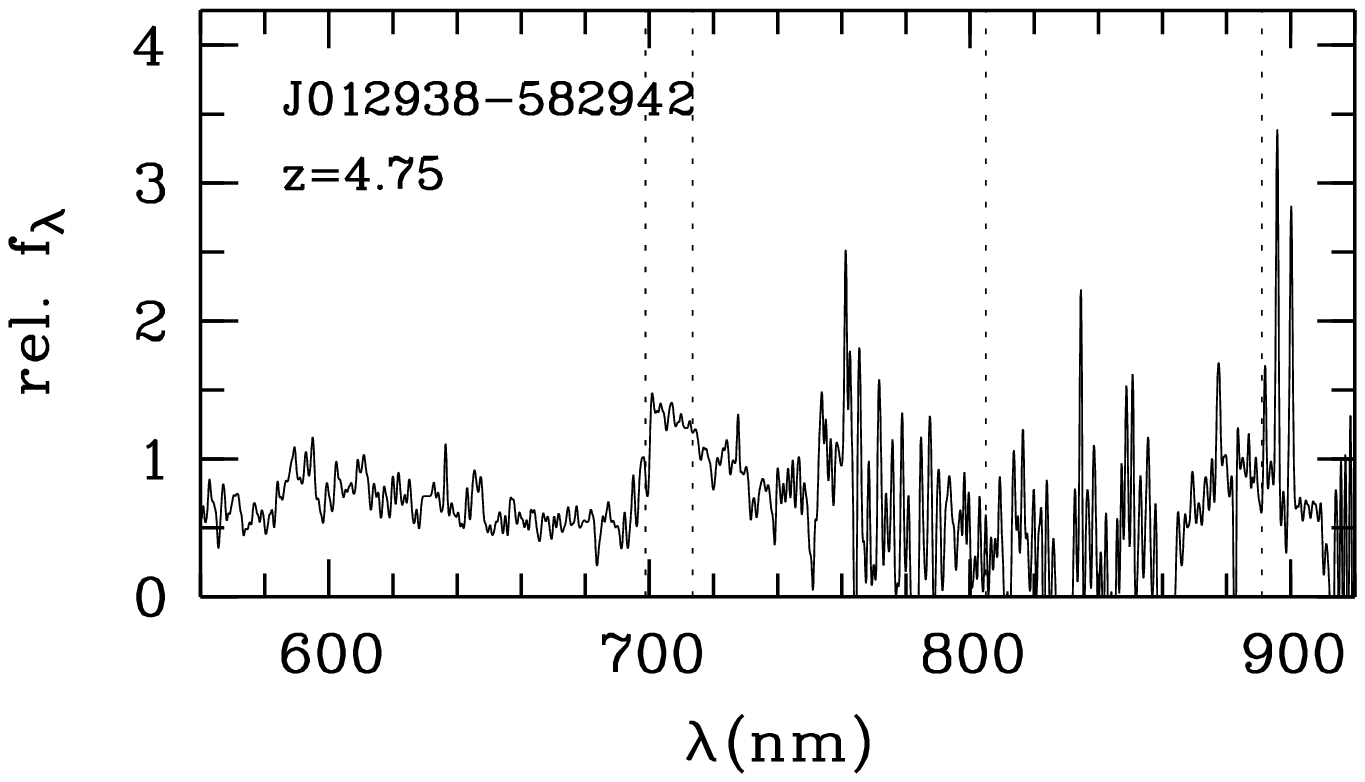}
\includegraphics[angle=270,width=0.32\textwidth,clip=true]{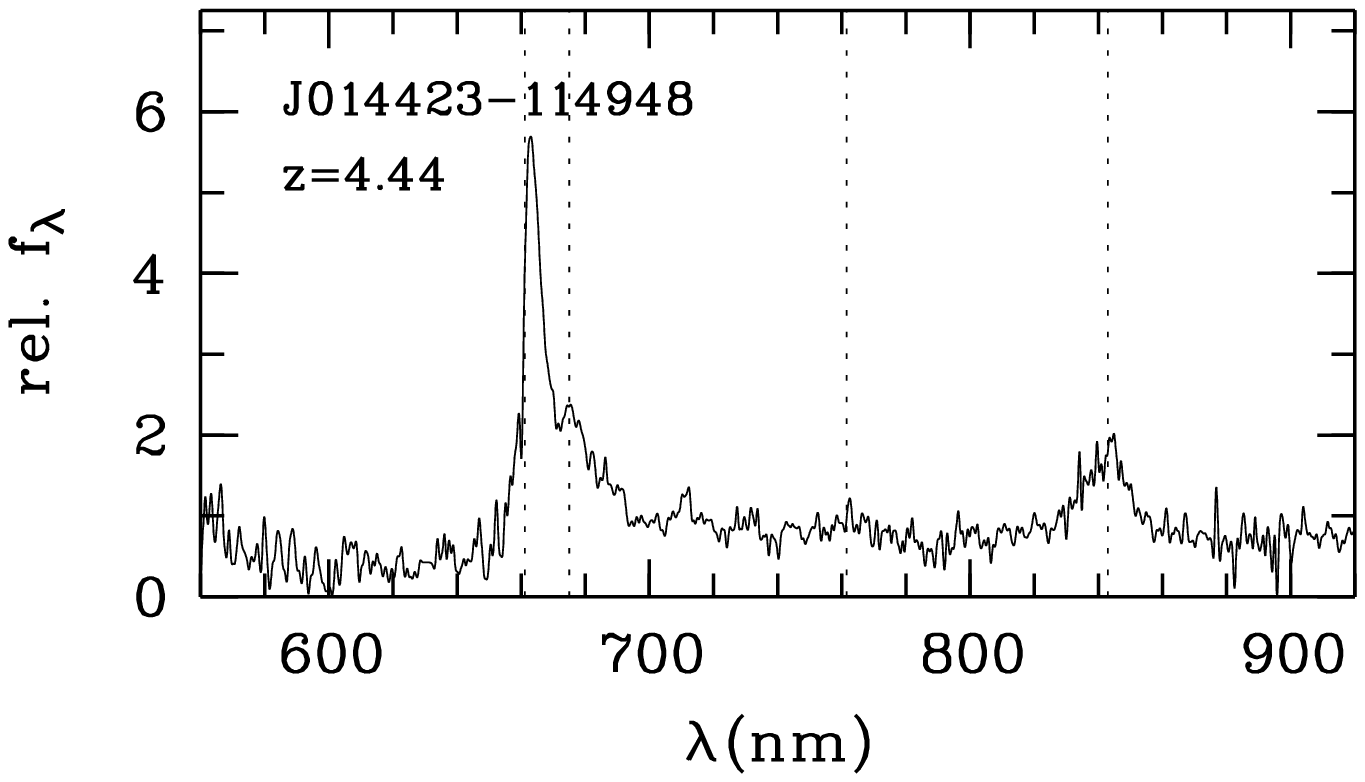}
\includegraphics[angle=270,width=0.32\textwidth,clip=true]{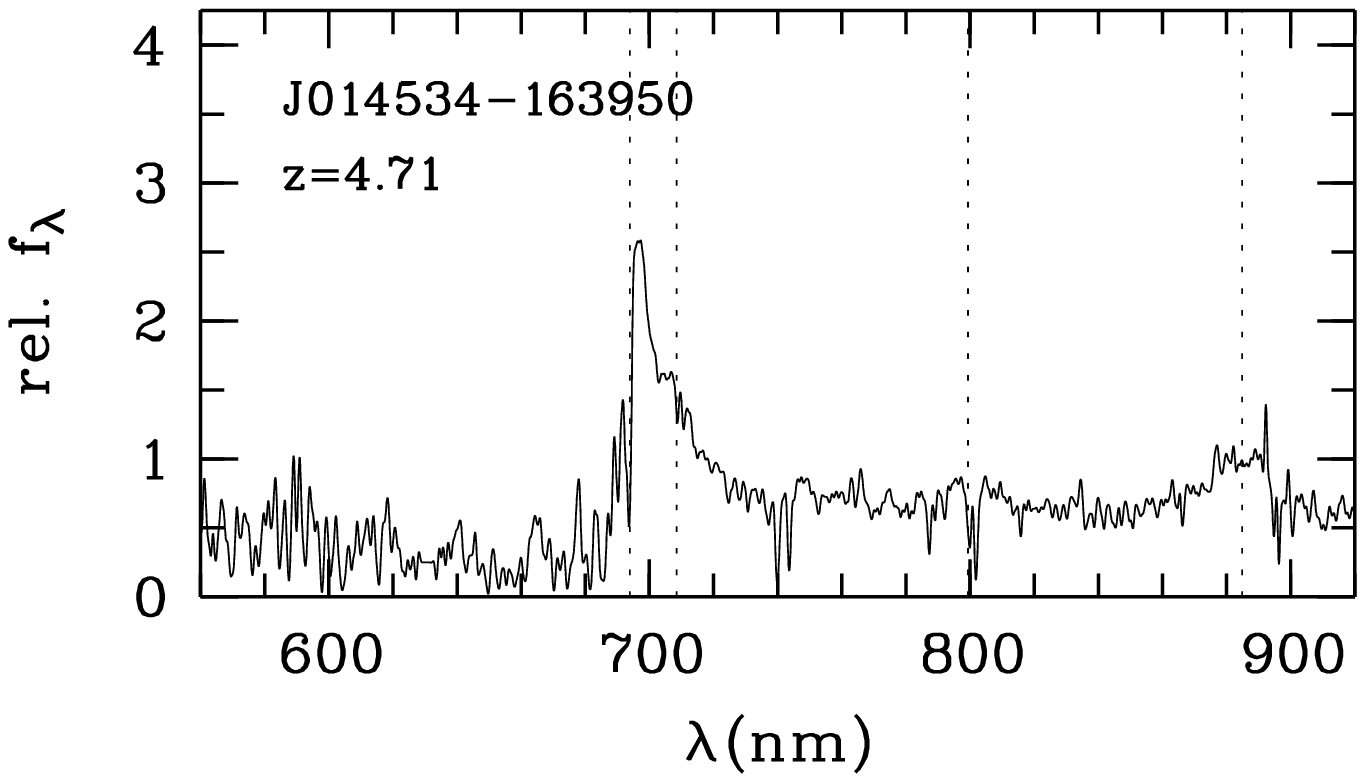}
\includegraphics[angle=270,width=0.32\textwidth,clip=true]{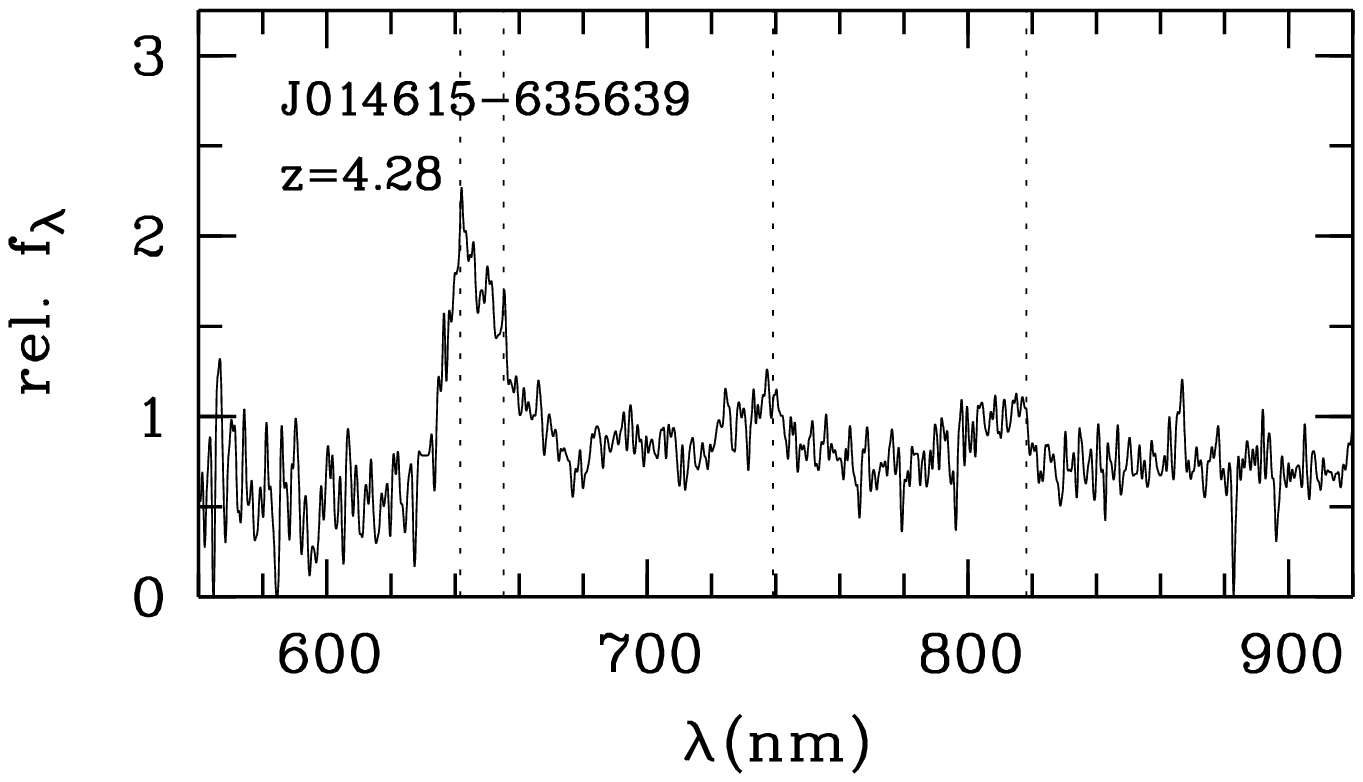}
\includegraphics[angle=270,width=0.32\textwidth,clip=true]{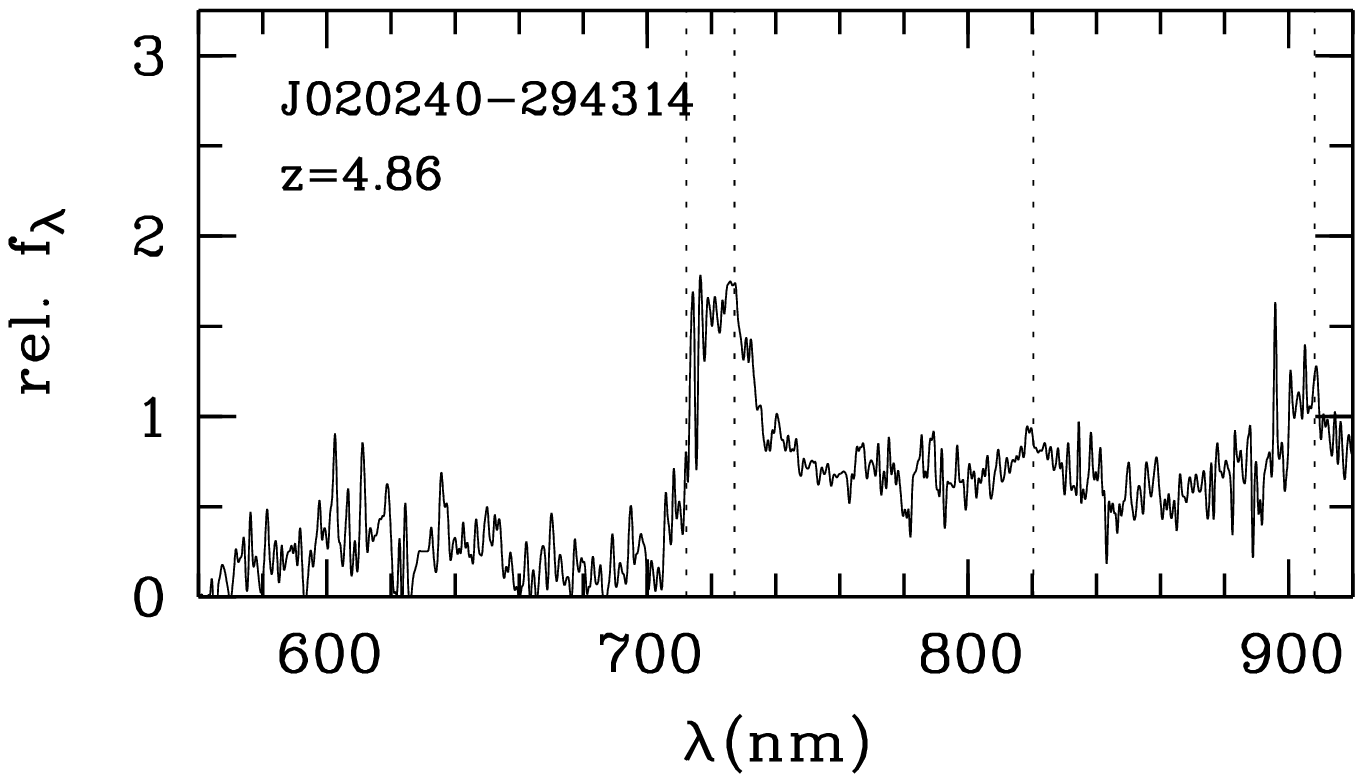}
\includegraphics[angle=270,width=0.32\textwidth,clip=true]{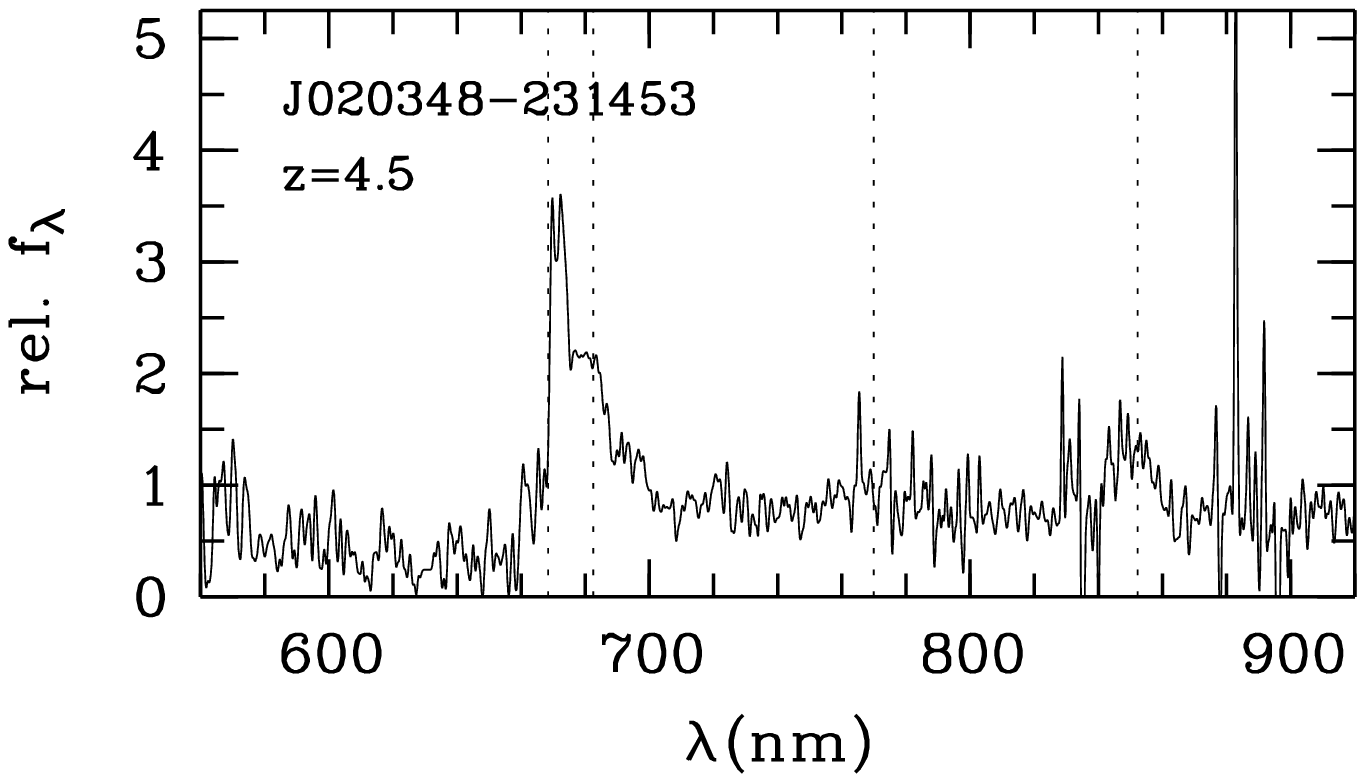}
\includegraphics[angle=270,width=0.32\textwidth,clip=true]{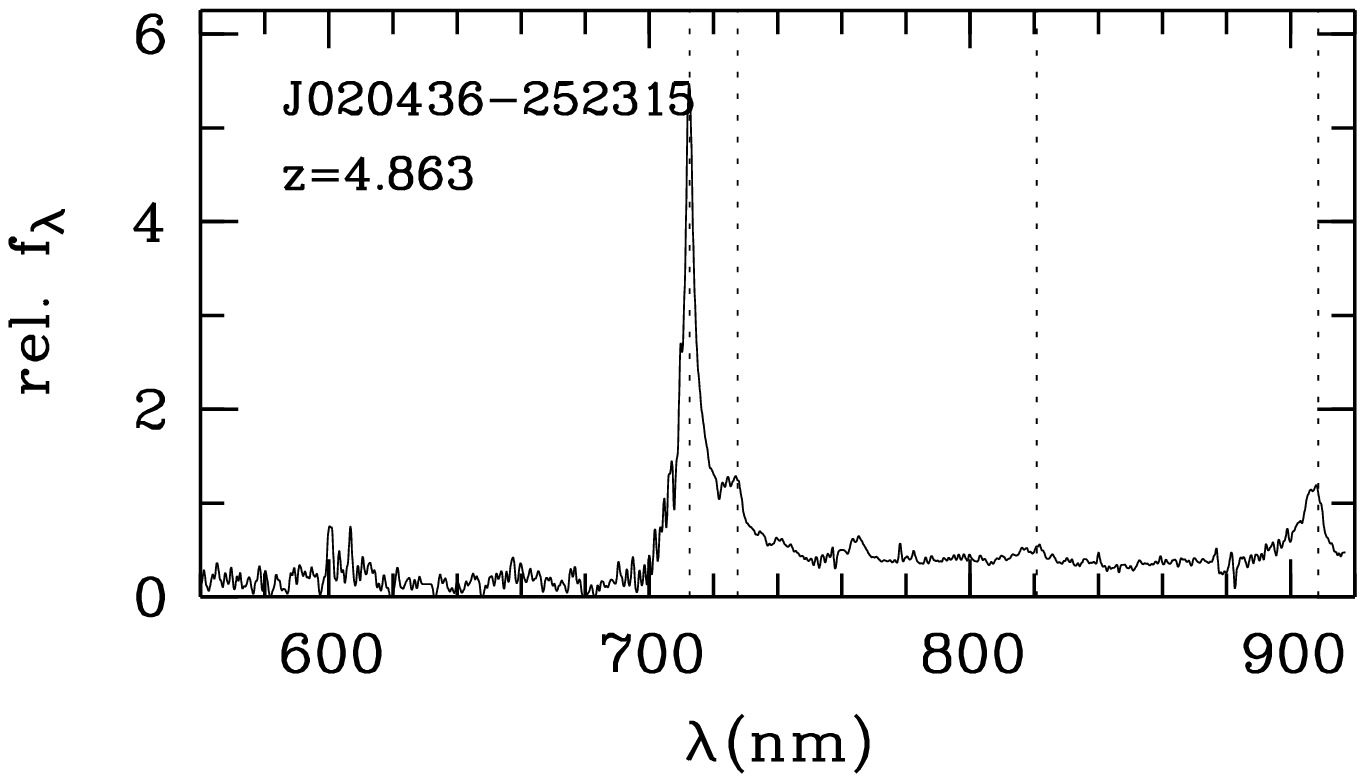}
\includegraphics[angle=270,width=0.32\textwidth,clip=true]{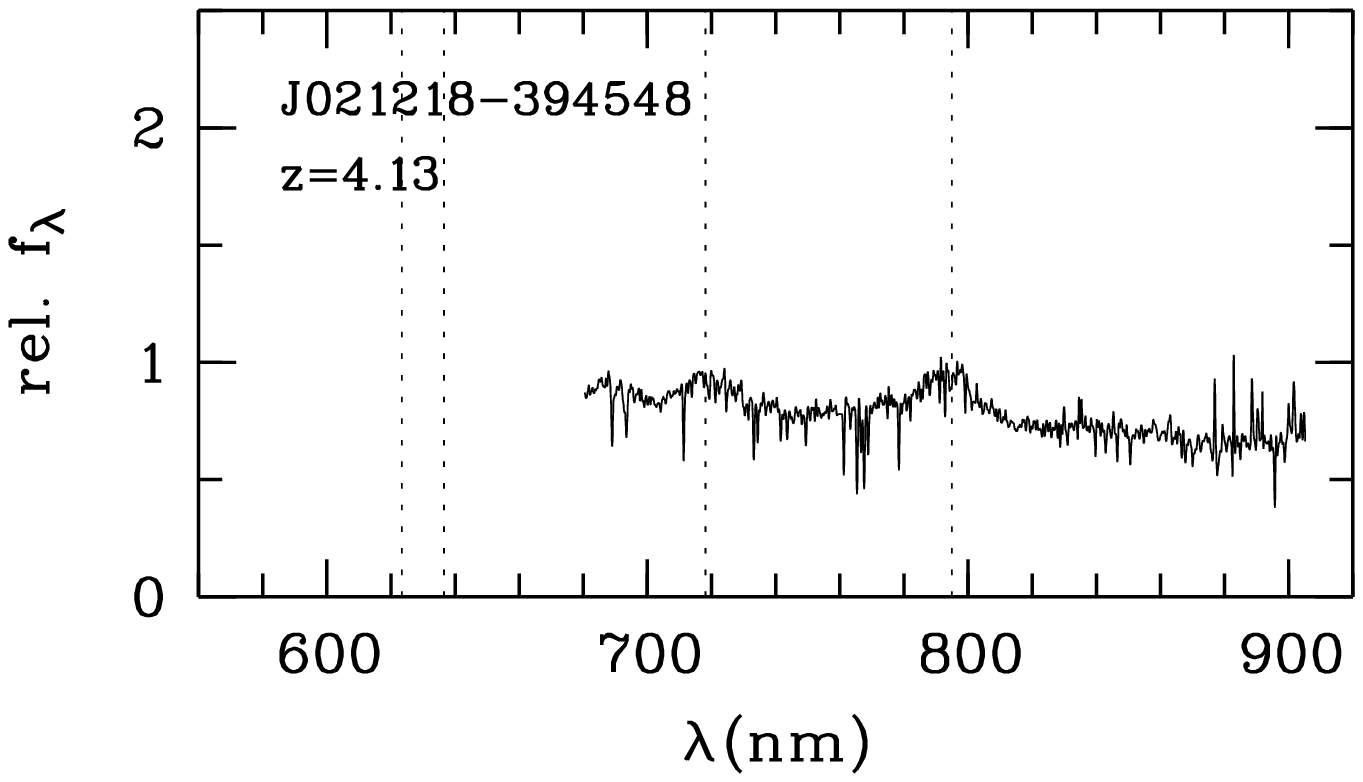}
\caption{Gallery of $3.8>z>5.5$ QSO spectra obtained in this work, ordered by RA, page 1. Vertical dashed lines indicate the wavelengths of Ly\,$\alpha$, \ion{N}{v}, \ion{Si}{iv}, and \ion{C}{iv} at our adopted redshift.
\label{gallery1}}
\end{center}
\end{figure*}

\begin{figure*}
\begin{center}
\includegraphics[angle=270,width=0.32\textwidth,clip=true]{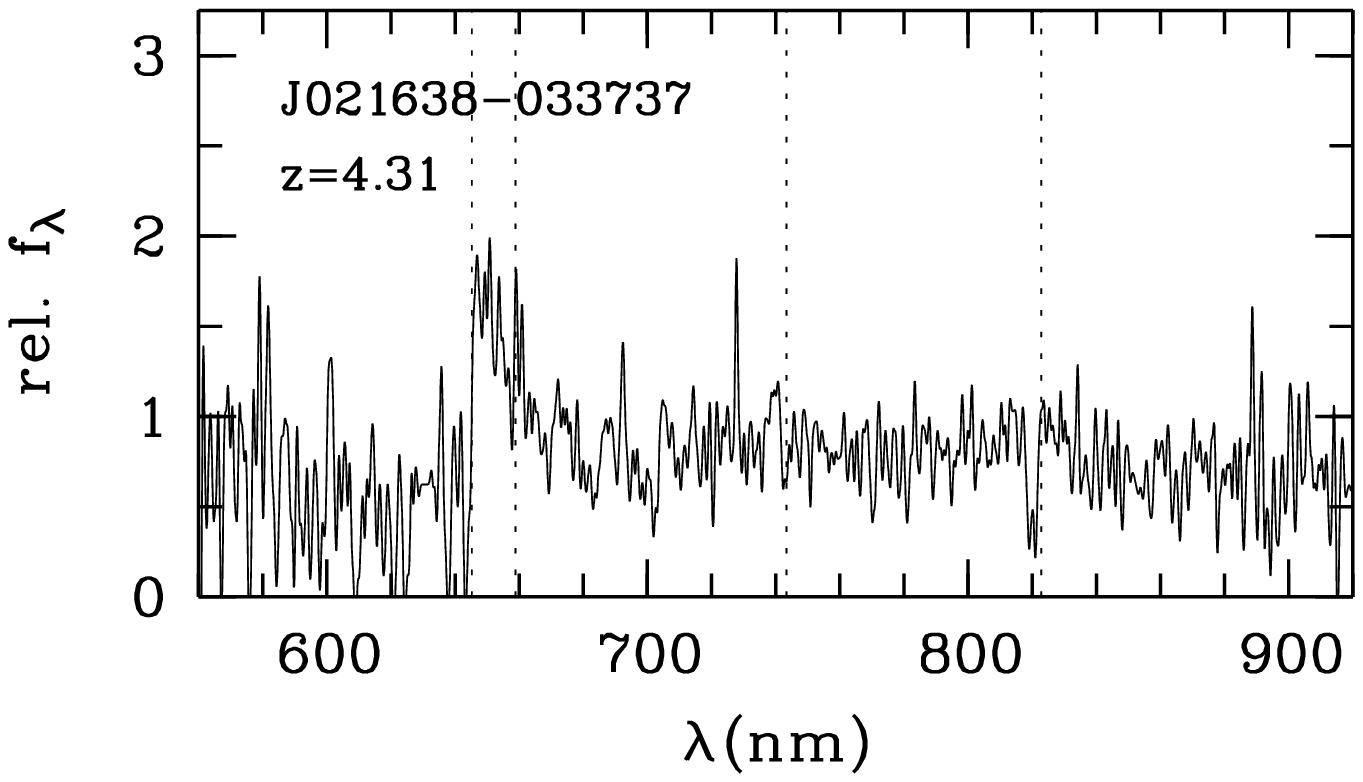}
\includegraphics[angle=270,width=0.32\textwidth,clip=true]{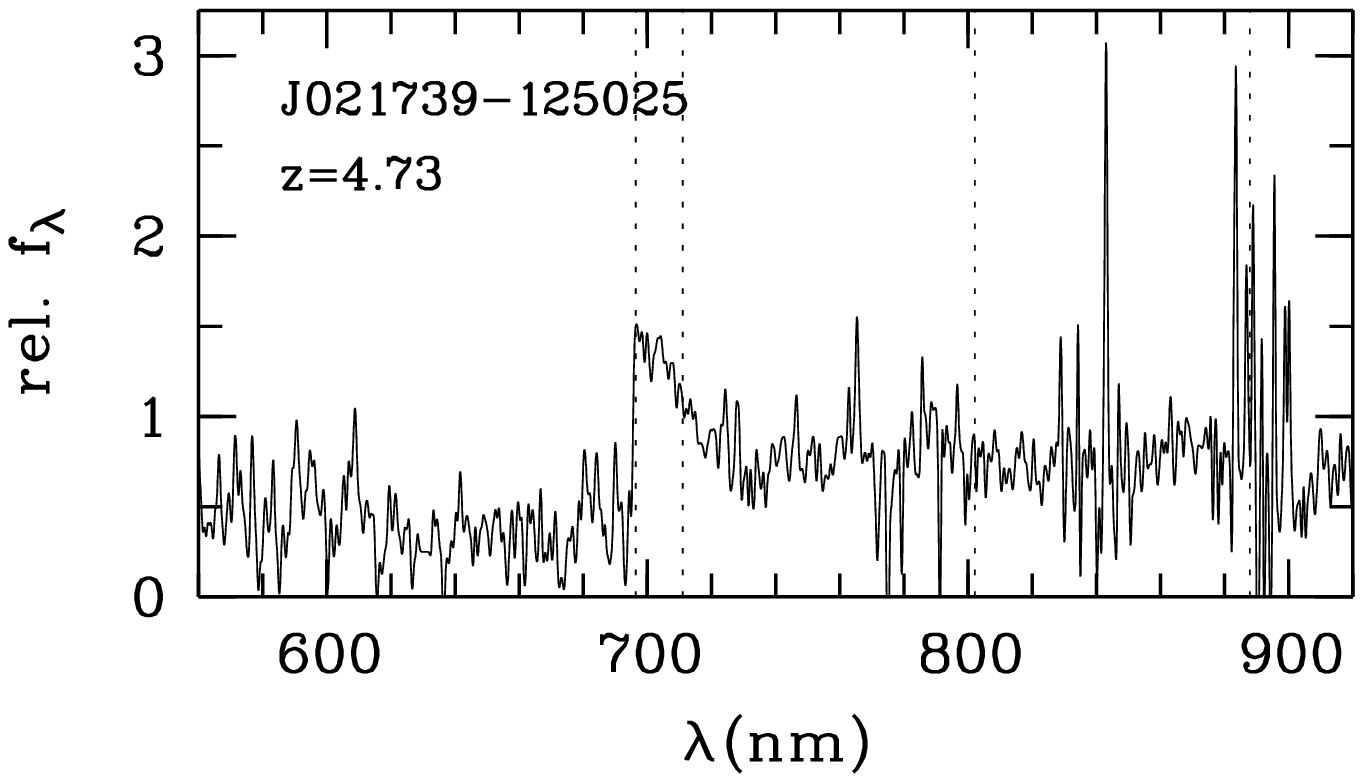}
\includegraphics[angle=270,width=0.32\textwidth,clip=true]{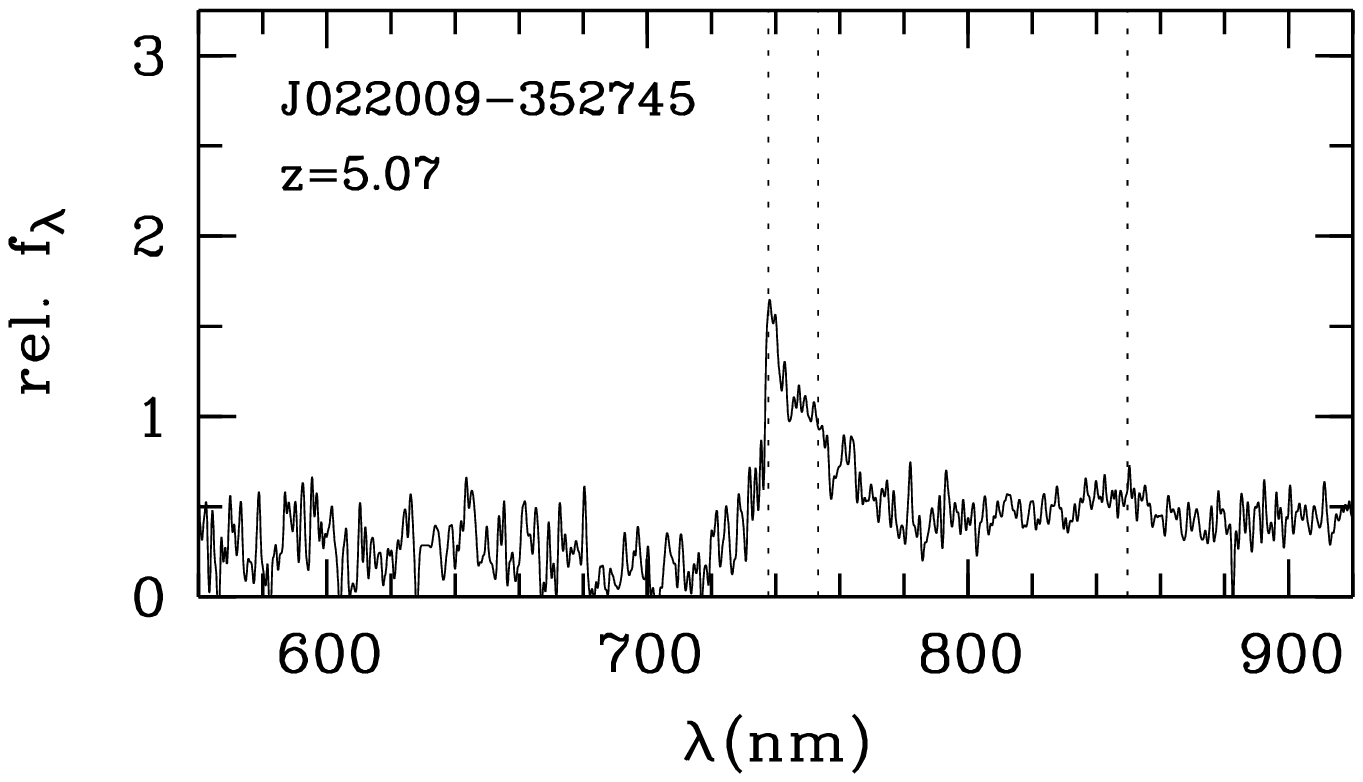}
\includegraphics[angle=270,width=0.32\textwidth,clip=true]{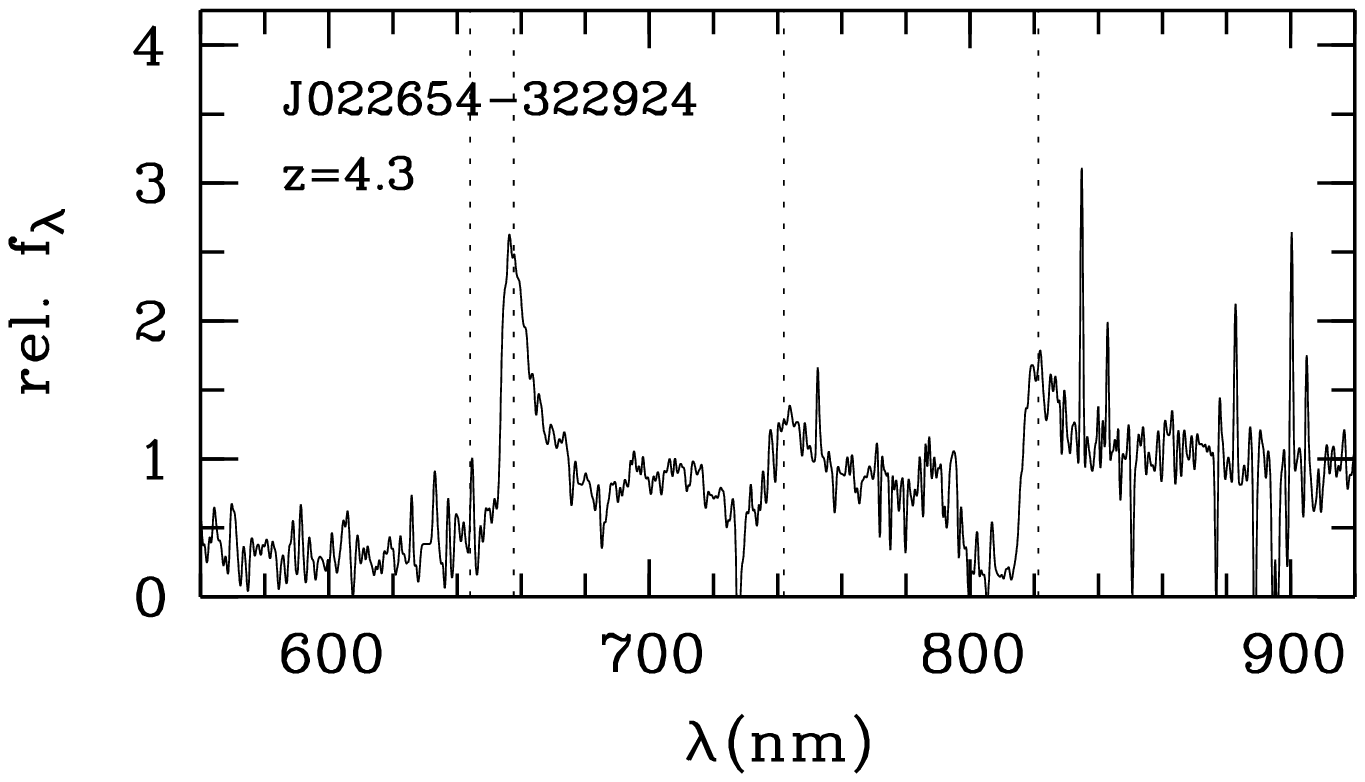}
\includegraphics[angle=270,width=0.32\textwidth,clip=true]{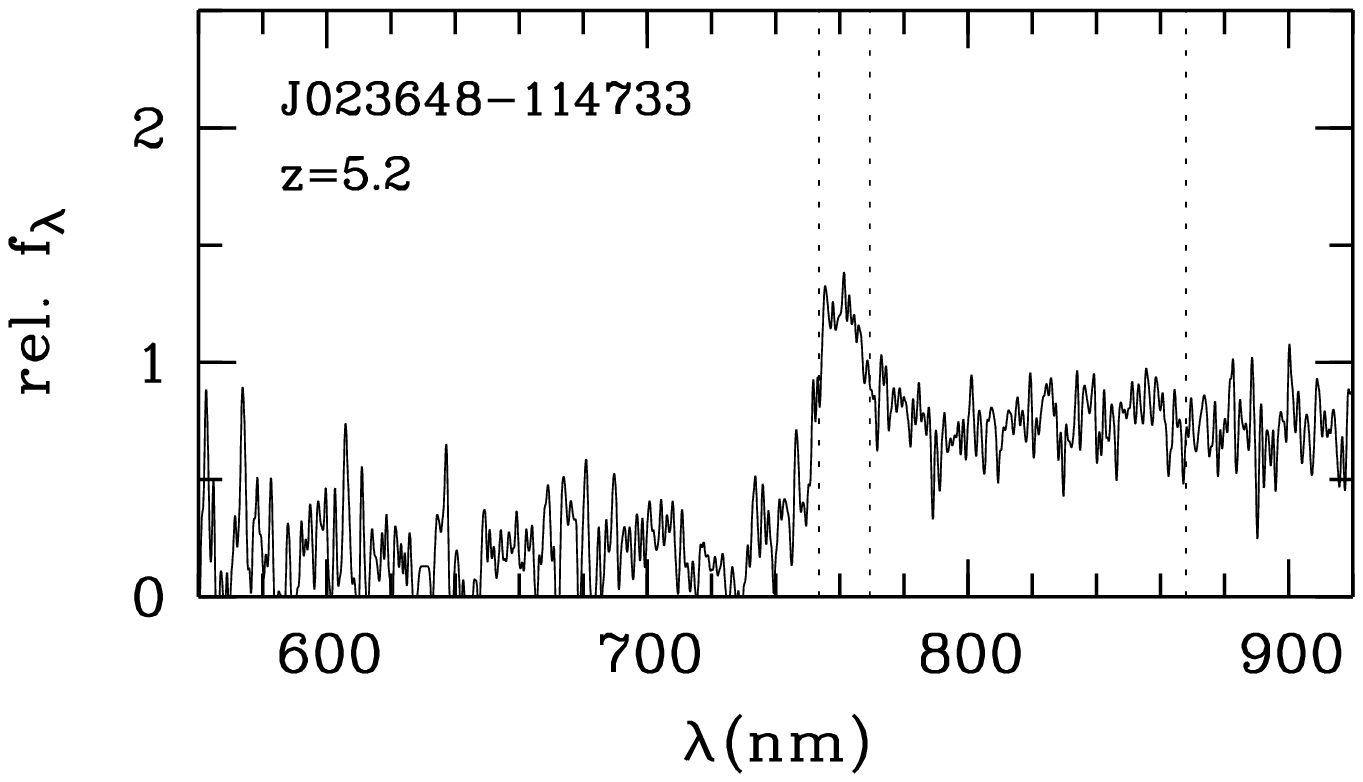}
\includegraphics[angle=270,width=0.32\textwidth,clip=true]{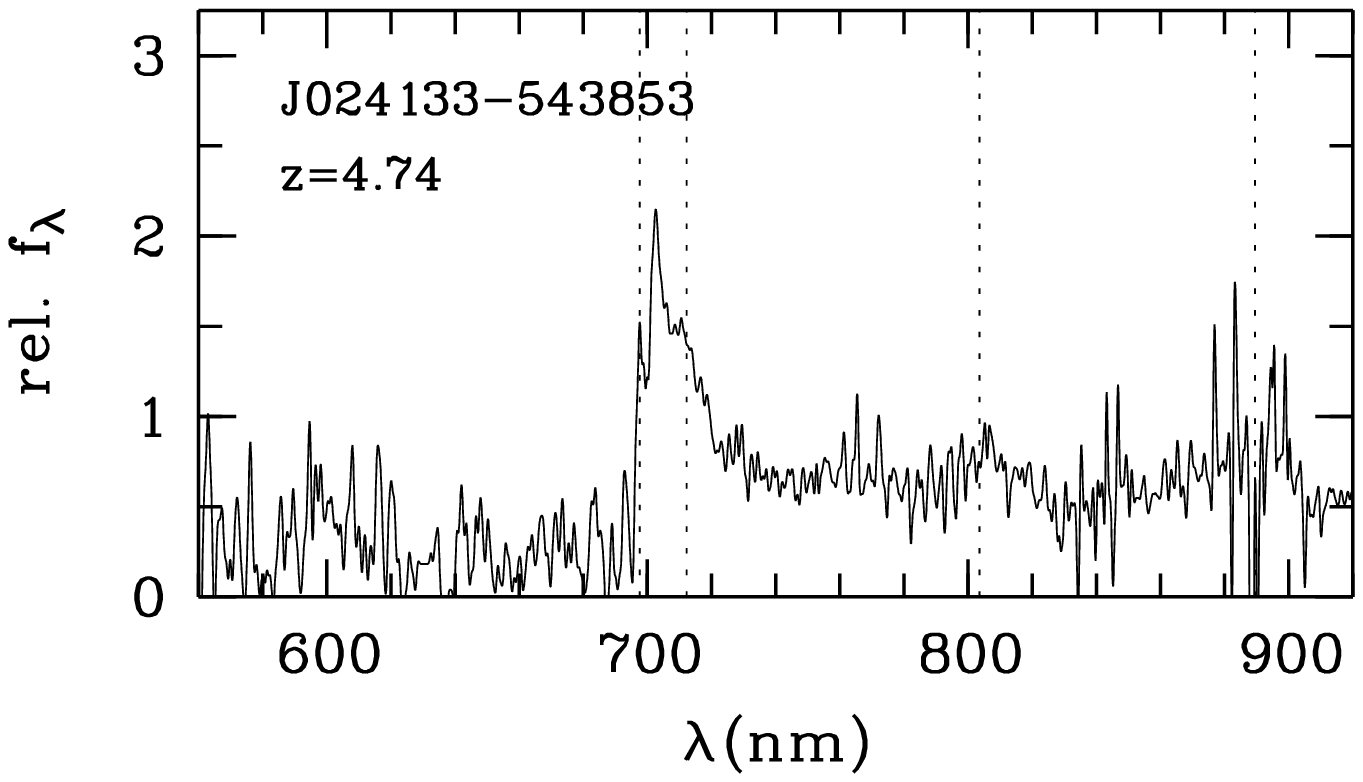}
\includegraphics[angle=270,width=0.32\textwidth,clip=true]{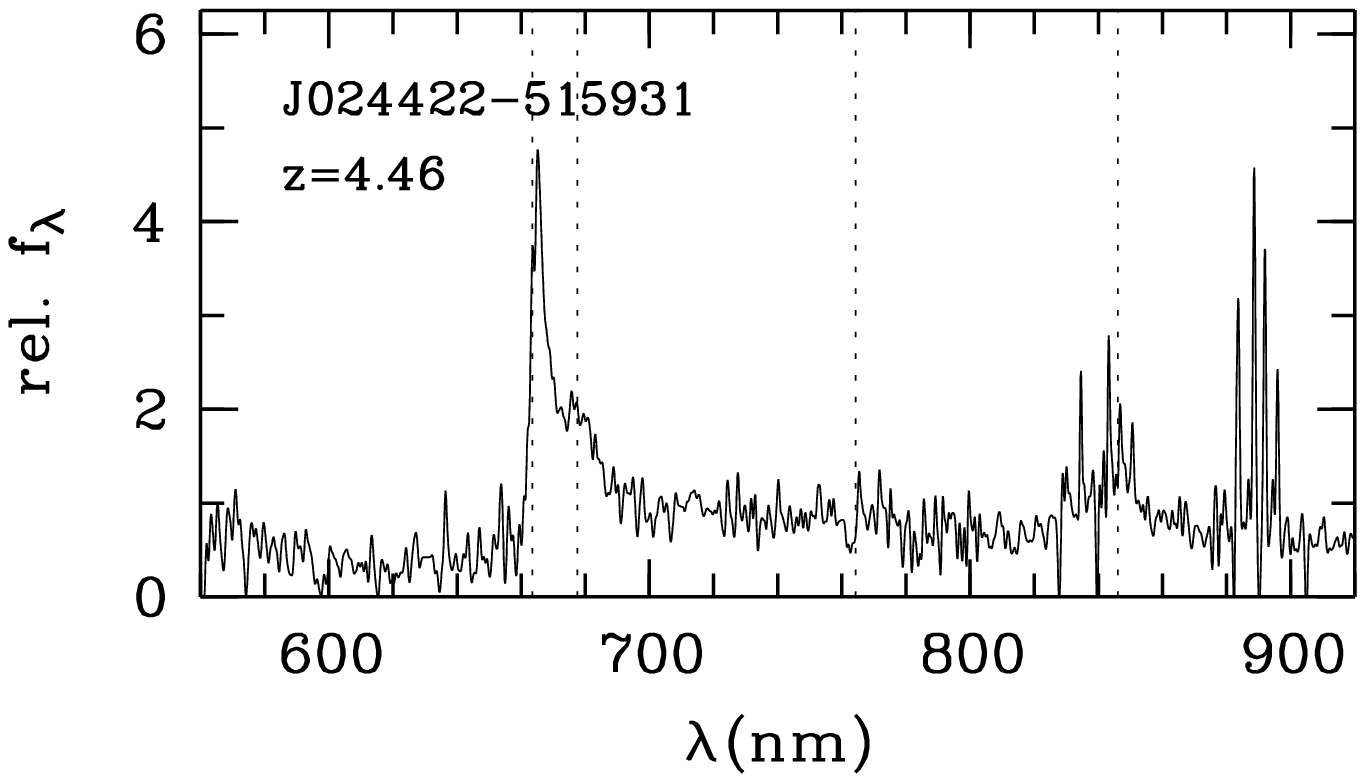}
\includegraphics[angle=270,width=0.32\textwidth,clip=true]{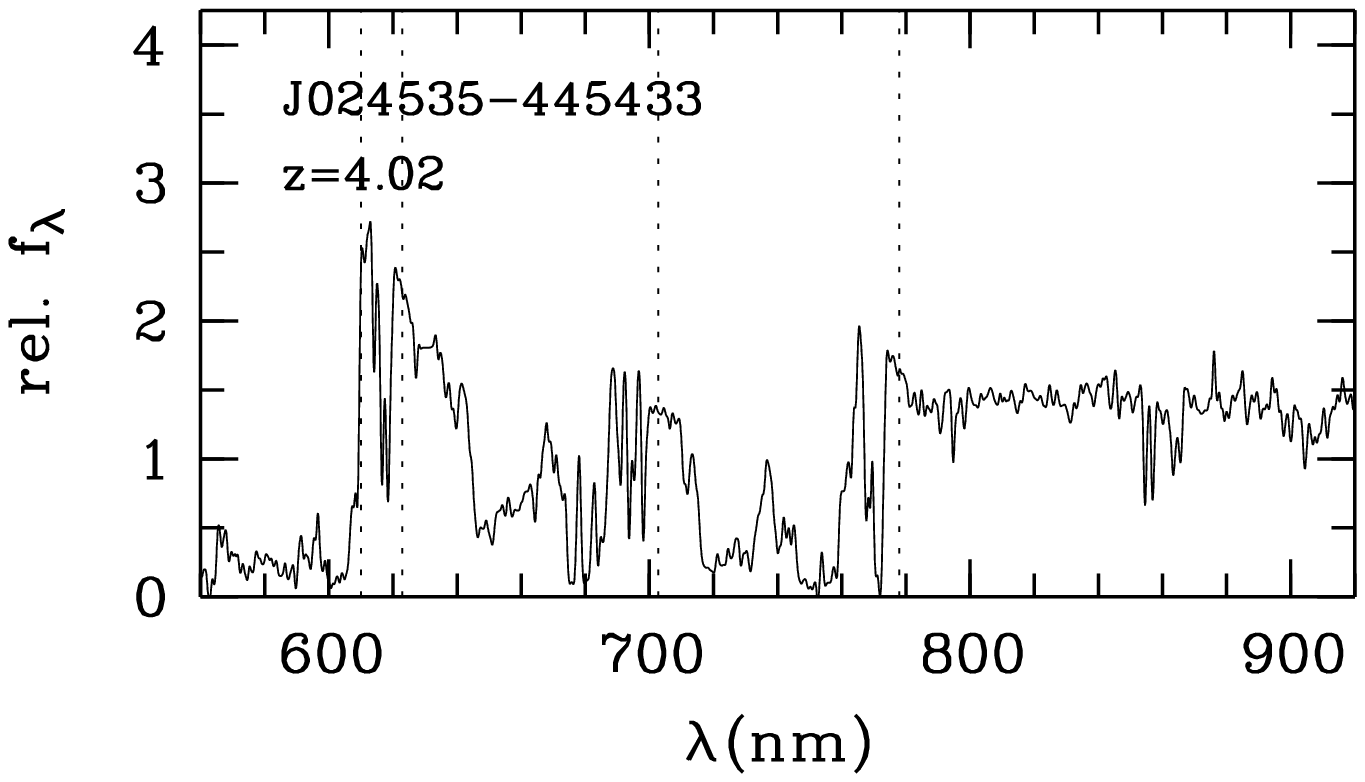}
\includegraphics[angle=270,width=0.32\textwidth,clip=true]{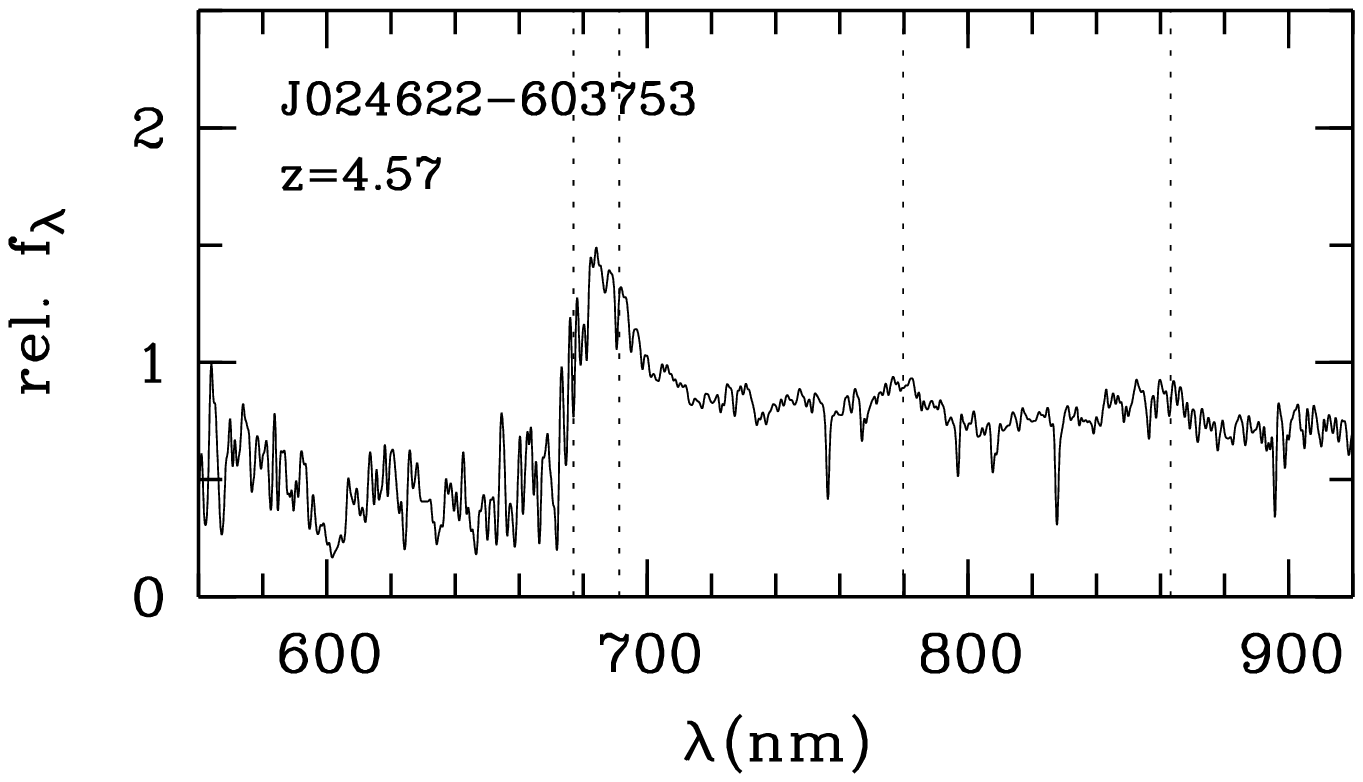}
\includegraphics[angle=270,width=0.32\textwidth,clip=true]{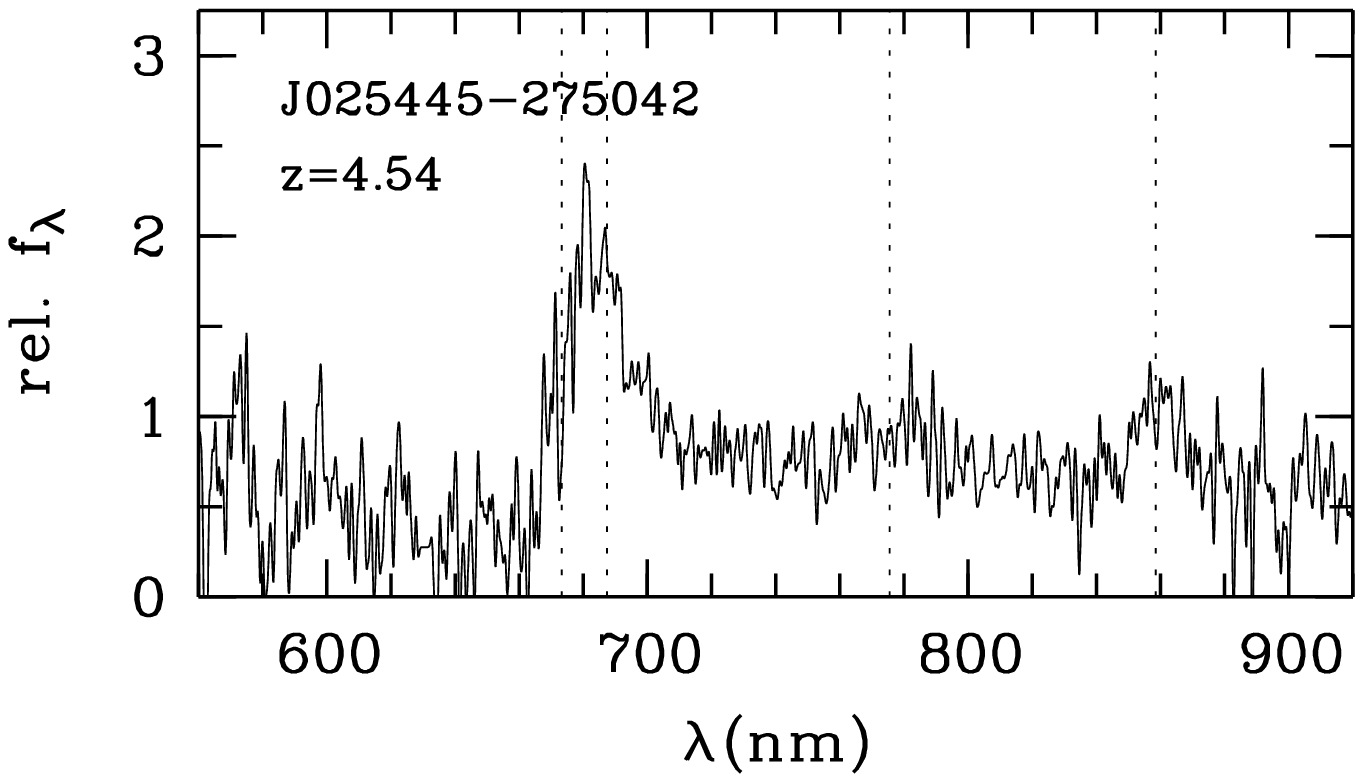}
\includegraphics[angle=270,width=0.32\textwidth,clip=true]{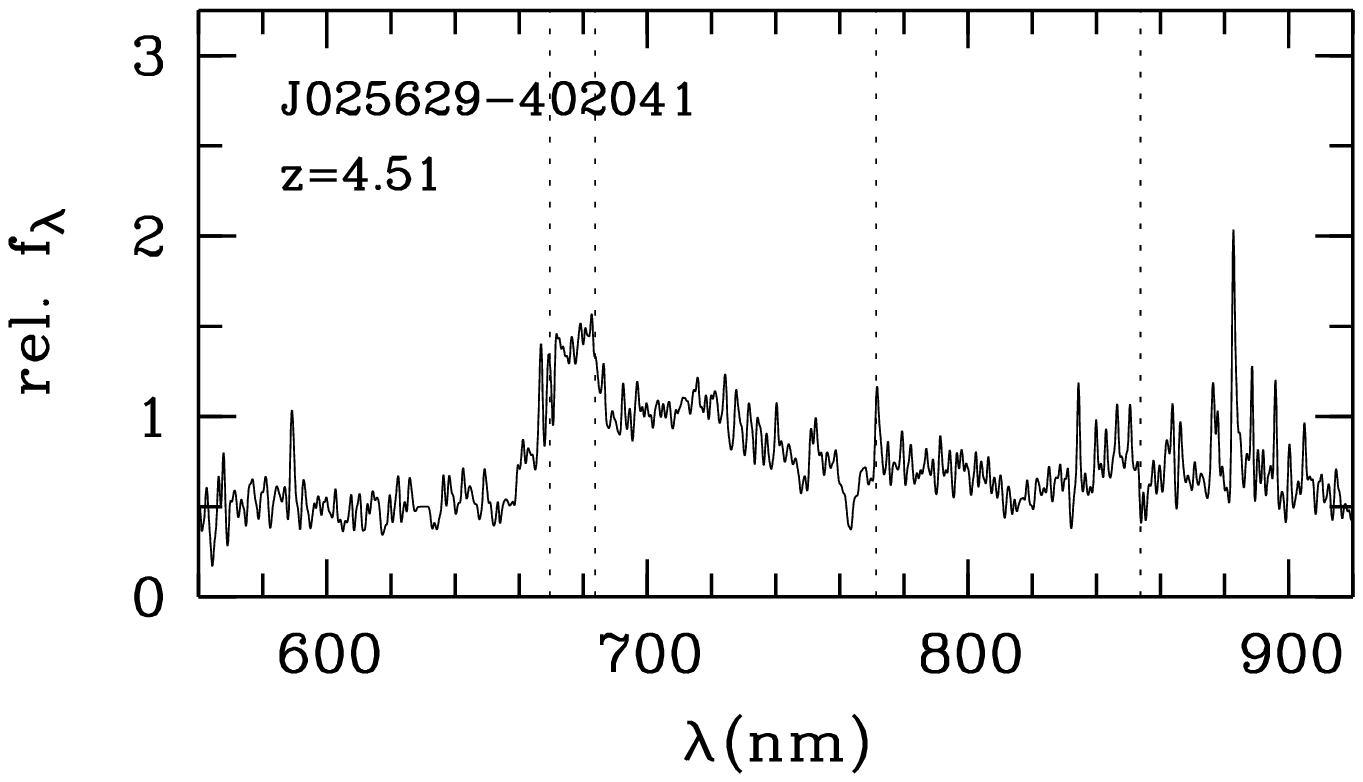}
\includegraphics[angle=270,width=0.32\textwidth,clip=true]{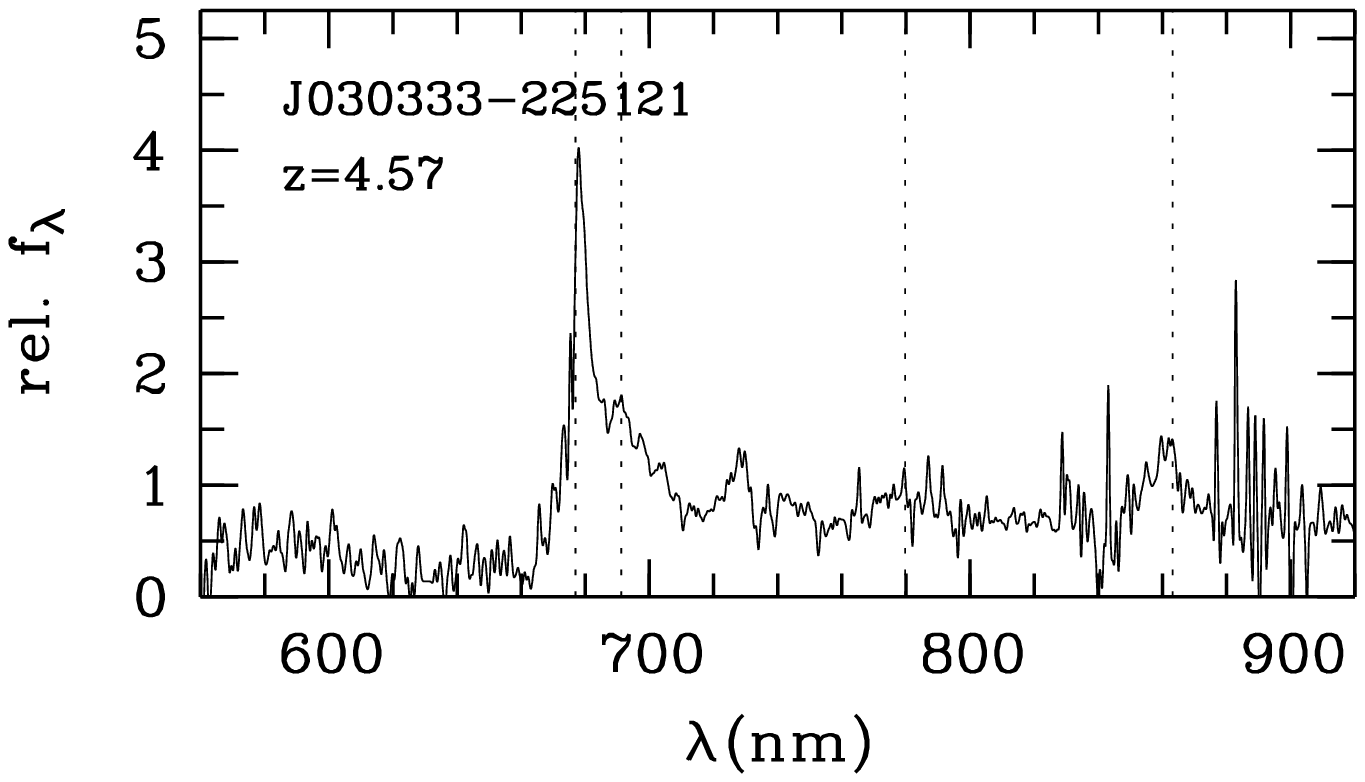}
\includegraphics[angle=270,width=0.32\textwidth,clip=true]{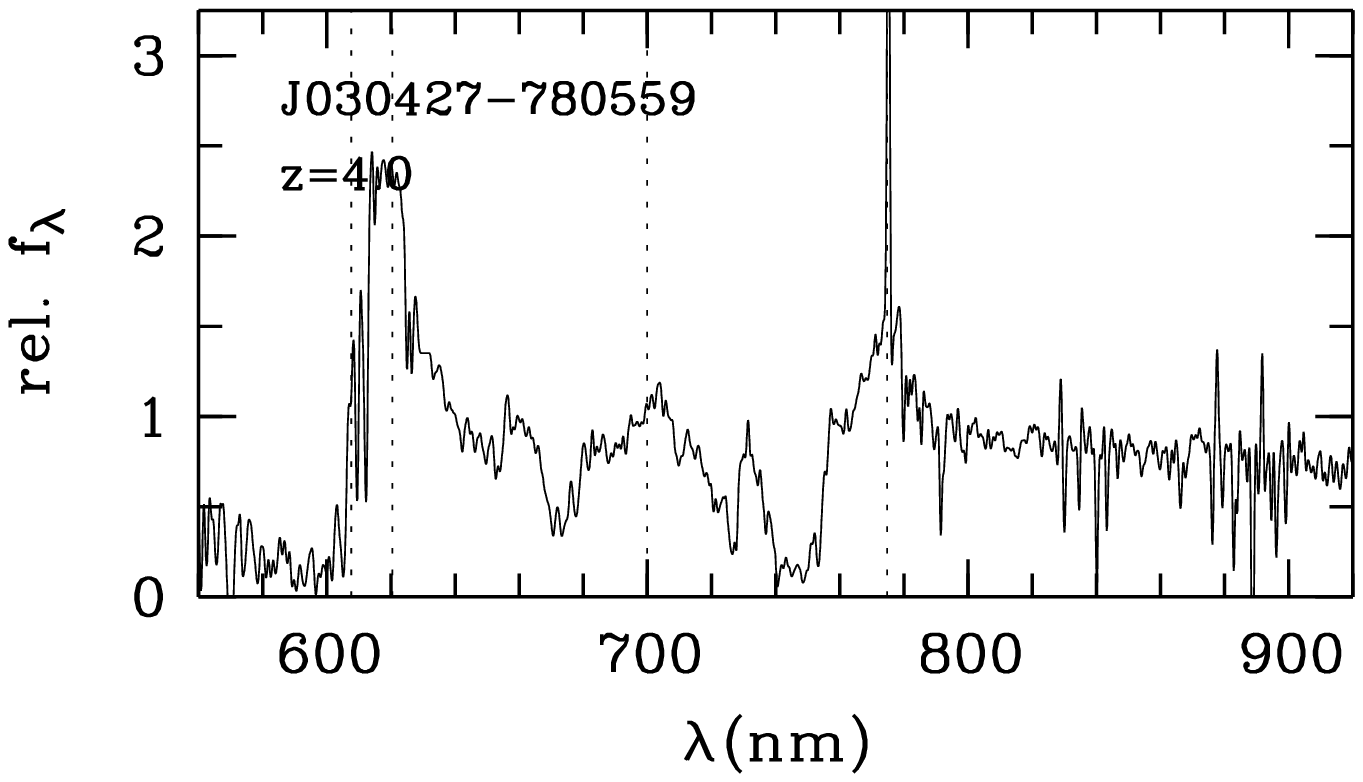}
\includegraphics[angle=270,width=0.32\textwidth,clip=true]{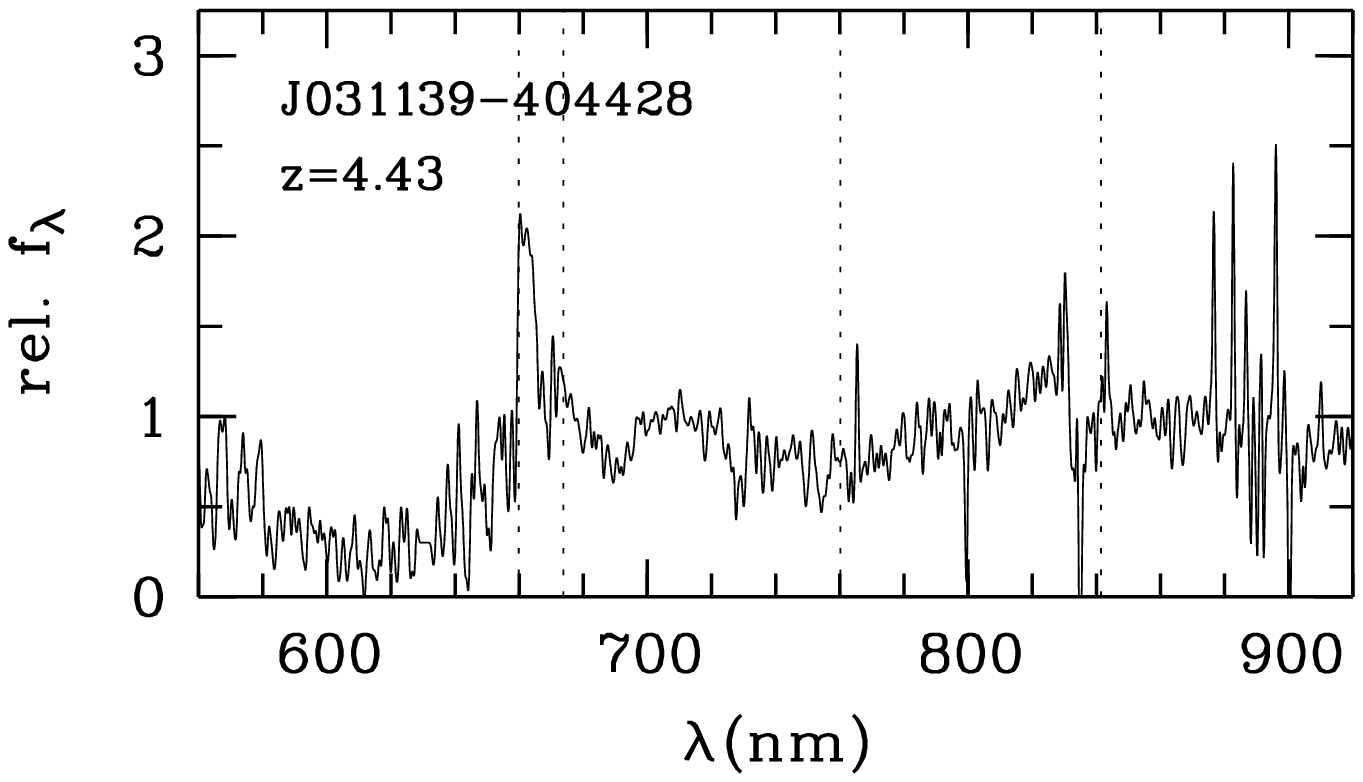}
\includegraphics[angle=270,width=0.32\textwidth,clip=true]{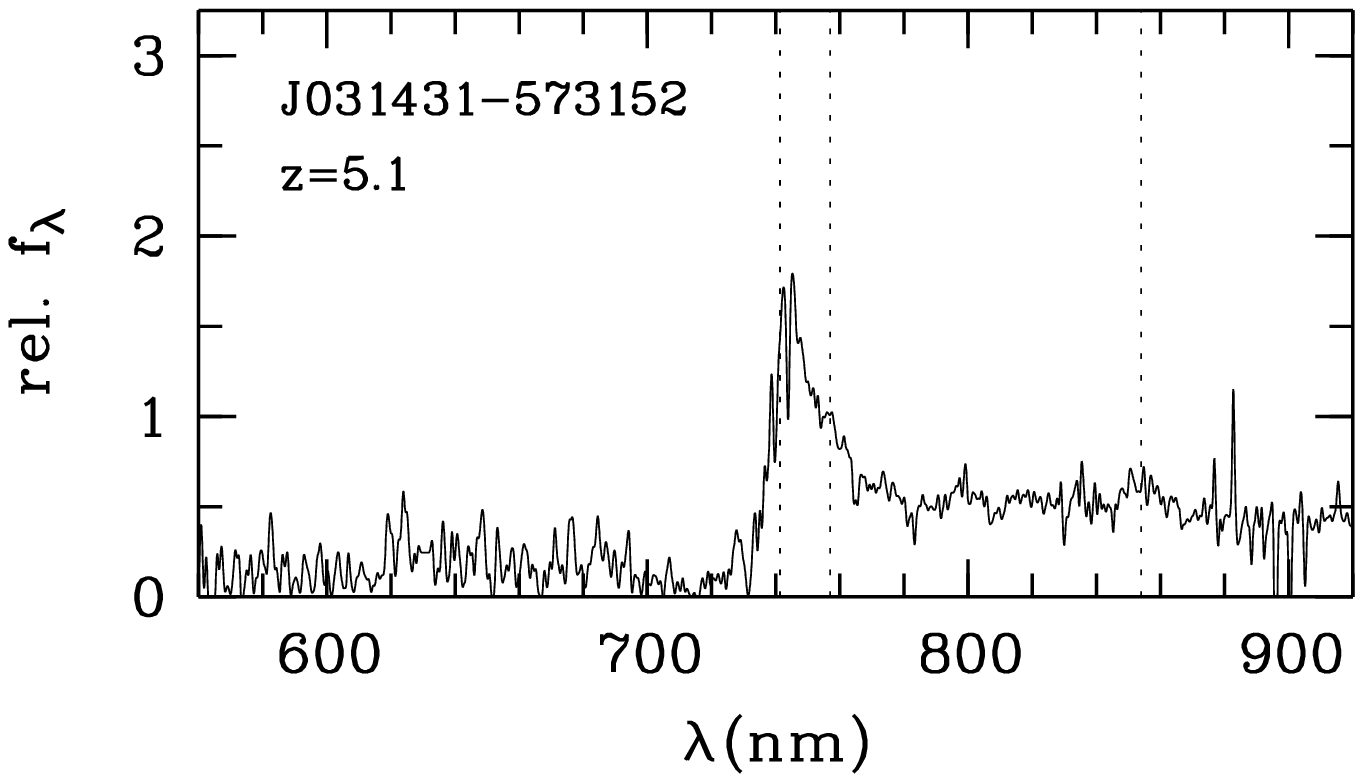}
\includegraphics[angle=270,width=0.32\textwidth,clip=true]{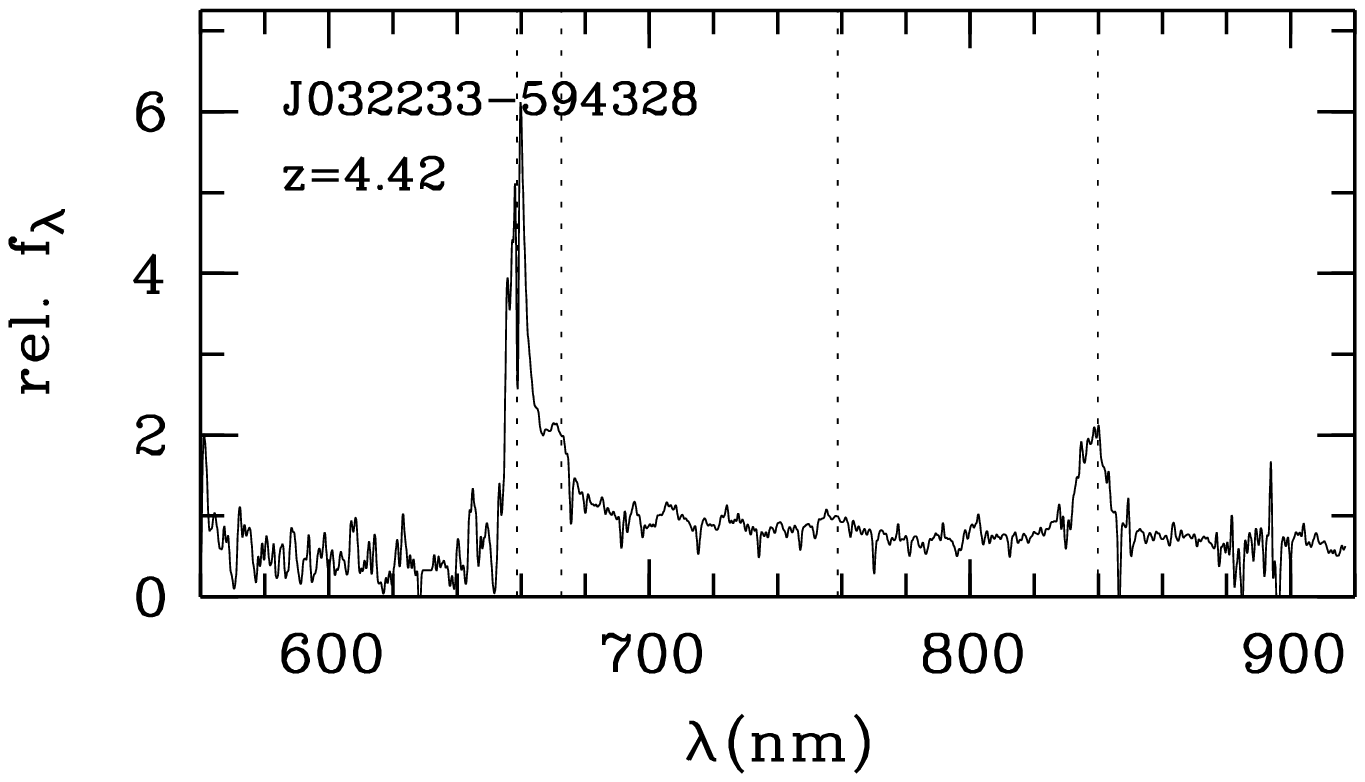}
\includegraphics[angle=270,width=0.32\textwidth,clip=true]{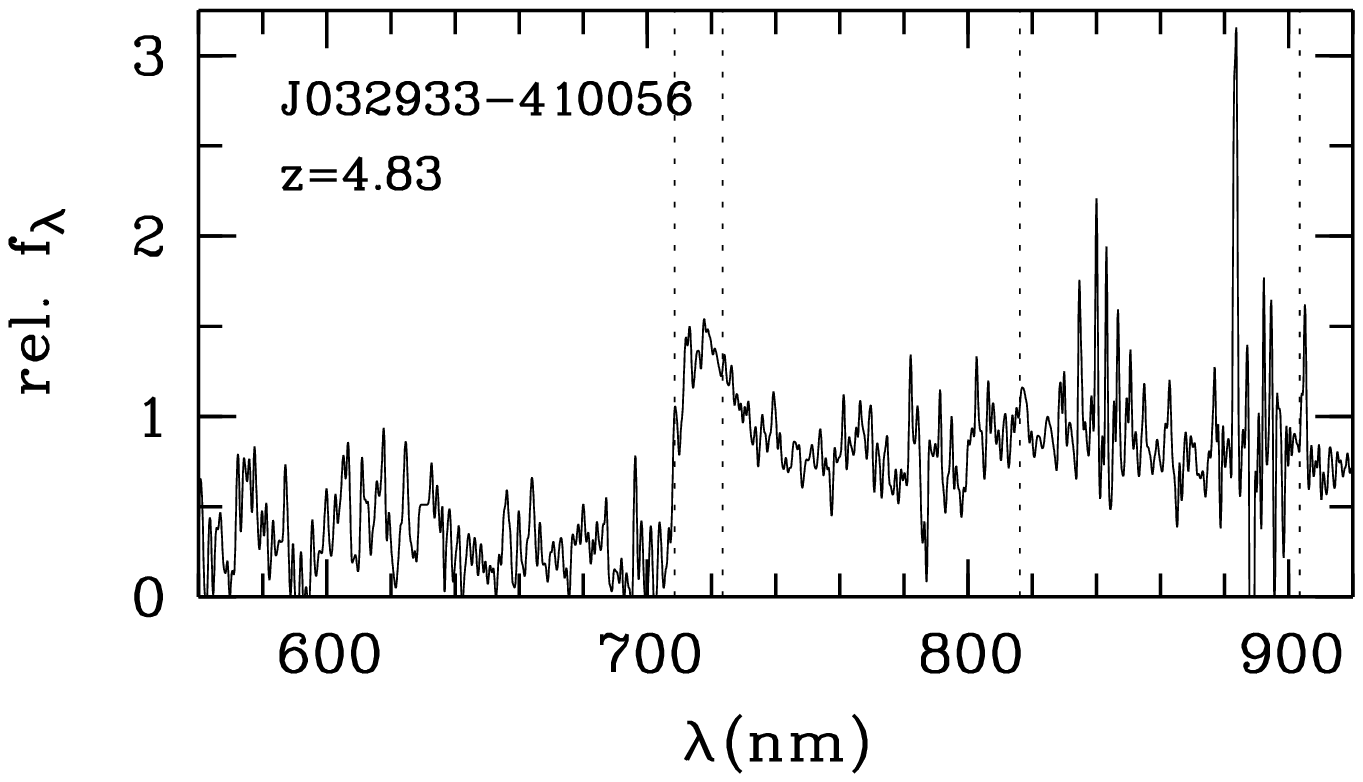}
\includegraphics[angle=270,width=0.32\textwidth,clip=true]{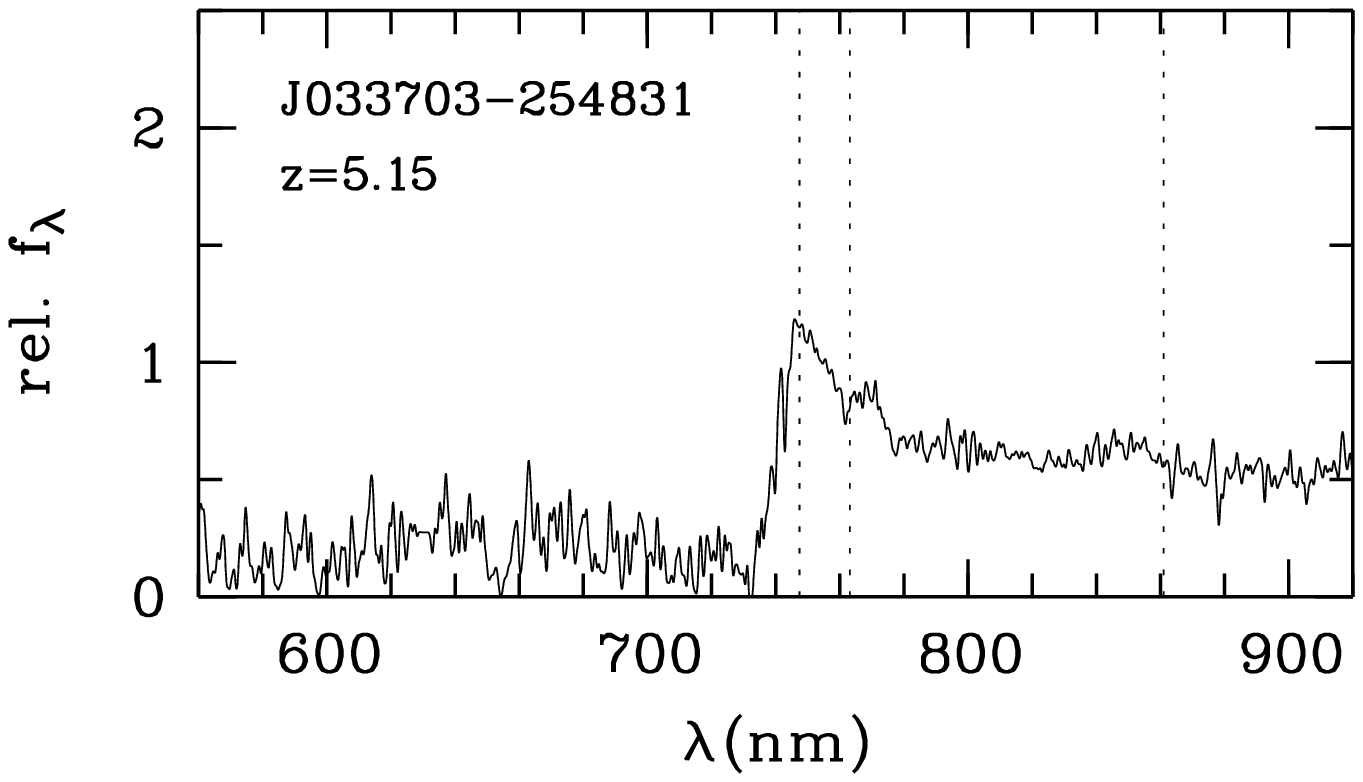}
\includegraphics[angle=270,width=0.32\textwidth,clip=true]{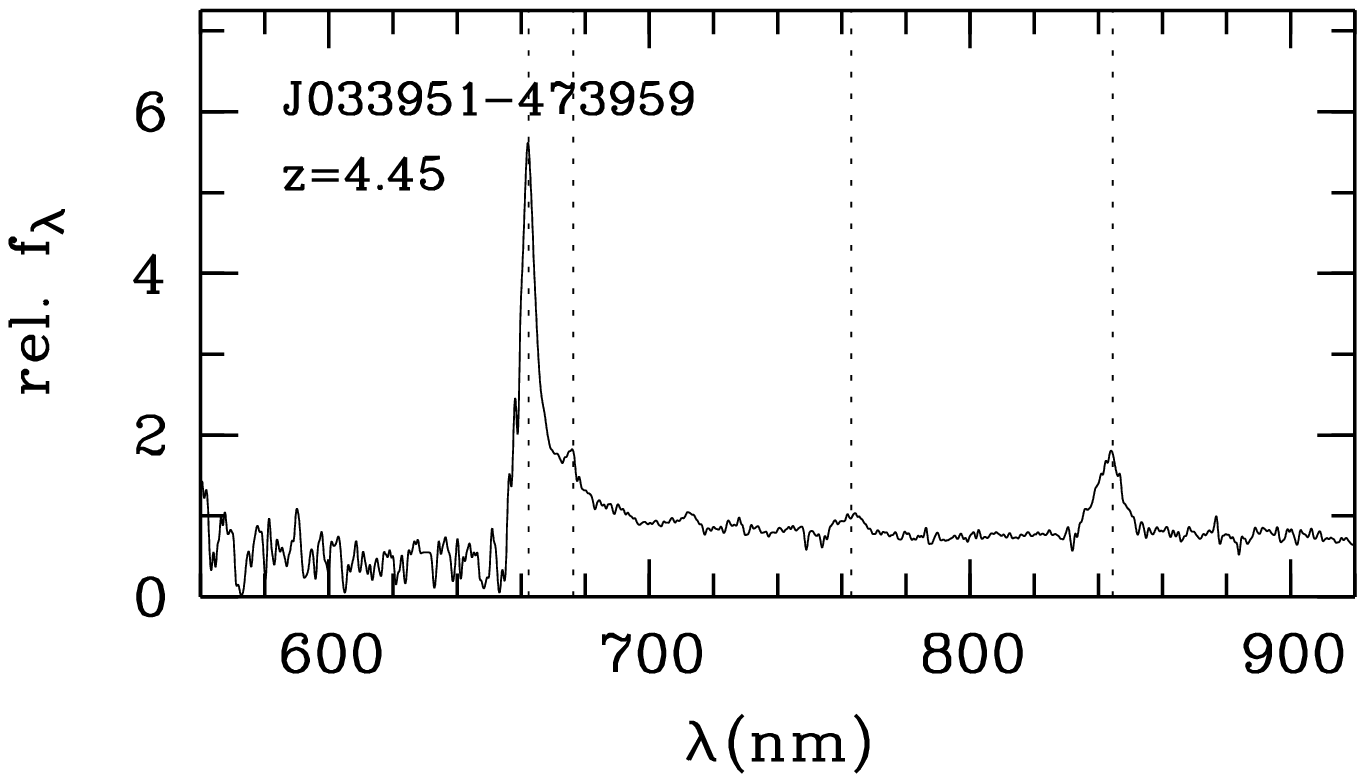}
\includegraphics[angle=270,width=0.32\textwidth,clip=true]{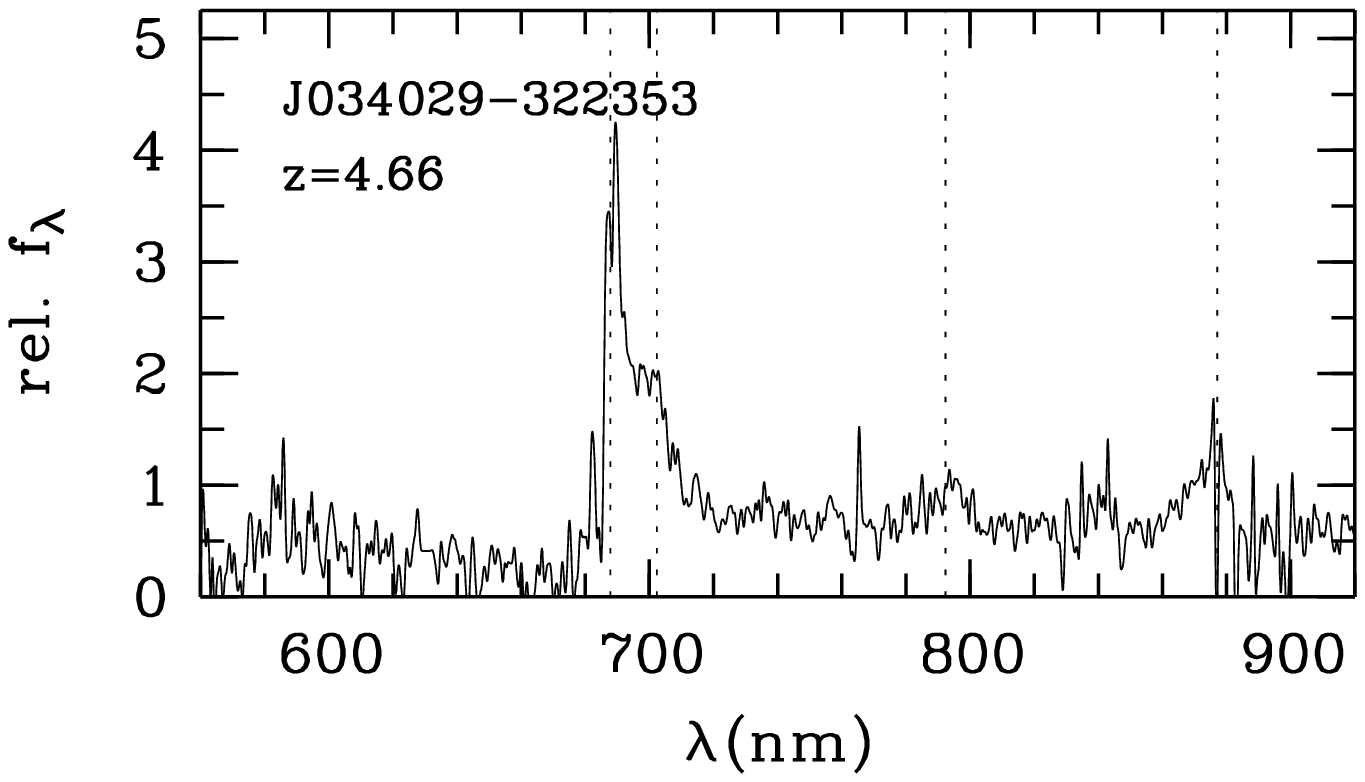}
\includegraphics[angle=270,width=0.32\textwidth,clip=true]{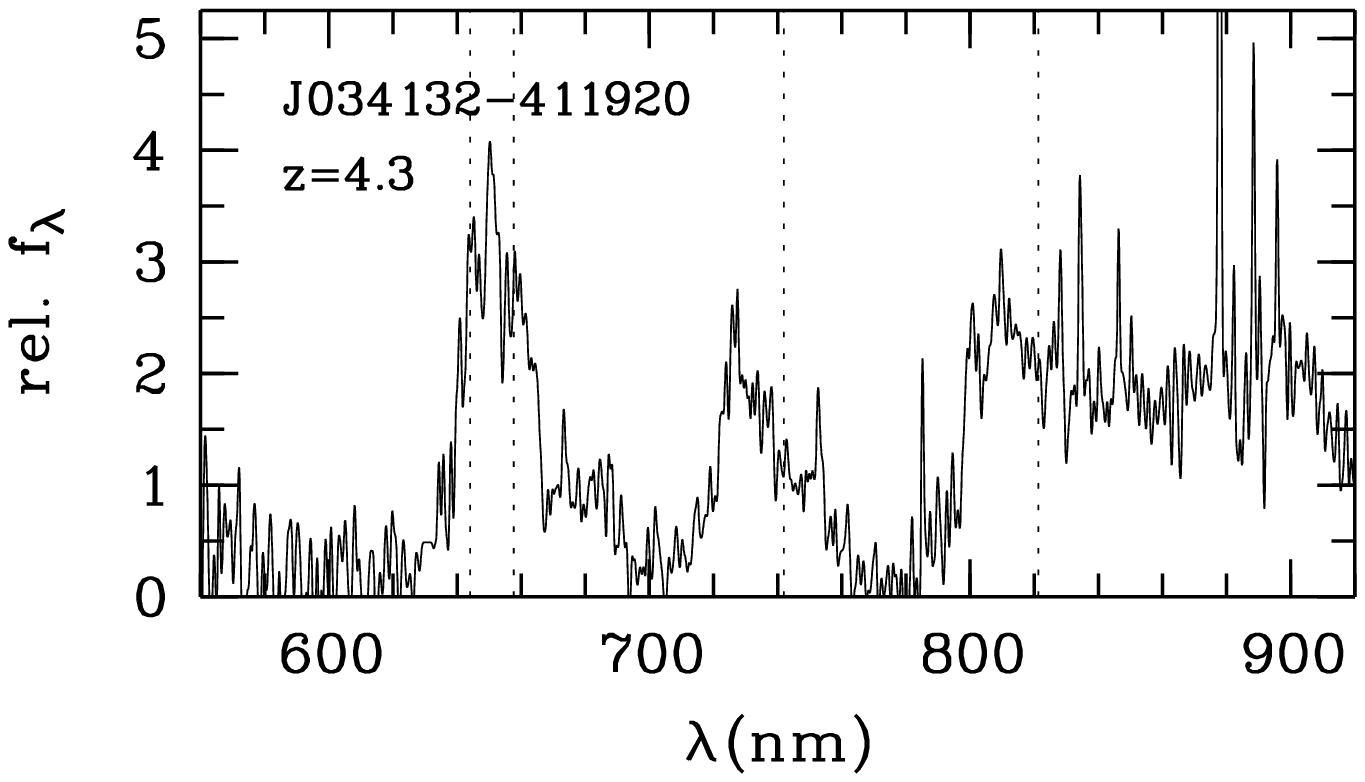}
\caption{Gallery of $3.8>z>5.5$ QSO spectra obtained in this work, ordered by RA, page 2.
\label{gallery2}}
\end{center}
\end{figure*}

\begin{figure*}
\begin{center}
\includegraphics[angle=270,width=0.32\textwidth,clip=true]{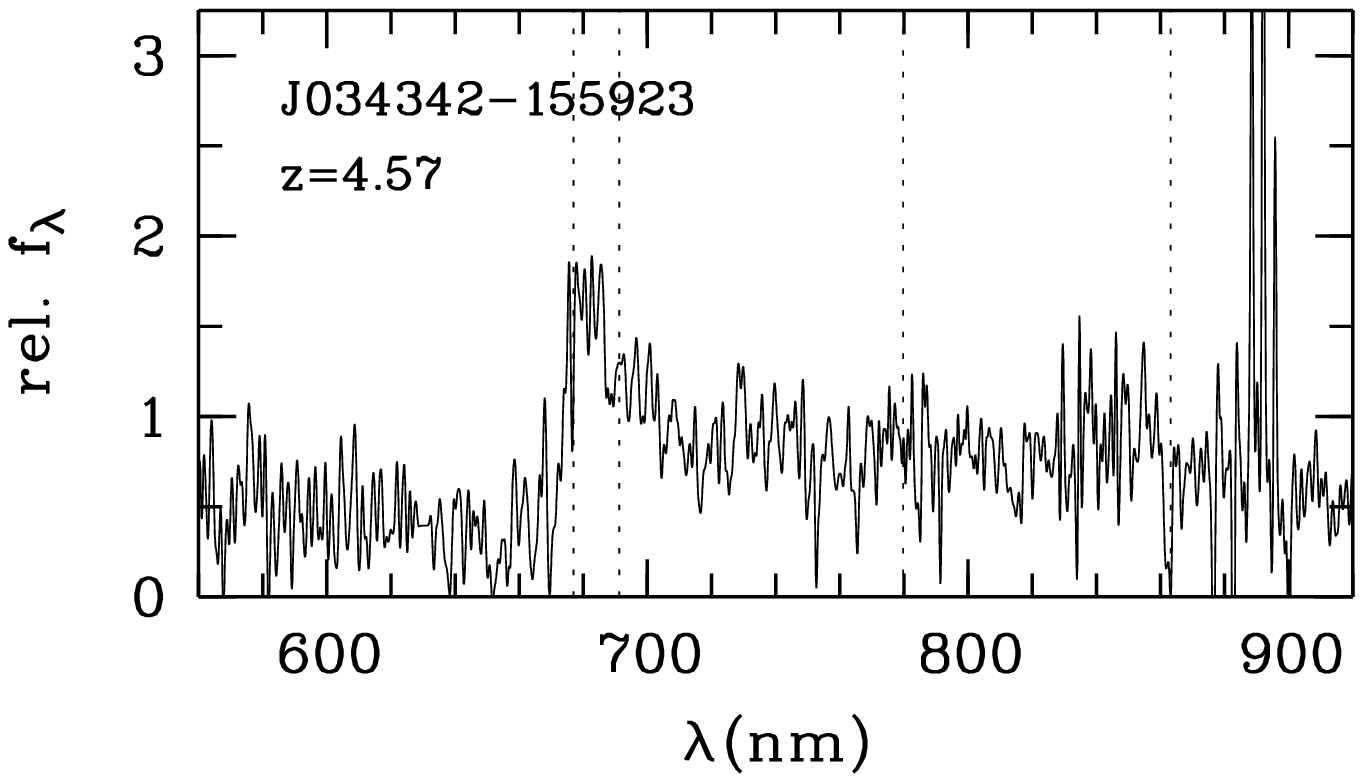}
\includegraphics[angle=270,width=0.32\textwidth,clip=true]{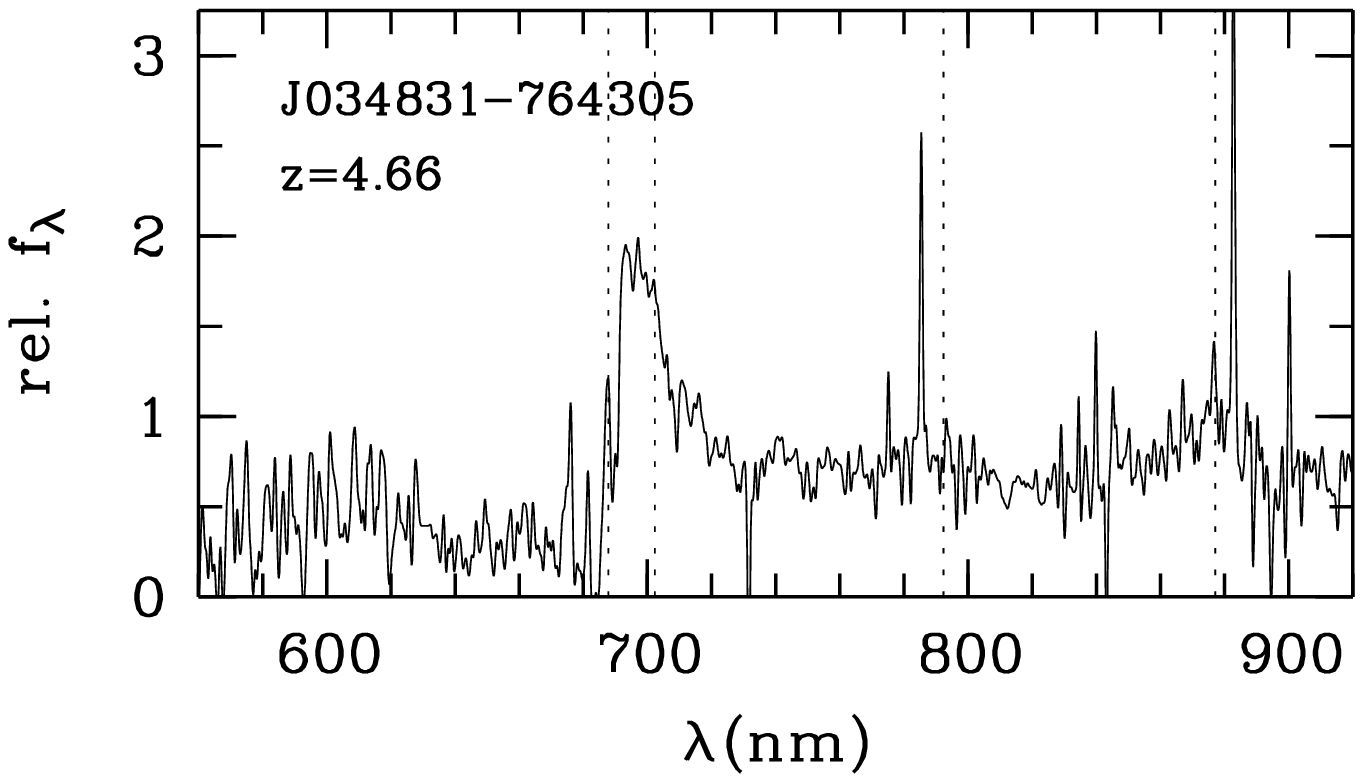}
\includegraphics[angle=270,width=0.32\textwidth,clip=true]{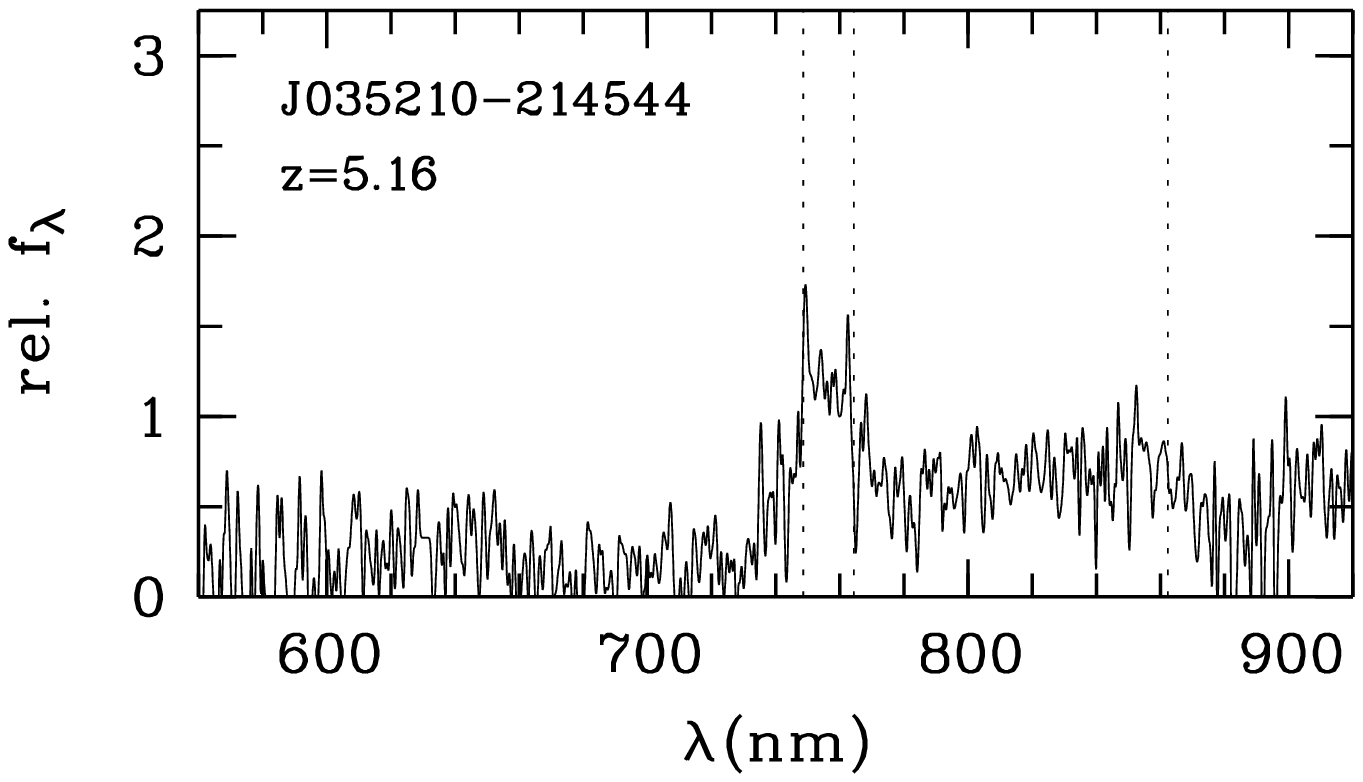}
\includegraphics[angle=270,width=0.32\textwidth,clip=true]{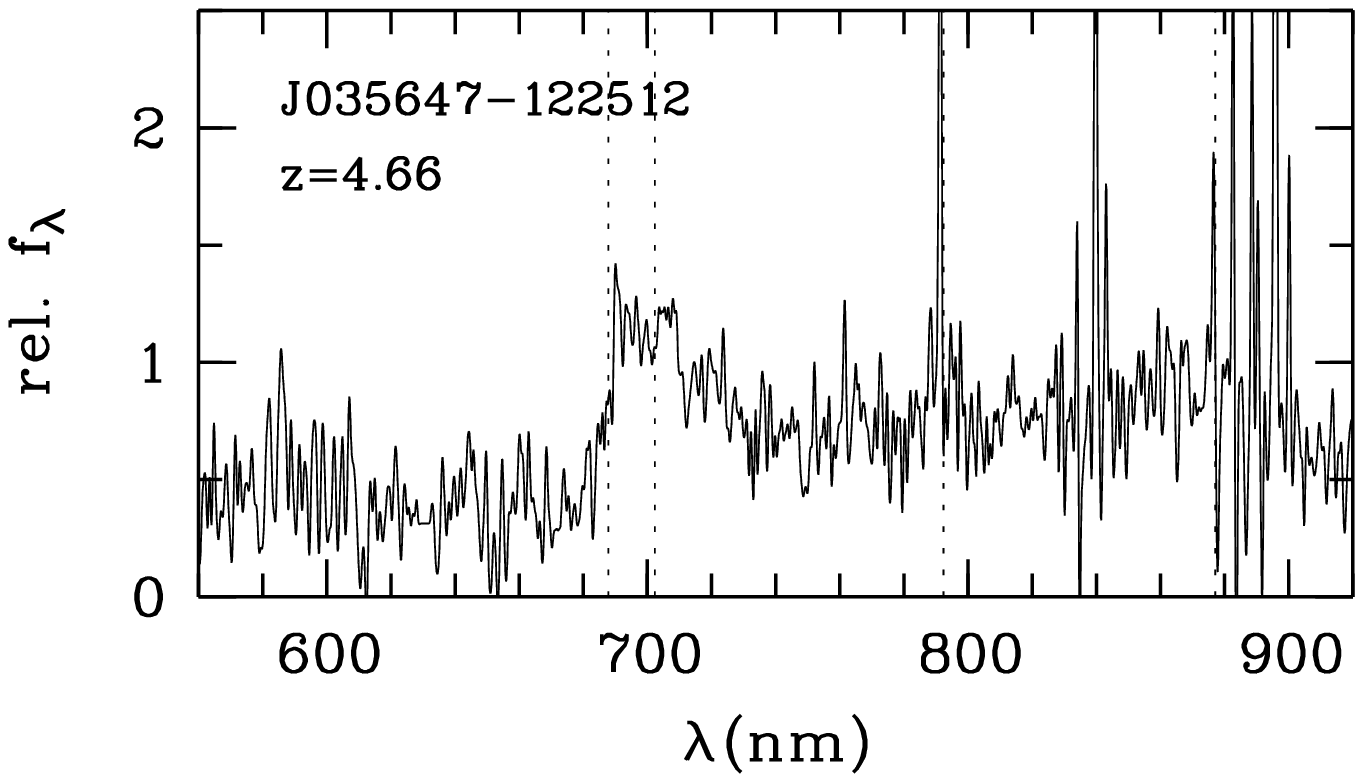}
\includegraphics[angle=270,width=0.32\textwidth,clip=true]{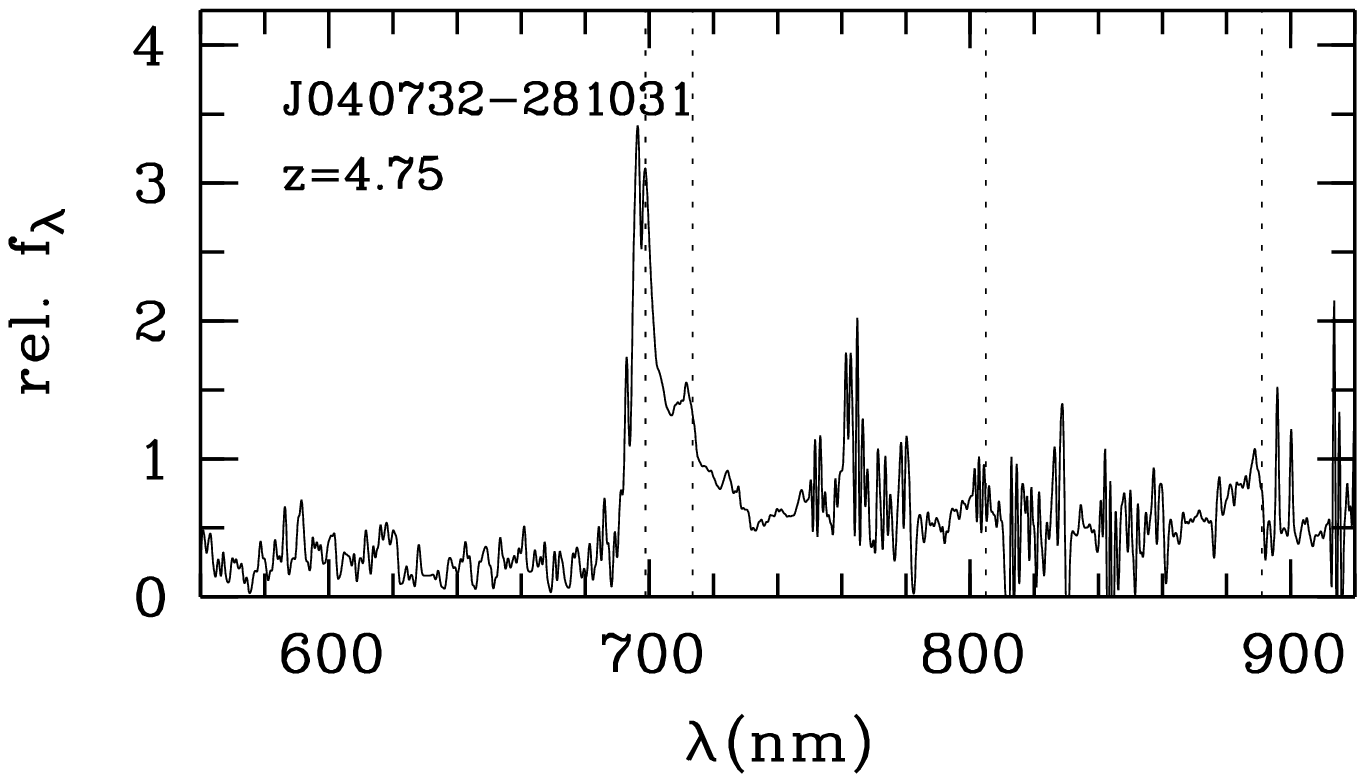}
\includegraphics[angle=270,width=0.32\textwidth,clip=true]{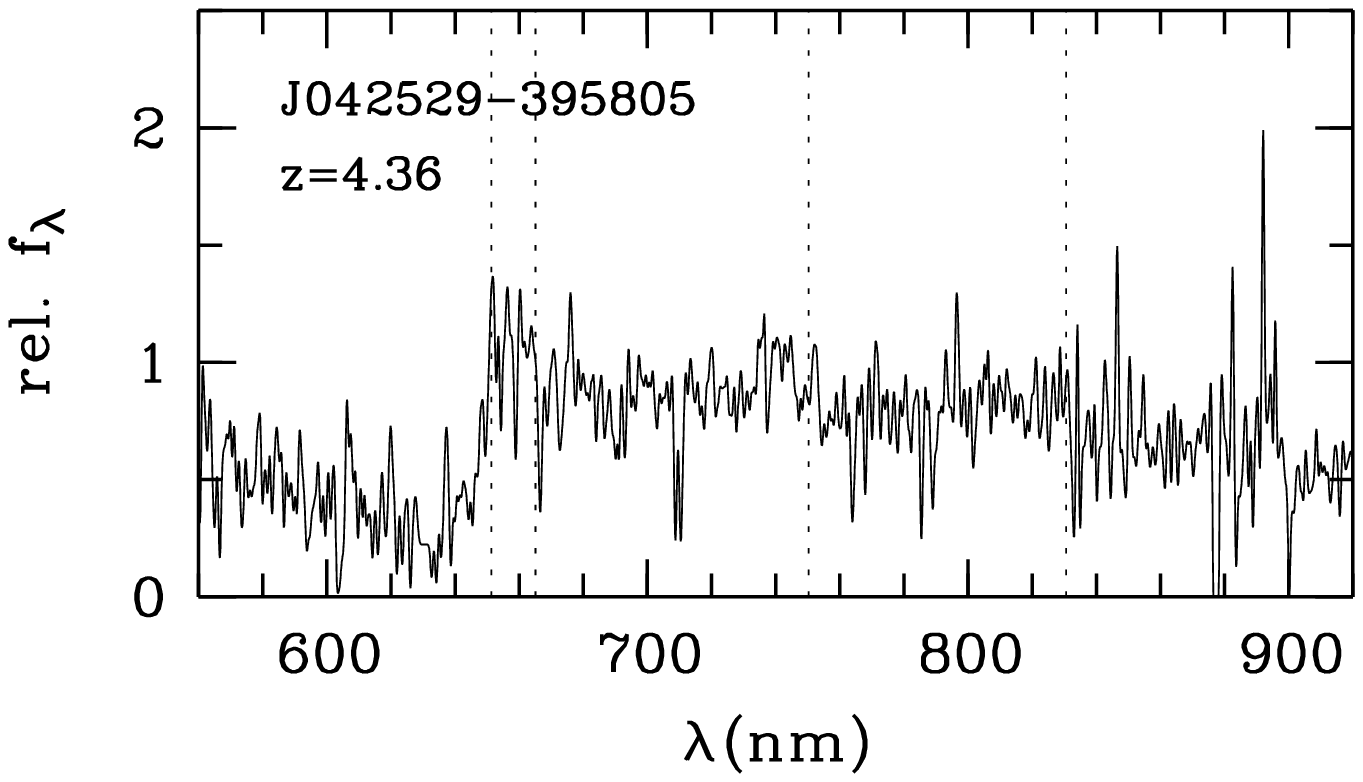}
\includegraphics[angle=270,width=0.32\textwidth,clip=true]{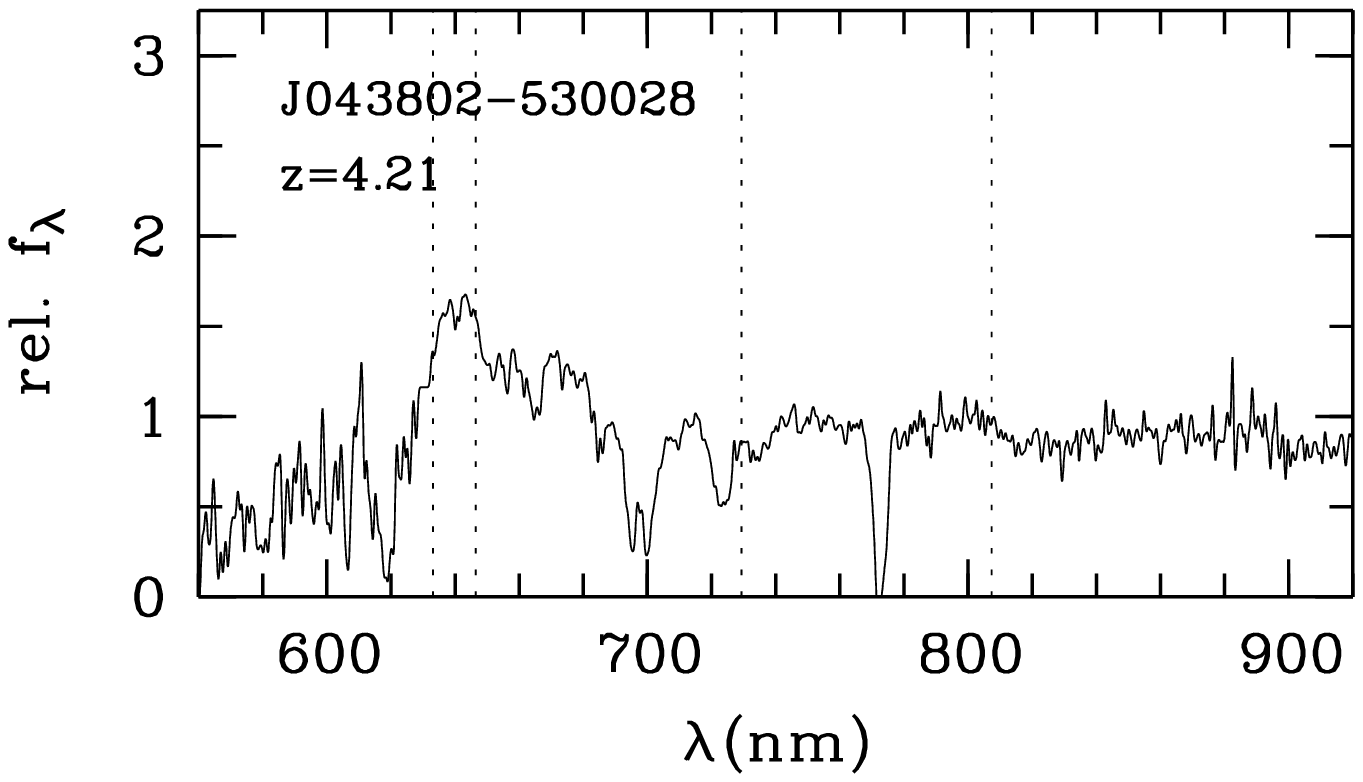}
\includegraphics[angle=270,width=0.32\textwidth,clip=true]{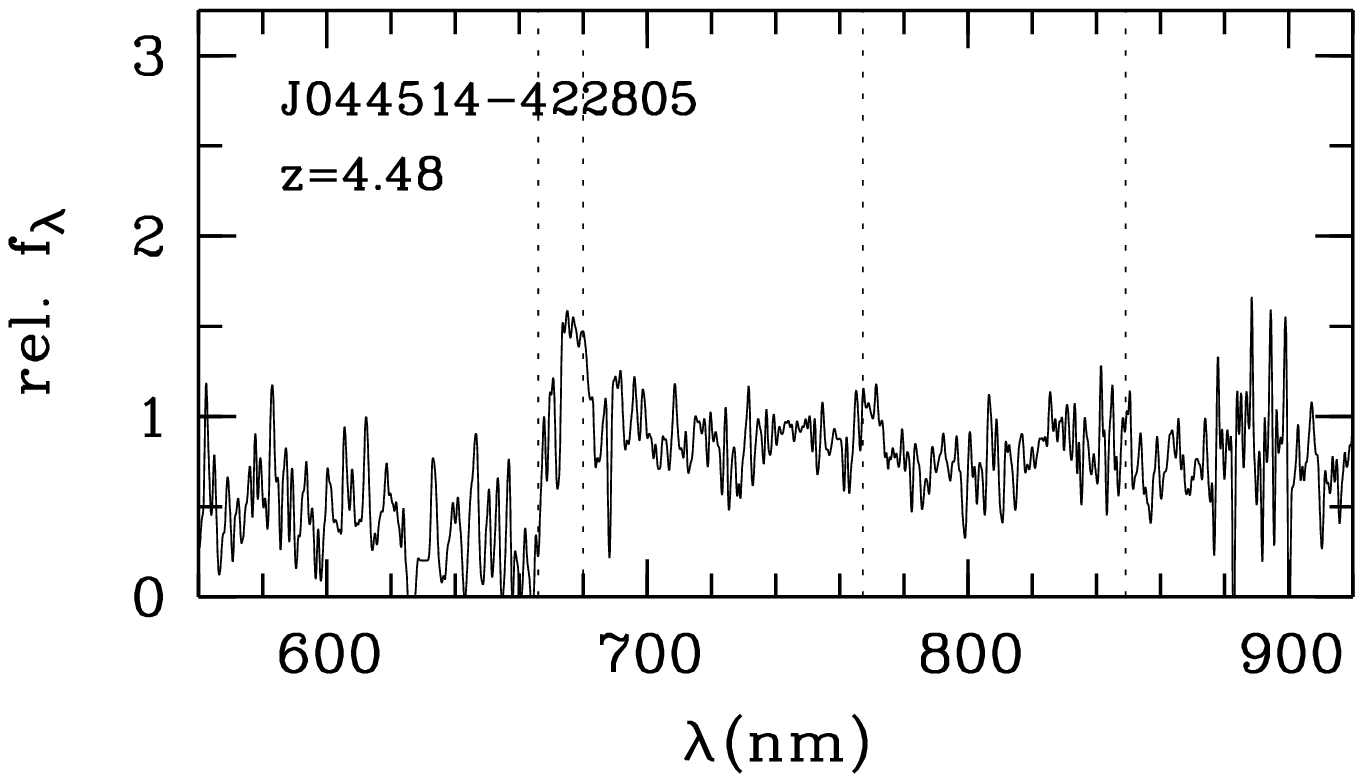}
\includegraphics[angle=270,width=0.32\textwidth,clip=true]{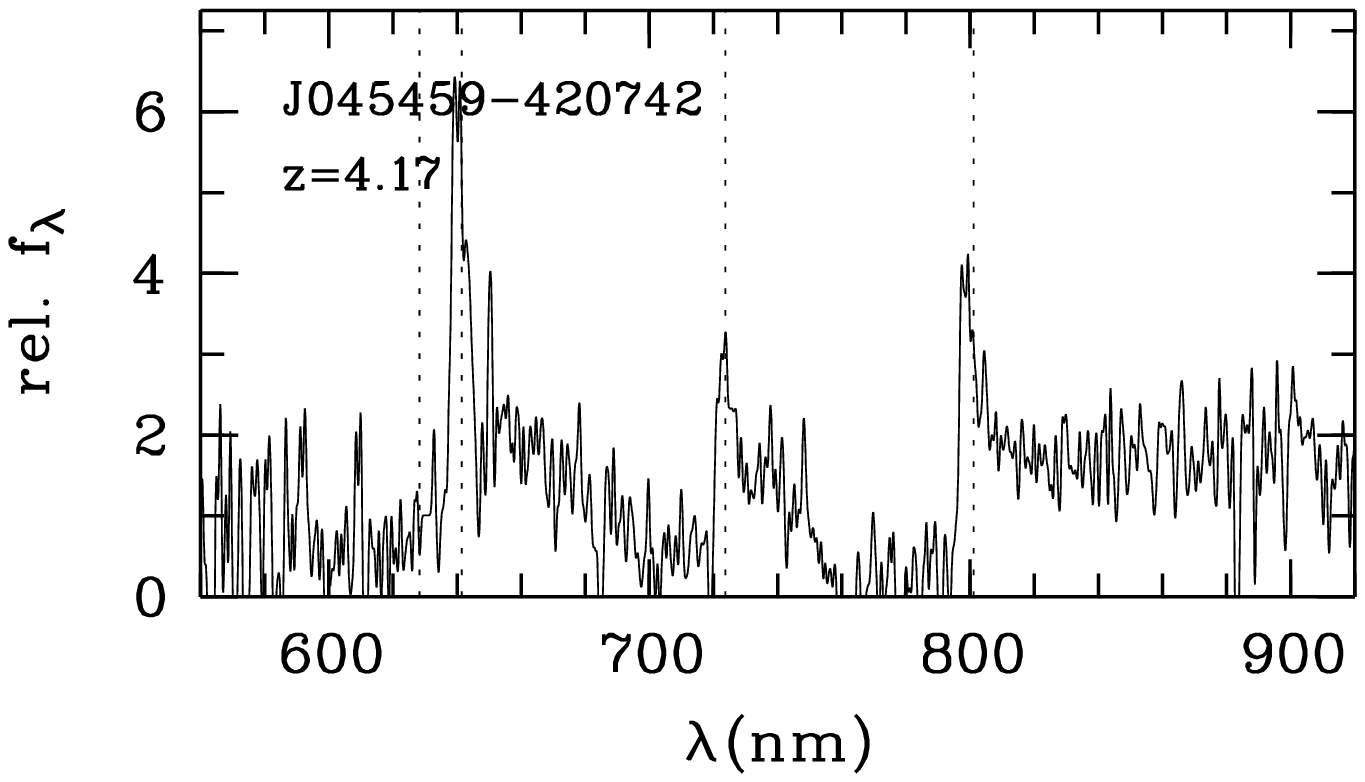}
\includegraphics[angle=270,width=0.32\textwidth,clip=true]{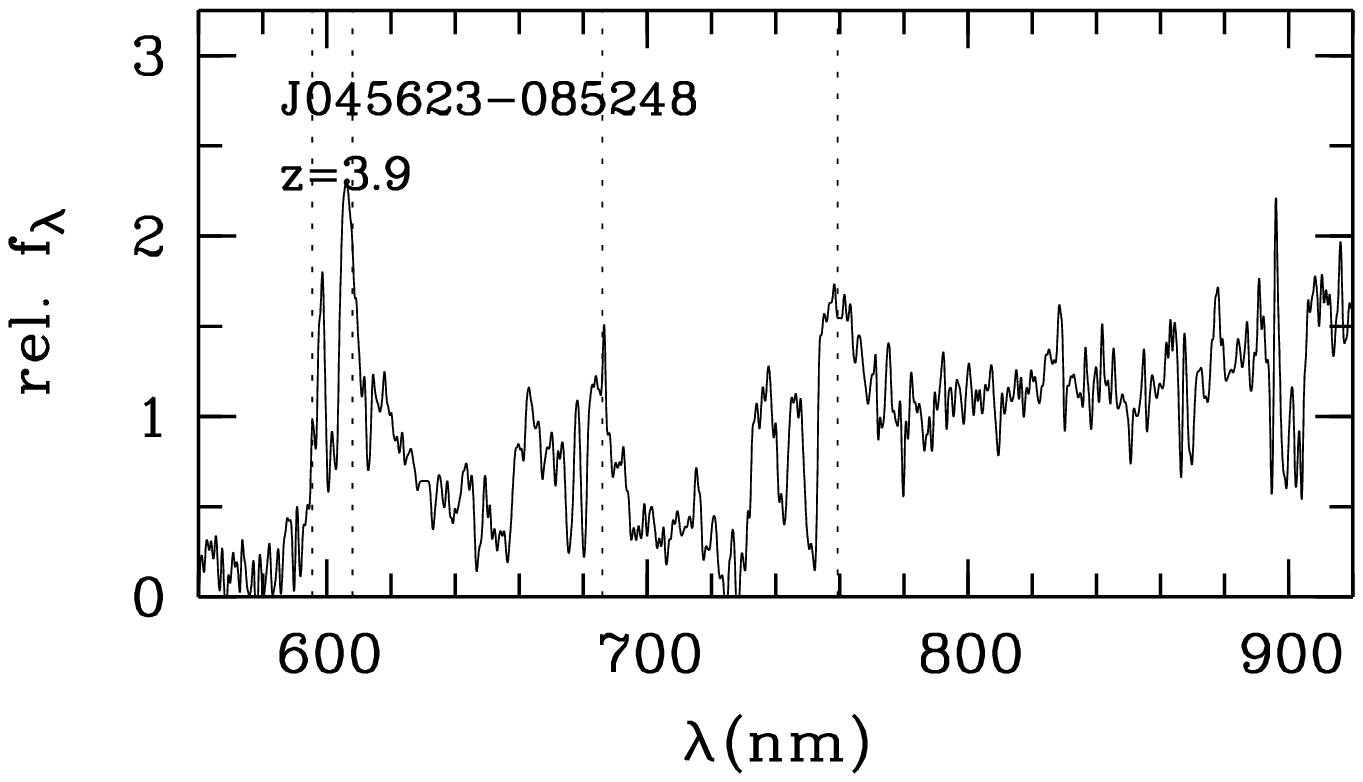}
\includegraphics[angle=270,width=0.32\textwidth,clip=true]{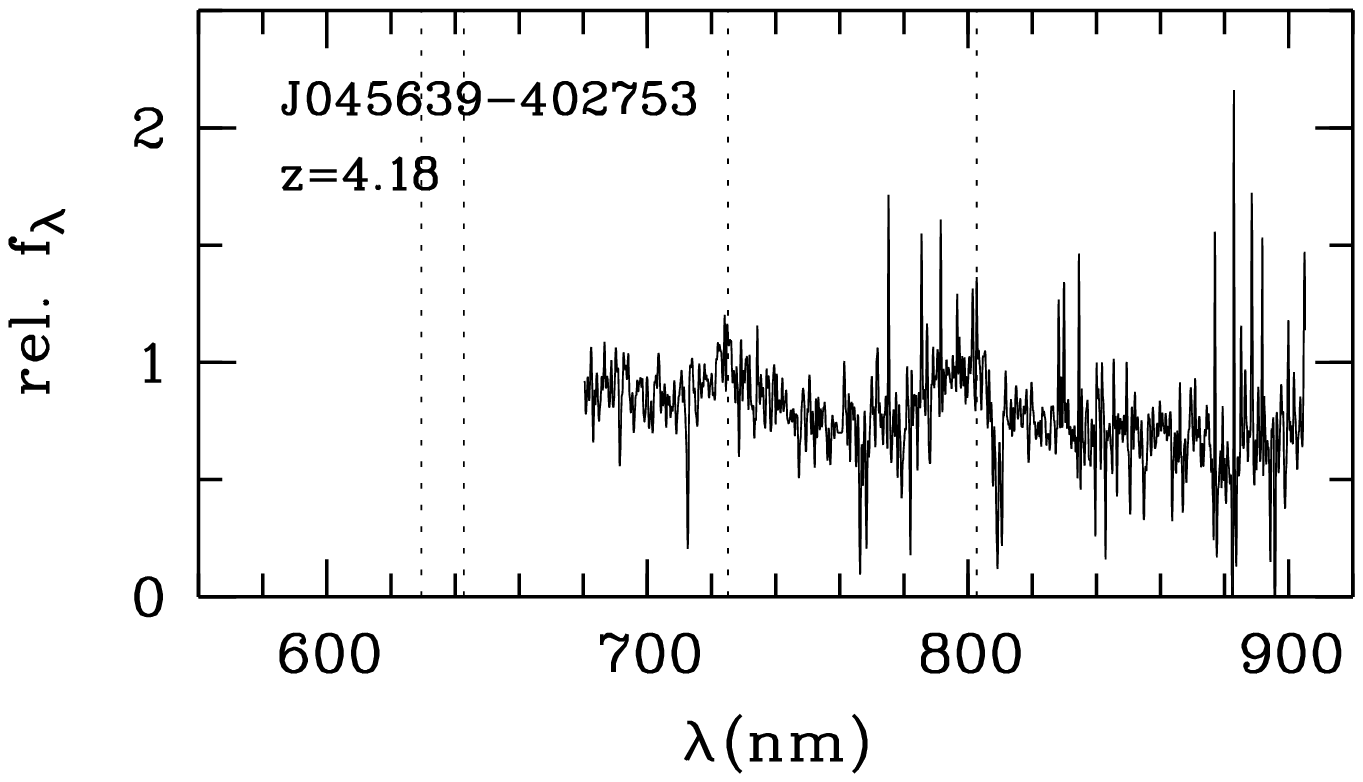}
\includegraphics[angle=270,width=0.32\textwidth,clip=true]{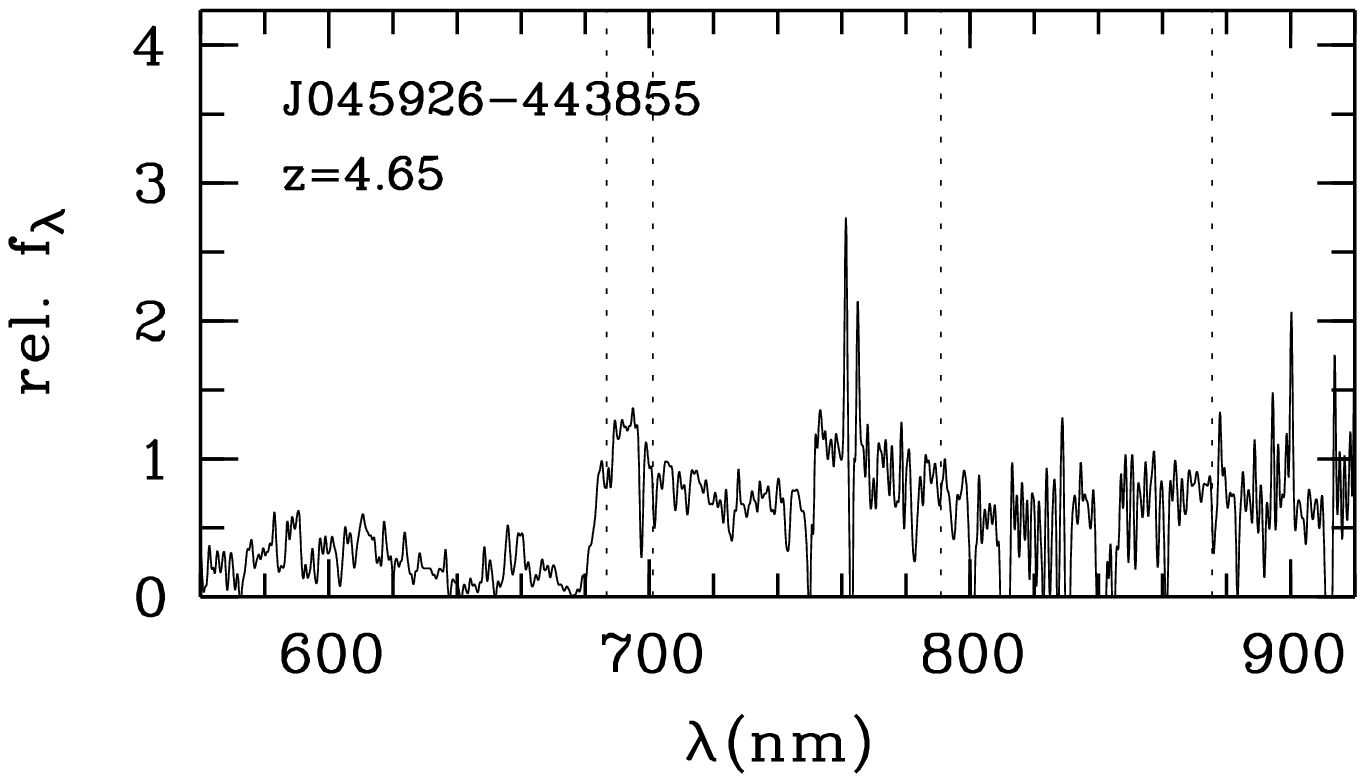}
\includegraphics[angle=270,width=0.32\textwidth,clip=true]{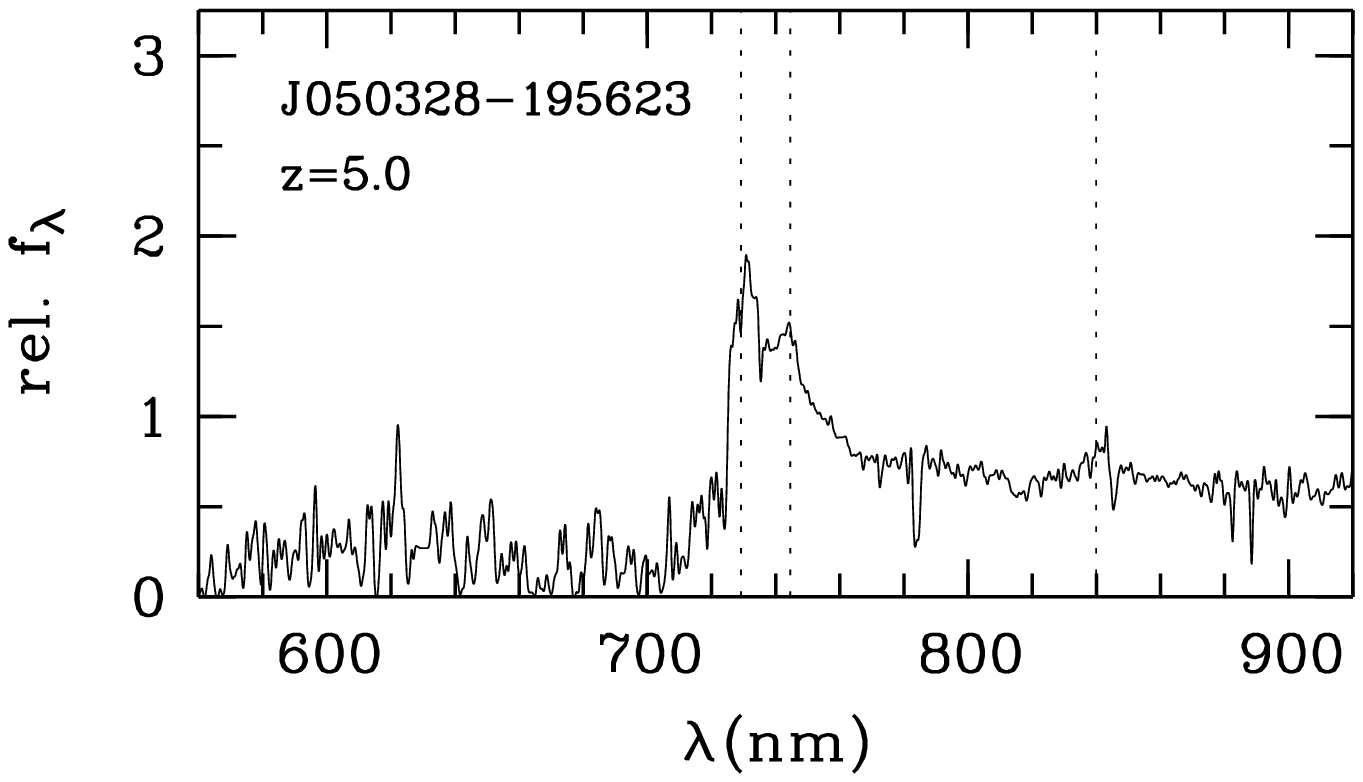}
\includegraphics[angle=270,width=0.32\textwidth,clip=true]{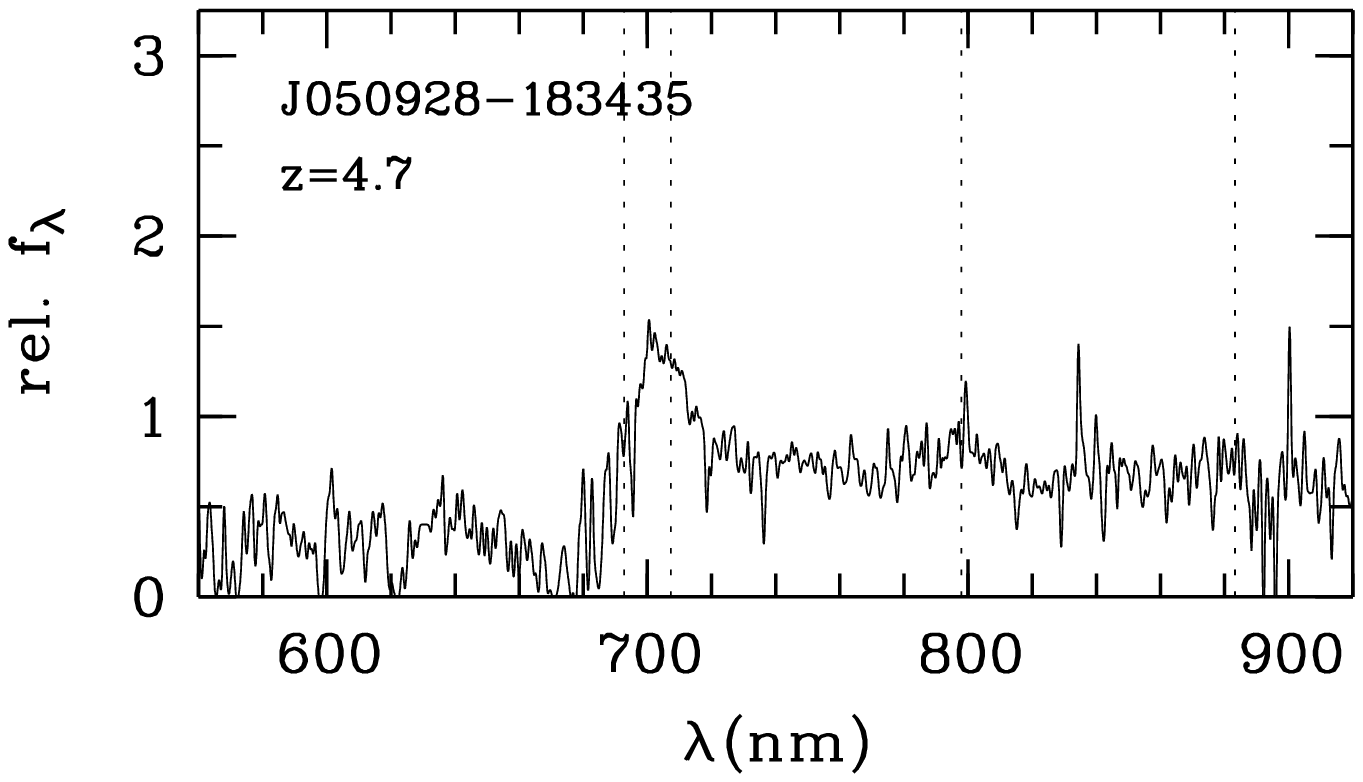}
\includegraphics[angle=270,width=0.32\textwidth,clip=true]{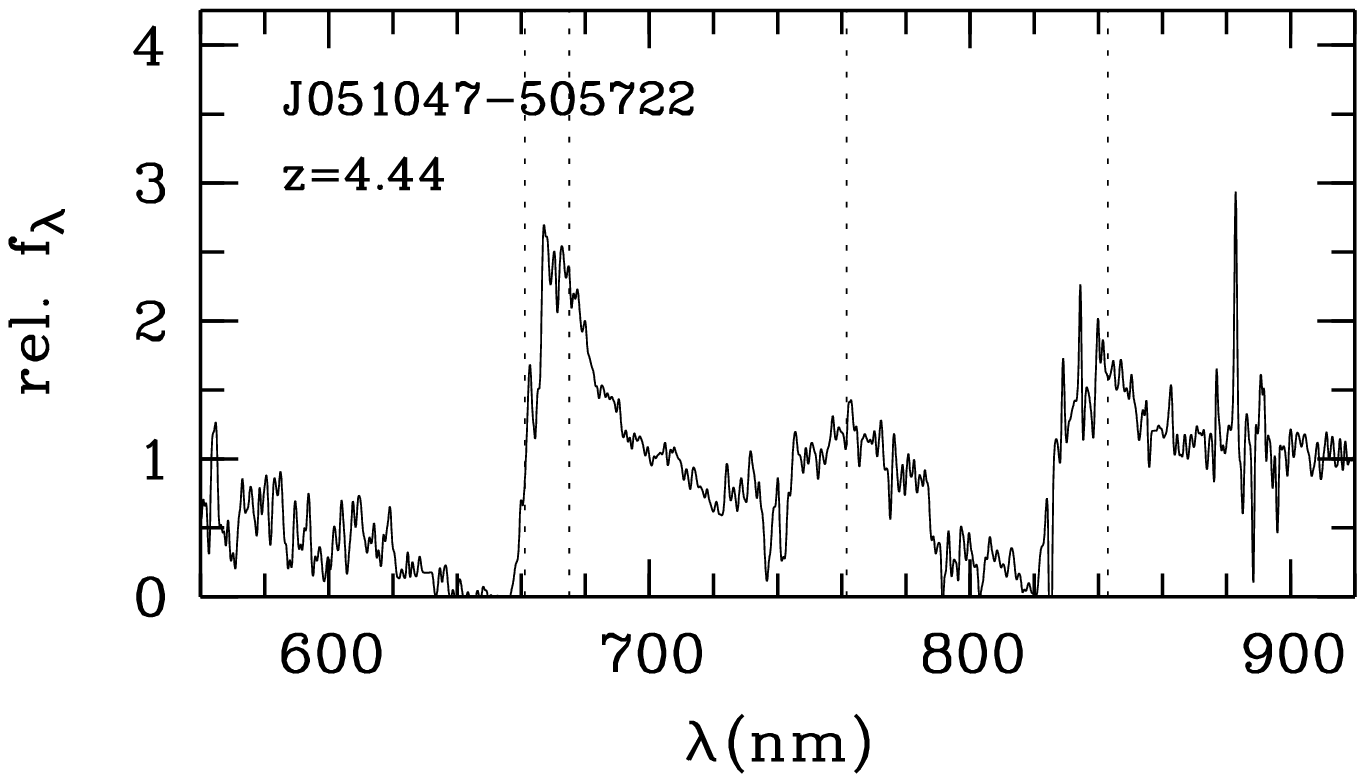}
\includegraphics[angle=270,width=0.32\textwidth,clip=true]{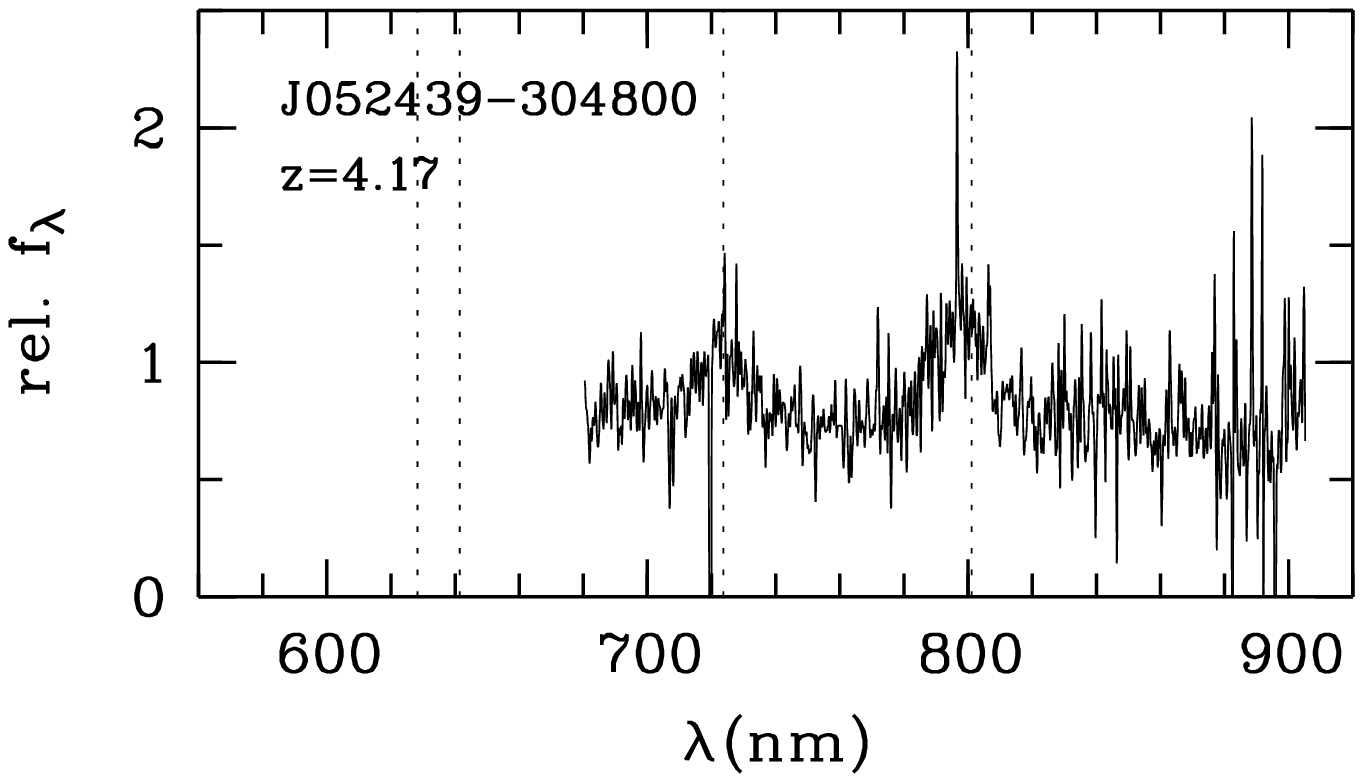}
\includegraphics[angle=270,width=0.32\textwidth,clip=true]{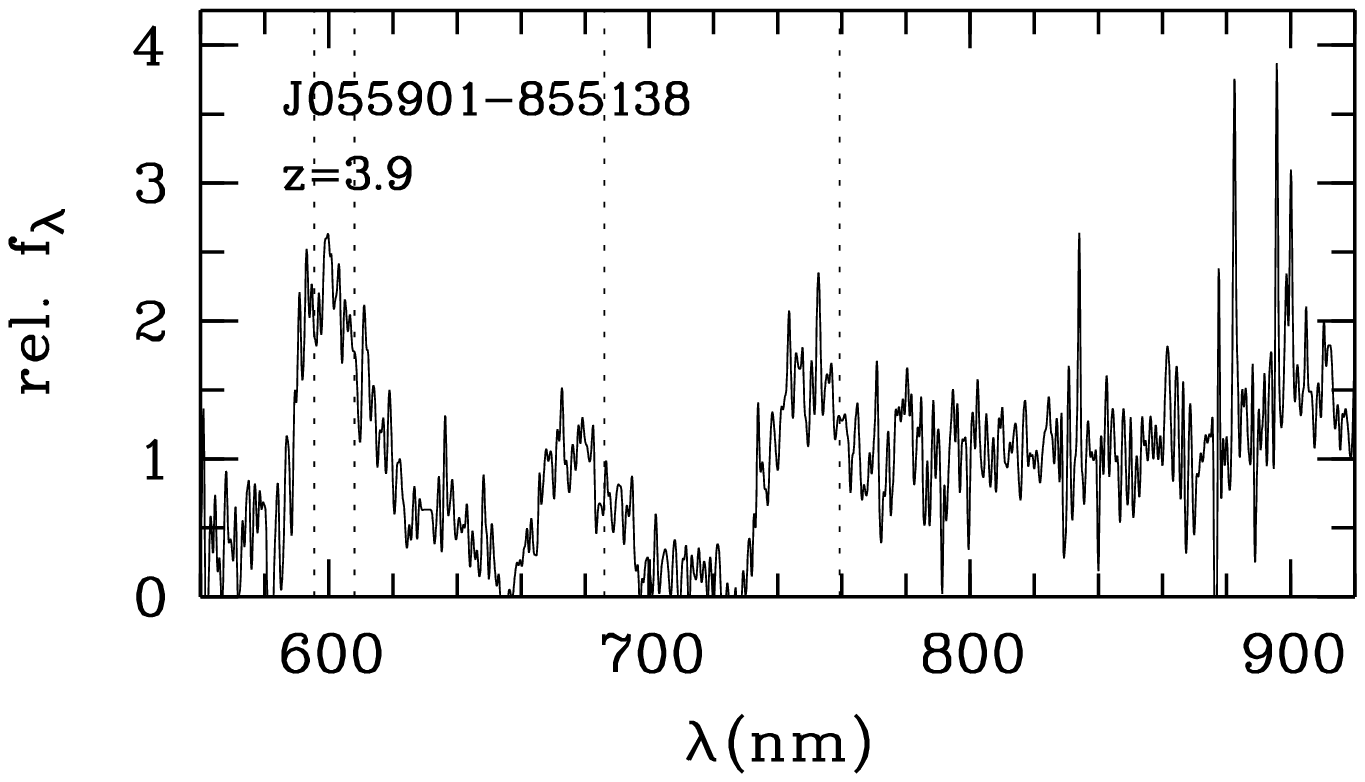}
\includegraphics[angle=270,width=0.32\textwidth,clip=true]{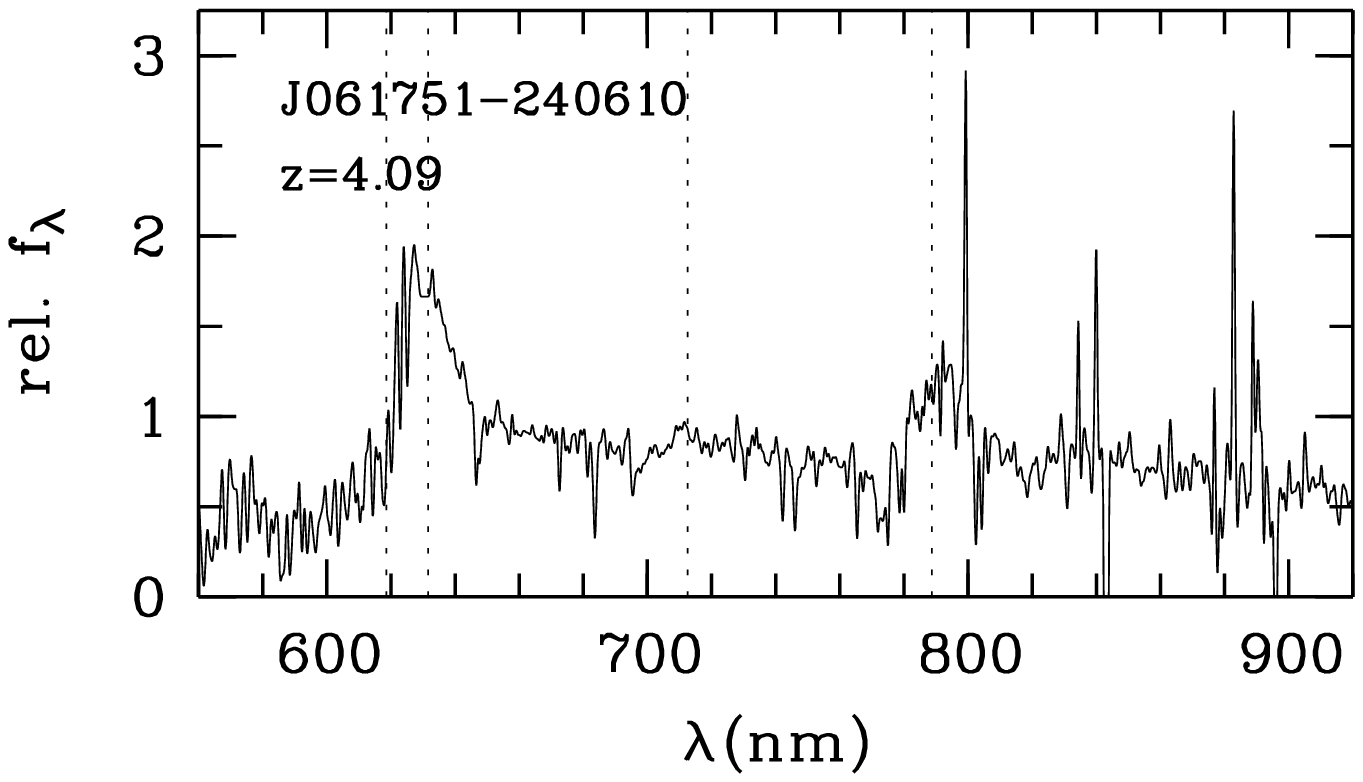}
\includegraphics[angle=270,width=0.32\textwidth,clip=true]{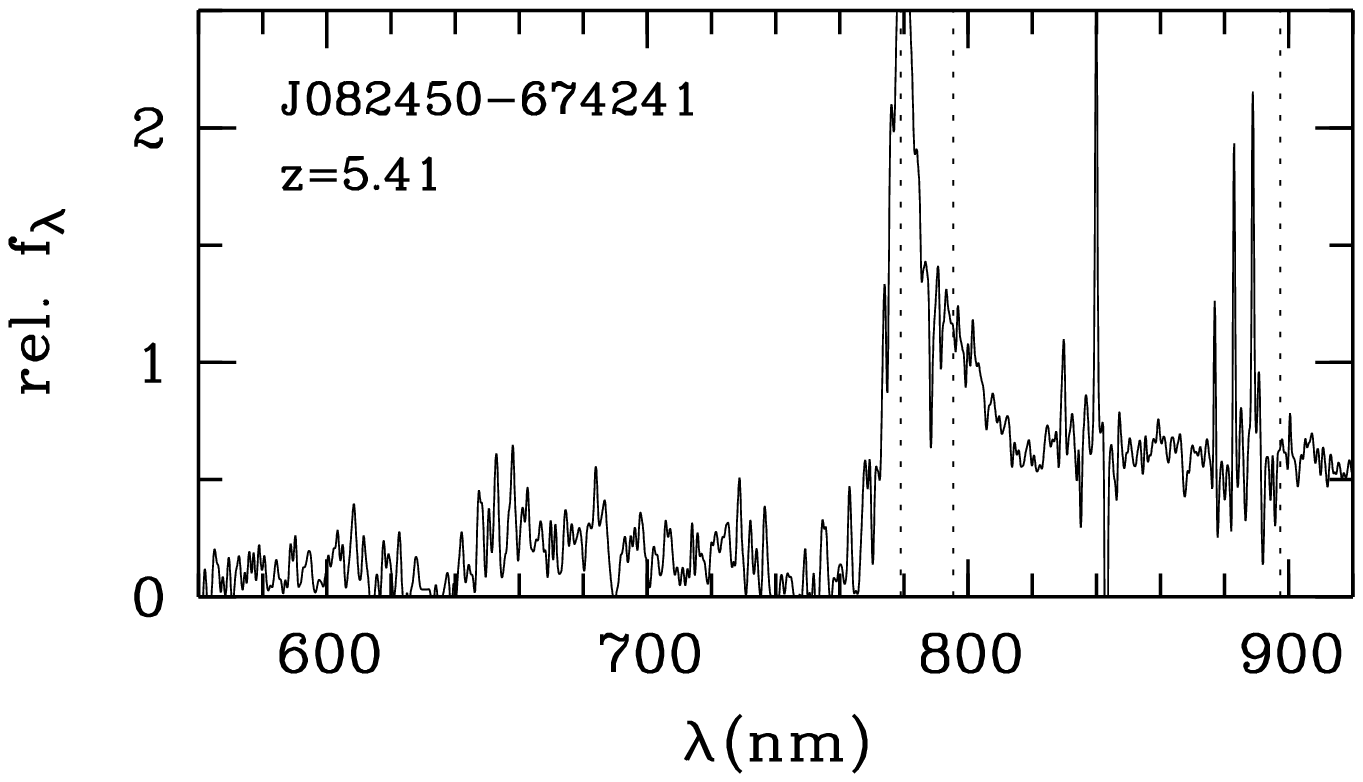}
\includegraphics[angle=270,width=0.32\textwidth,clip=true]{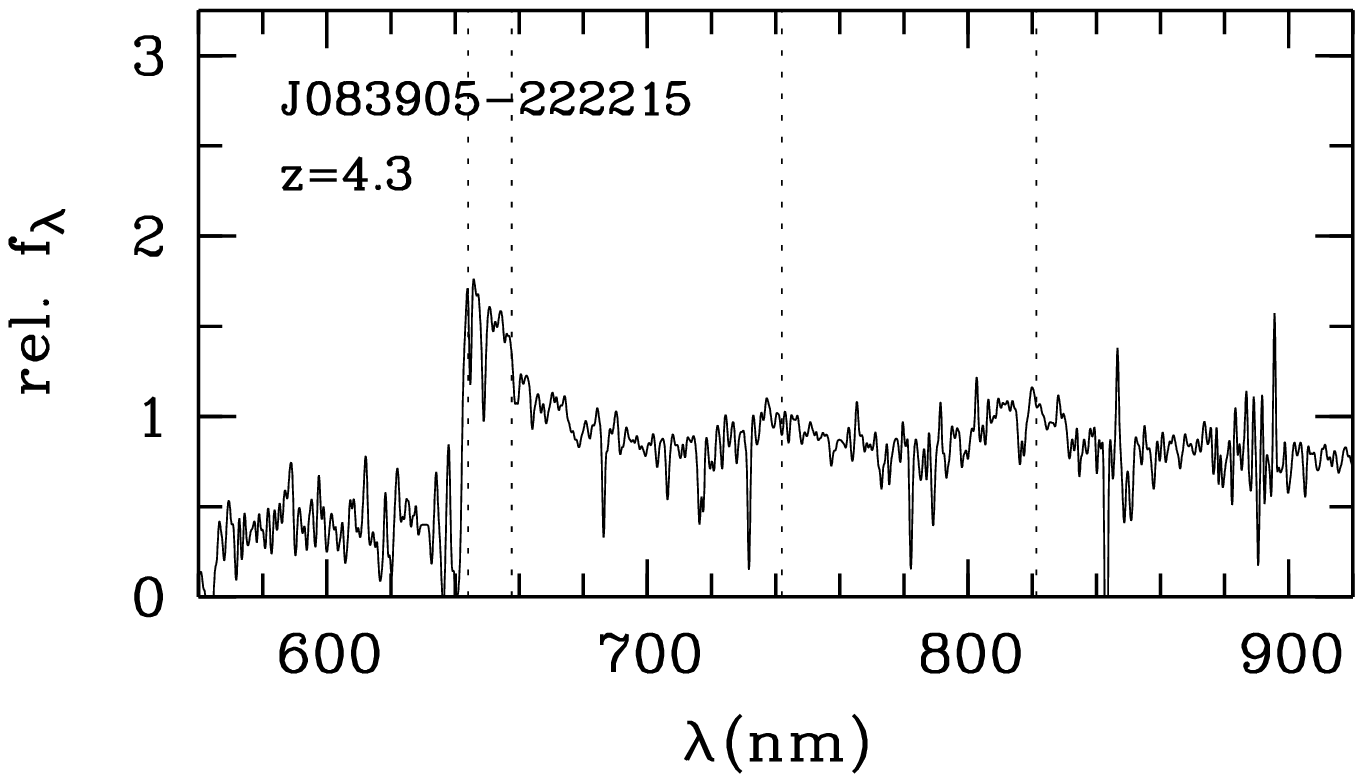}
\includegraphics[angle=270,width=0.32\textwidth,clip=true]{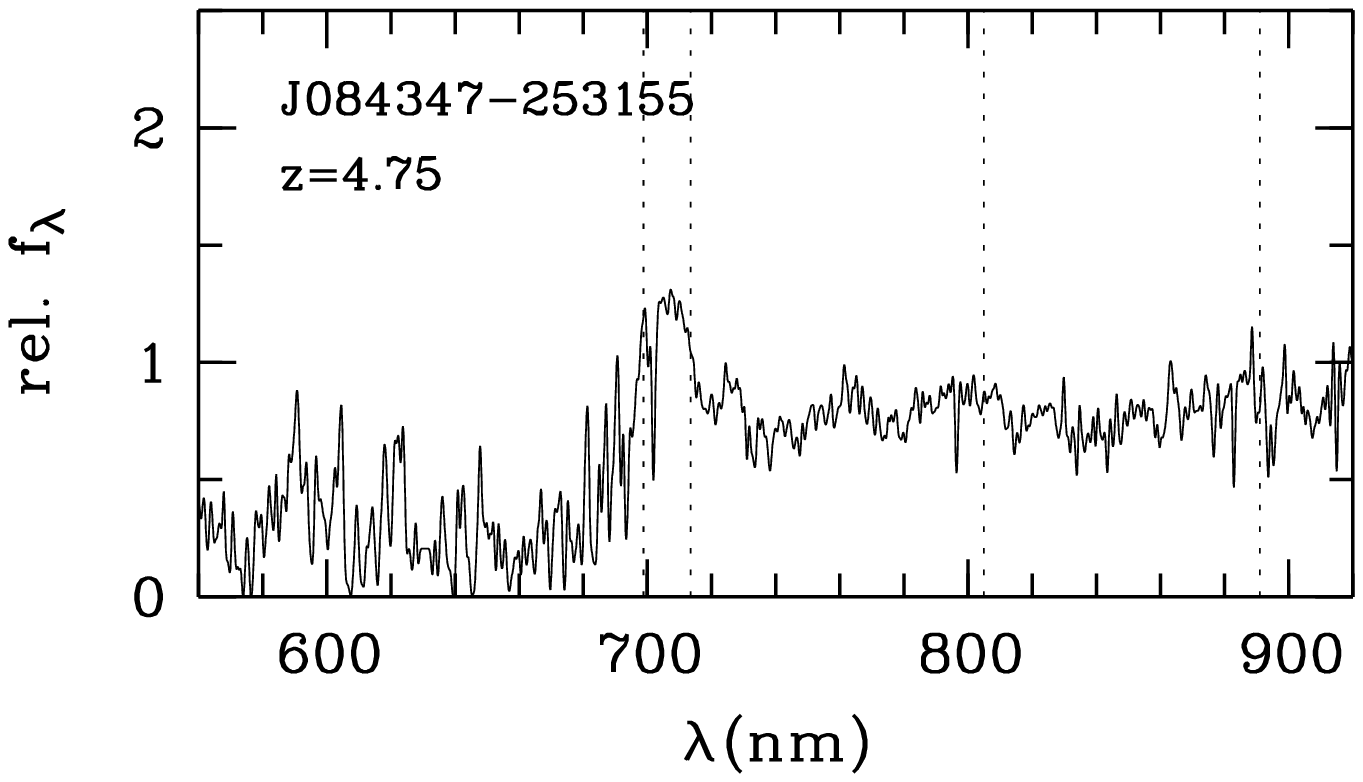}
\caption{Gallery of $3.8>z>5.5$ QSO spectra obtained in this work, ordered by RA, page 3.
\label{gallery3}}
\end{center}
\end{figure*}

\begin{figure*}
\begin{center}
\includegraphics[angle=270,width=0.32\textwidth,clip=true]{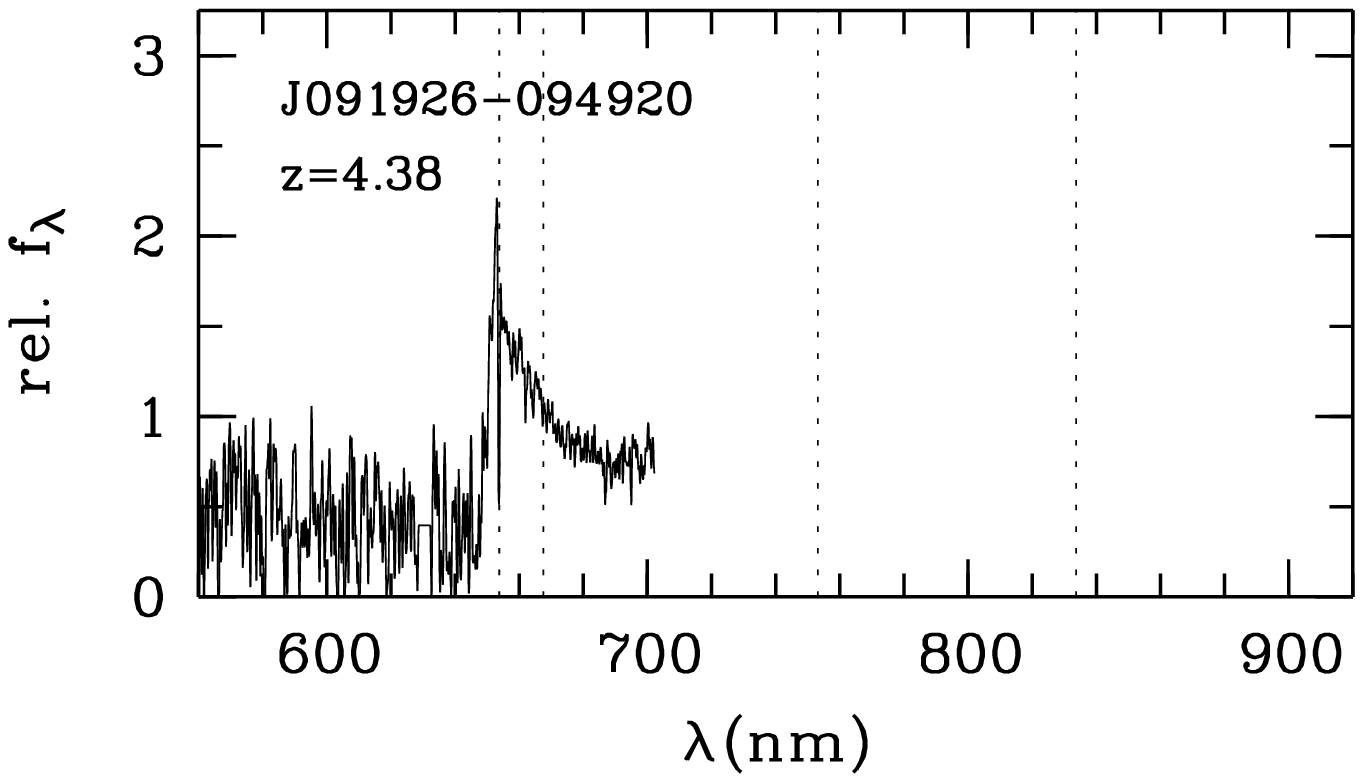}
\includegraphics[angle=270,width=0.32\textwidth,clip=true]{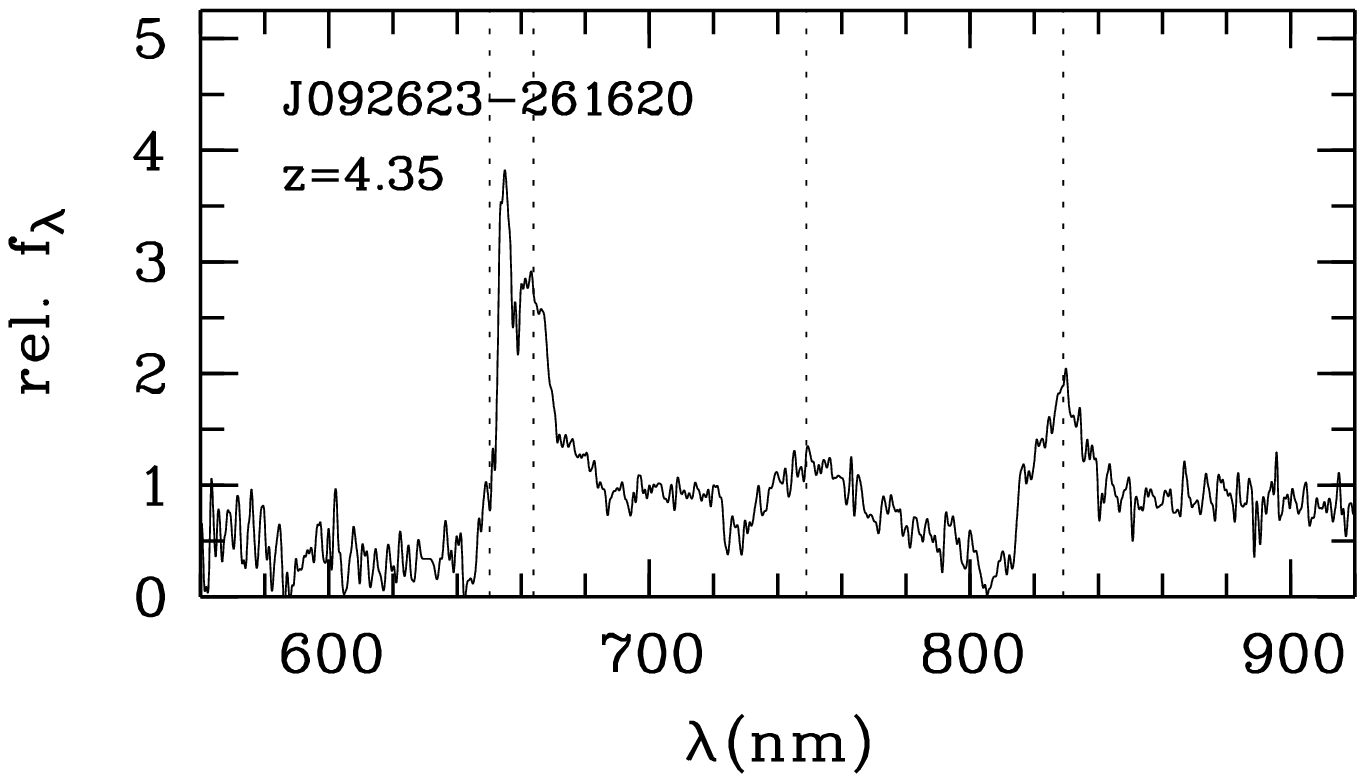}
\includegraphics[angle=270,width=0.32\textwidth,clip=true]{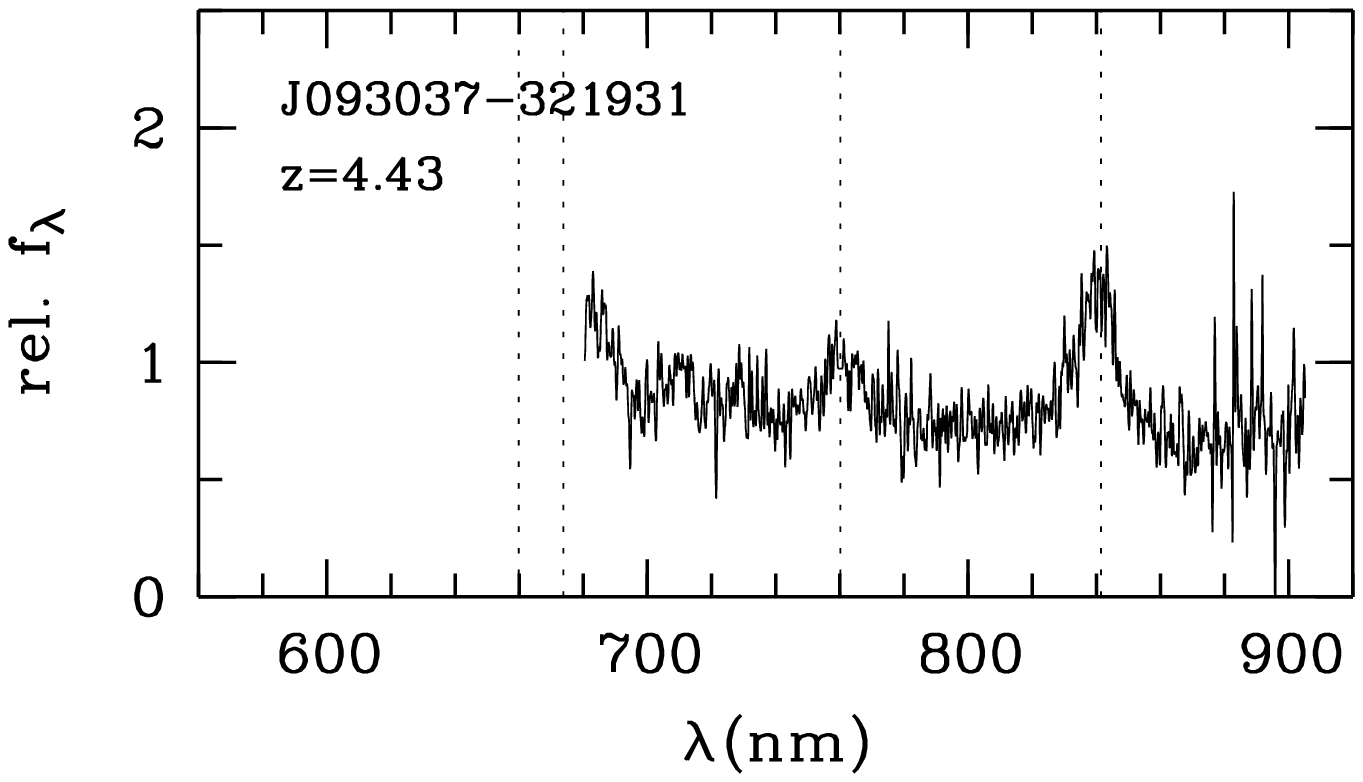}
\includegraphics[angle=270,width=0.32\textwidth,clip=true]{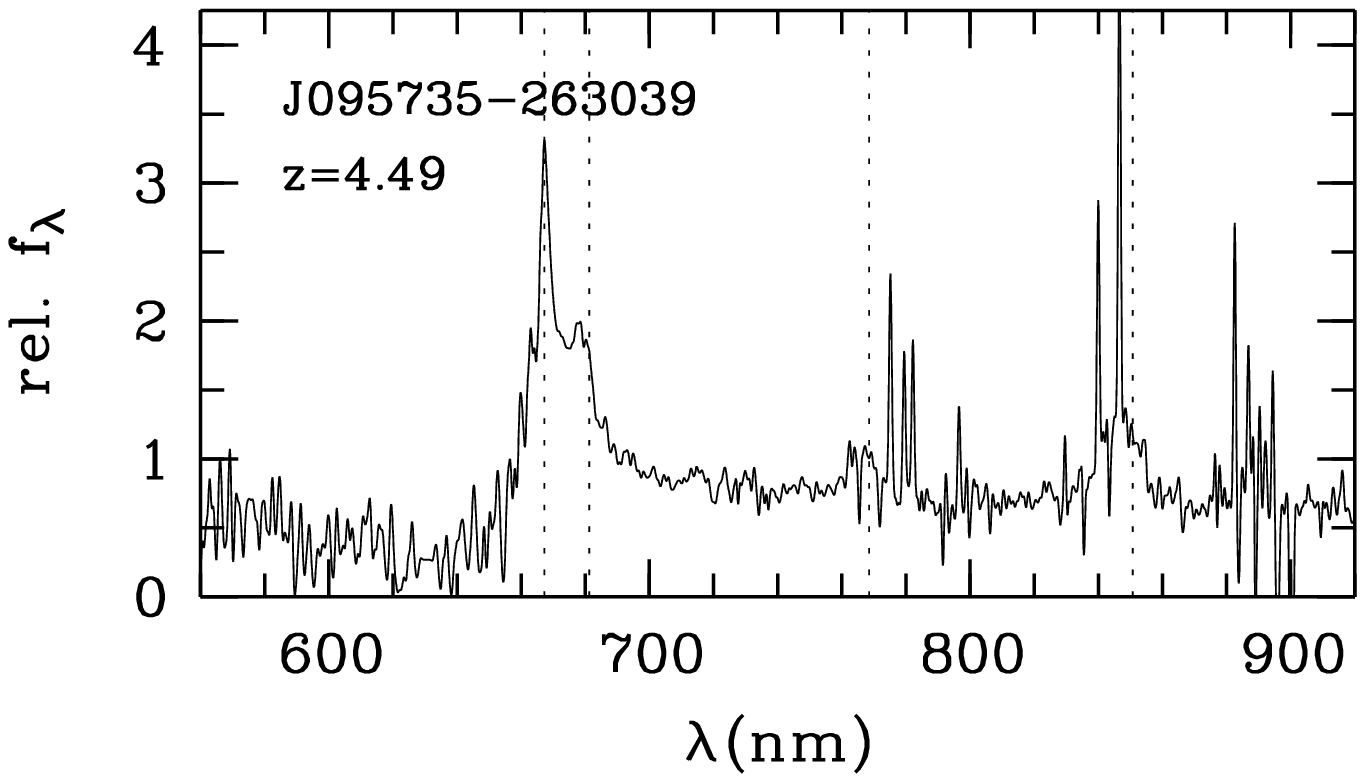}
\includegraphics[angle=270,width=0.32\textwidth,clip=true]{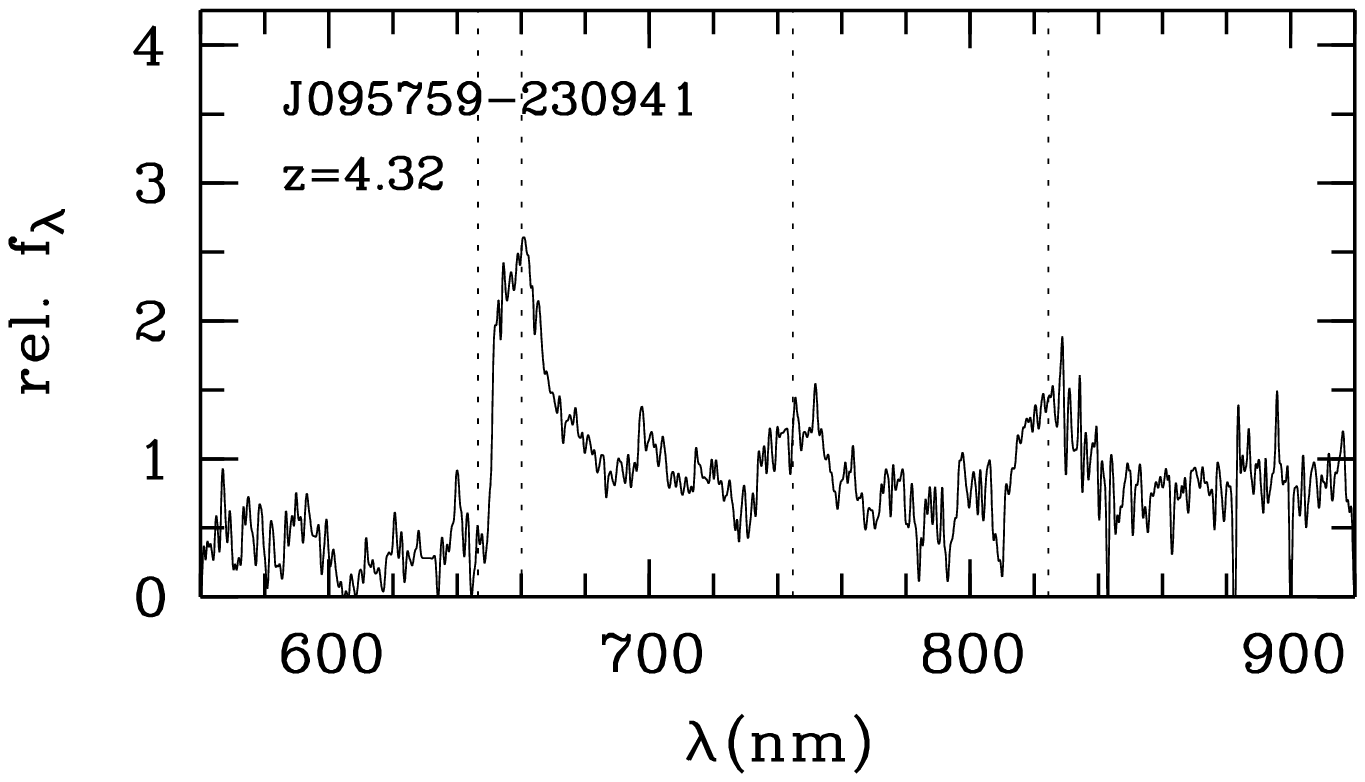}
\includegraphics[angle=270,width=0.32\textwidth,clip=true]{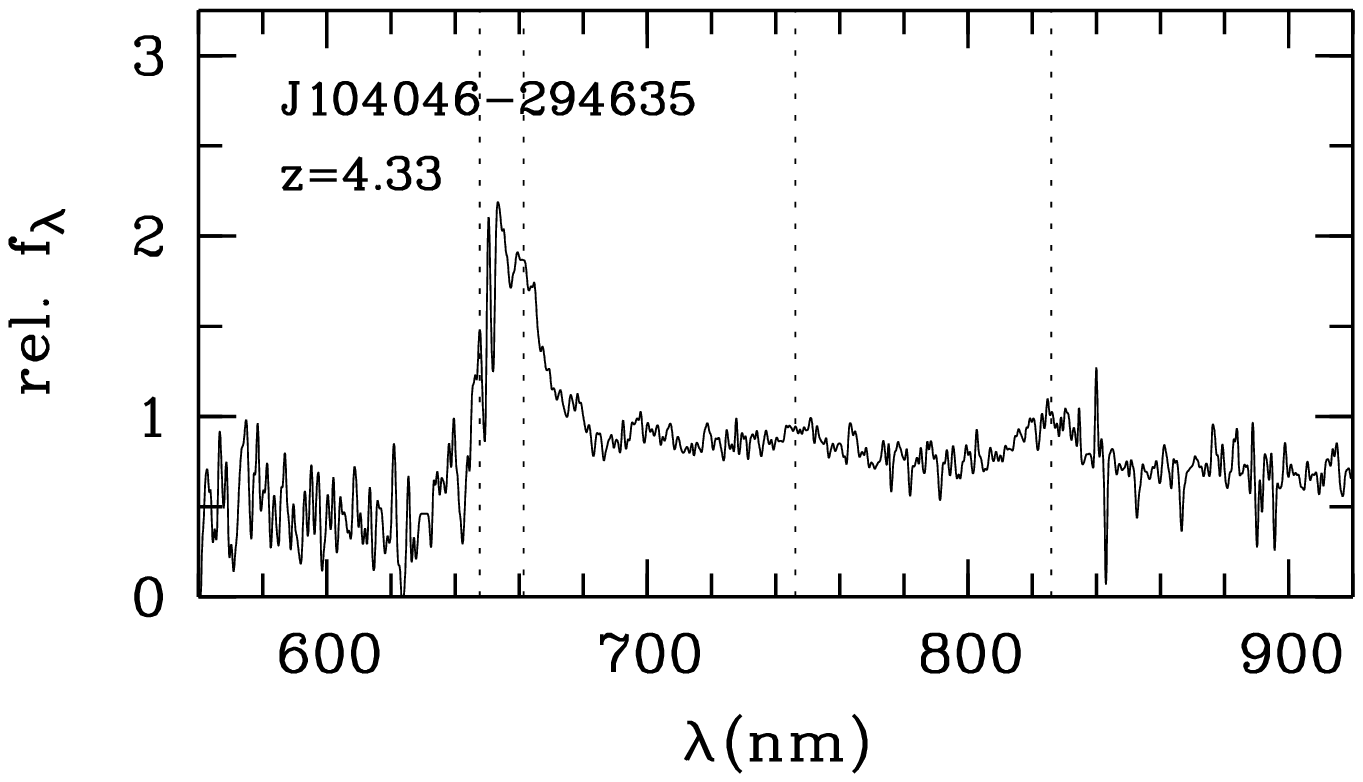}
\includegraphics[angle=270,width=0.32\textwidth,clip=true]{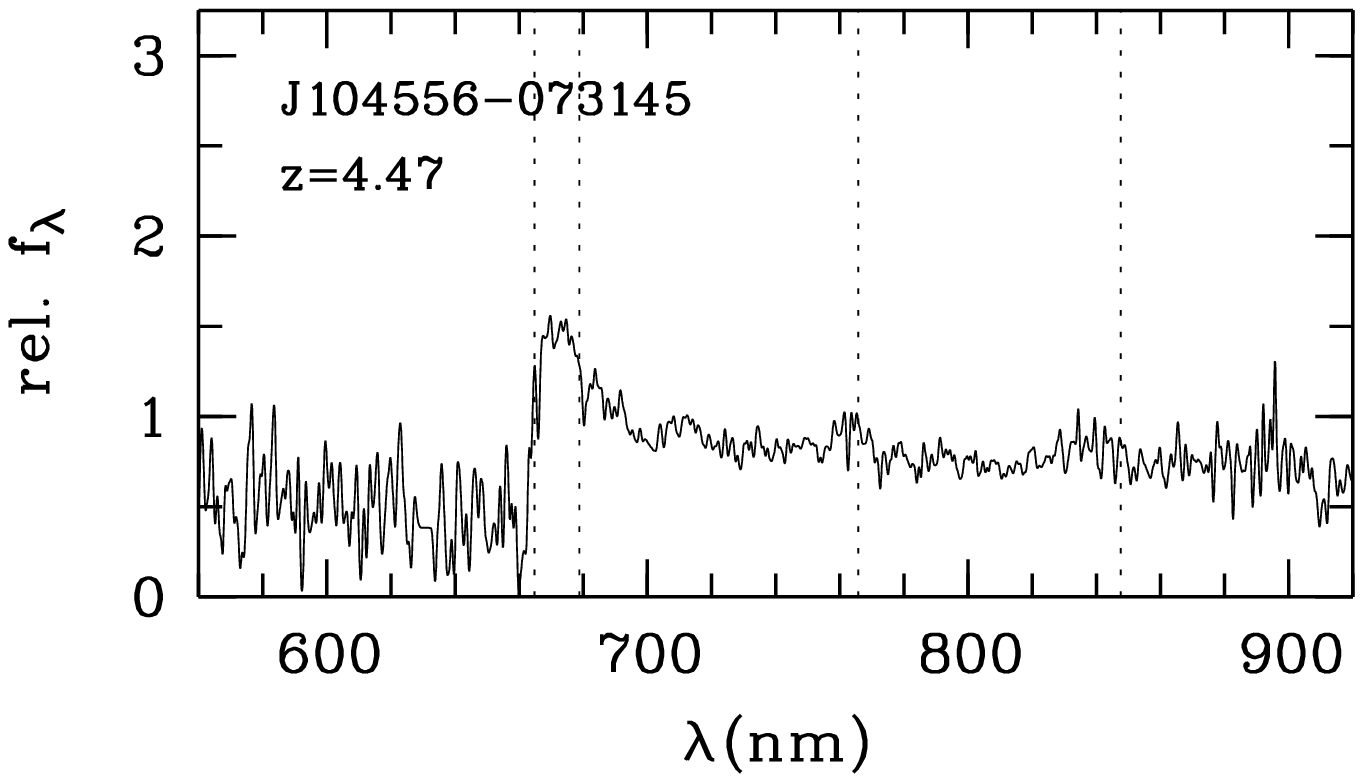}
\includegraphics[angle=270,width=0.32\textwidth,clip=true]{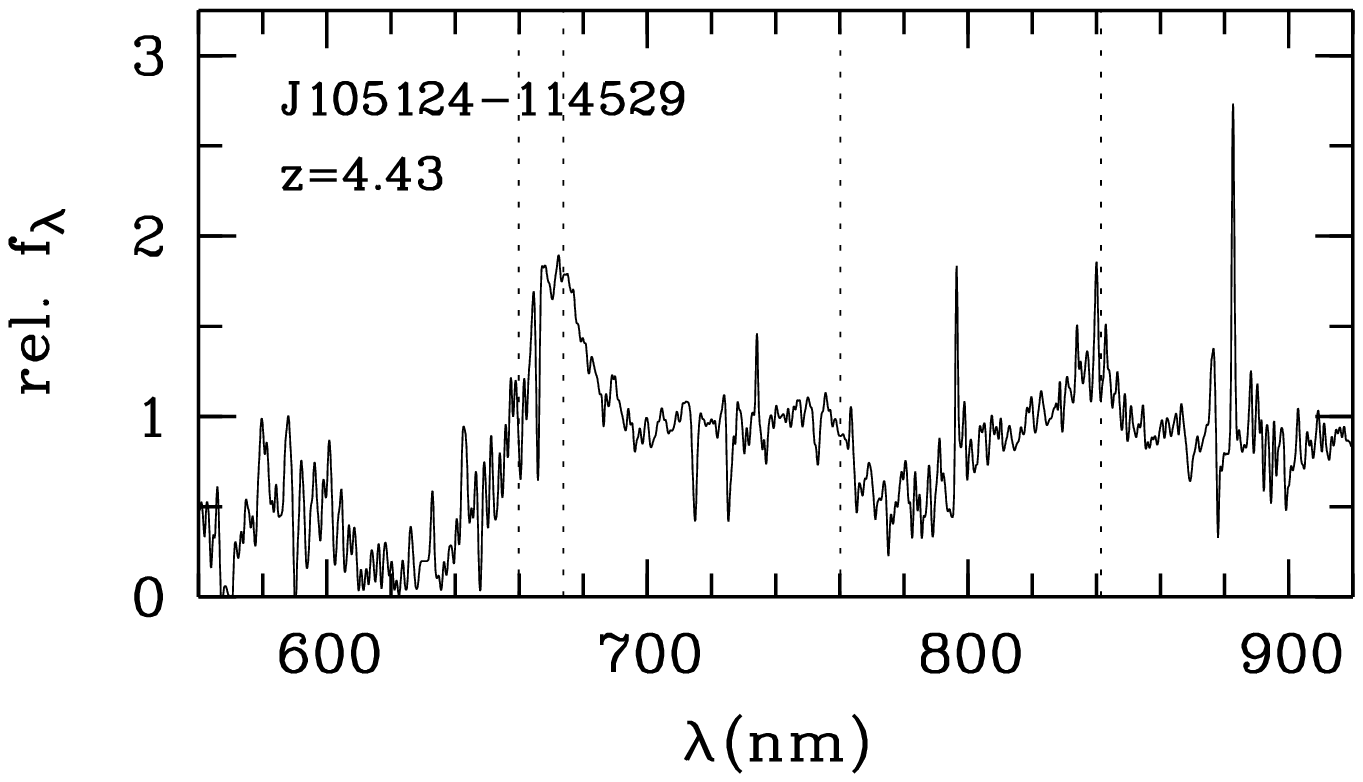}
\includegraphics[angle=270,width=0.32\textwidth,clip=true]{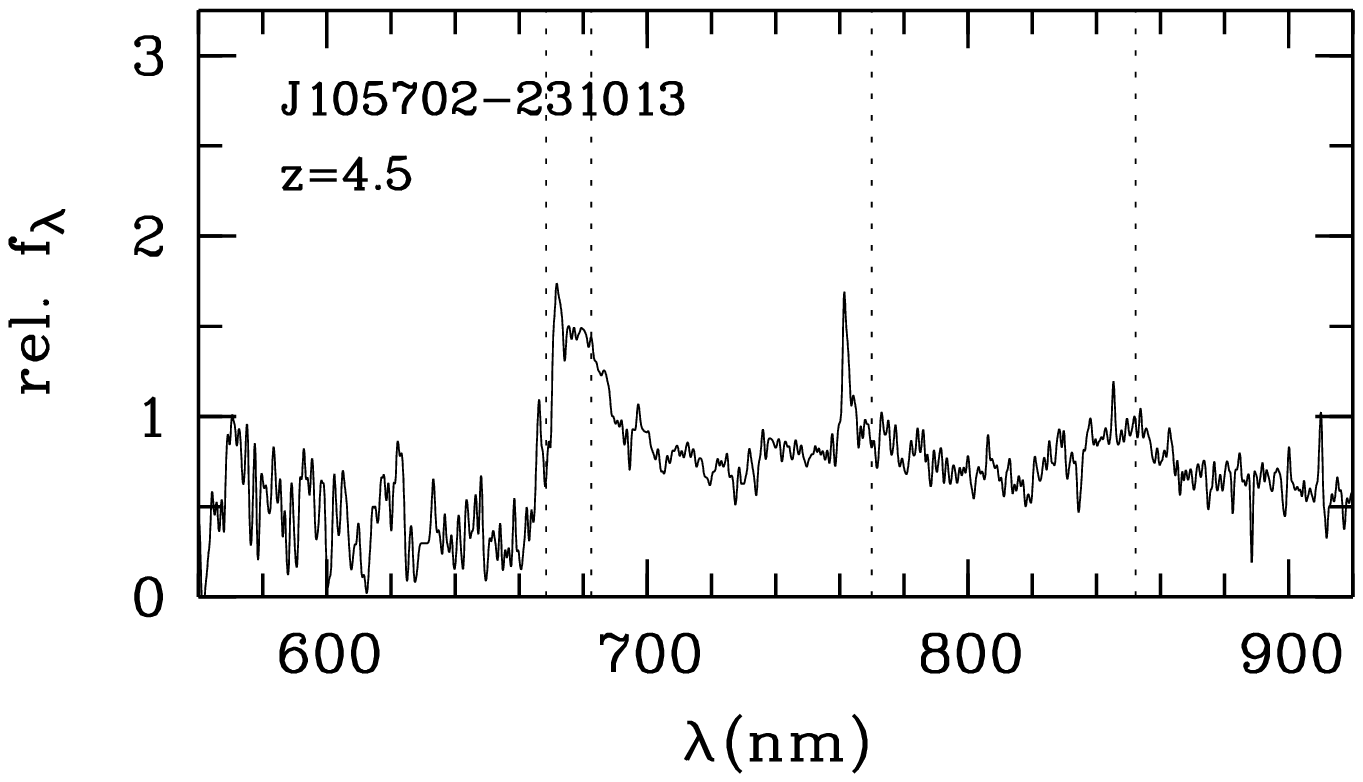}
\includegraphics[angle=270,width=0.32\textwidth,clip=true]{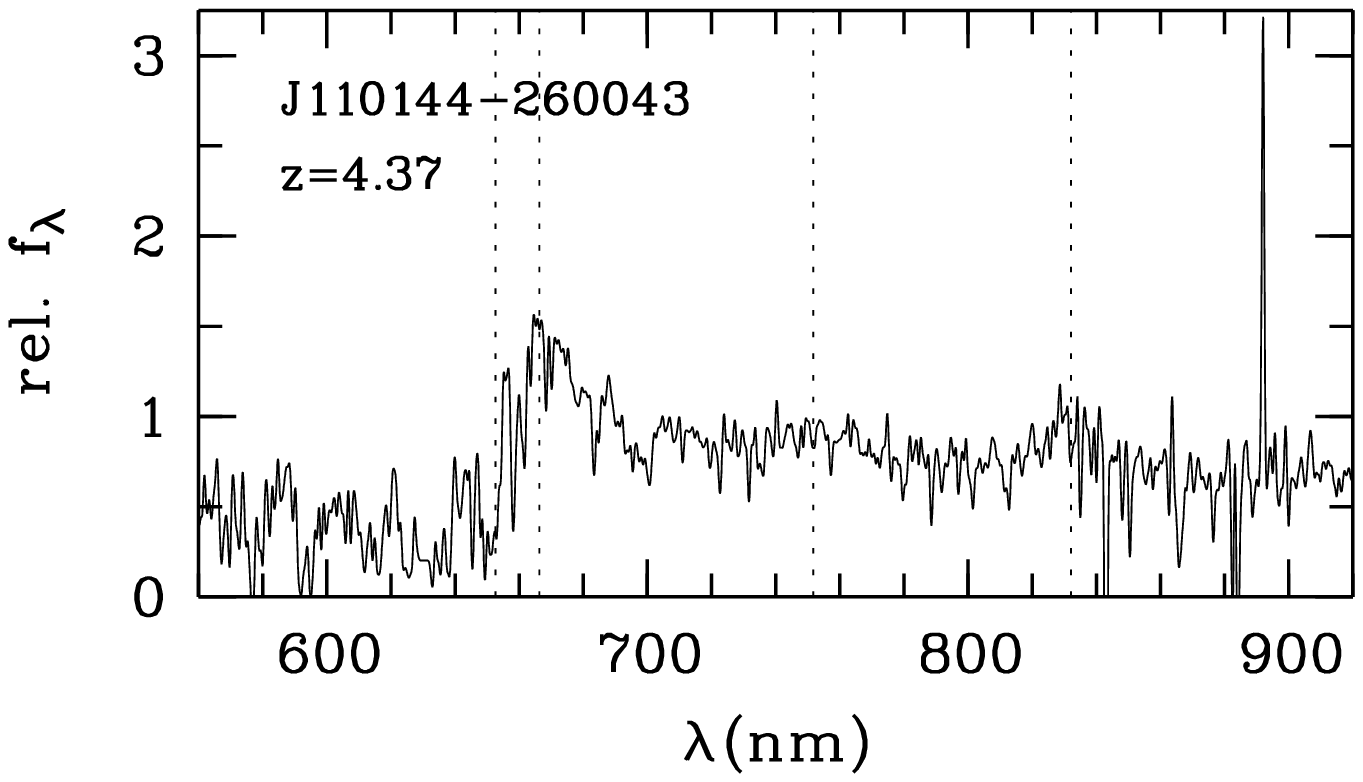}
\includegraphics[angle=270,width=0.32\textwidth,clip=true]{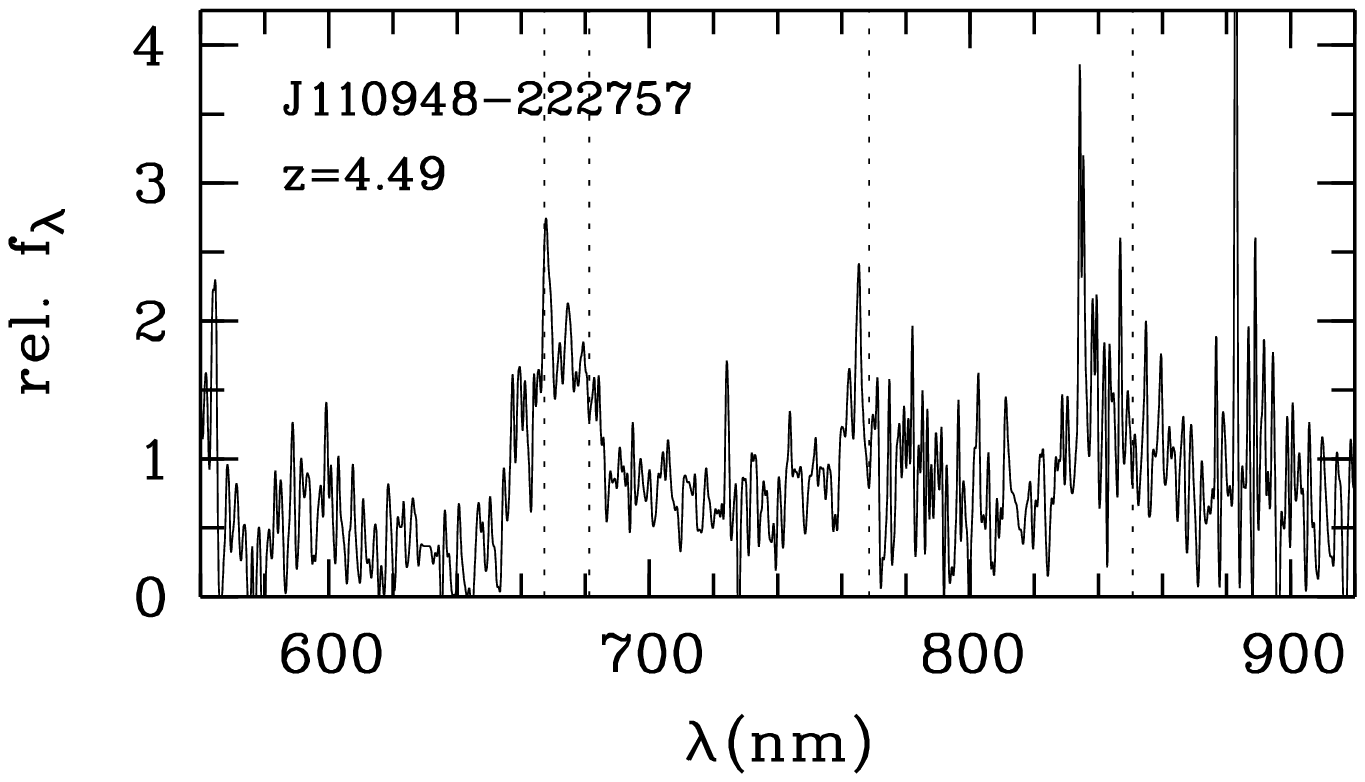}
\includegraphics[angle=270,width=0.32\textwidth,clip=true]{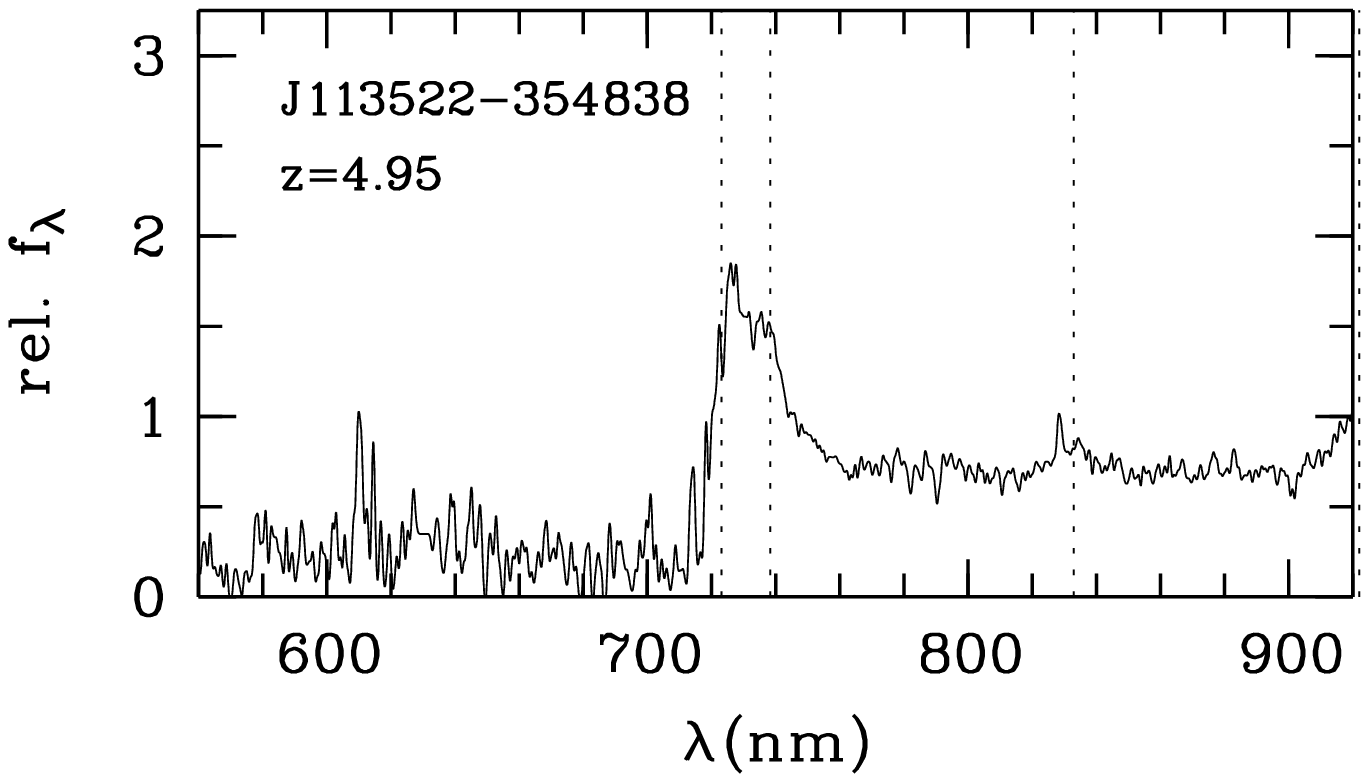}
\includegraphics[angle=270,width=0.32\textwidth,clip=true]{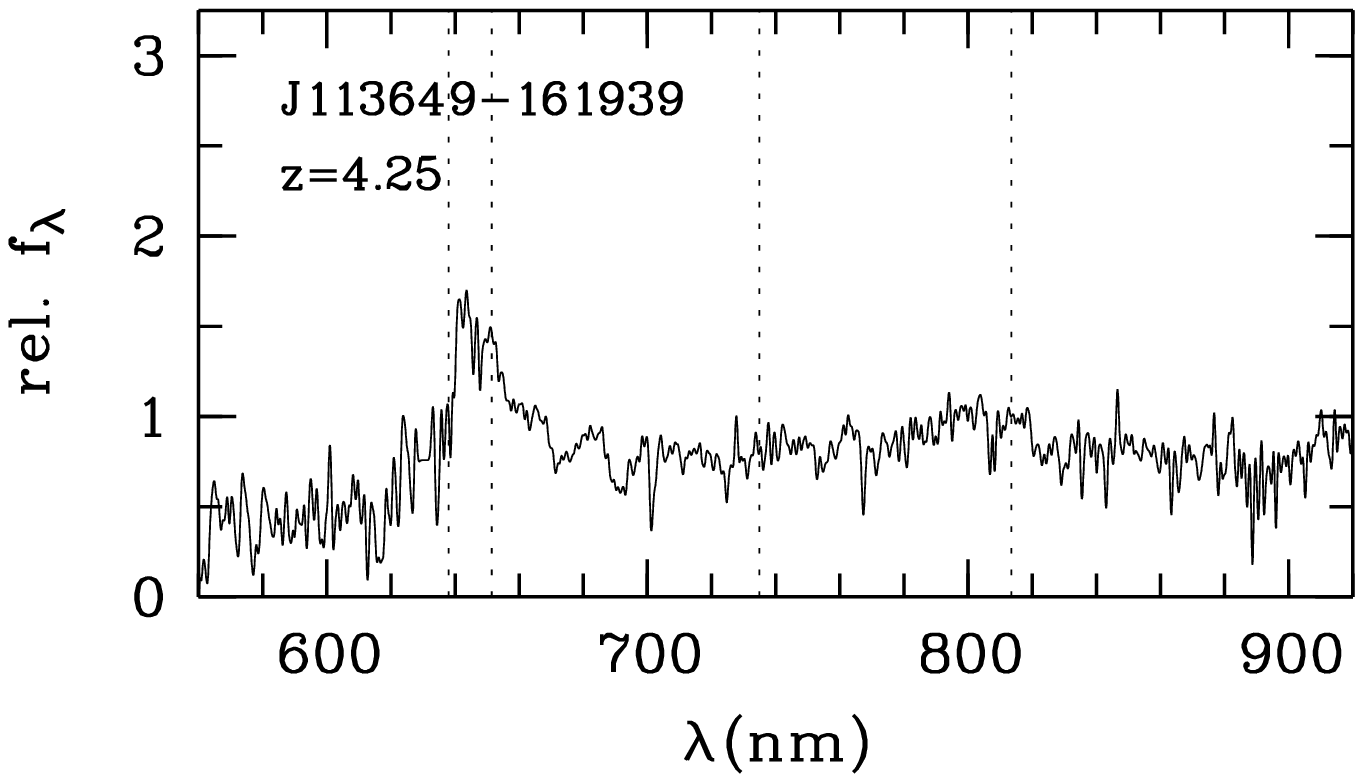}
\includegraphics[angle=270,width=0.32\textwidth,clip=true]{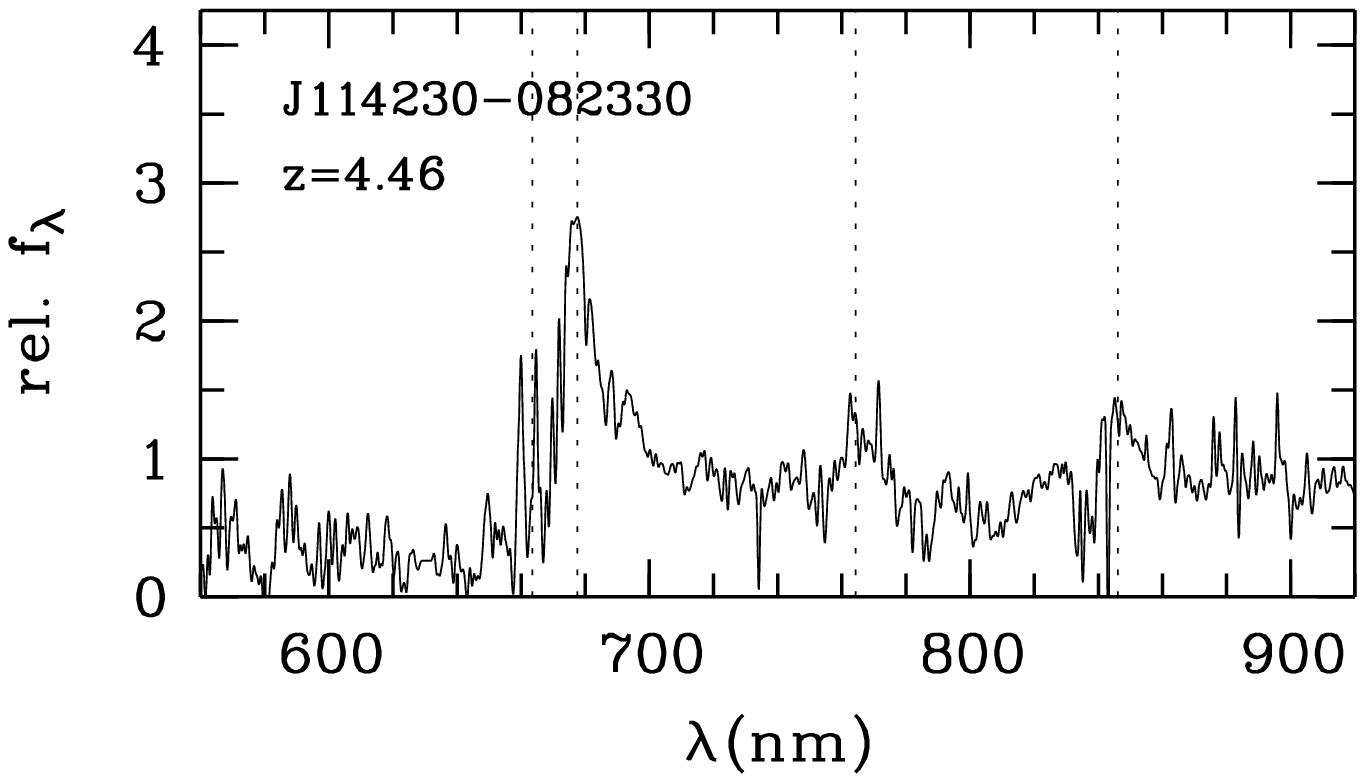}
\includegraphics[angle=270,width=0.32\textwidth,clip=true]{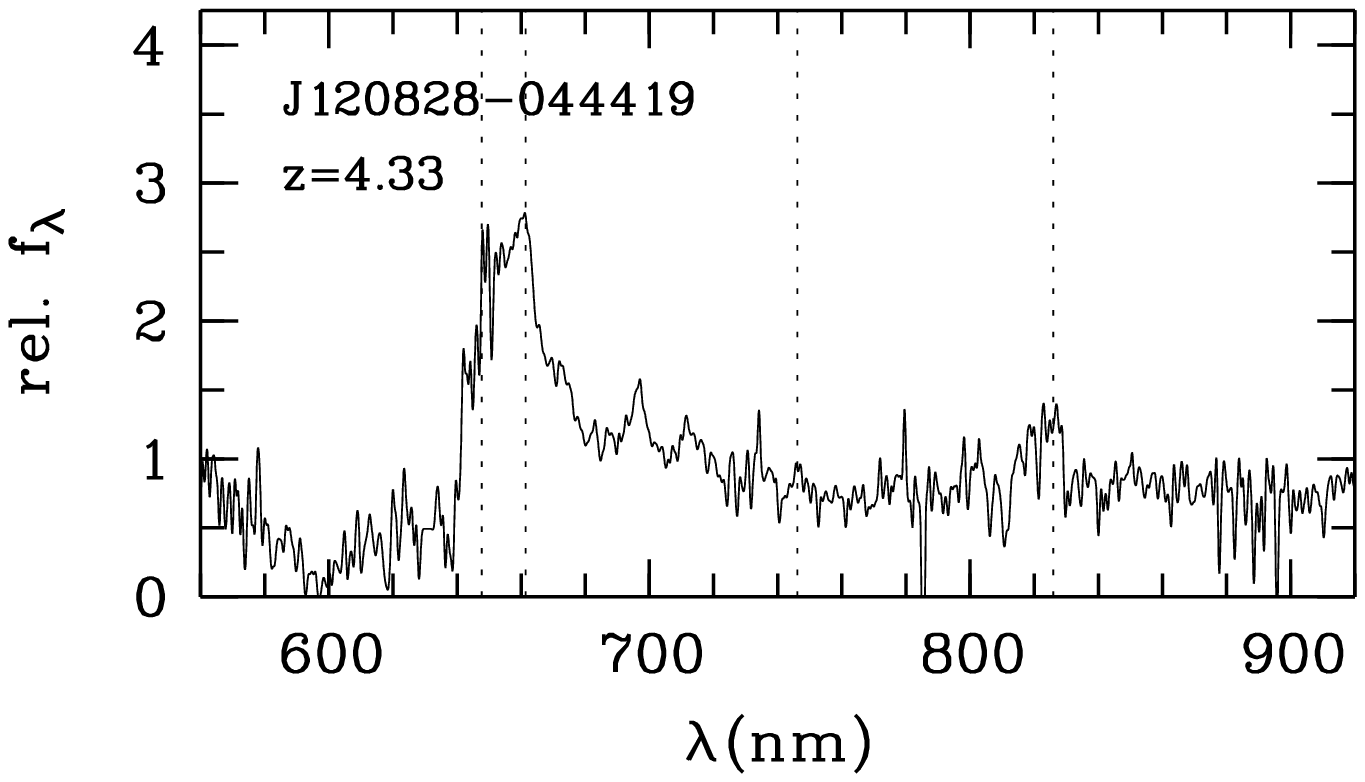}
\includegraphics[angle=270,width=0.32\textwidth,clip=true]{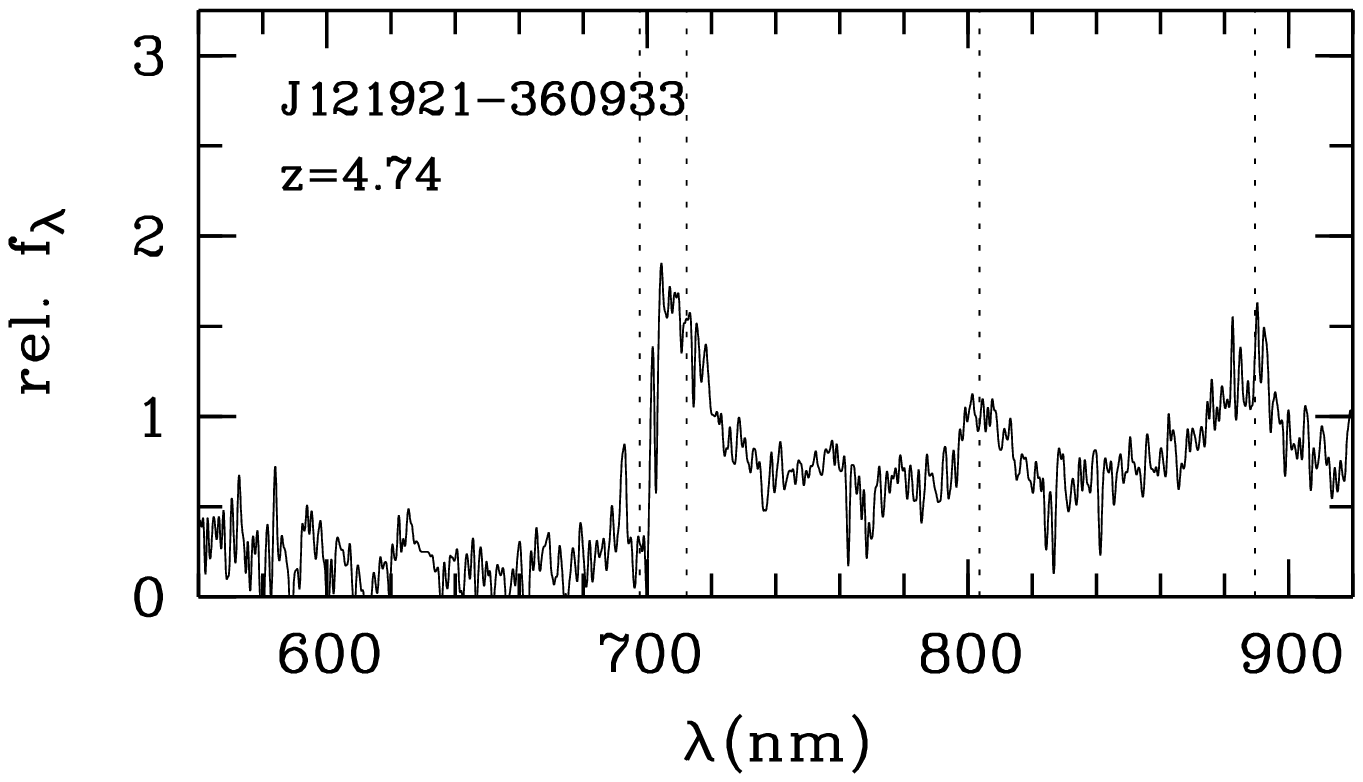}
\includegraphics[angle=270,width=0.32\textwidth,clip=true]{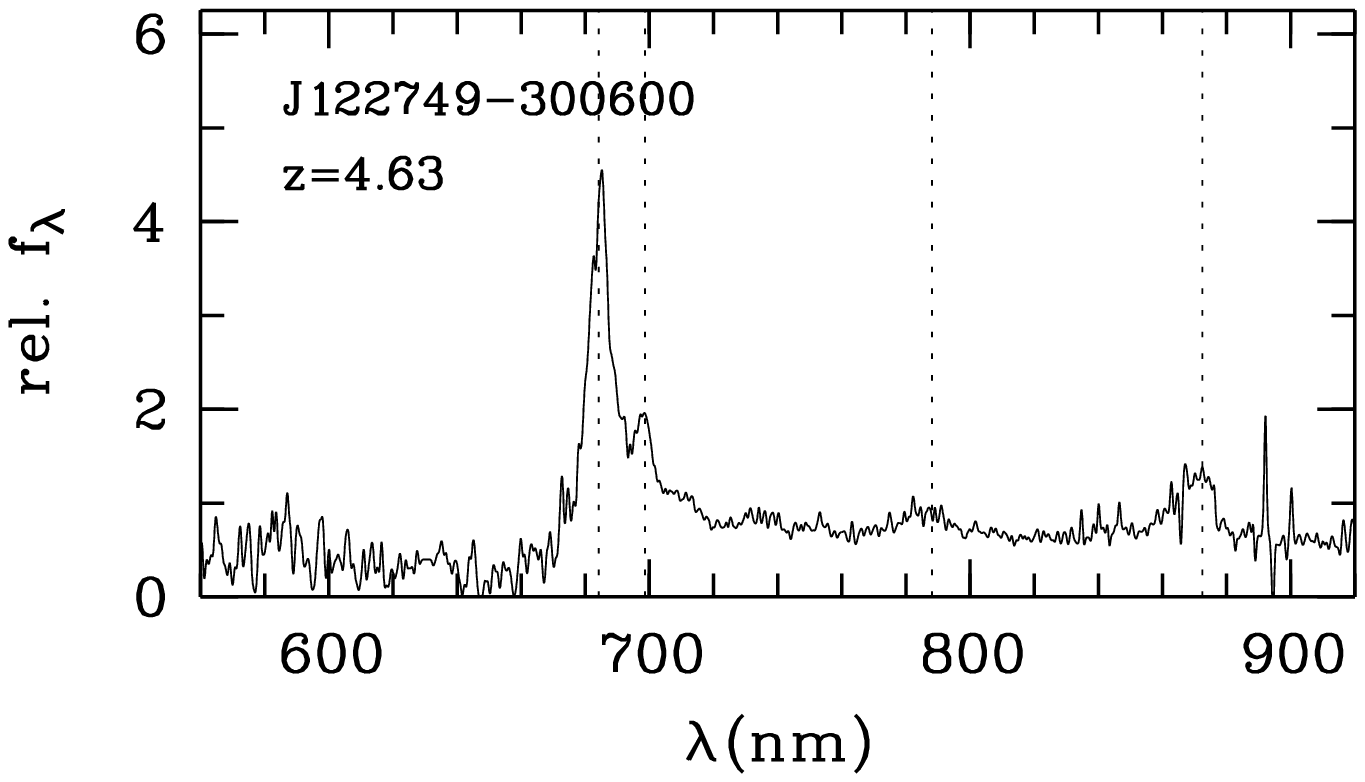}
\includegraphics[angle=270,width=0.32\textwidth,clip=true]{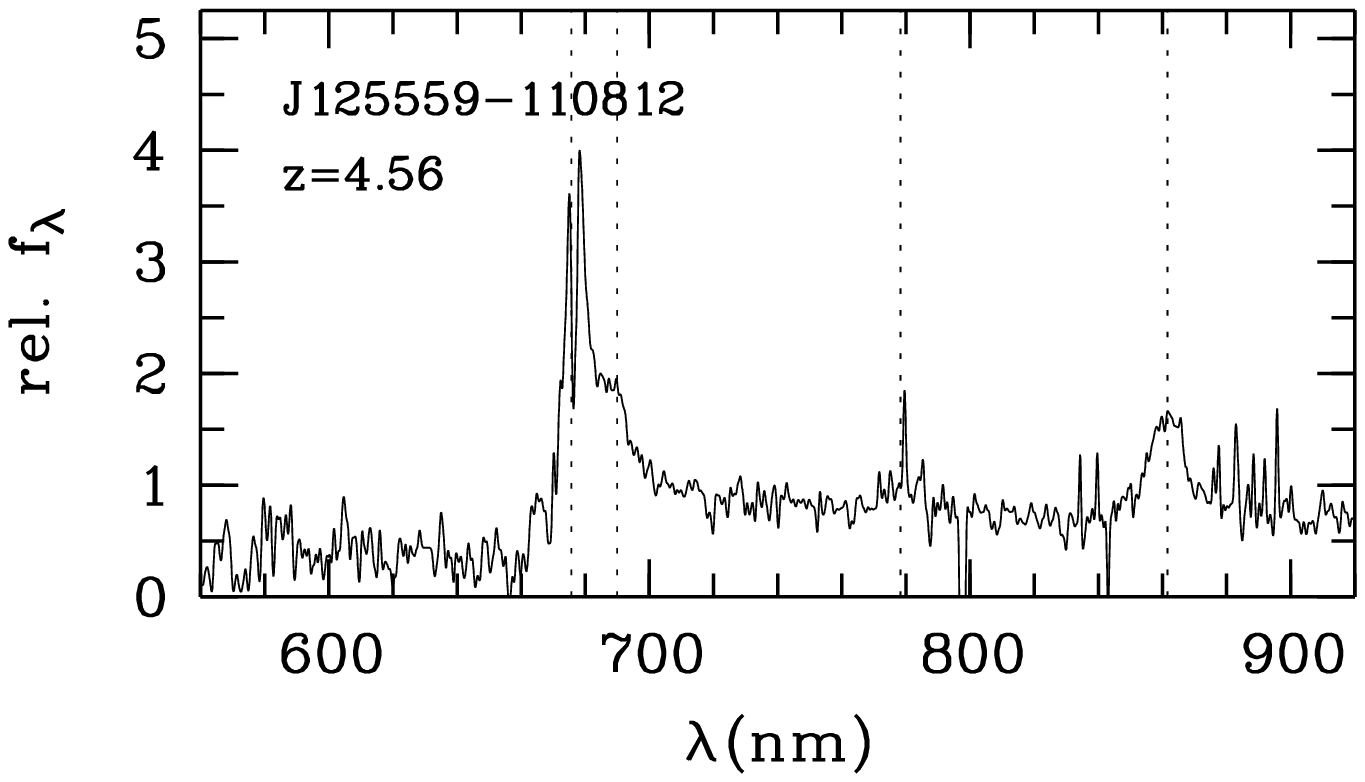}
\includegraphics[angle=270,width=0.32\textwidth,clip=true]{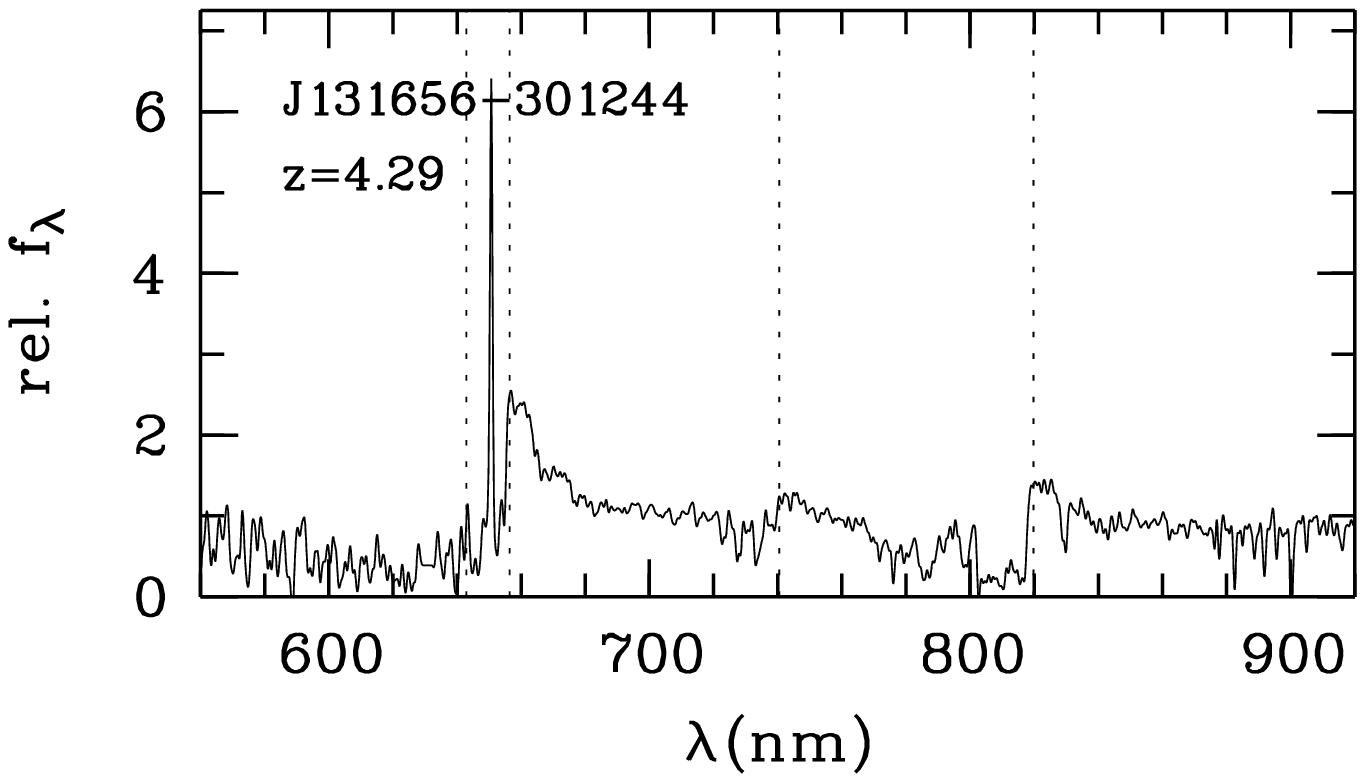}
\includegraphics[angle=270,width=0.32\textwidth,clip=true]{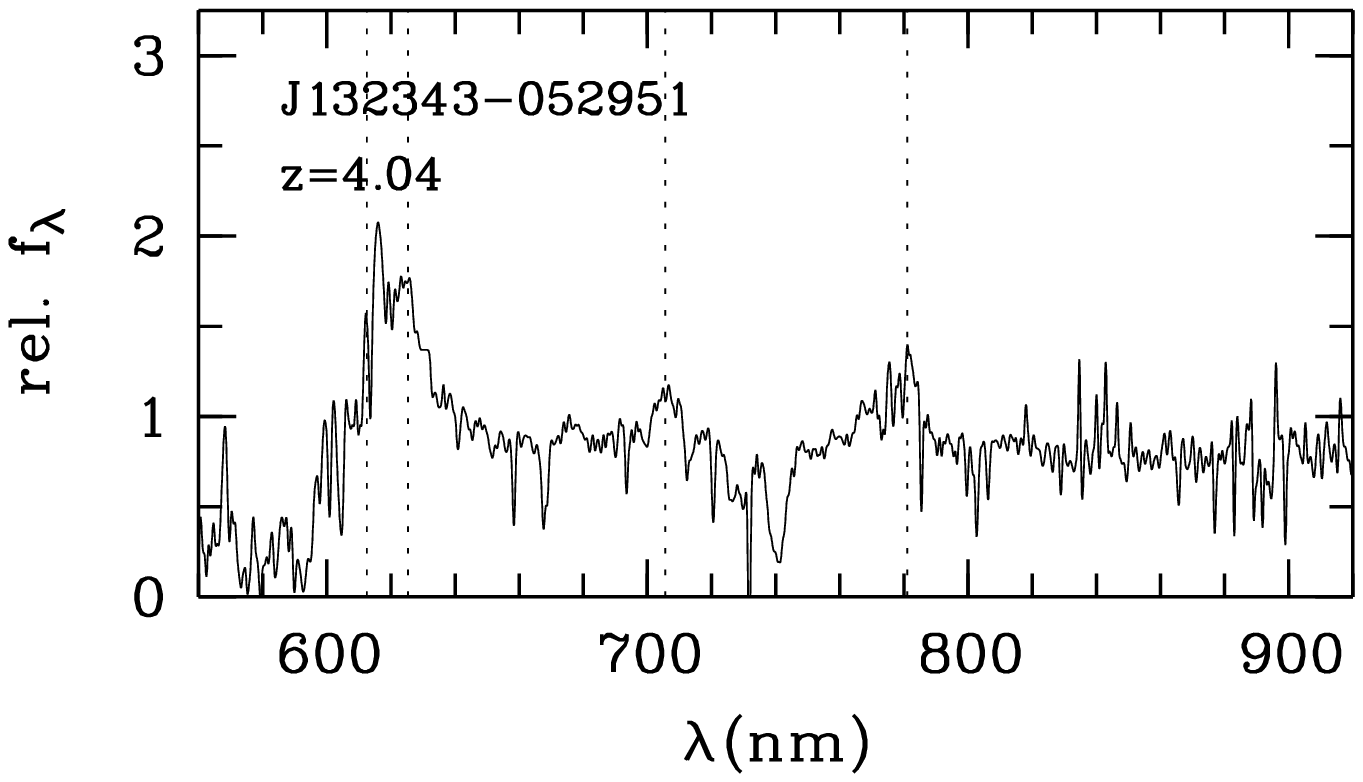}
\includegraphics[angle=270,width=0.32\textwidth,clip=true]{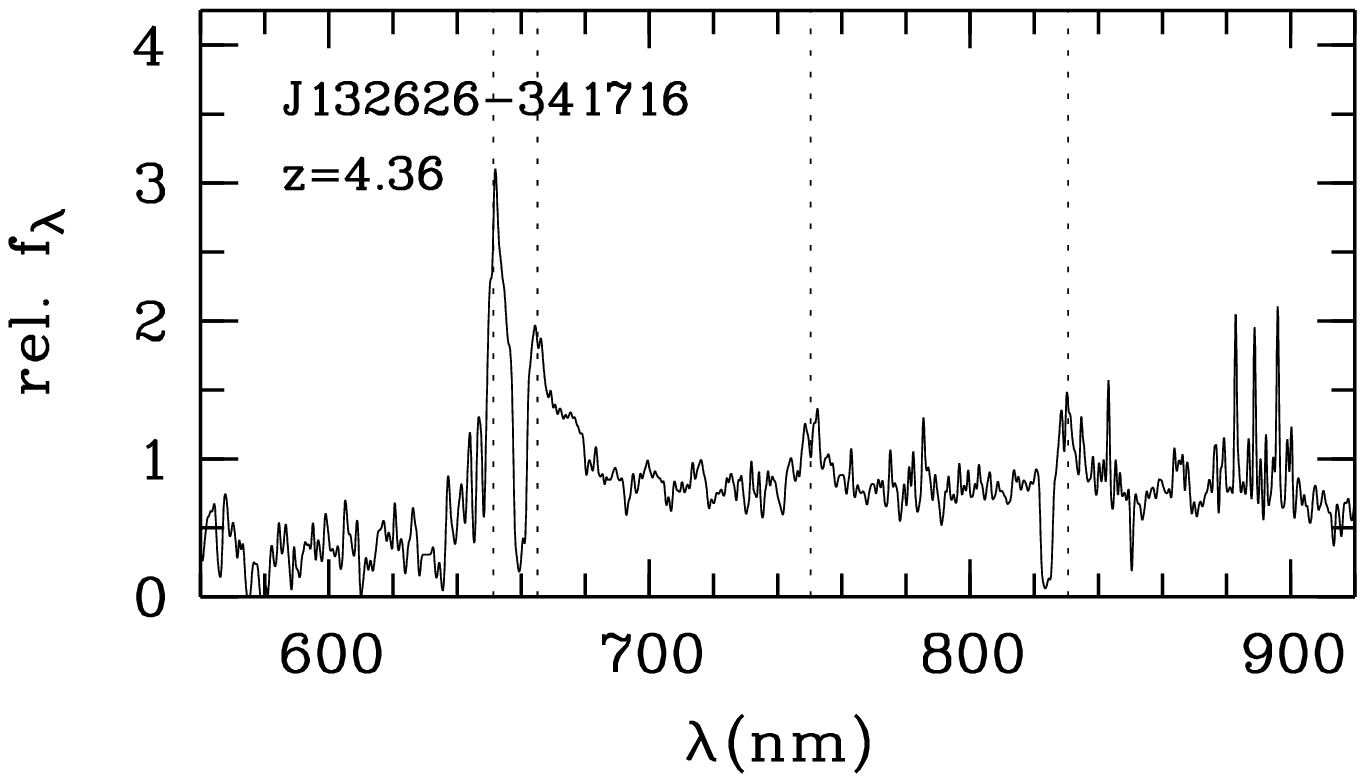}
\caption{Gallery of $3.8>z>5.5$ QSO spectra obtained in this work, ordered by RA, page 4.
\label{gallery4}}
\end{center}
\end{figure*}

\begin{figure*}
\begin{center}
\includegraphics[angle=270,width=0.32\textwidth,clip=true]{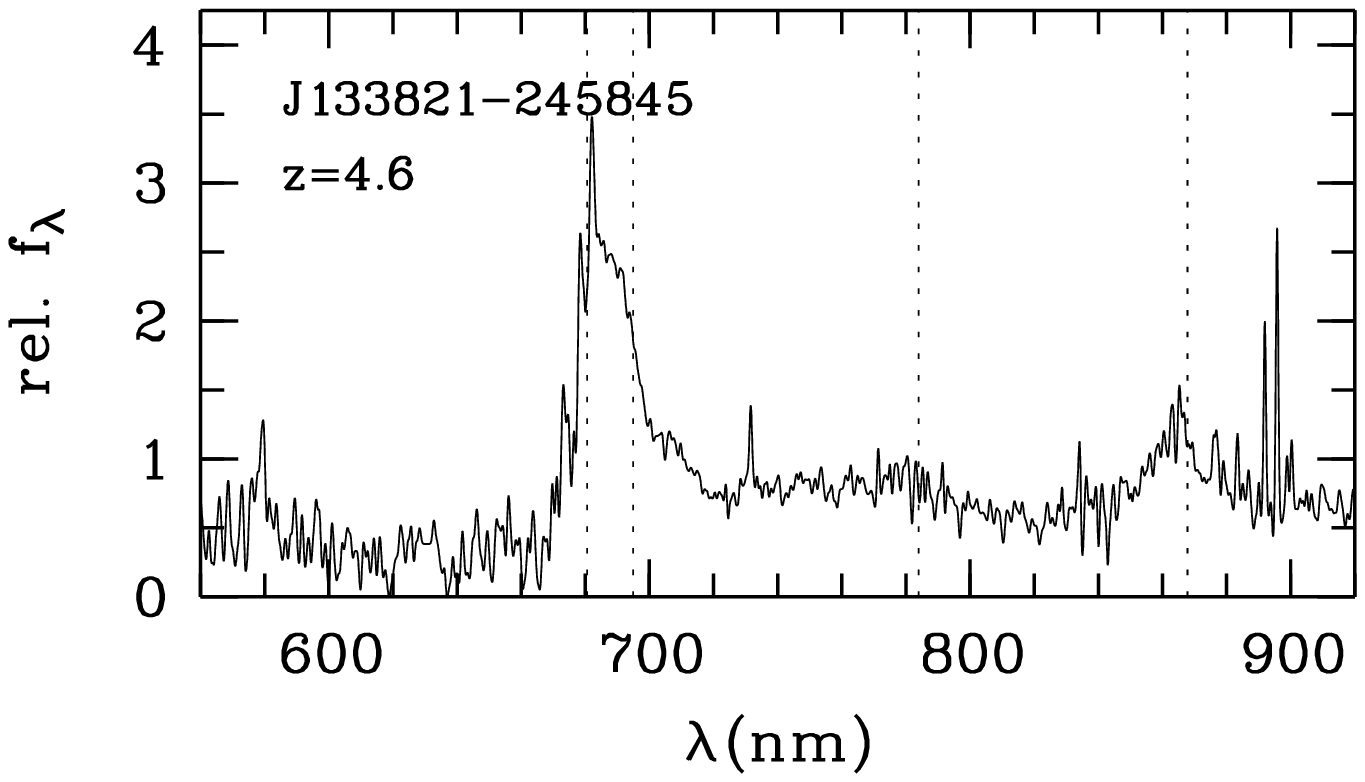}
\includegraphics[angle=270,width=0.32\textwidth,clip=true]{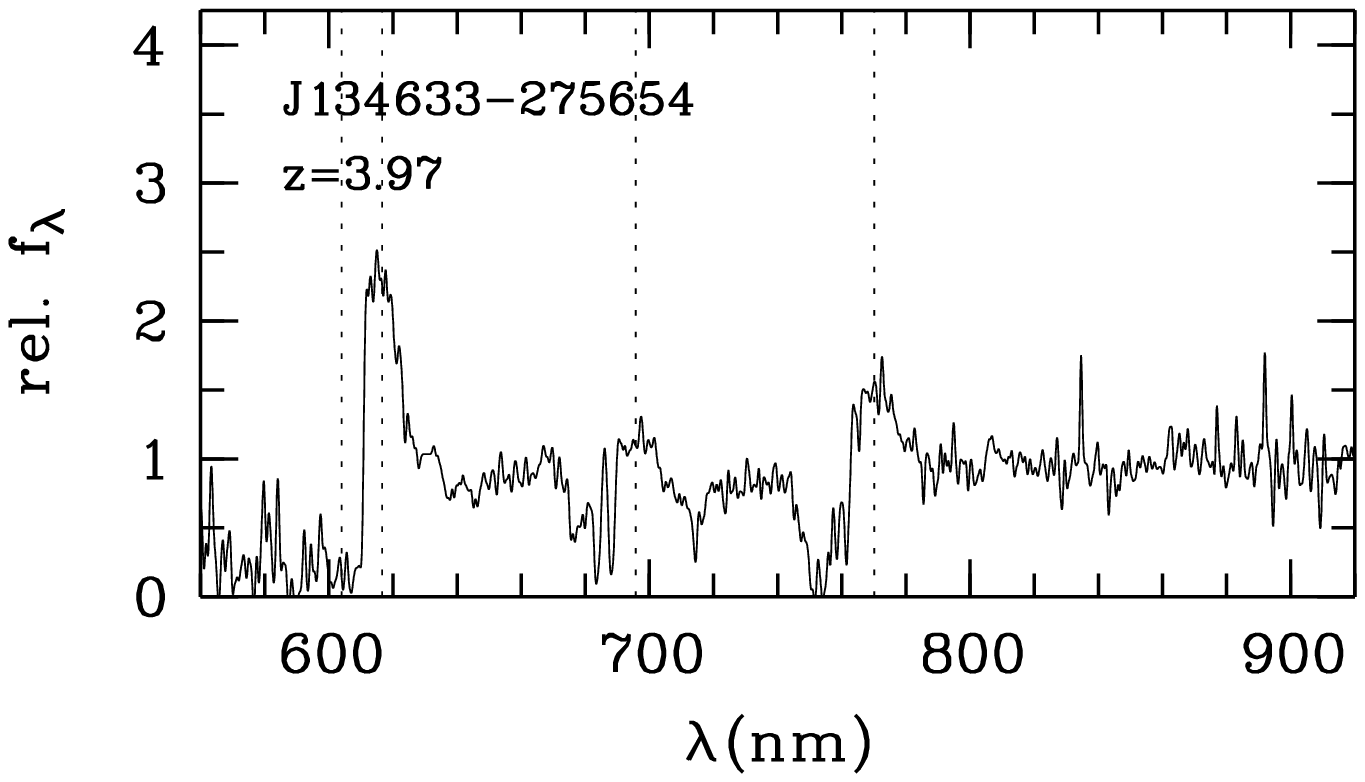}
\includegraphics[angle=270,width=0.32\textwidth,clip=true]{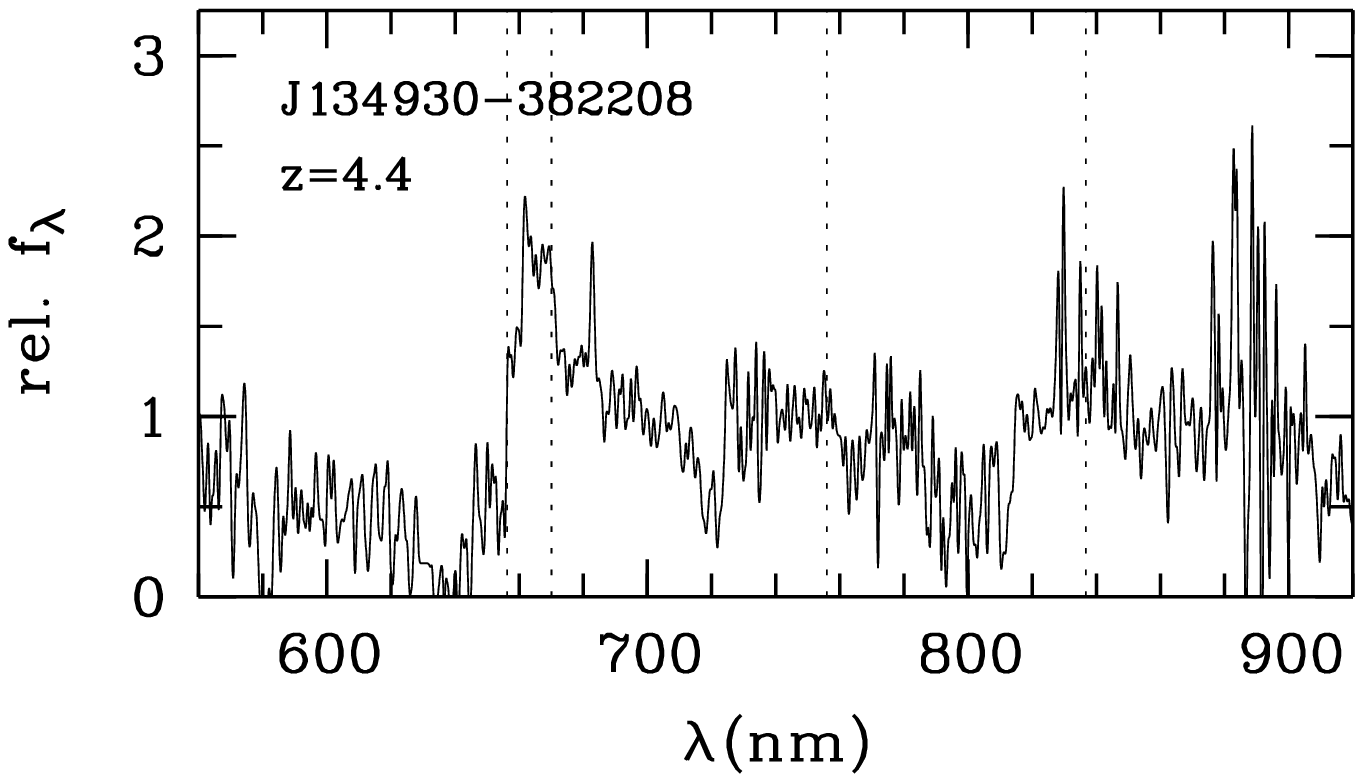}
\includegraphics[angle=270,width=0.32\textwidth,clip=true]{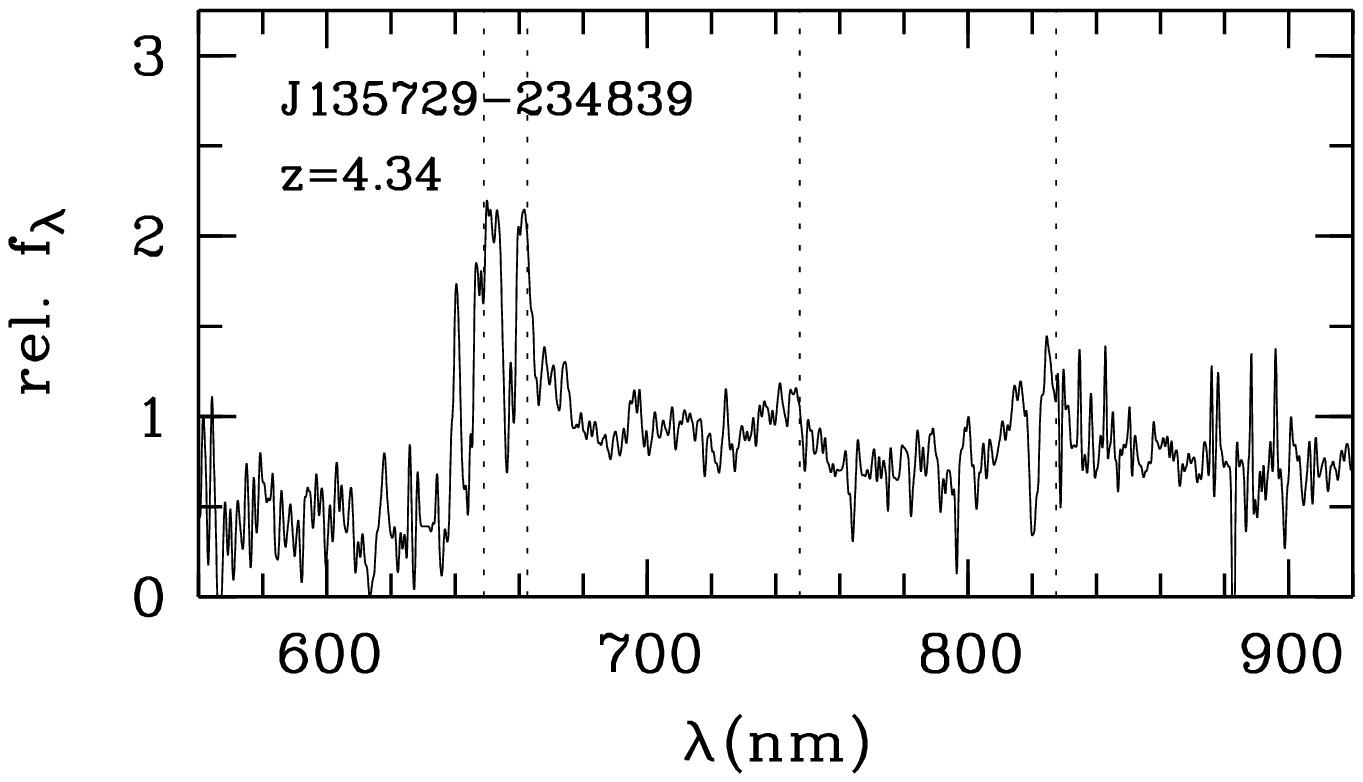}
\includegraphics[angle=270,width=0.32\textwidth,clip=true]{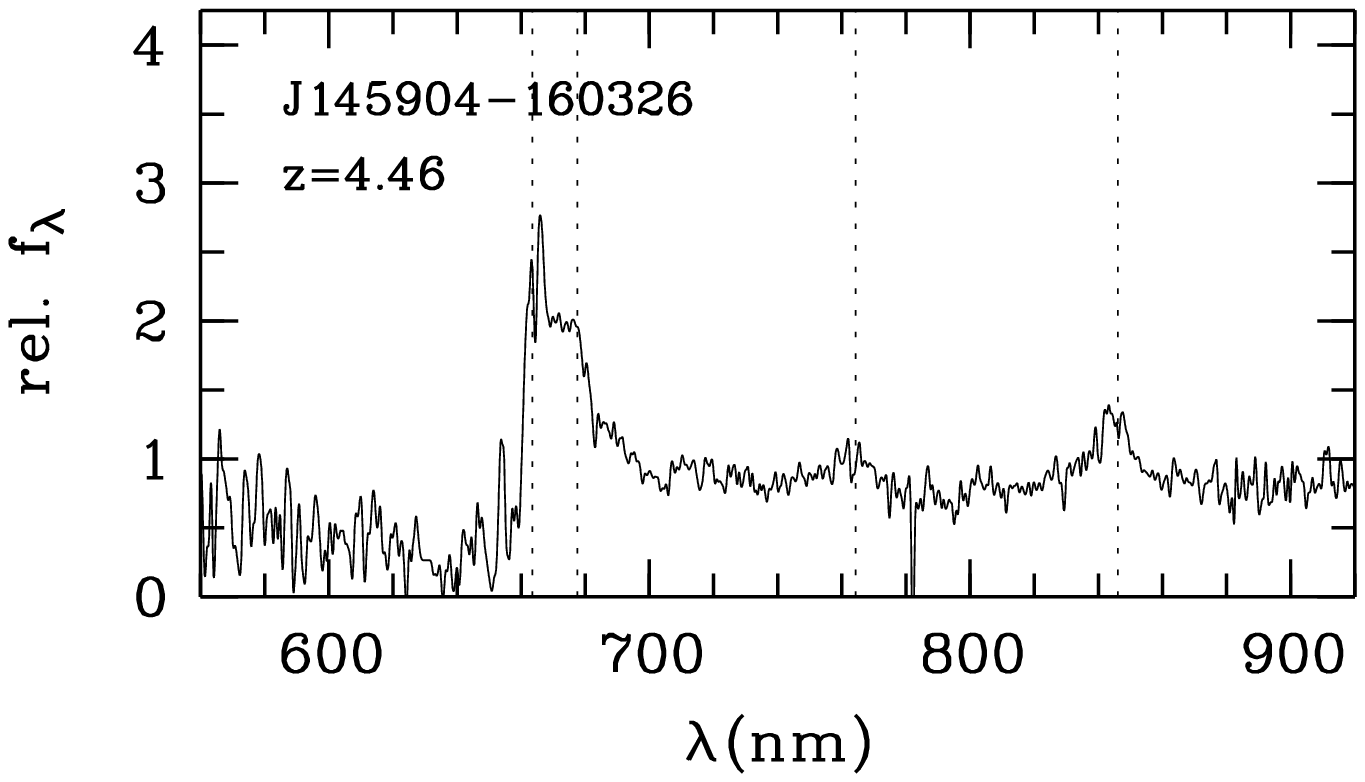}
\includegraphics[angle=270,width=0.32\textwidth,clip=true]{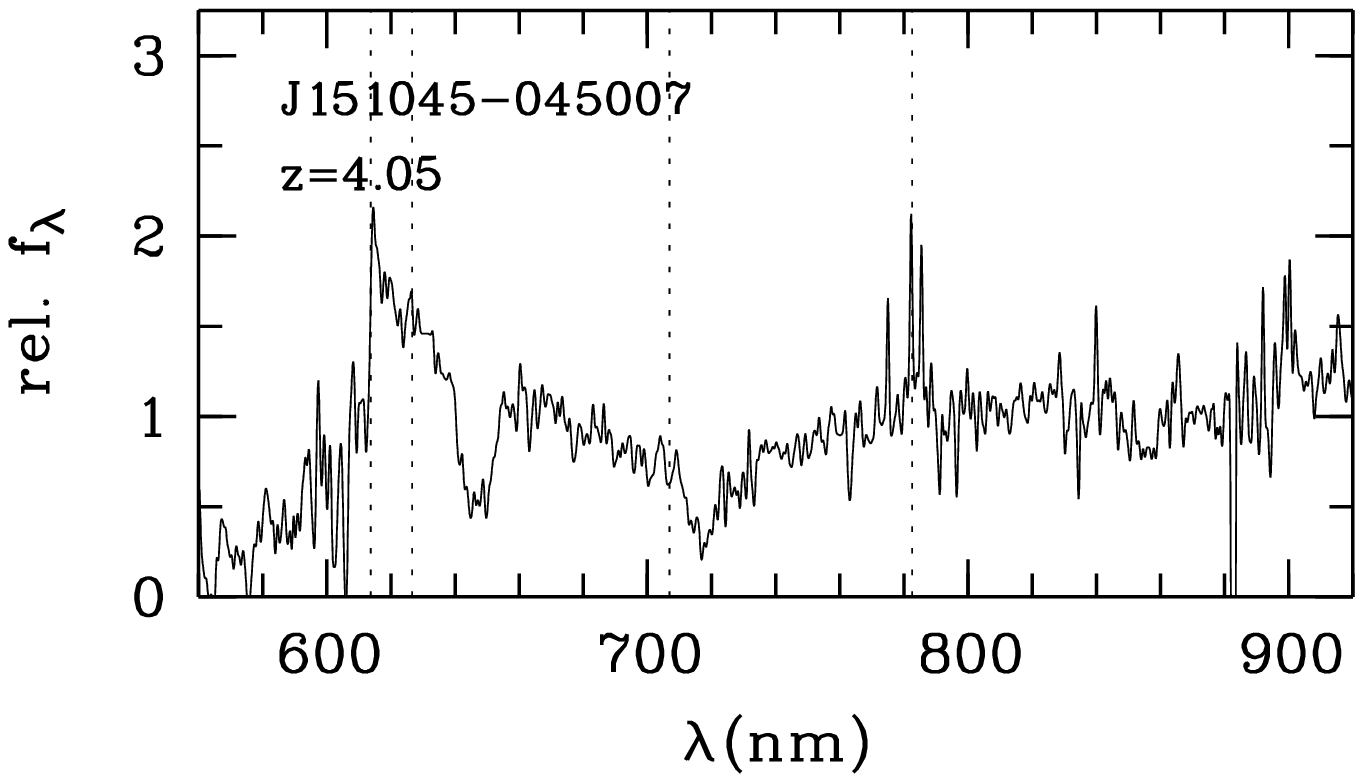}
\includegraphics[angle=270,width=0.32\textwidth,clip=true]{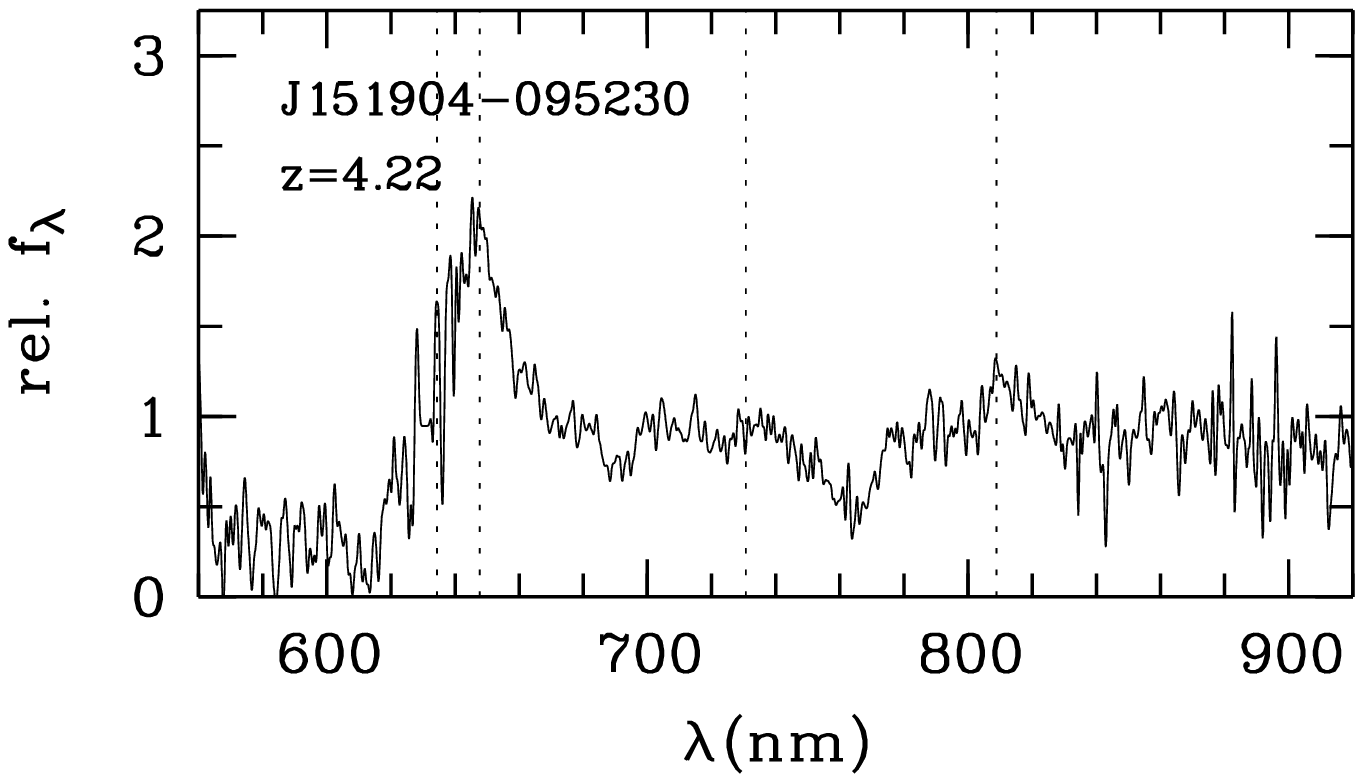}
\includegraphics[angle=270,width=0.32\textwidth,clip=true]{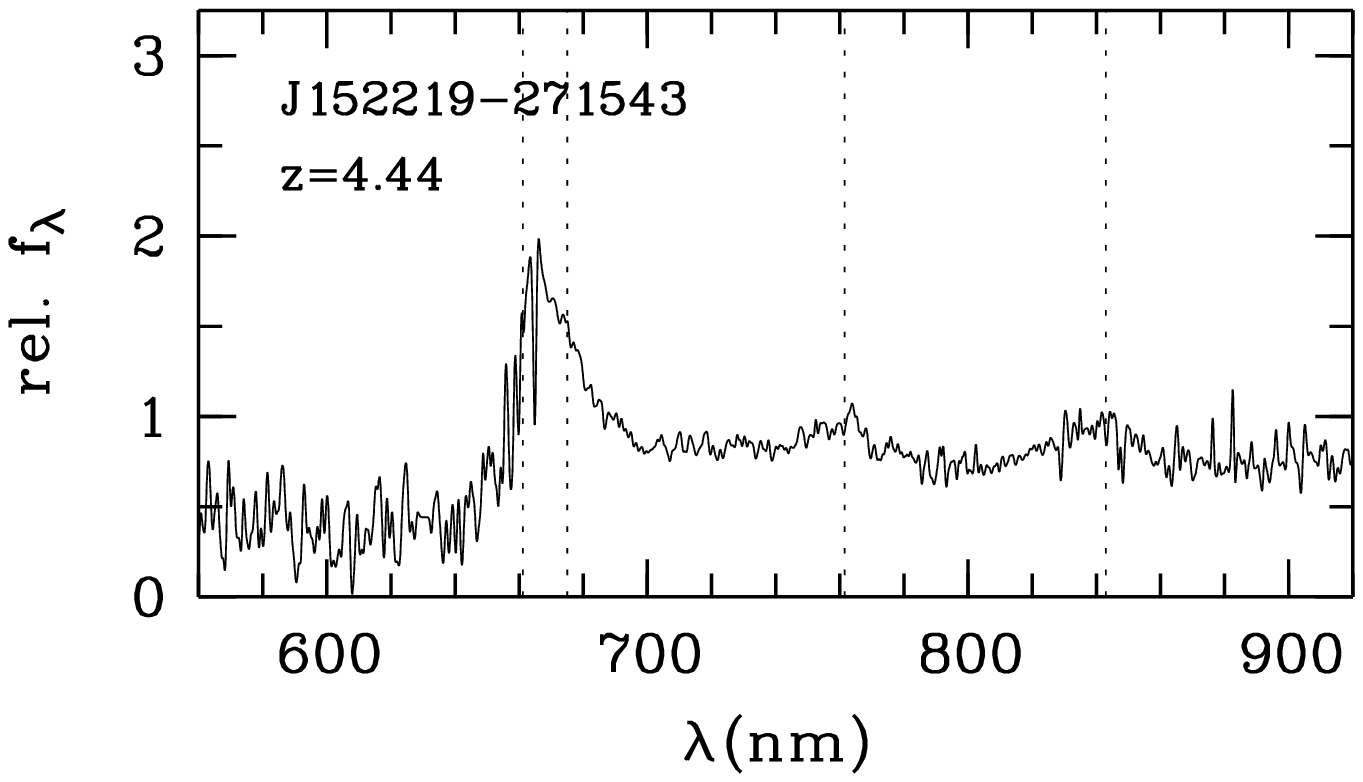}
\includegraphics[angle=270,width=0.32\textwidth,clip=true]{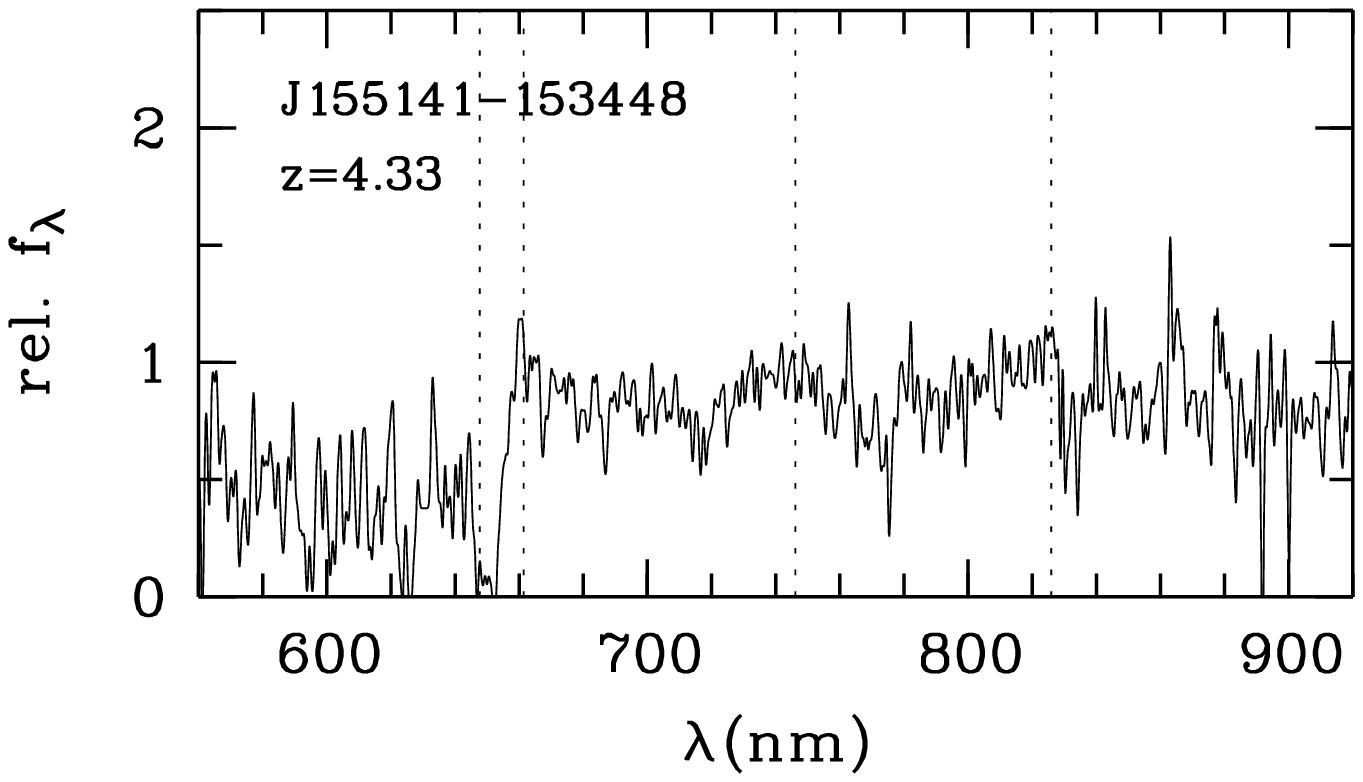}
\includegraphics[angle=270,width=0.32\textwidth,clip=true]{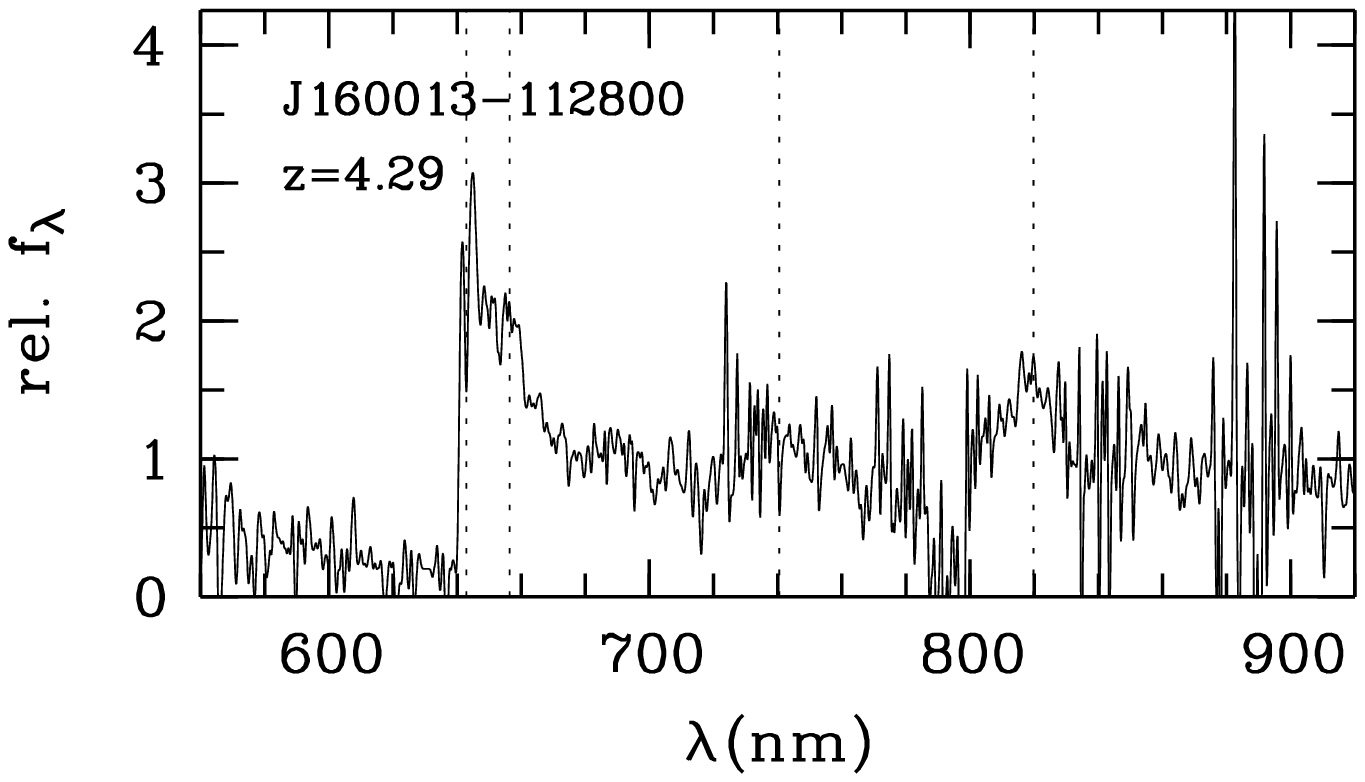}
\includegraphics[angle=270,width=0.32\textwidth,clip=true]{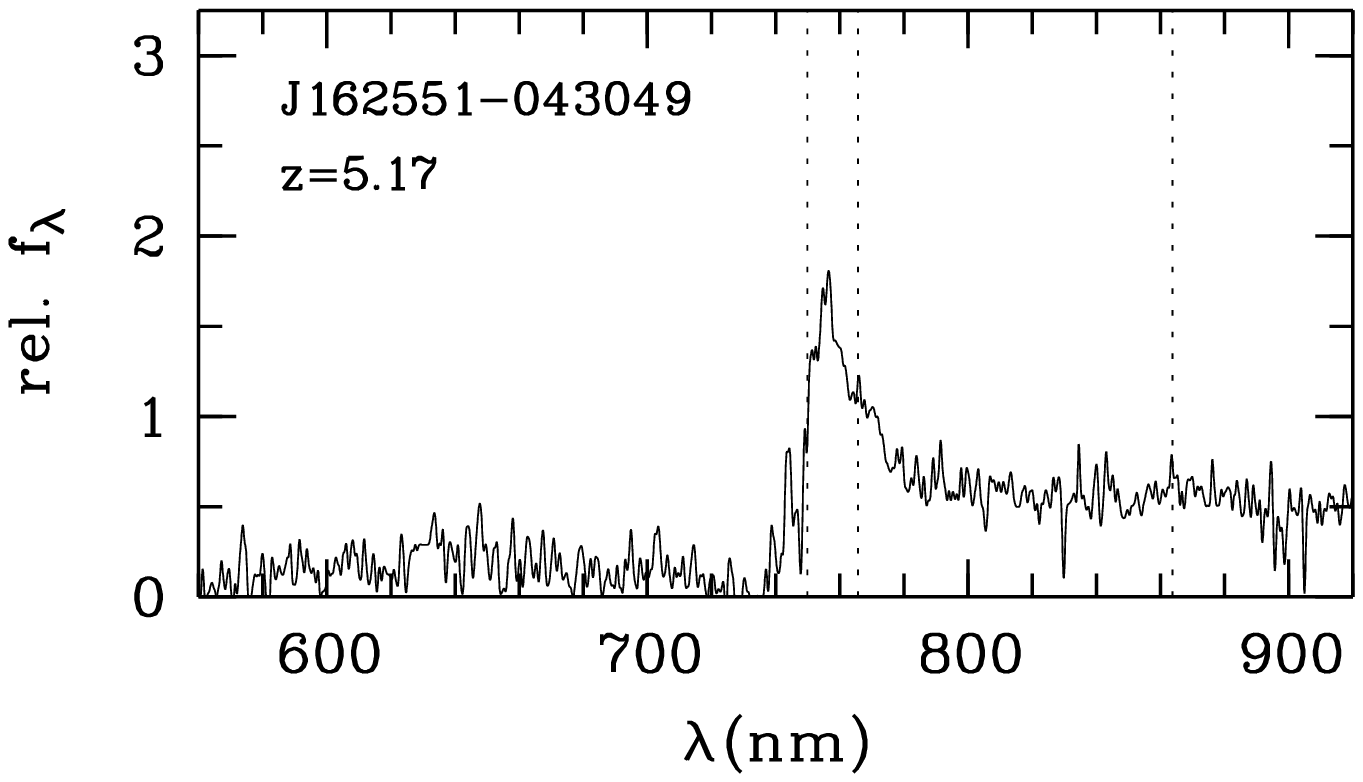}
\includegraphics[angle=270,width=0.32\textwidth,clip=true]{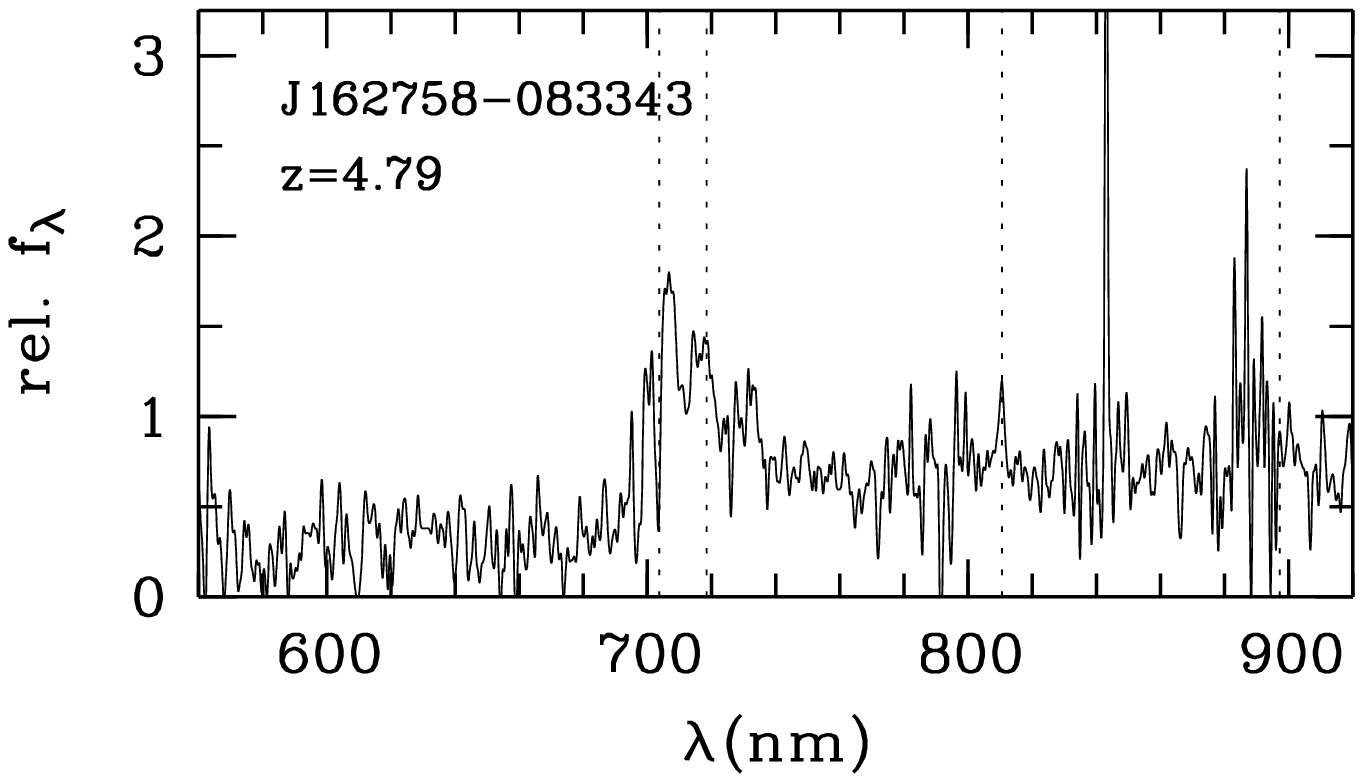}
\includegraphics[angle=270,width=0.32\textwidth,clip=true]{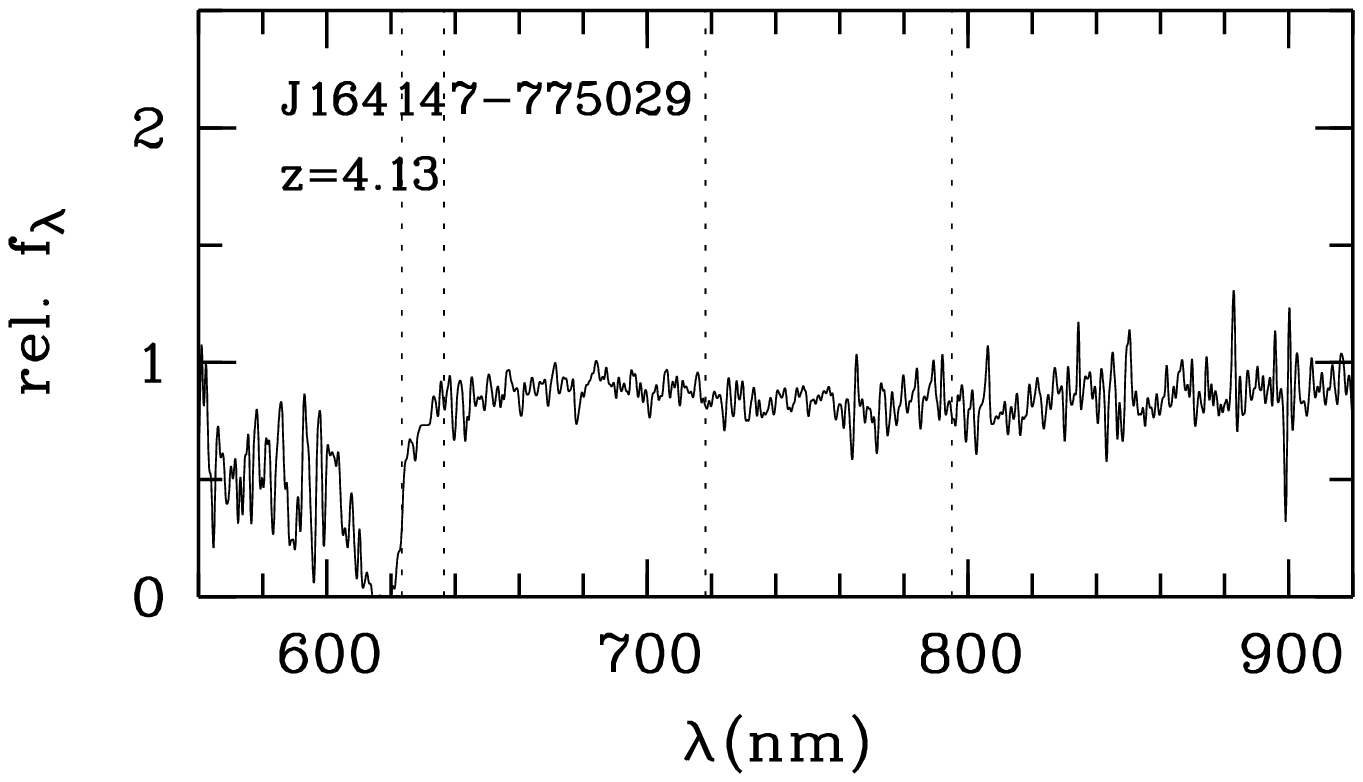}
\includegraphics[angle=270,width=0.32\textwidth,clip=true]{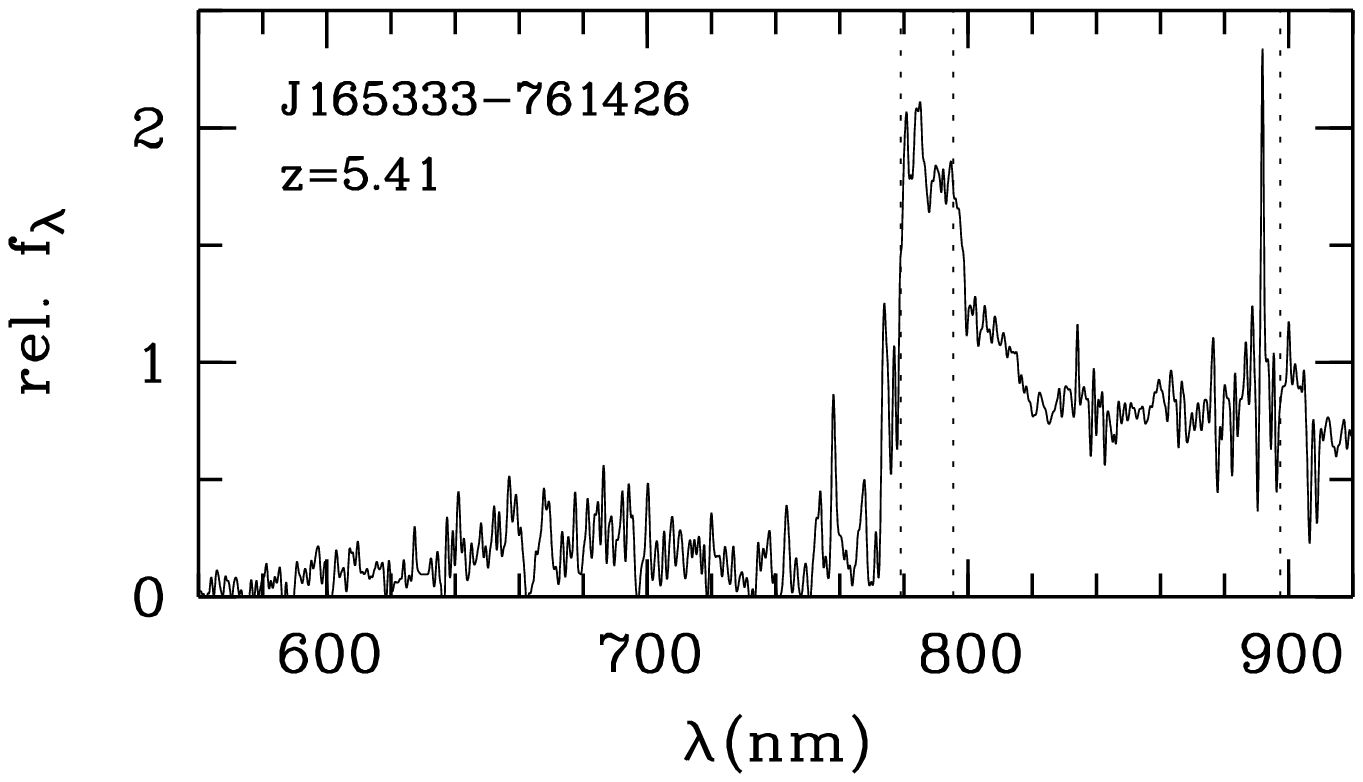}
\includegraphics[angle=270,width=0.32\textwidth,clip=true]{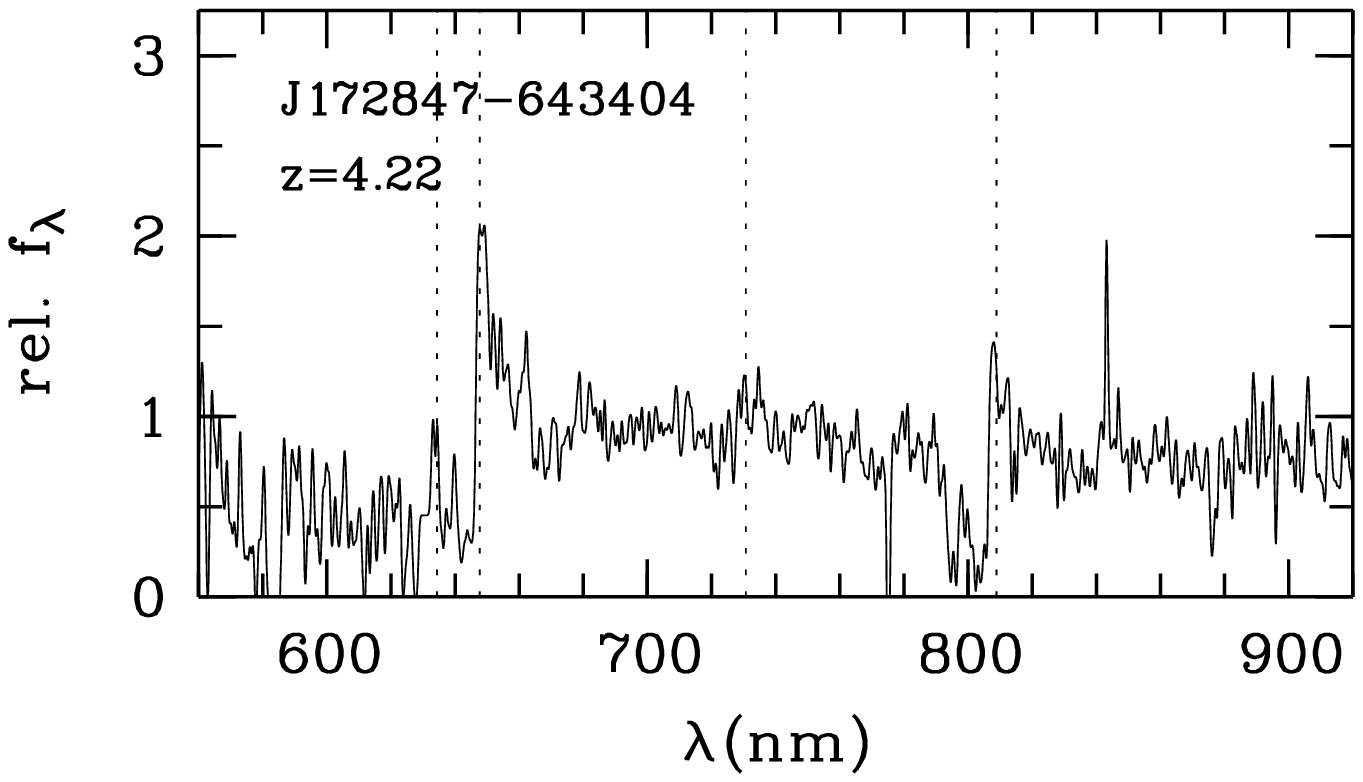}
\includegraphics[angle=270,width=0.32\textwidth,clip=true]{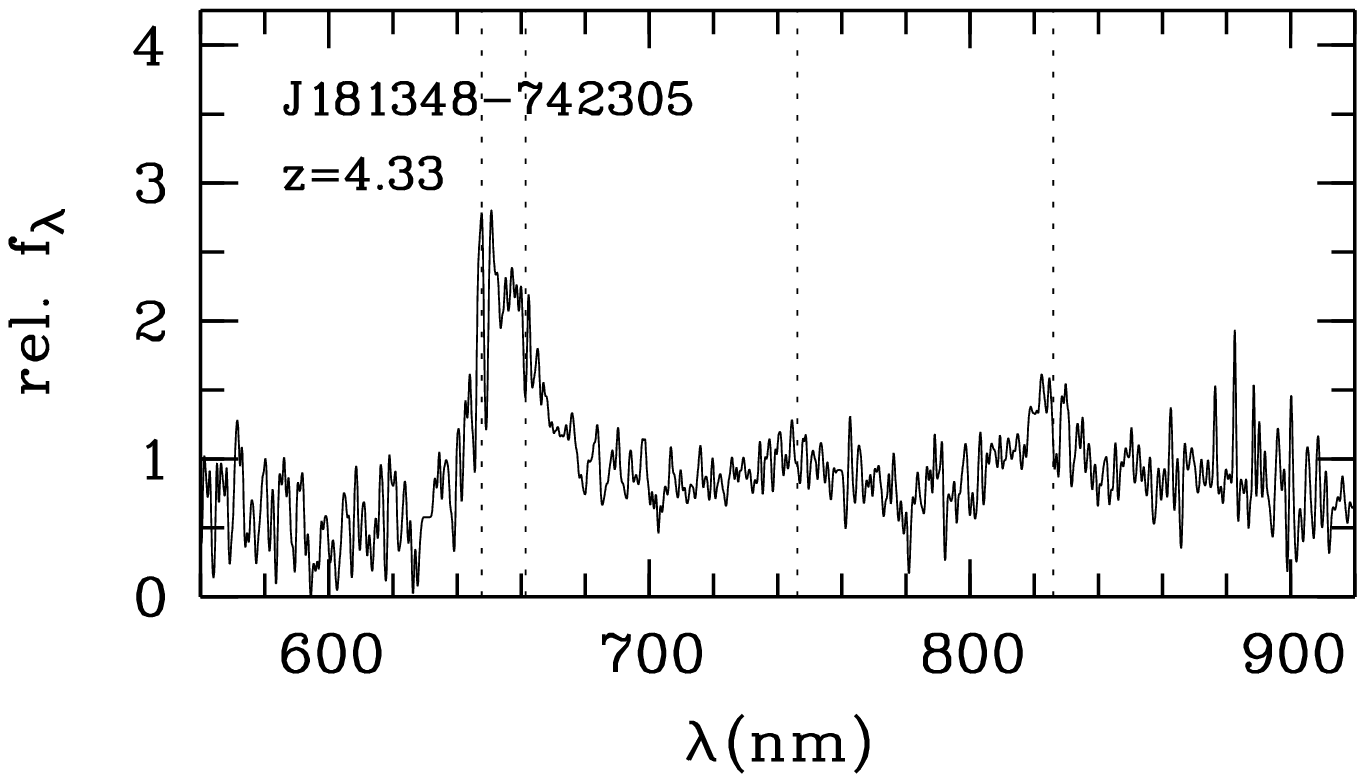}
\includegraphics[angle=270,width=0.32\textwidth,clip=true]{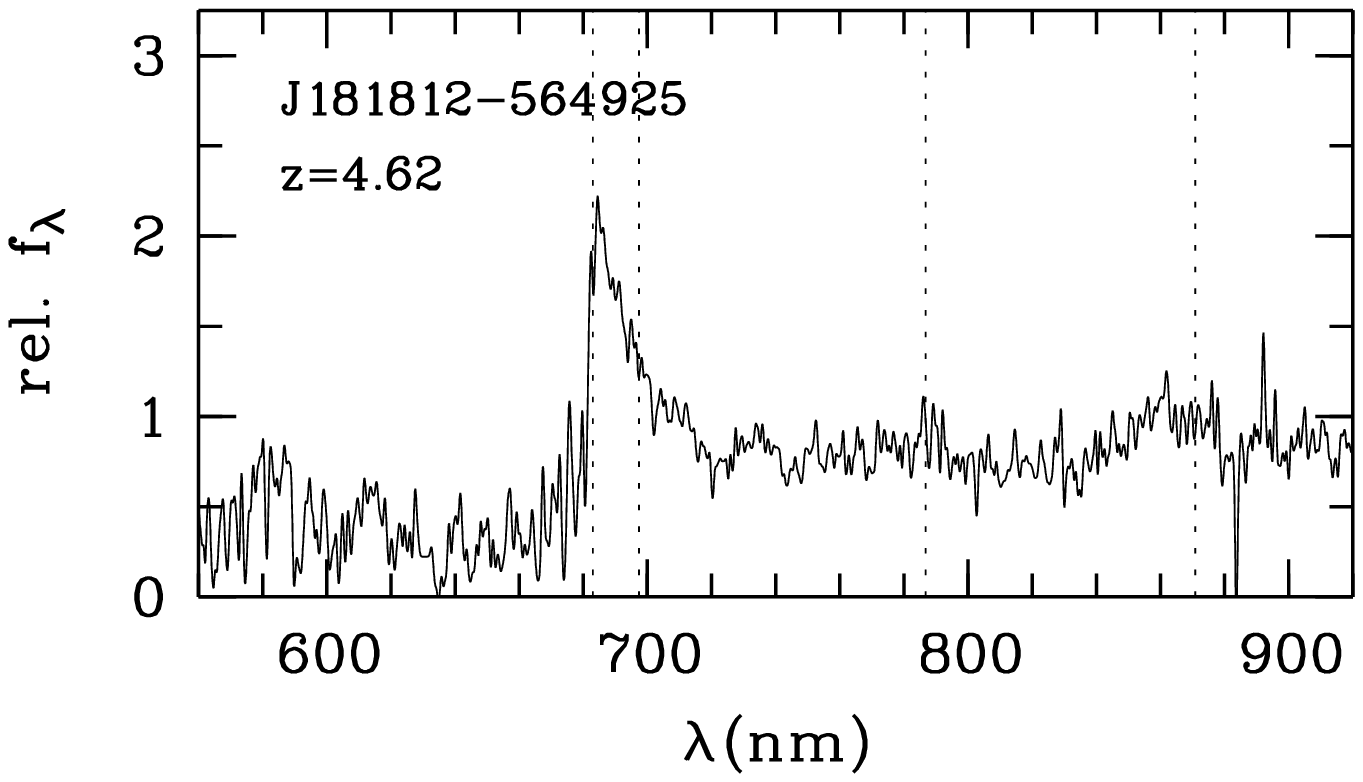}
\includegraphics[angle=270,width=0.32\textwidth,clip=true]{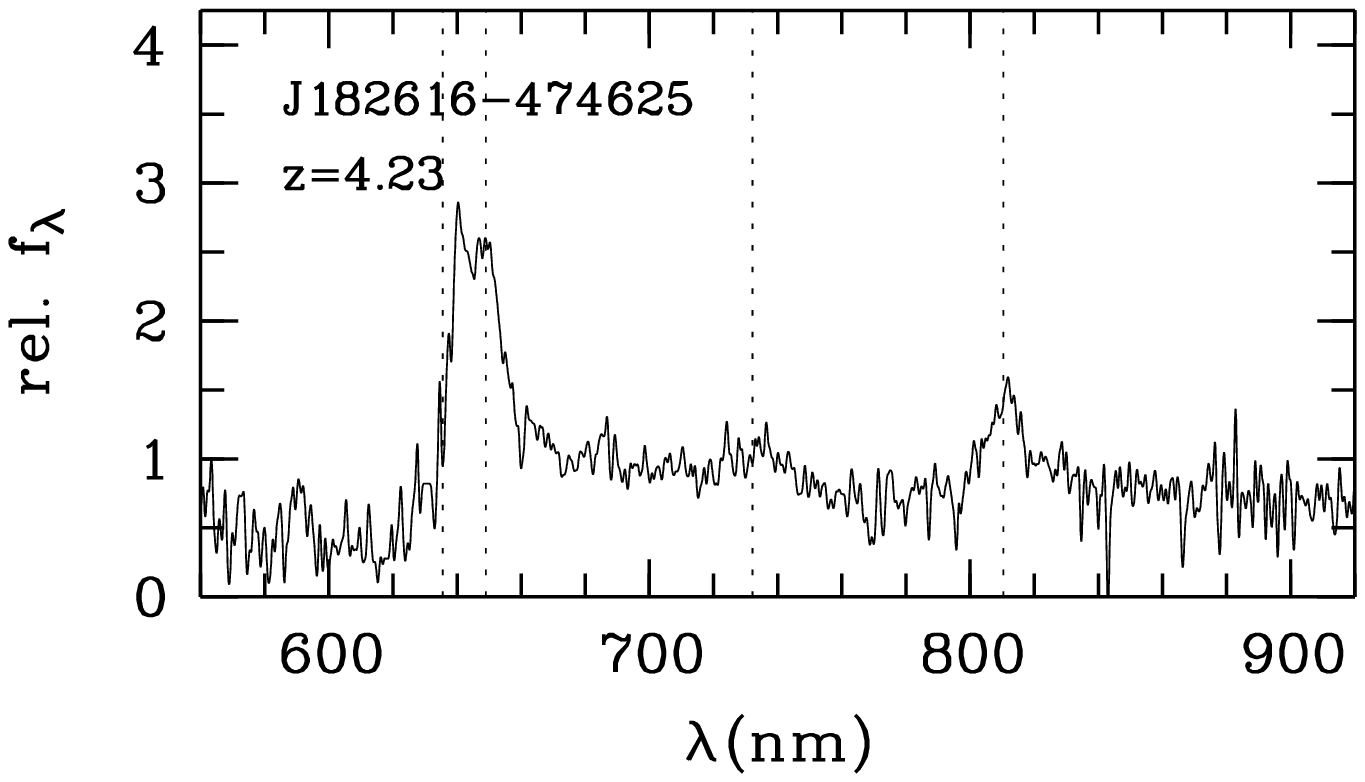}
\includegraphics[angle=270,width=0.32\textwidth,clip=true]{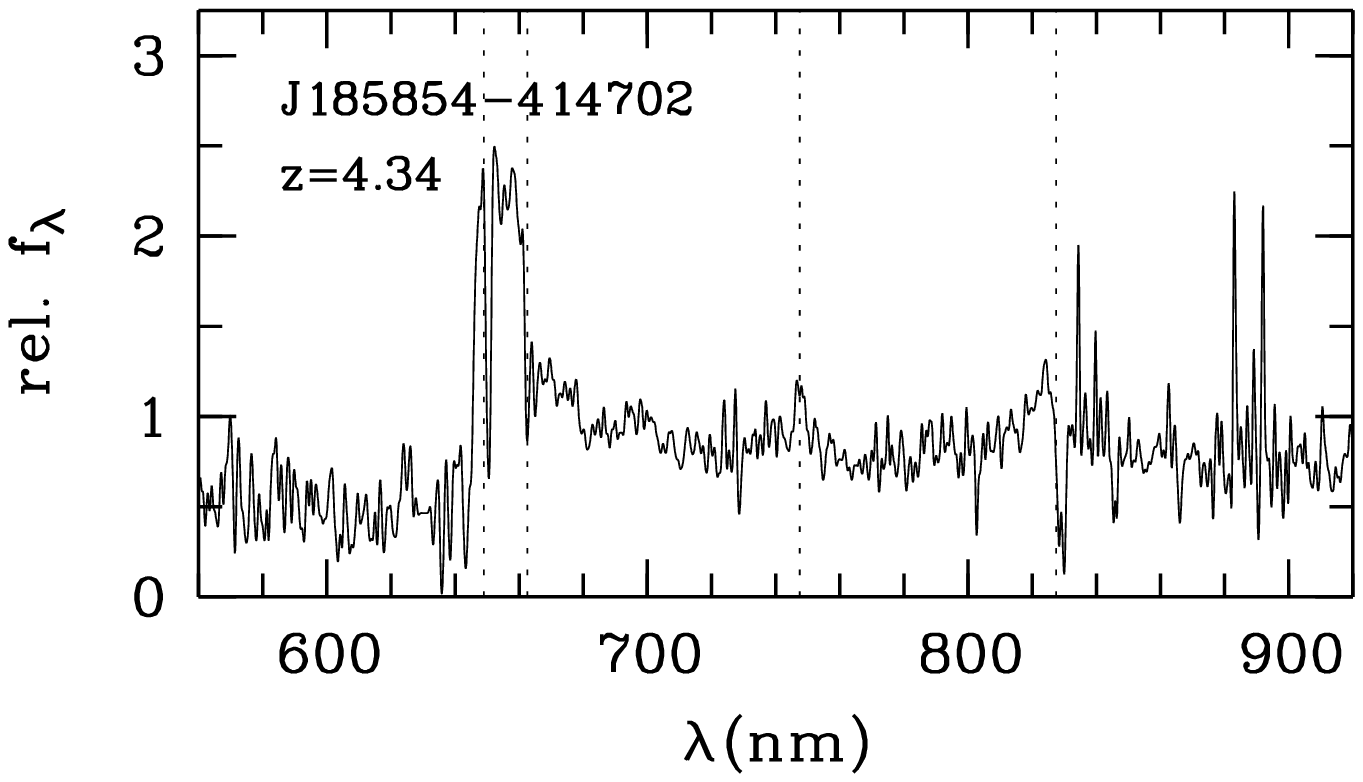}
\includegraphics[angle=270,width=0.32\textwidth,clip=true]{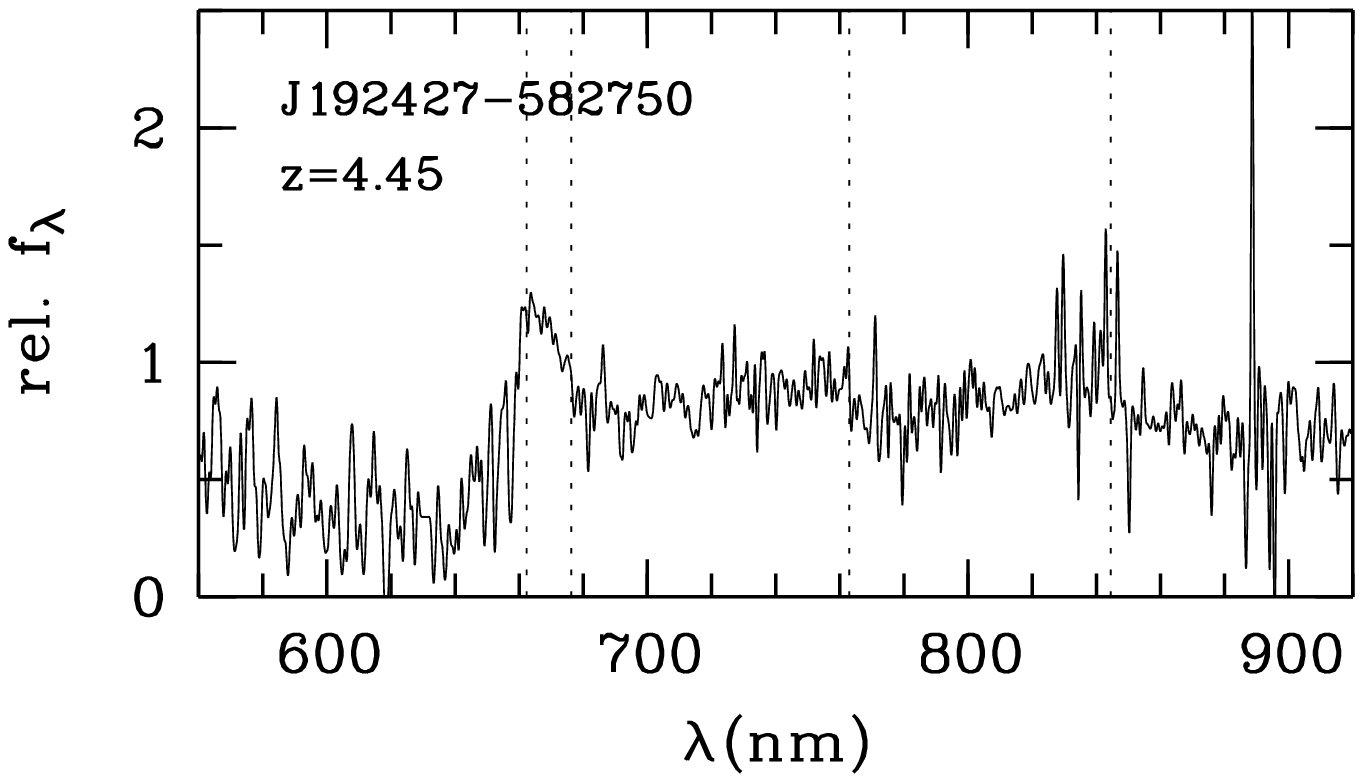}
\includegraphics[angle=270,width=0.32\textwidth,clip=true]{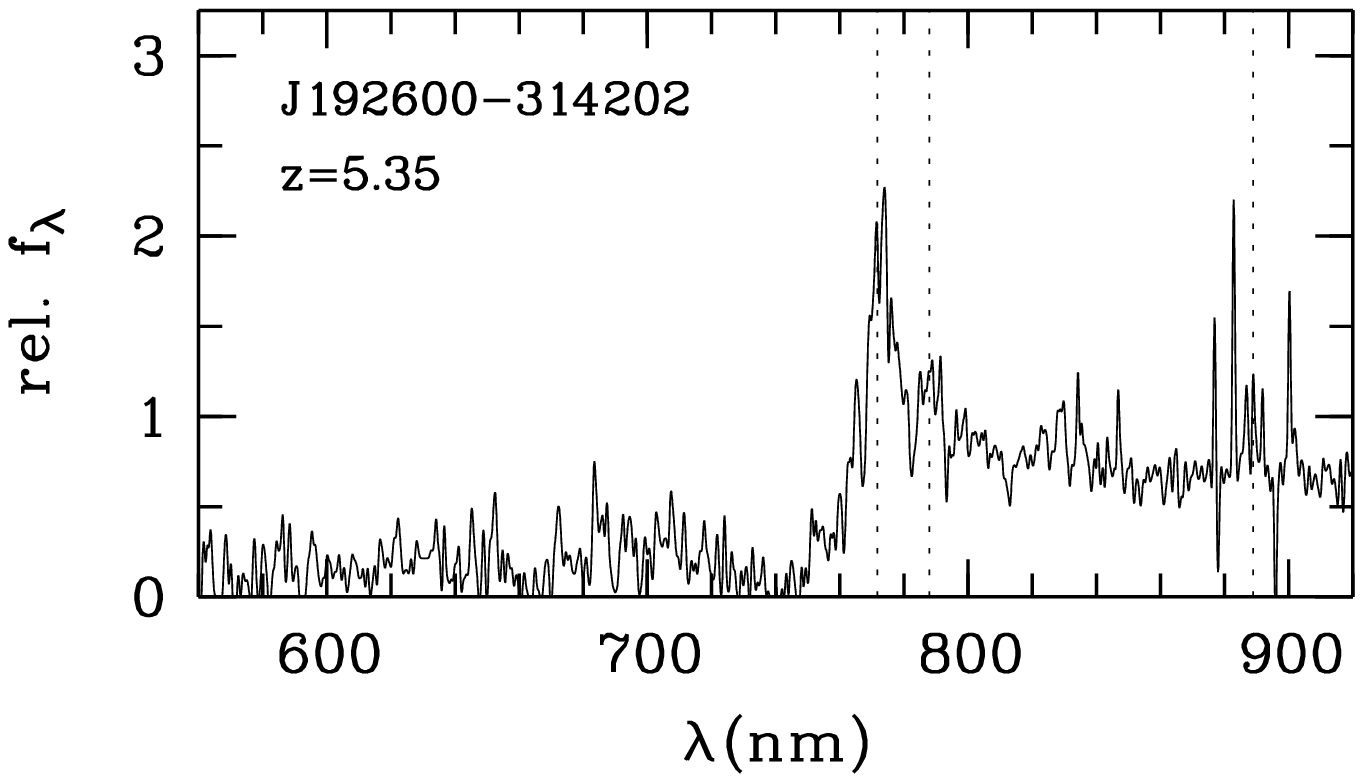}
\caption{Gallery of $3.8>z>5.5$ QSO spectra obtained in this work, ordered by RA, page 5.
\label{galler5y}}
\end{center}
\end{figure*}

\begin{figure*}
\begin{center}
\includegraphics[angle=270,width=0.32\textwidth,clip=true]{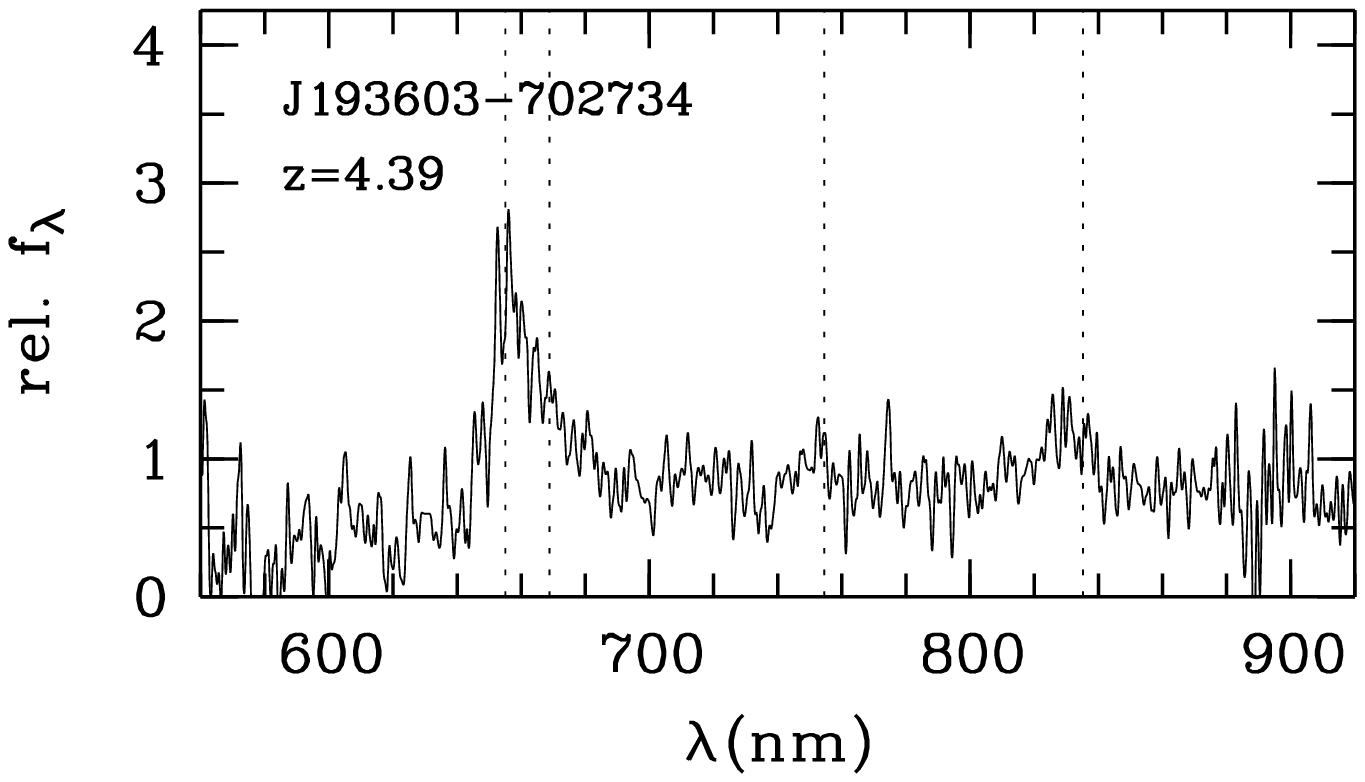}
\includegraphics[angle=270,width=0.32\textwidth,clip=true]{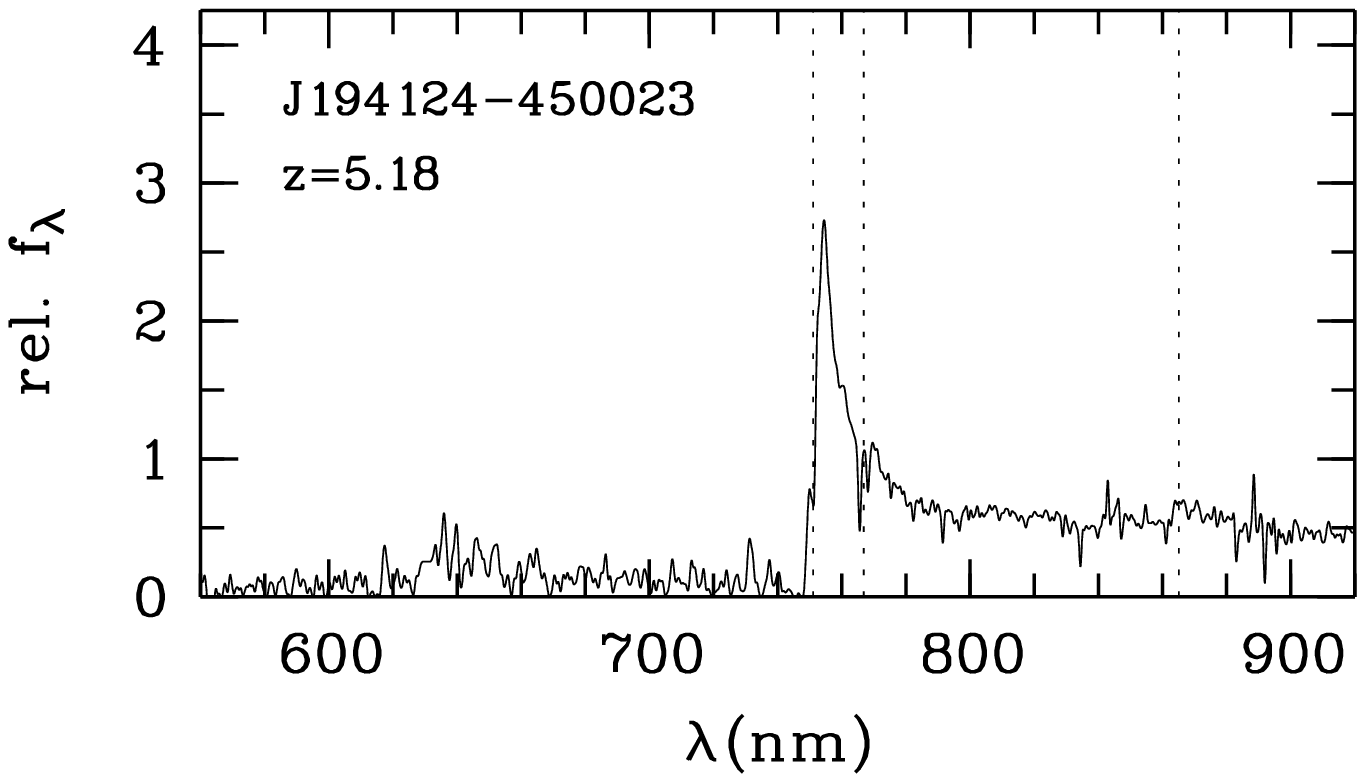}
\includegraphics[angle=270,width=0.32\textwidth,clip=true]{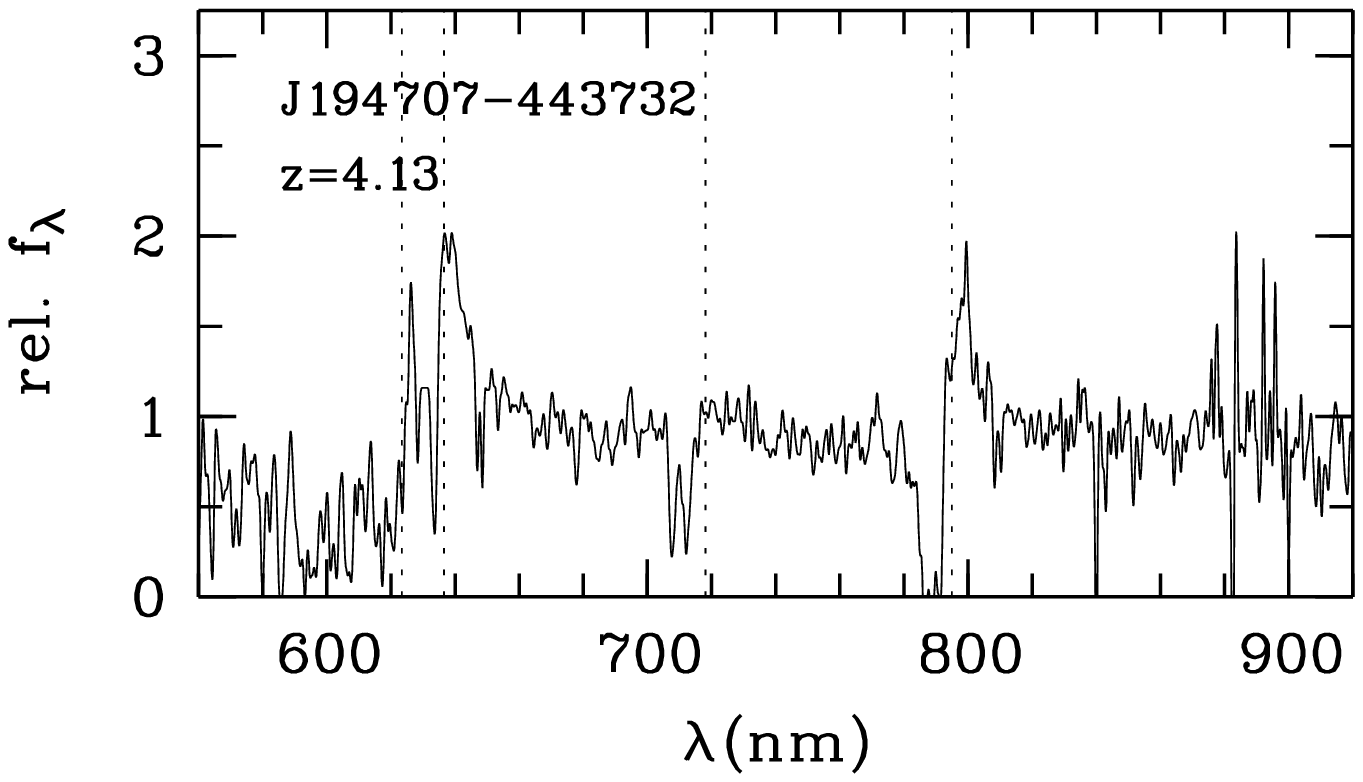}
\includegraphics[angle=270,width=0.32\textwidth,clip=true]{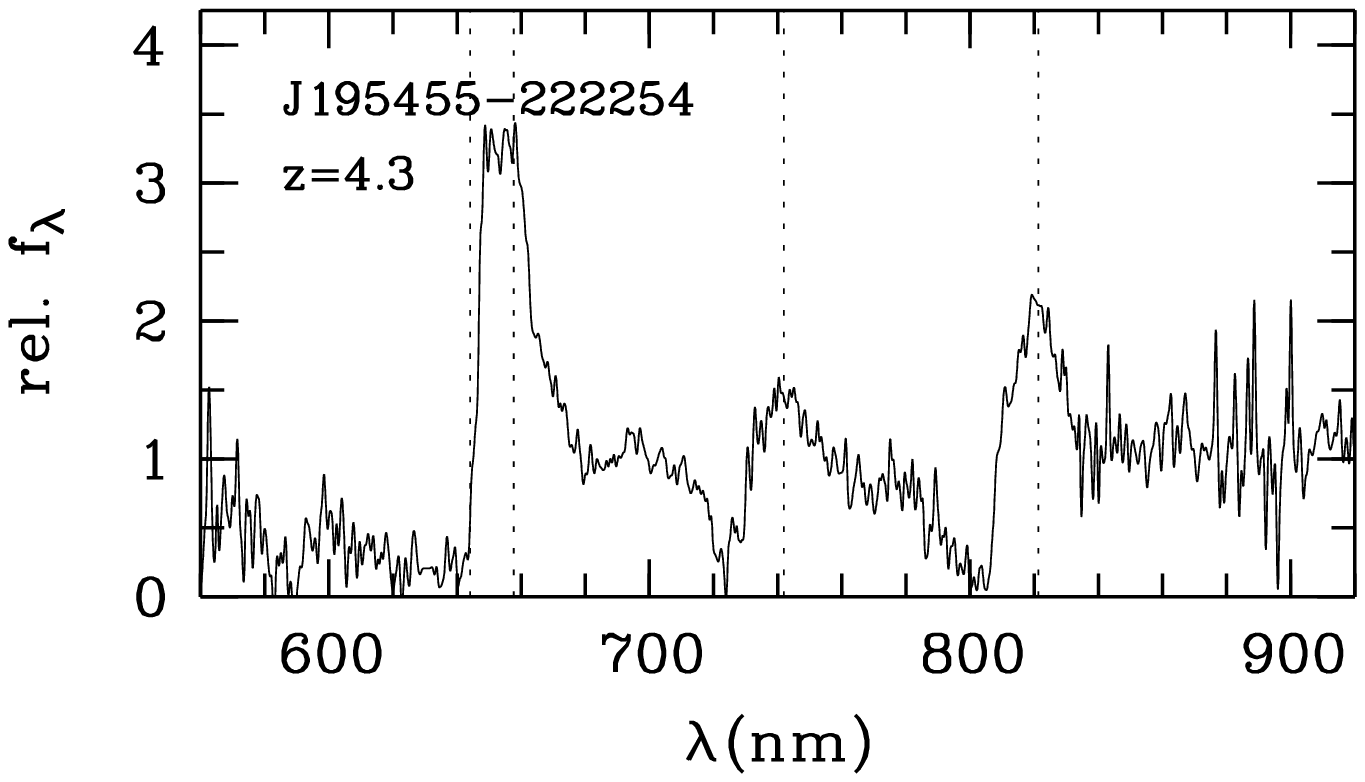}
\includegraphics[angle=270,width=0.32\textwidth,clip=true]{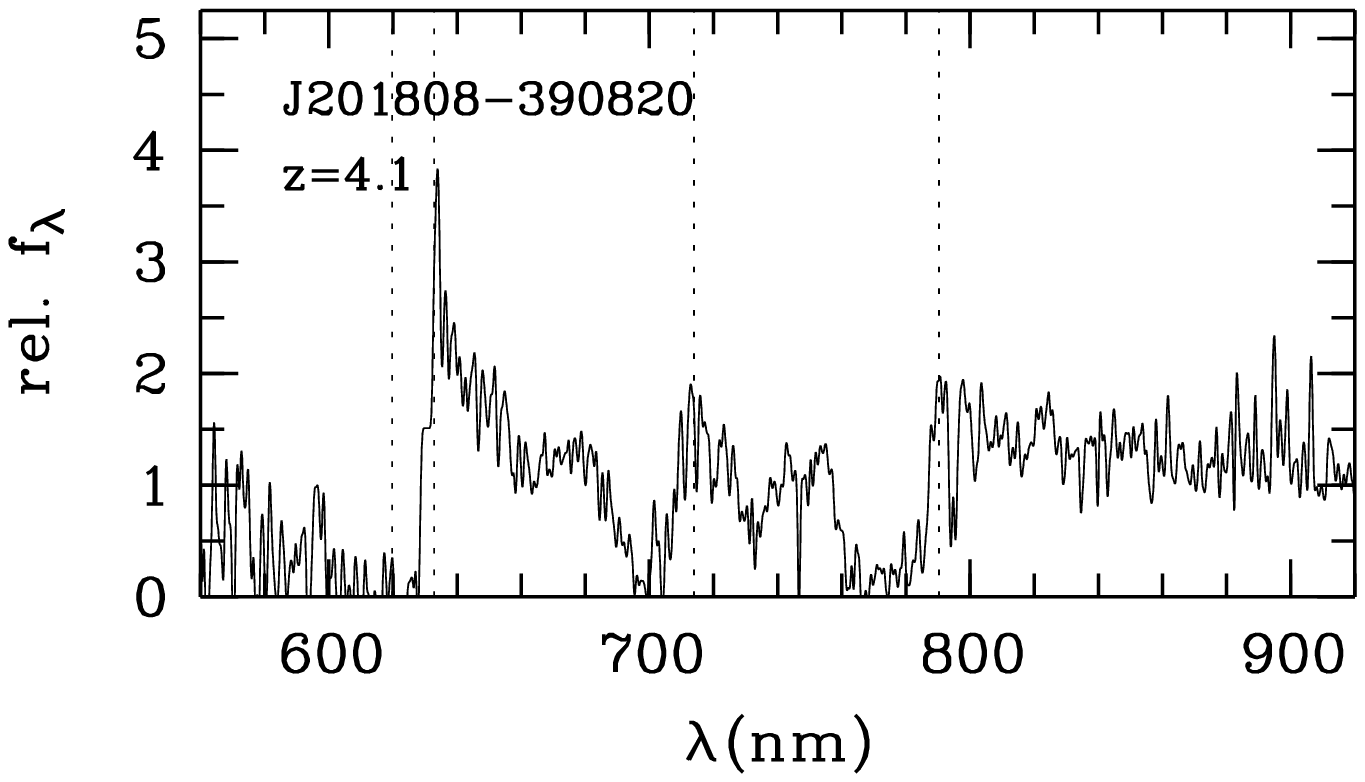}
\includegraphics[angle=270,width=0.32\textwidth,clip=true]{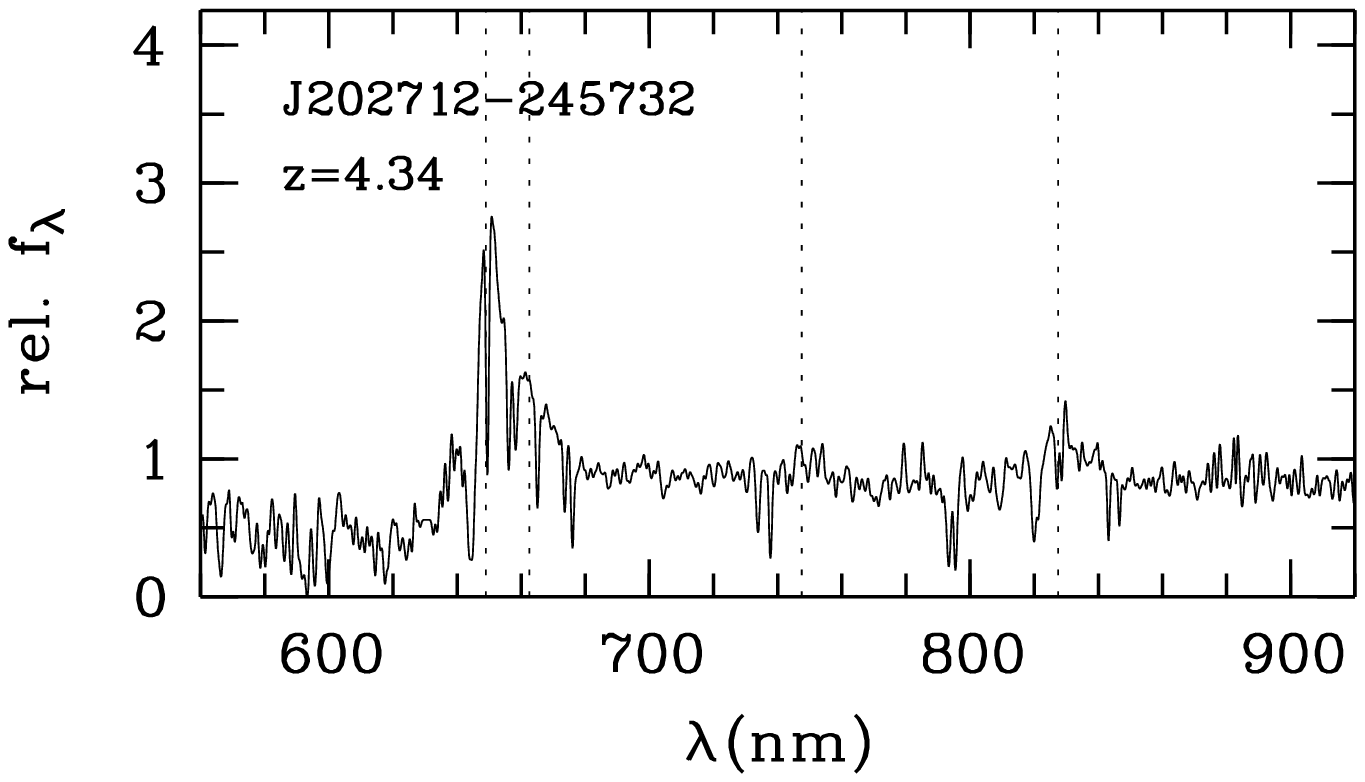}
\includegraphics[angle=270,width=0.32\textwidth,clip=true]{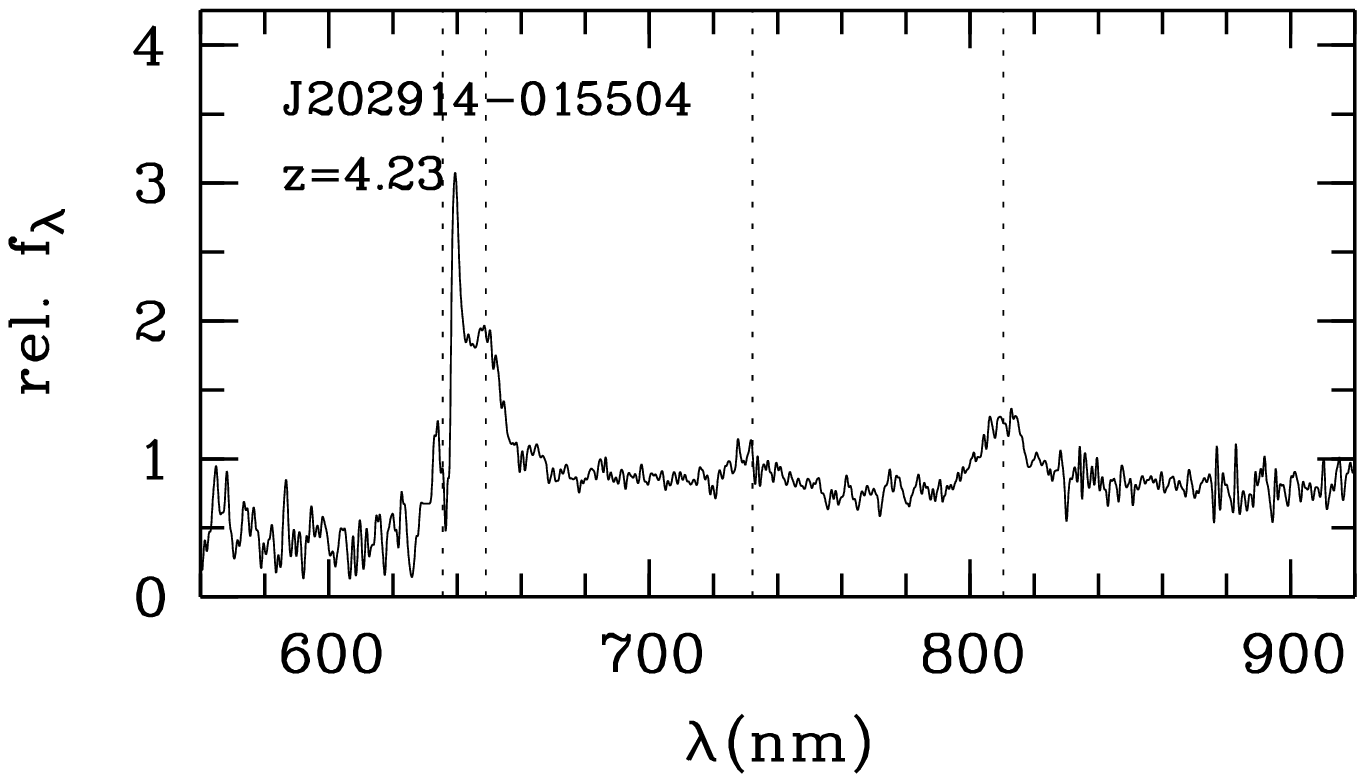}
\includegraphics[angle=270,width=0.32\textwidth,clip=true]{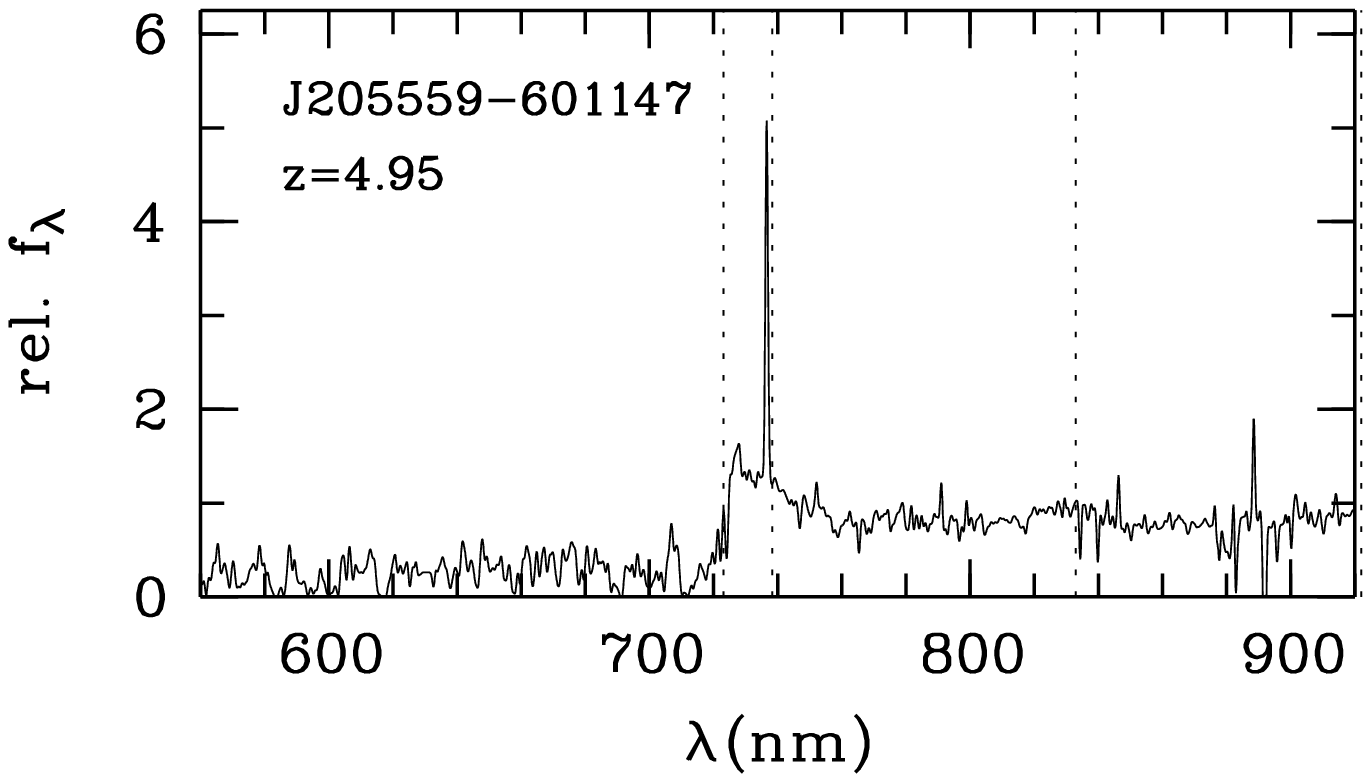}
\includegraphics[angle=270,width=0.32\textwidth,clip=true]{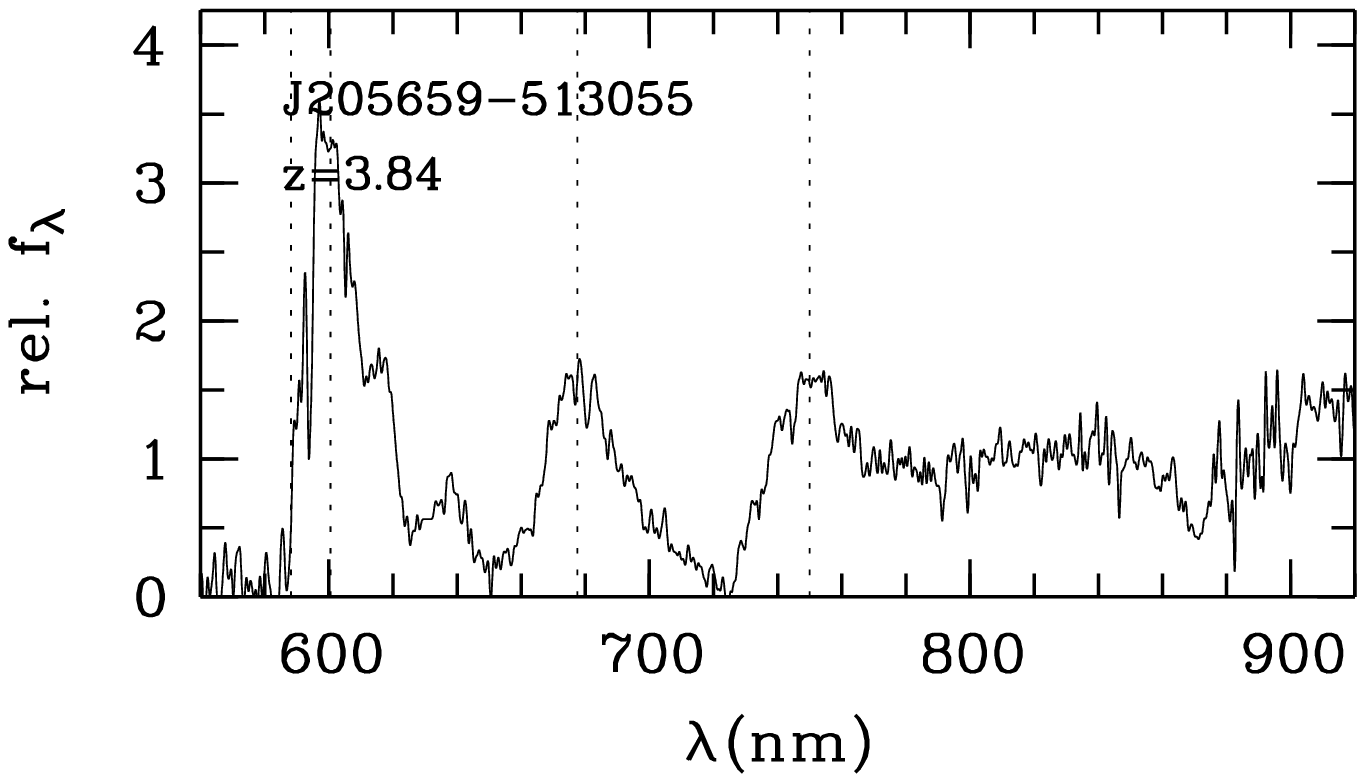}
\includegraphics[angle=270,width=0.32\textwidth,clip=true]{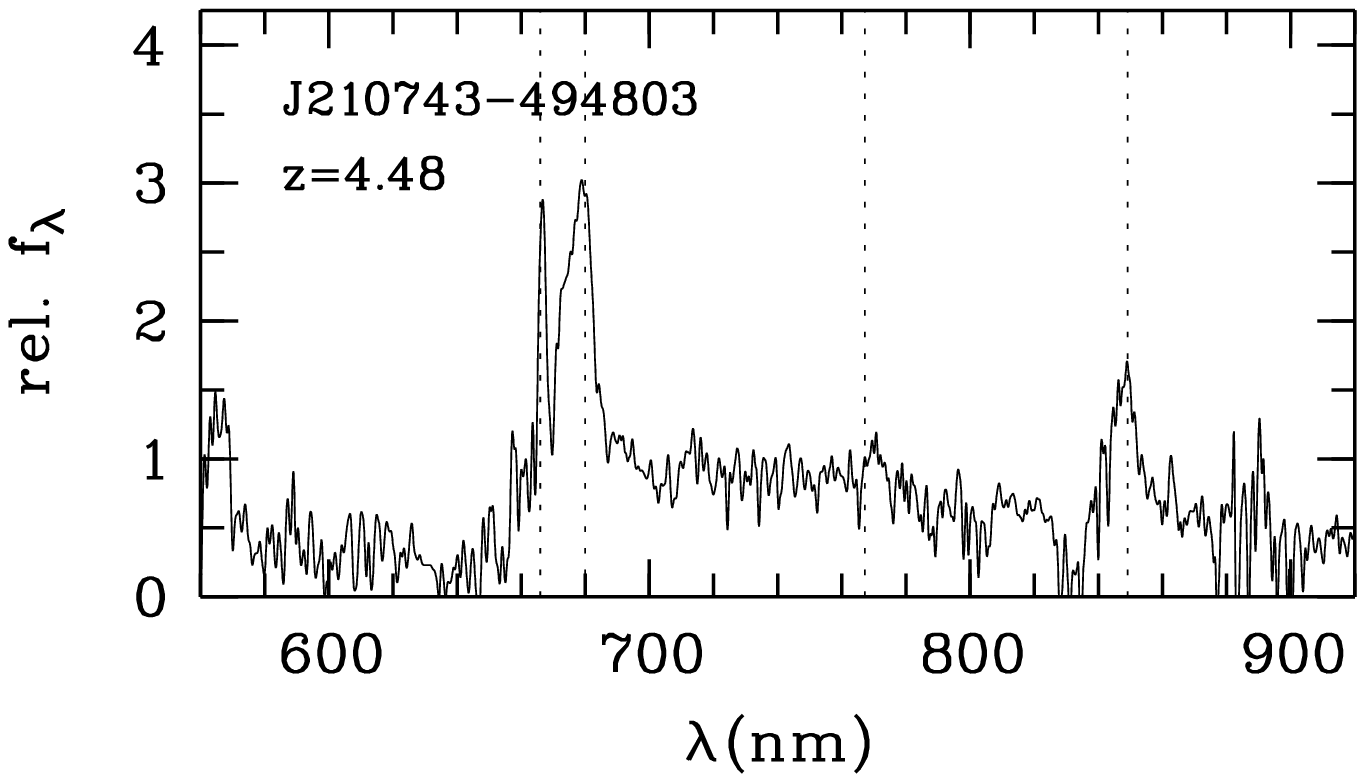}
\includegraphics[angle=270,width=0.32\textwidth,clip=true]{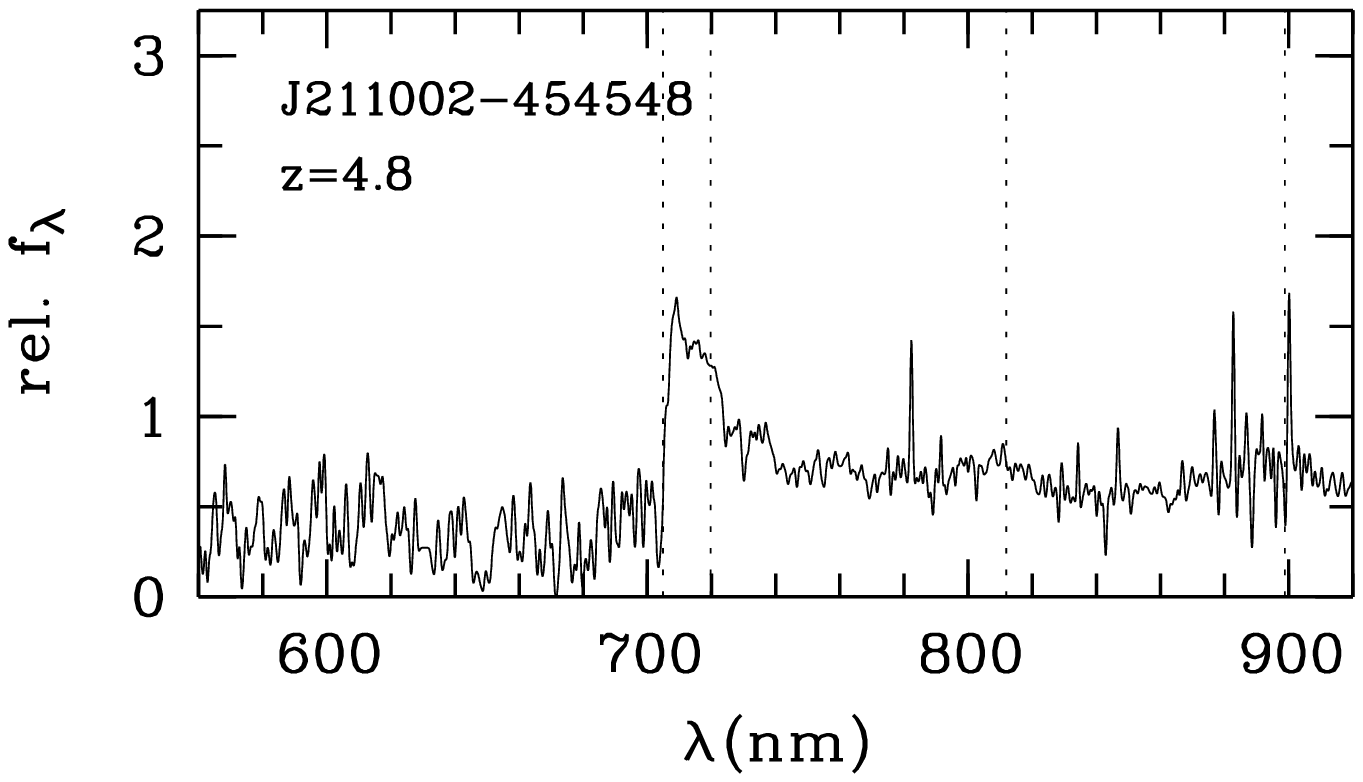}
\includegraphics[angle=270,width=0.32\textwidth,clip=true]{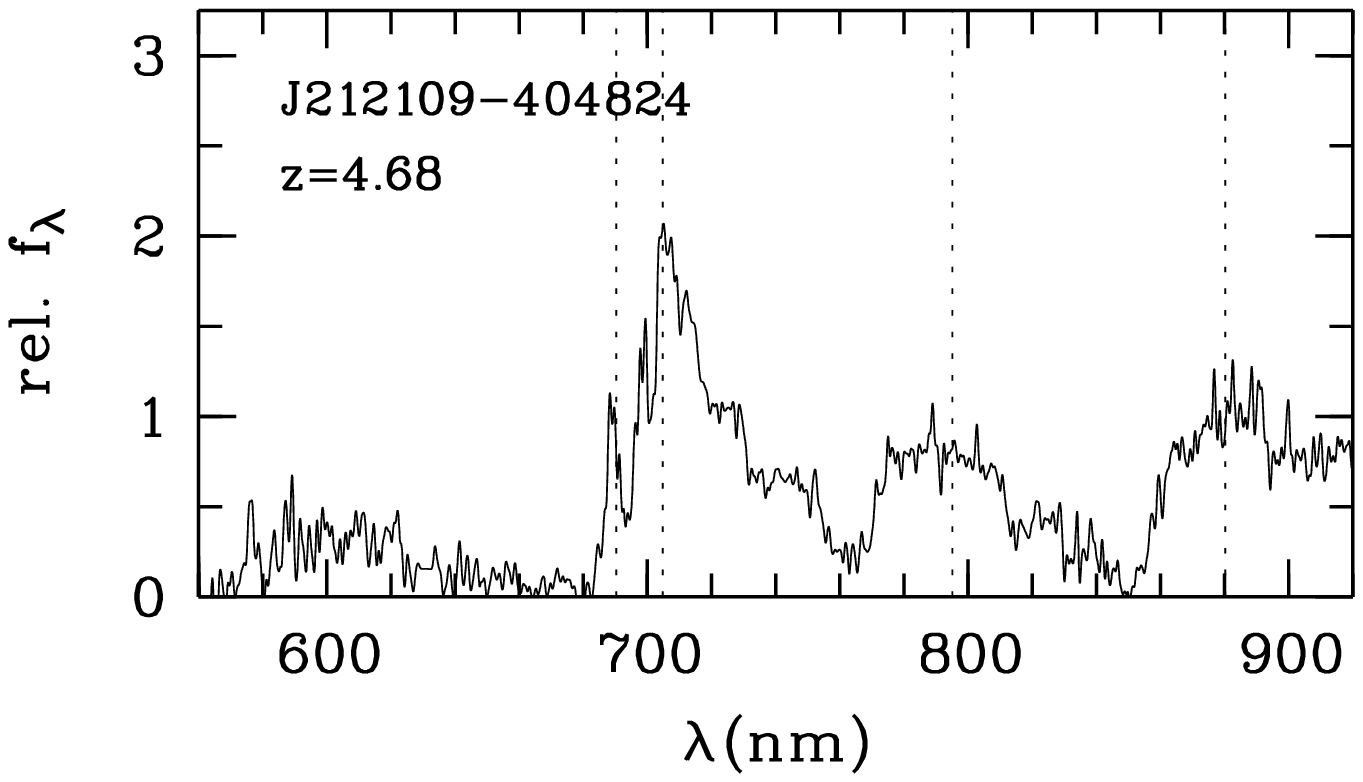}
\includegraphics[angle=270,width=0.32\textwidth,clip=true]{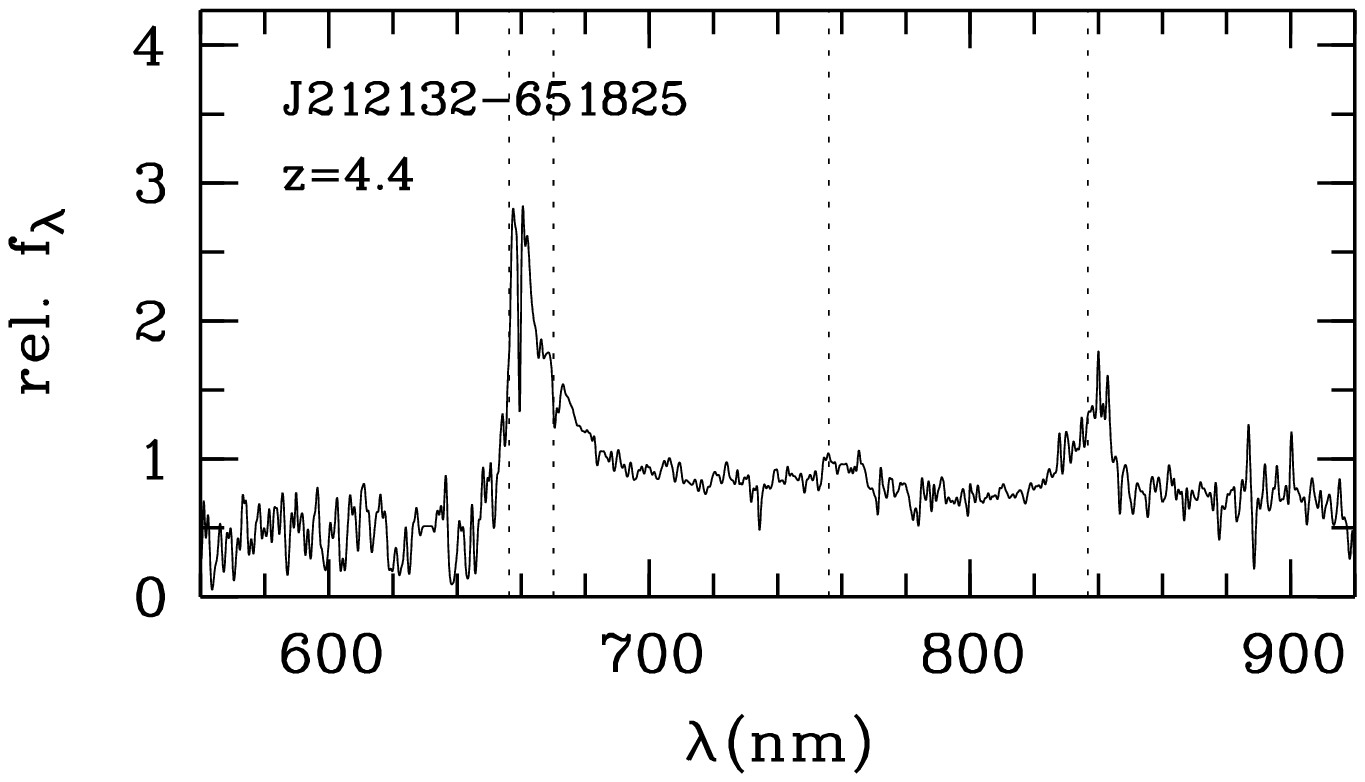}
\includegraphics[angle=270,width=0.32\textwidth,clip=true]{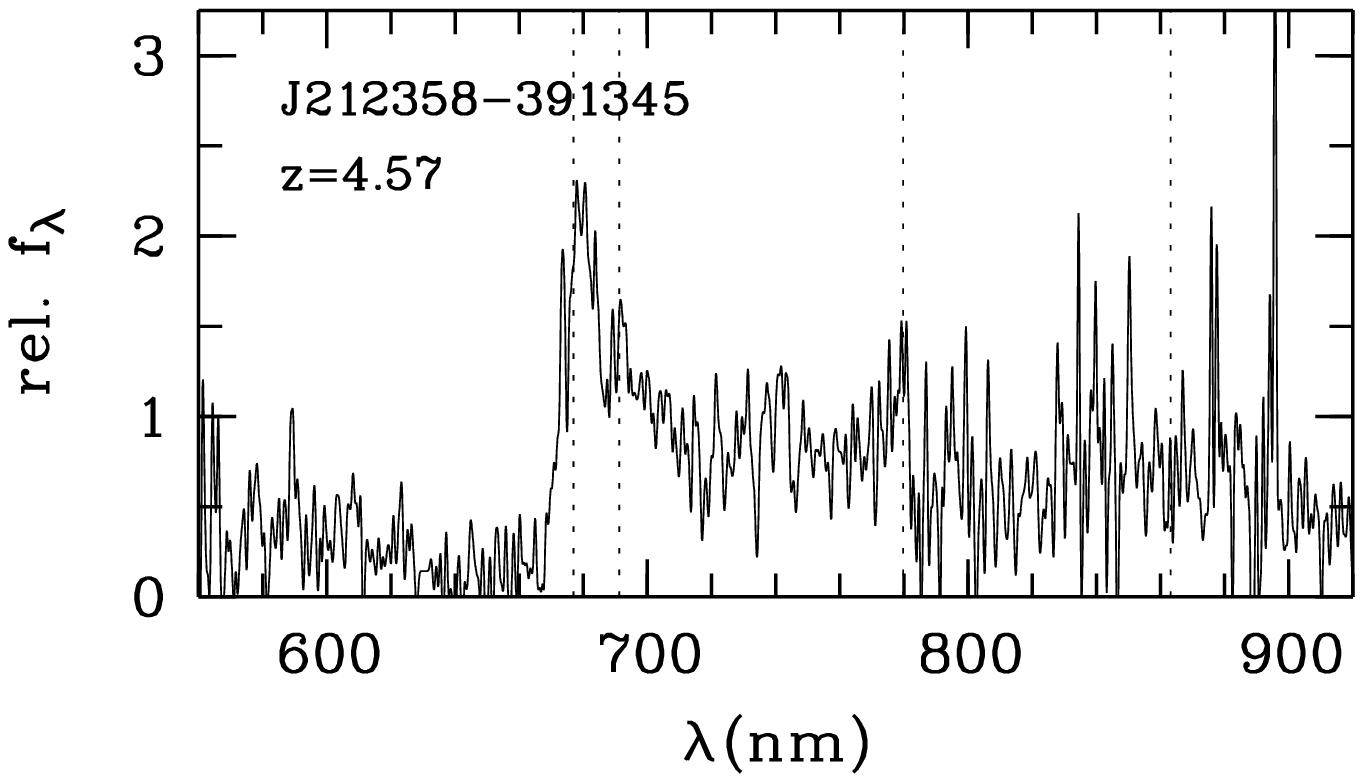}
\includegraphics[angle=270,width=0.32\textwidth,clip=true]{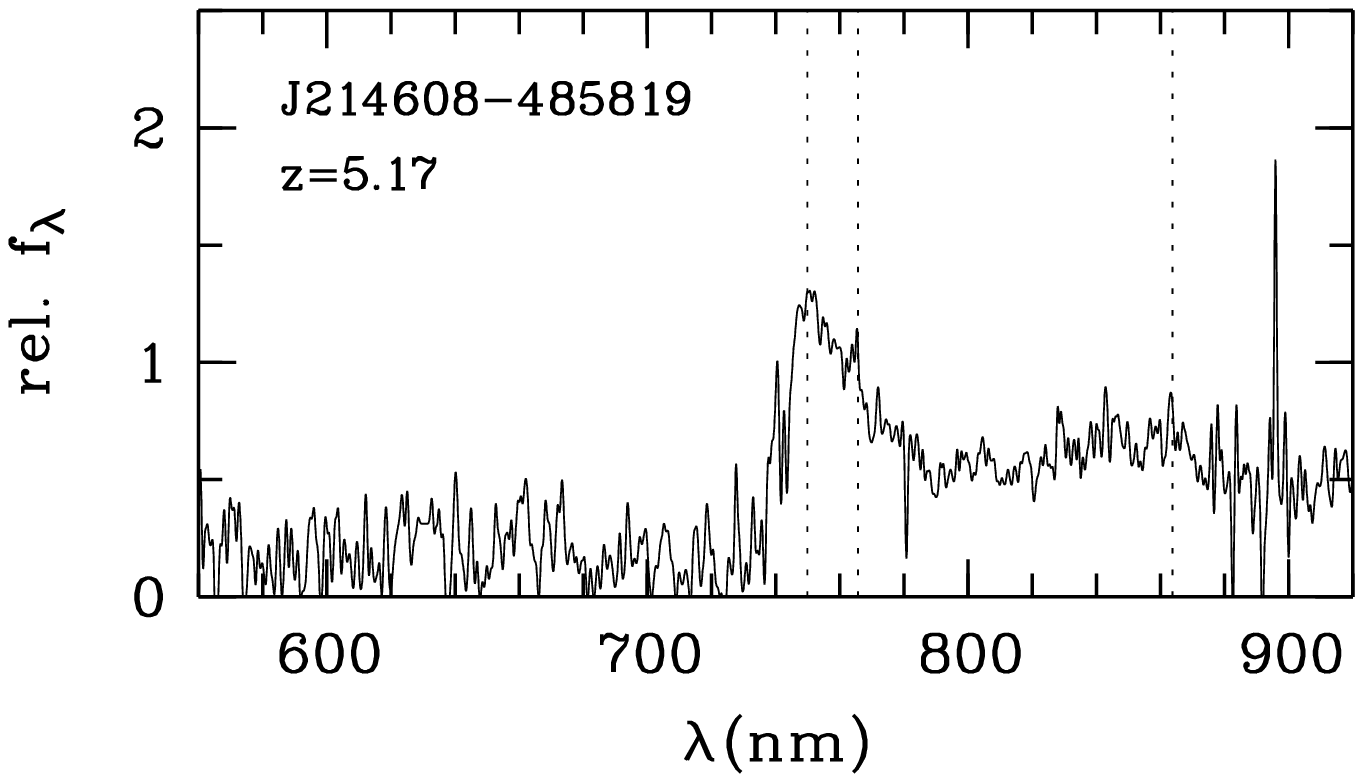}
\includegraphics[angle=270,width=0.32\textwidth,clip=true]{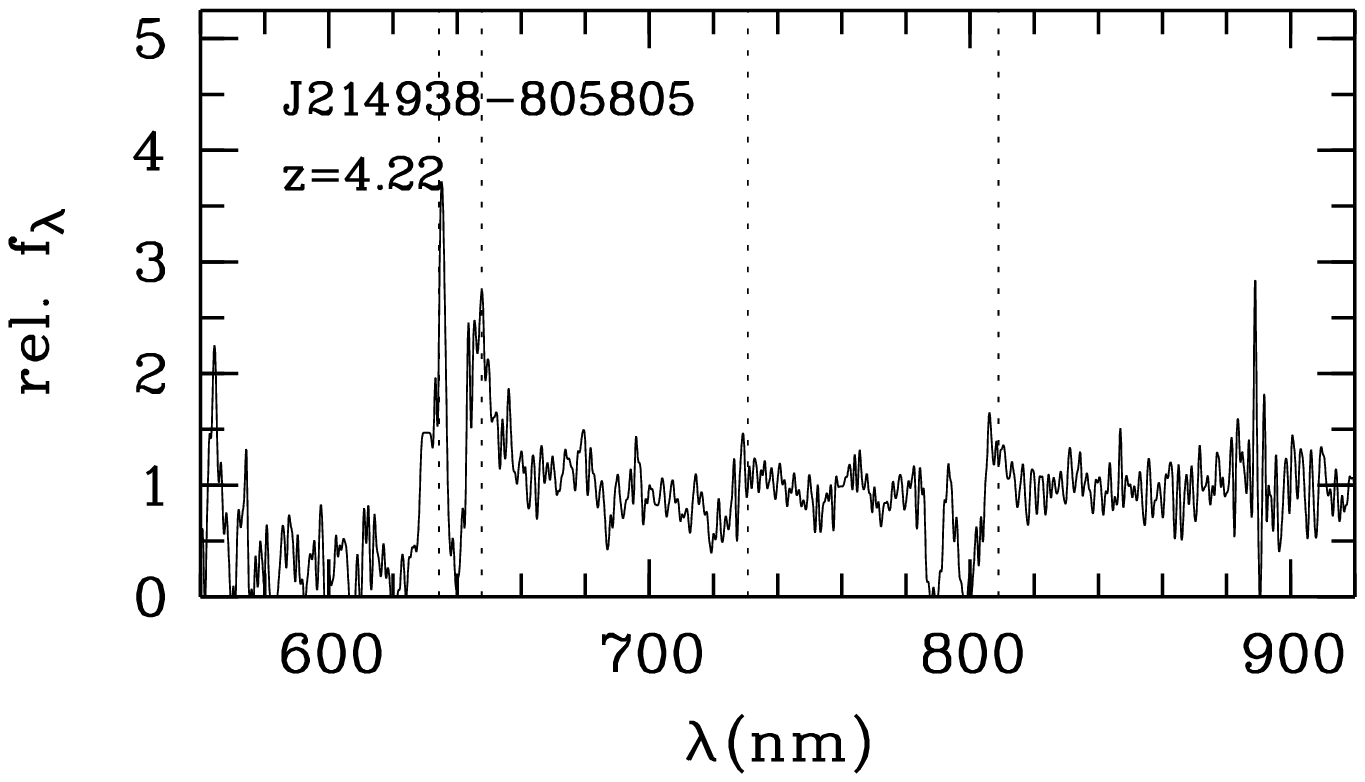}
\includegraphics[angle=270,width=0.32\textwidth,clip=true]{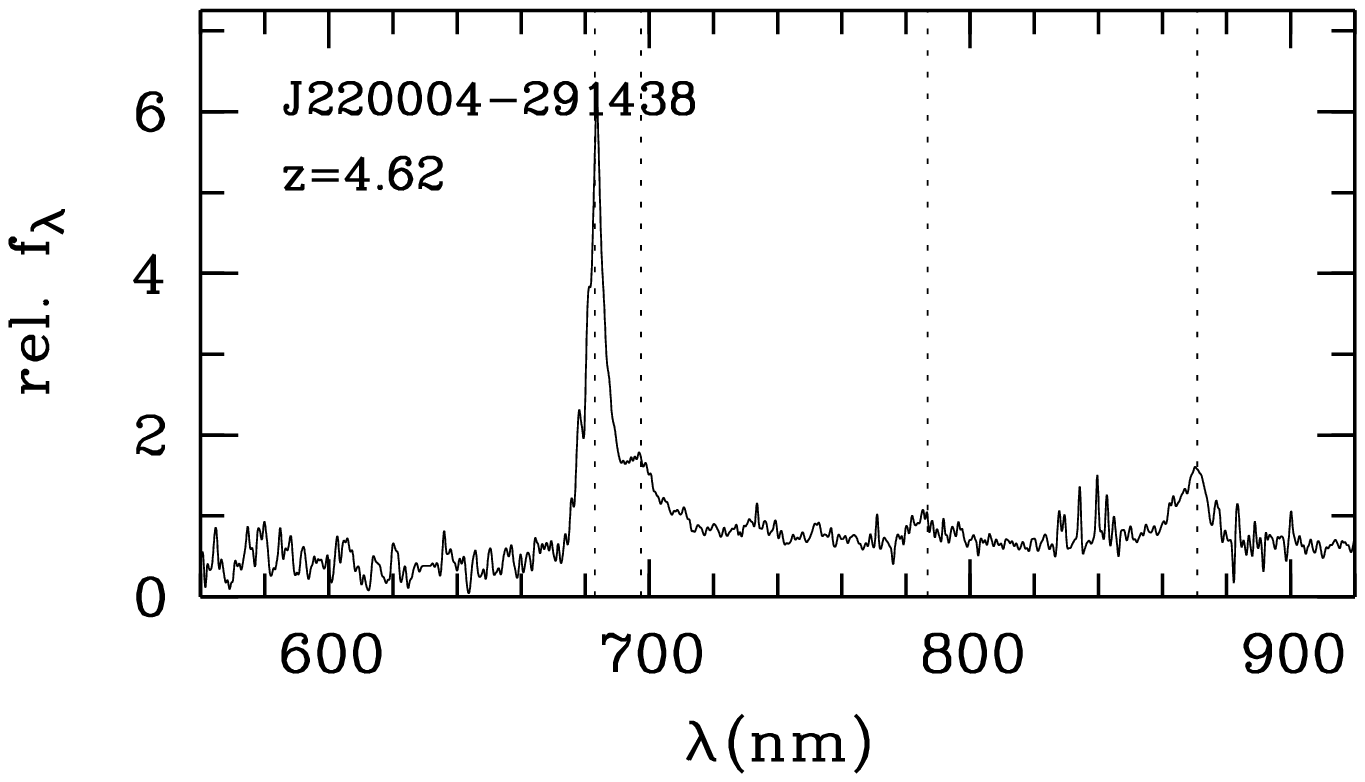}
\includegraphics[angle=270,width=0.32\textwidth,clip=true]{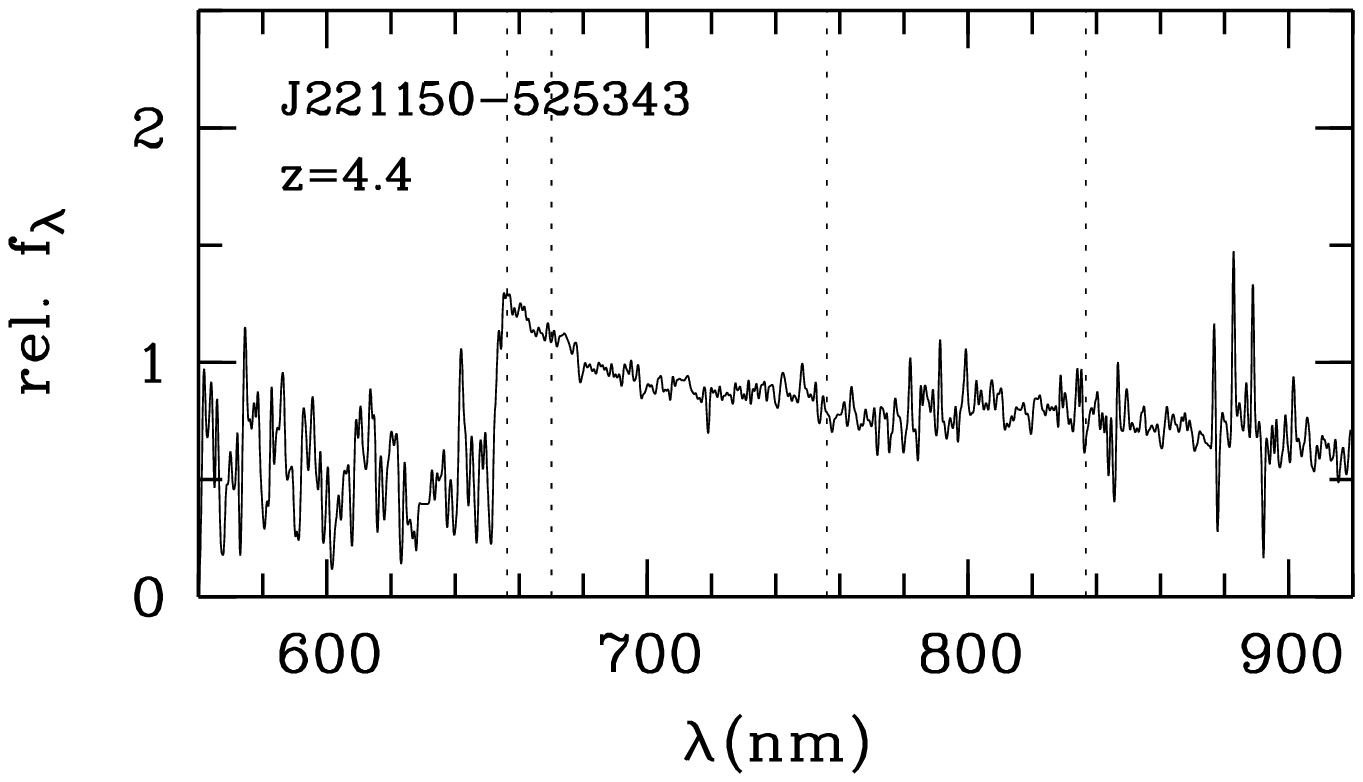}
\includegraphics[angle=270,width=0.32\textwidth,clip=true]{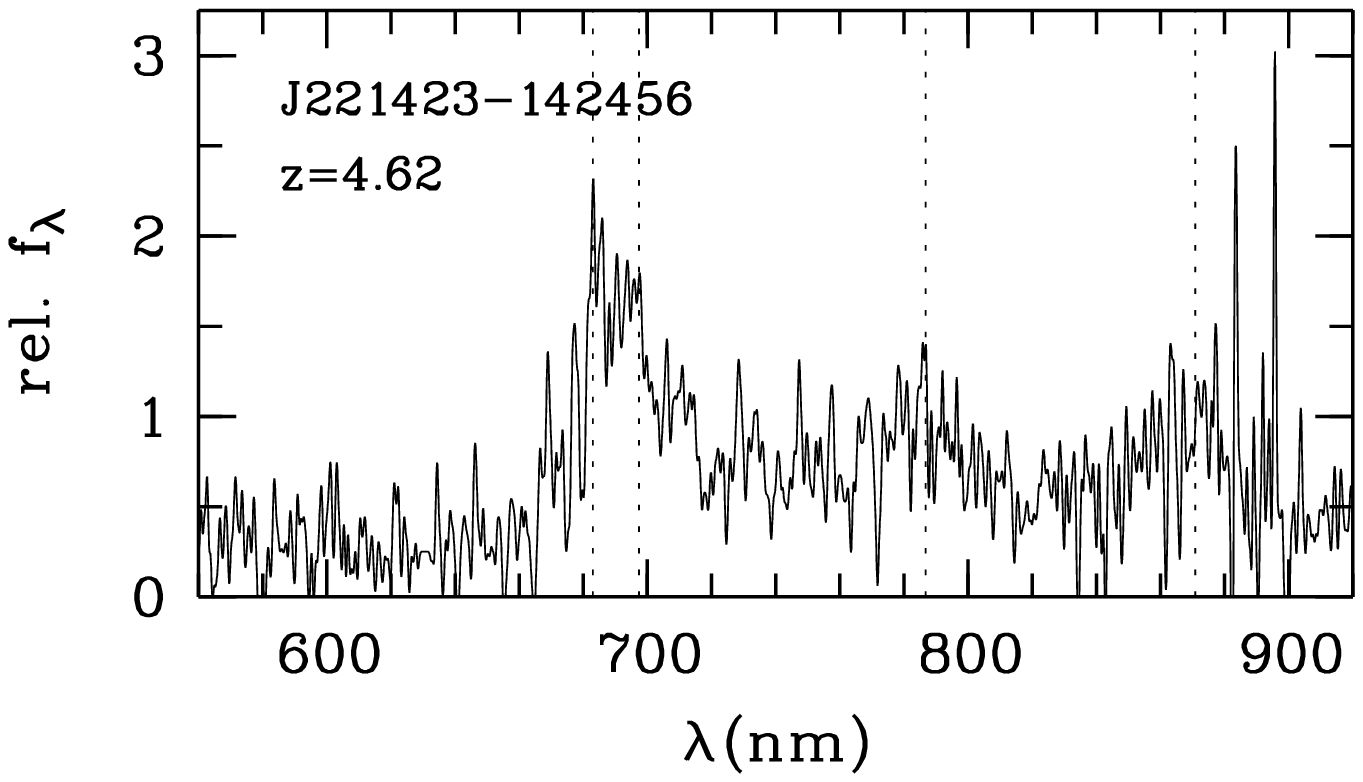}
\includegraphics[angle=270,width=0.32\textwidth,clip=true]{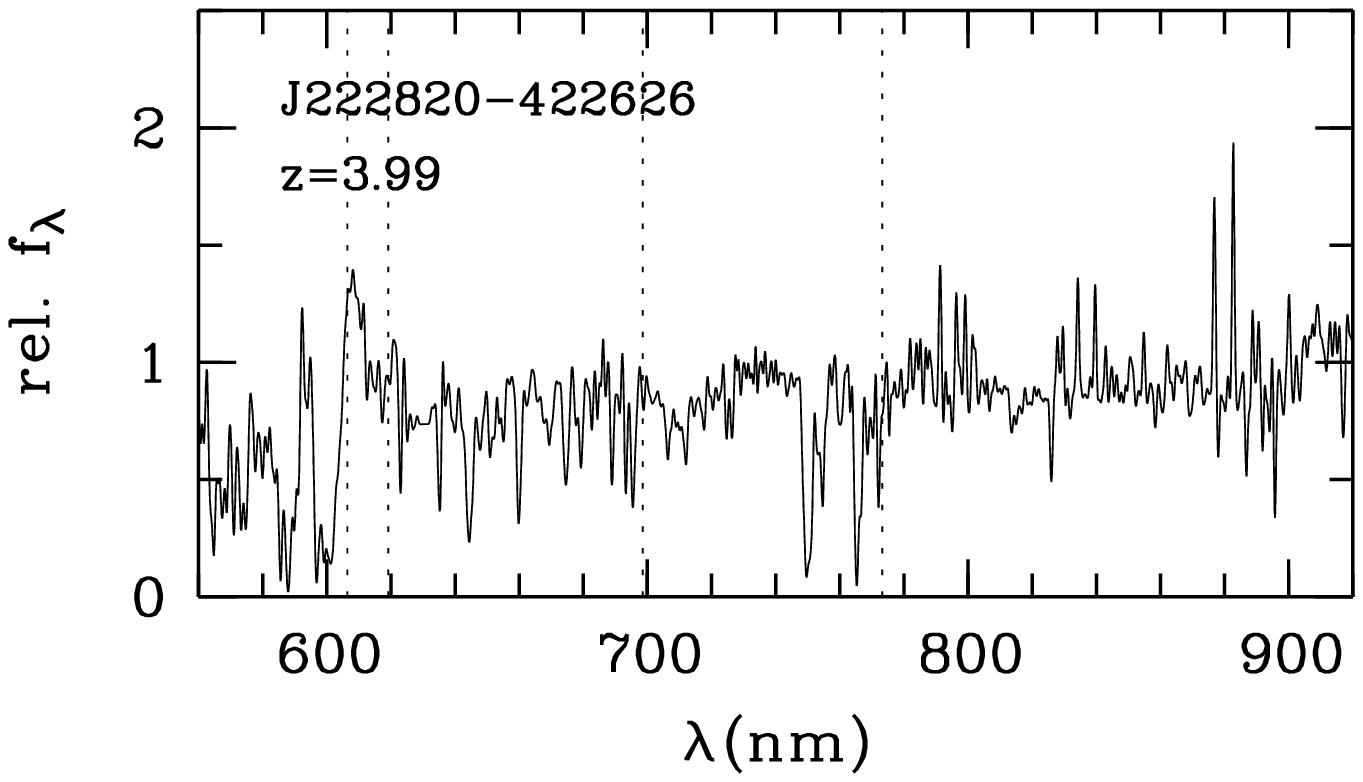}
\includegraphics[angle=270,width=0.32\textwidth,clip=true]{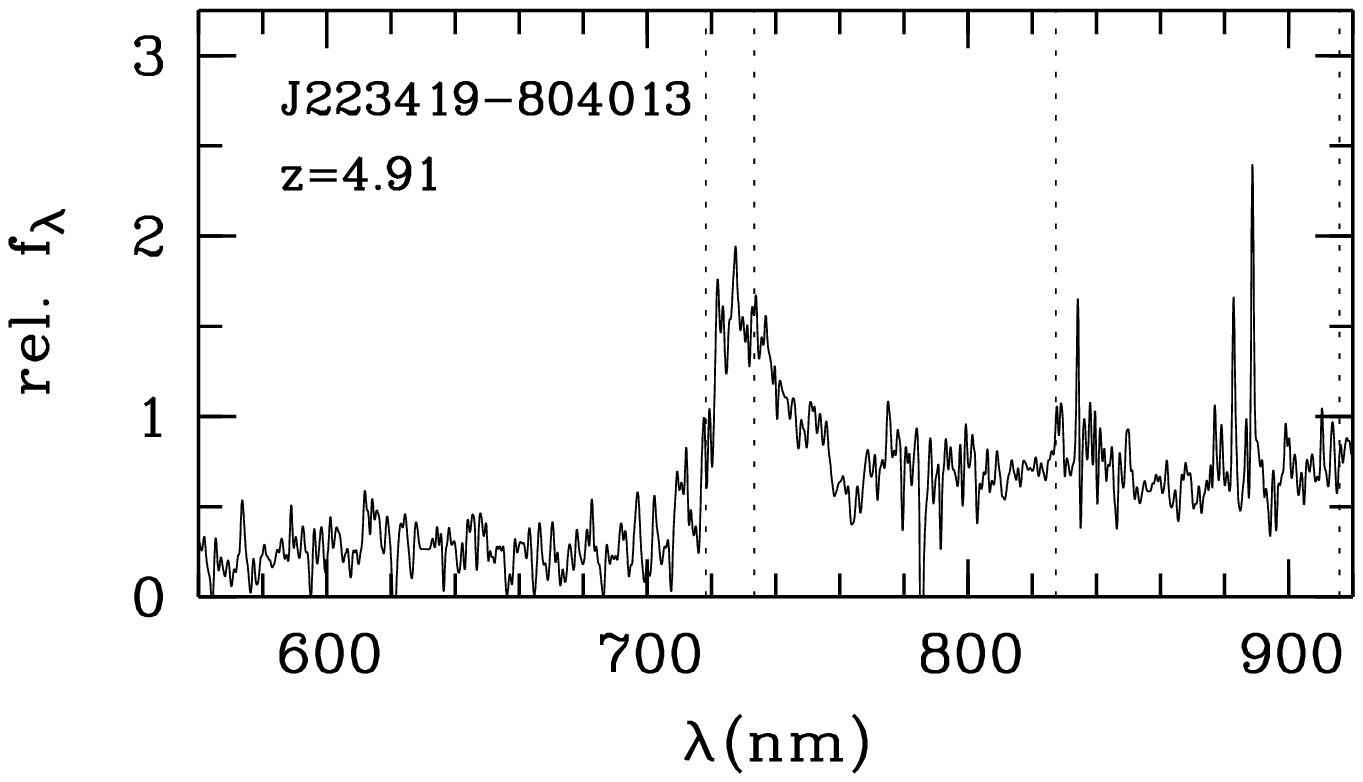}
\caption{Gallery of $3.8>z>5.5$ QSO spectra obtained in this work, ordered by RA, page 6.
\label{gallery6}}
\end{center}
\end{figure*}

\begin{figure*}
\begin{center}
\includegraphics[angle=270,width=0.32\textwidth,clip=true]{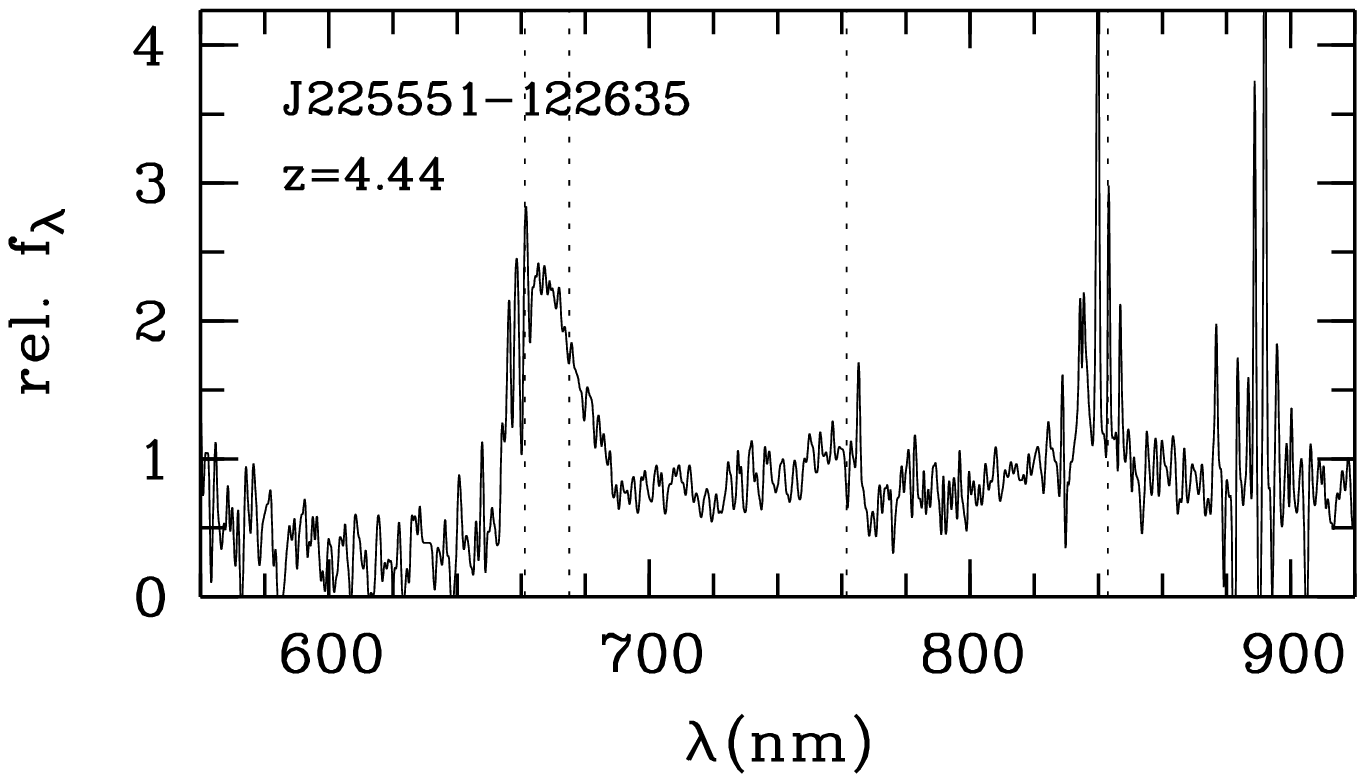}
\includegraphics[angle=270,width=0.32\textwidth,clip=true]{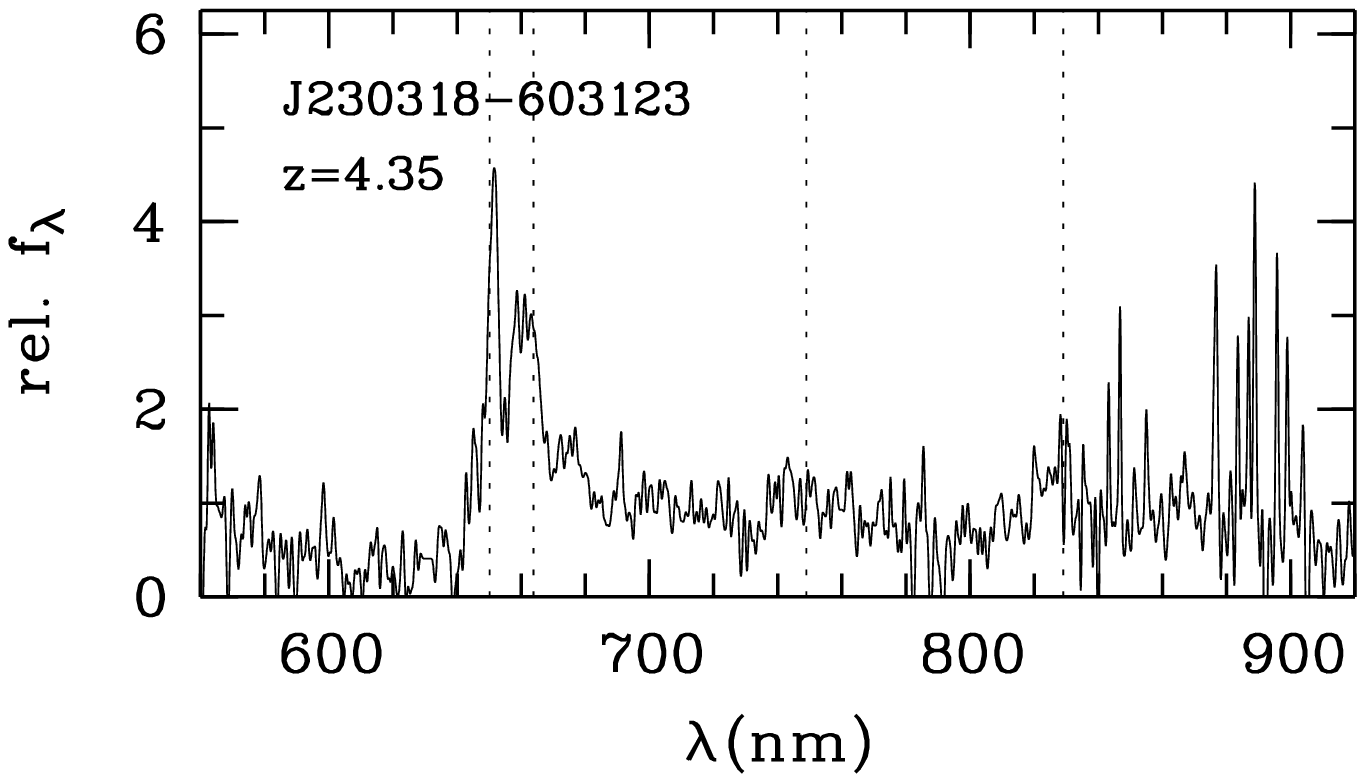}
\includegraphics[angle=270,width=0.32\textwidth,clip=true]{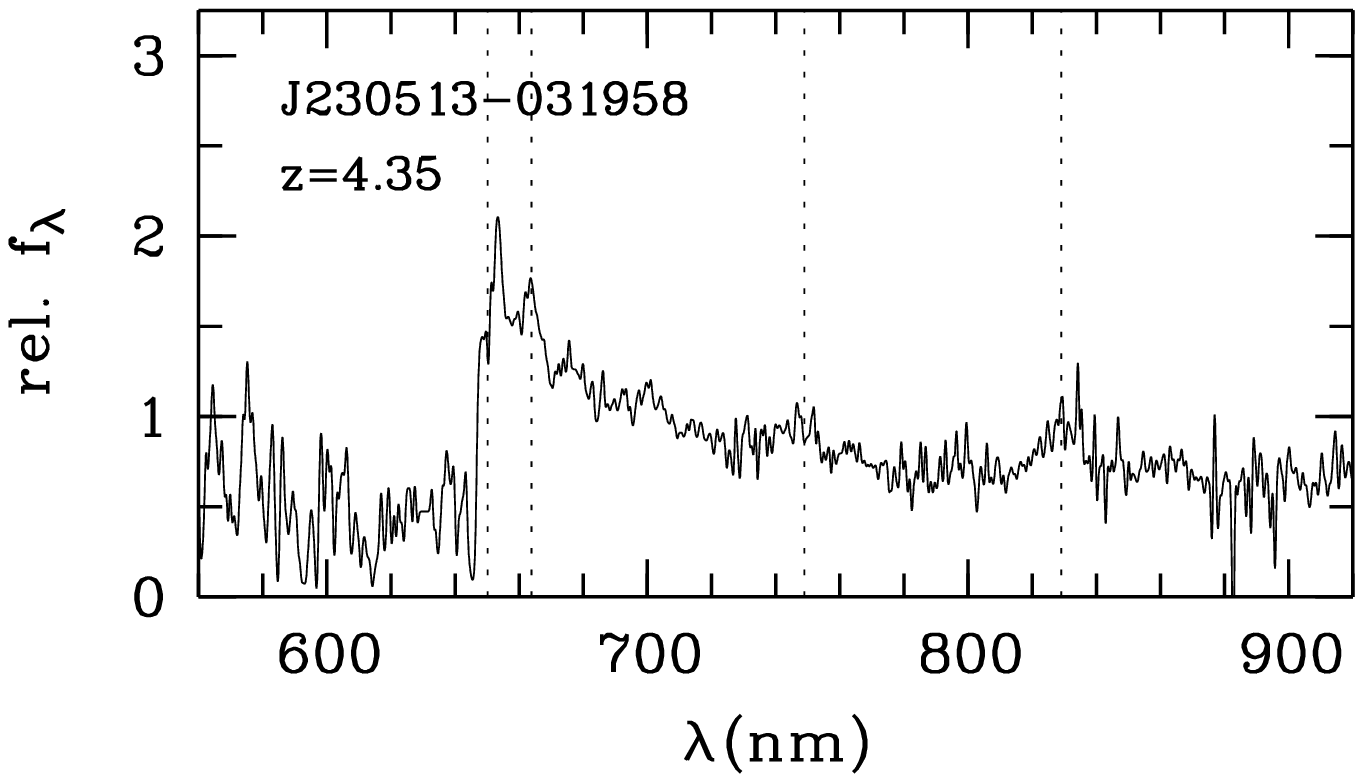}
\includegraphics[angle=270,width=0.32\textwidth,clip=true]{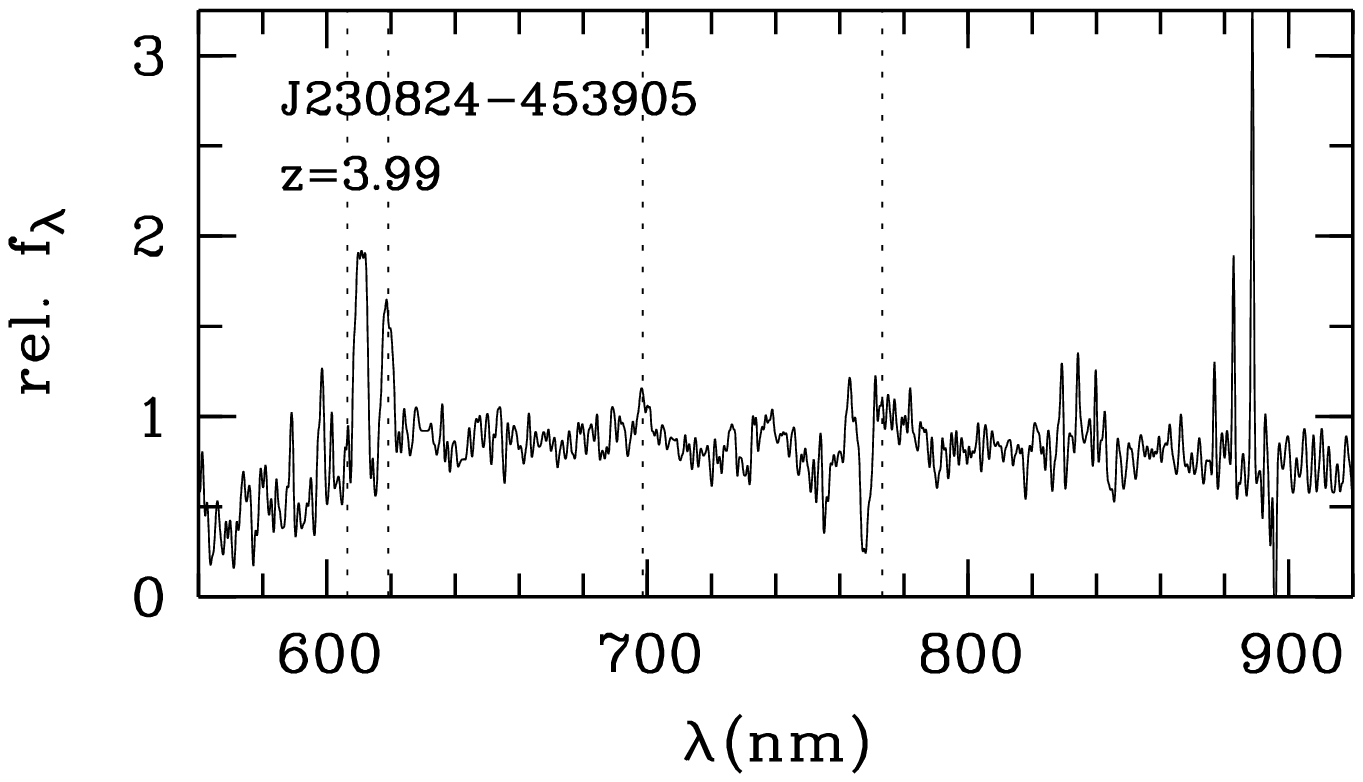}
\includegraphics[angle=270,width=0.32\textwidth,clip=true]{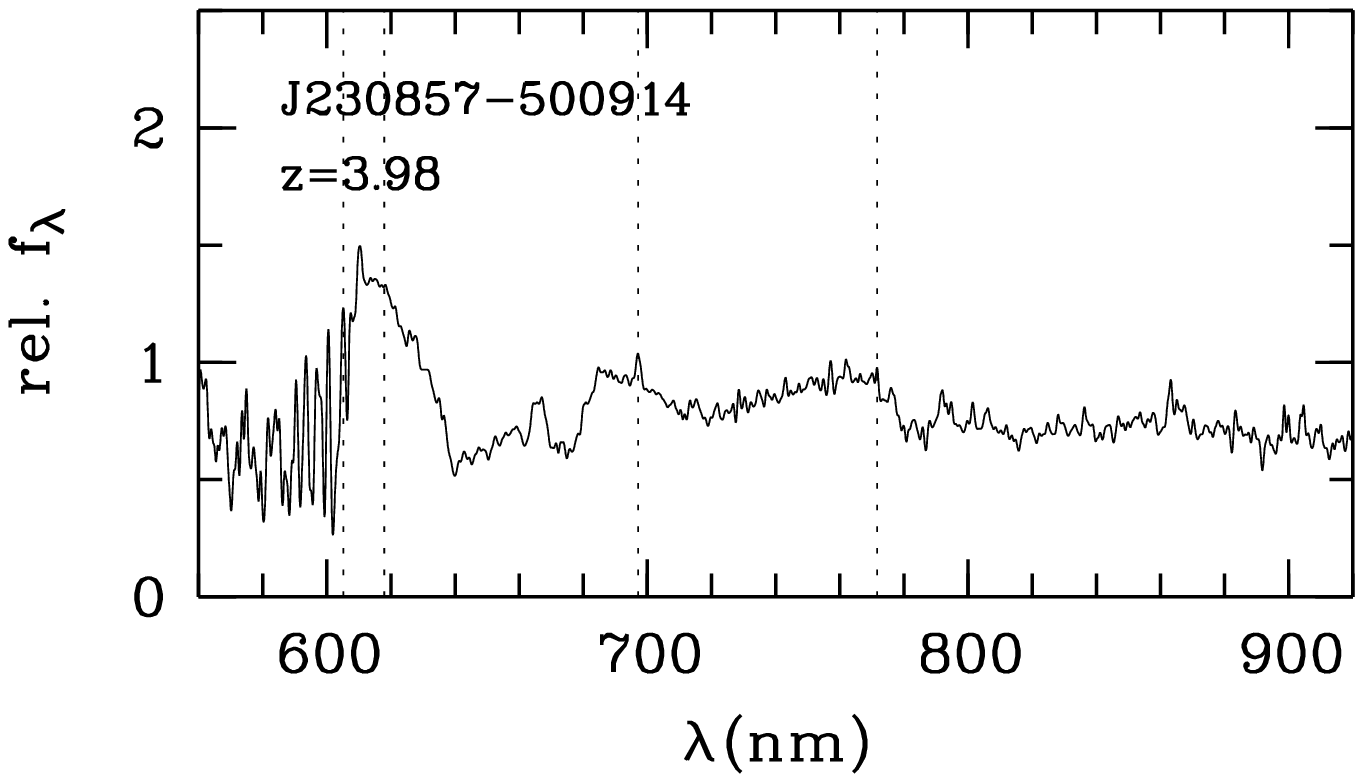}
\includegraphics[angle=270,width=0.32\textwidth,clip=true]{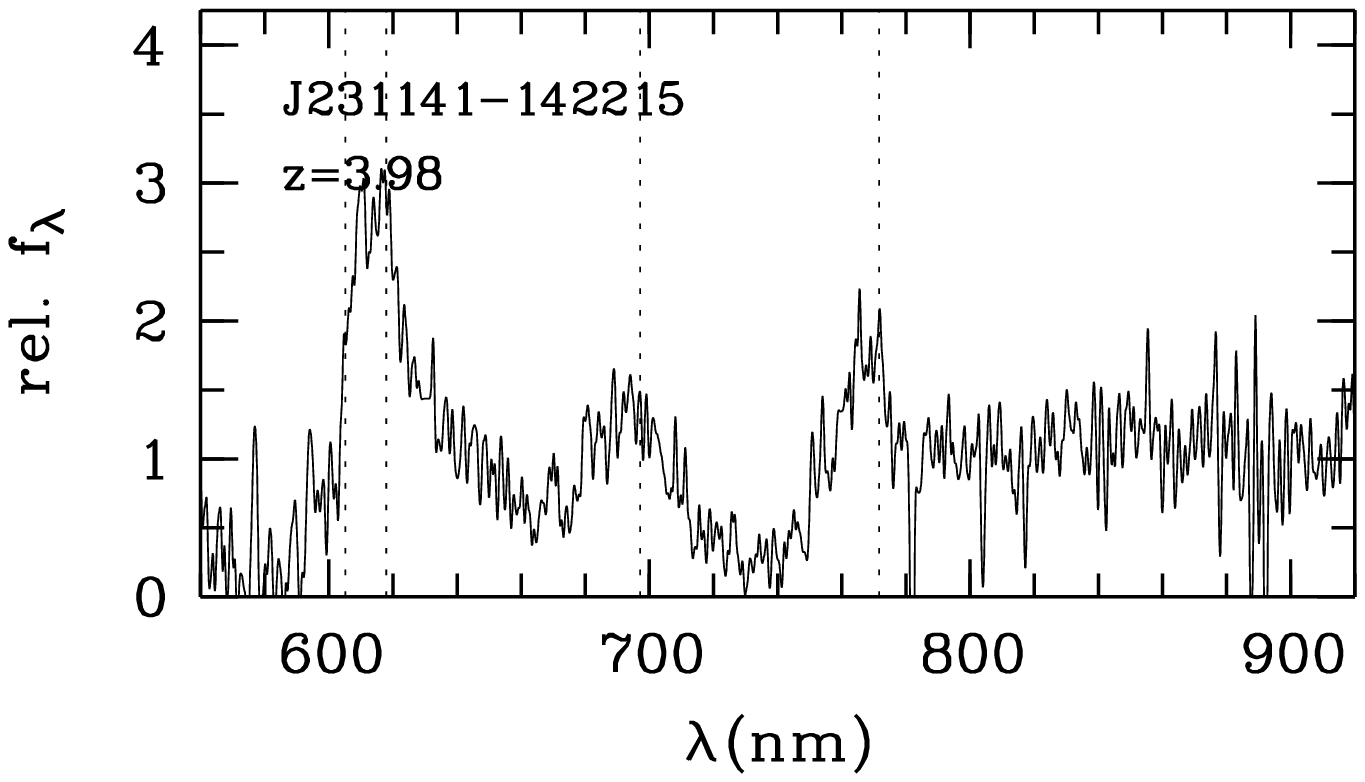}
\includegraphics[angle=270,width=0.32\textwidth,clip=true]{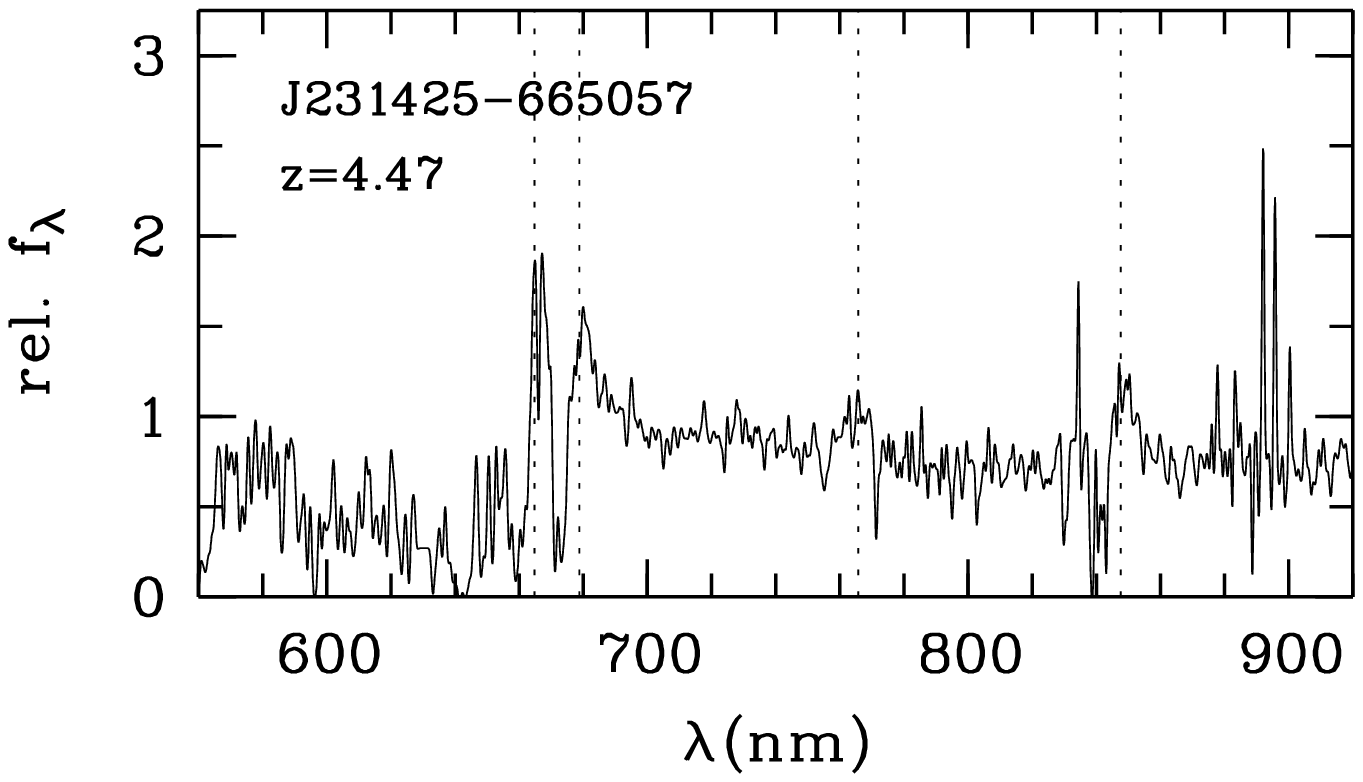}
\includegraphics[angle=270,width=0.32\textwidth,clip=true]{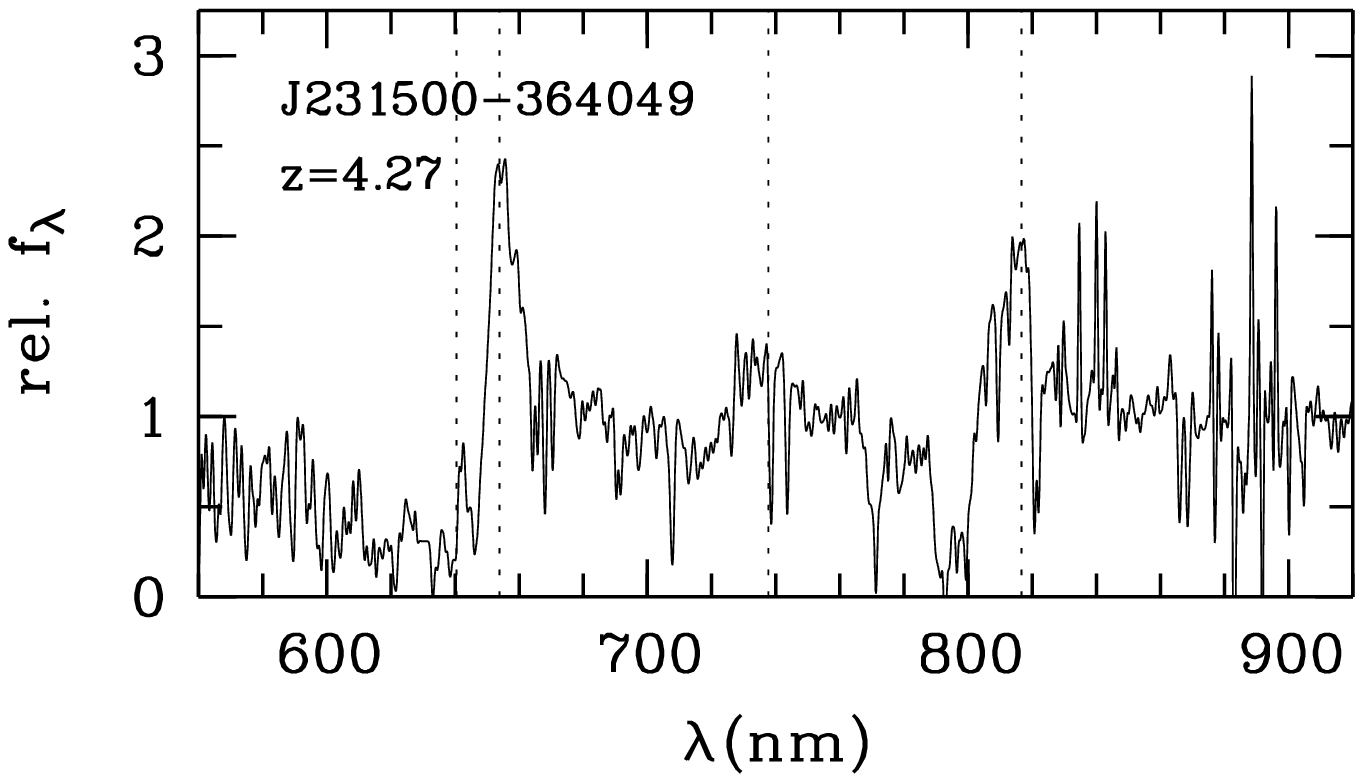}
\includegraphics[angle=270,width=0.32\textwidth,clip=true]{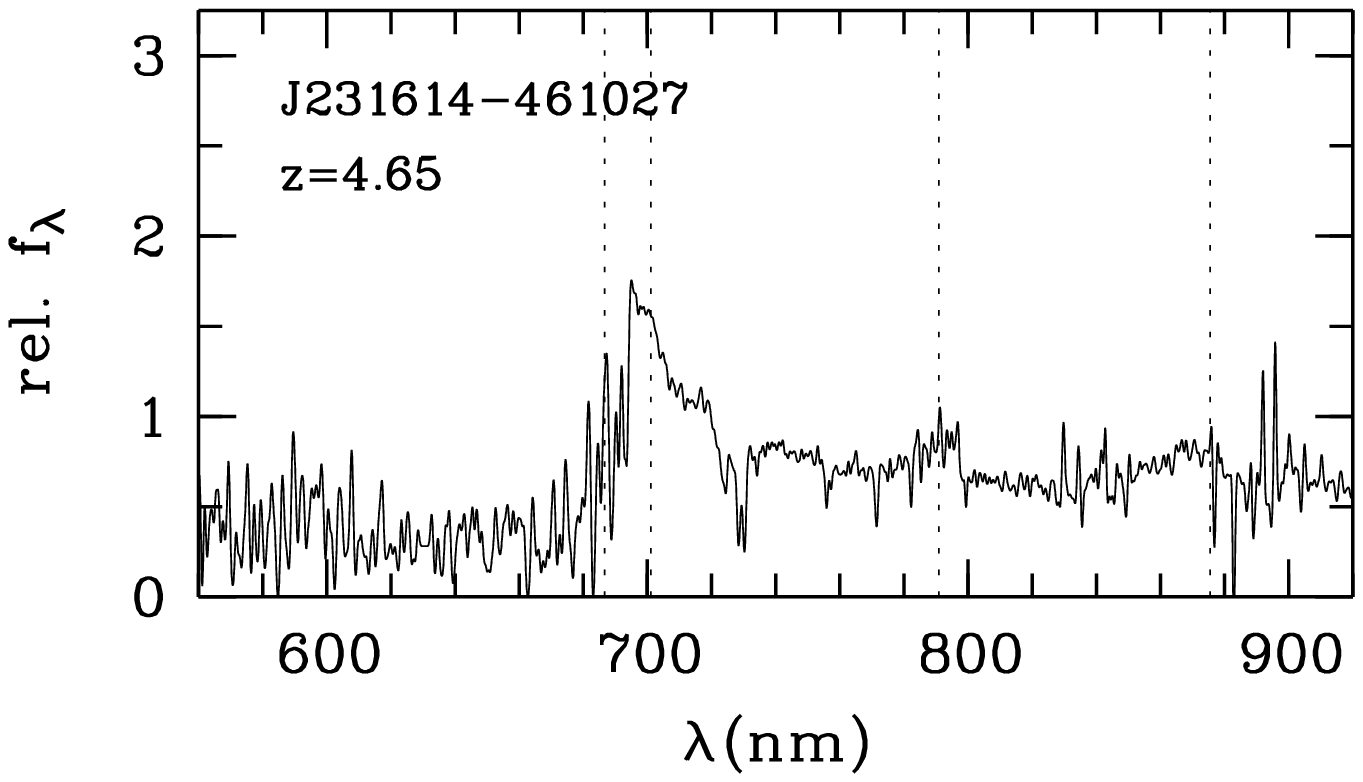}
\includegraphics[angle=270,width=0.32\textwidth,clip=true]{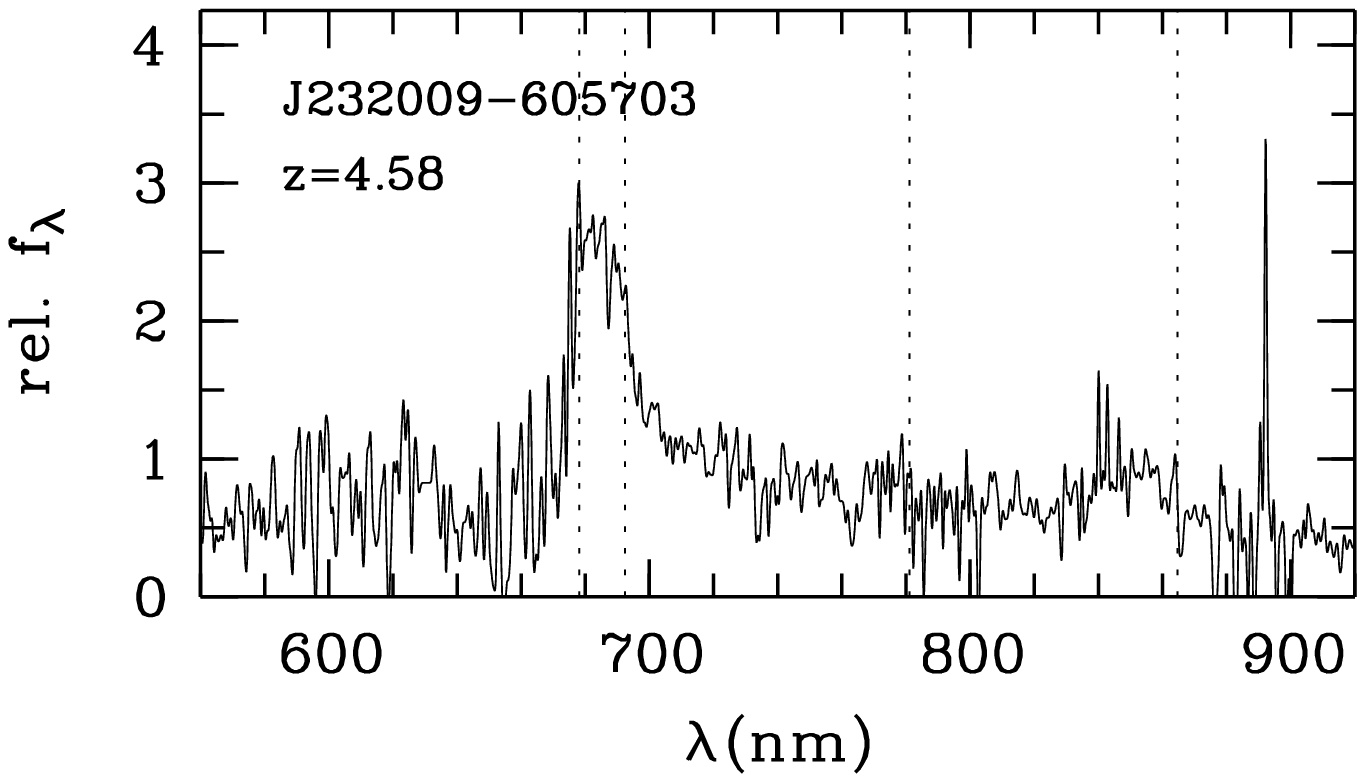}
\includegraphics[angle=270,width=0.32\textwidth,clip=true]{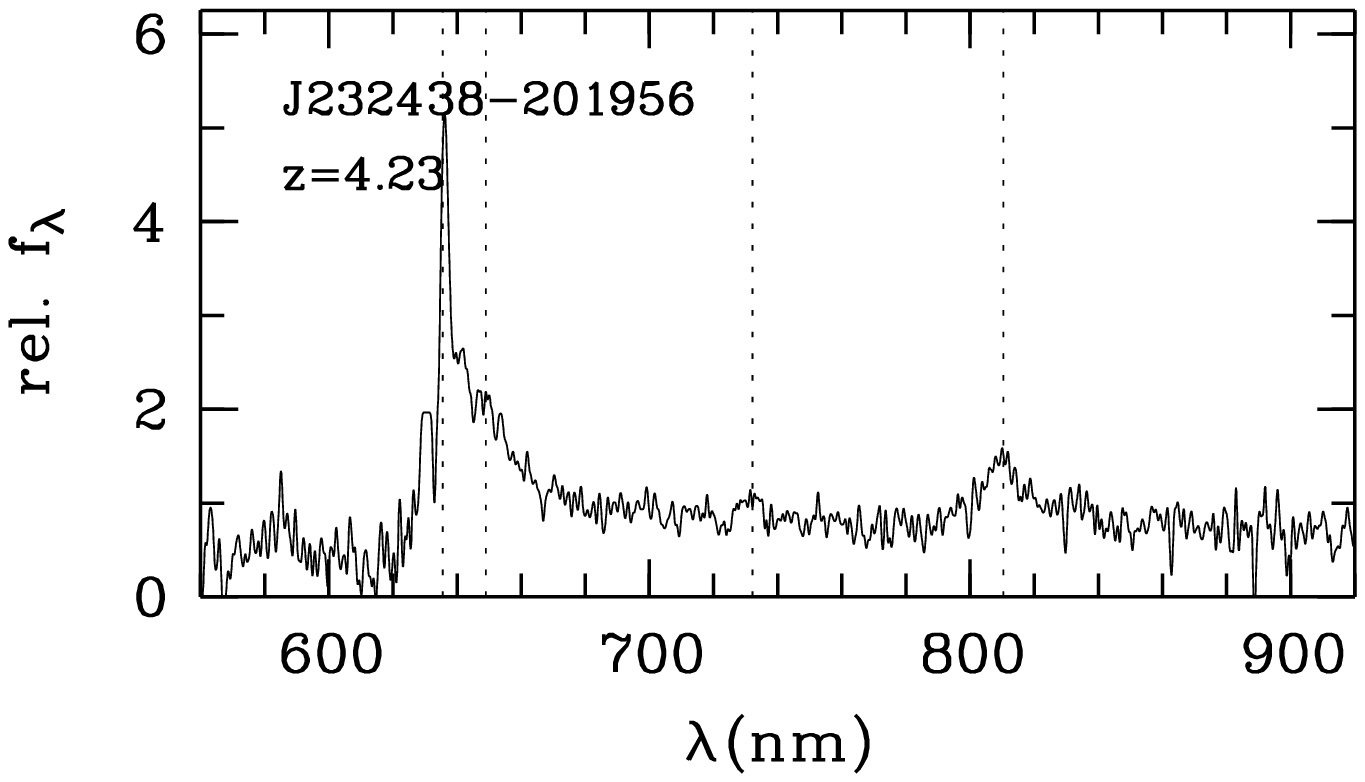}
\includegraphics[angle=270,width=0.32\textwidth,clip=true]{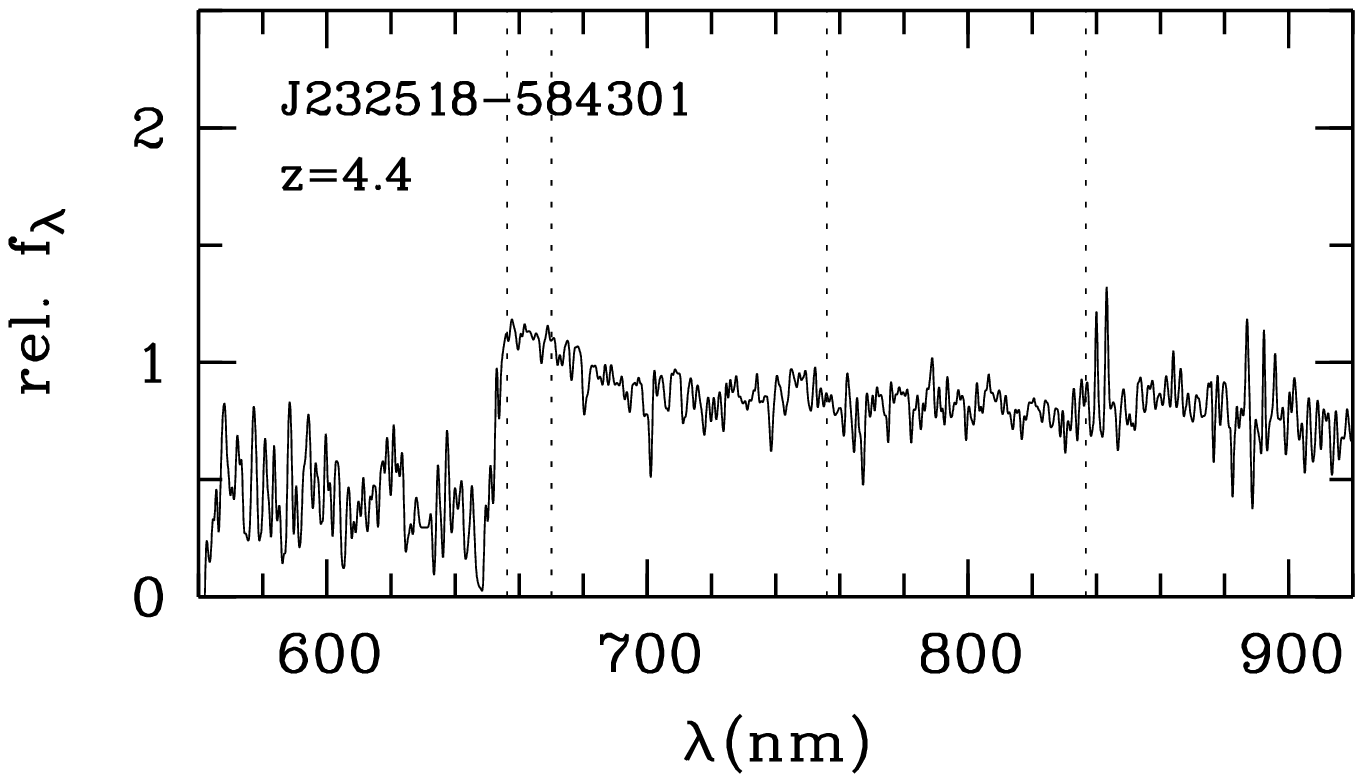}
\includegraphics[angle=270,width=0.32\textwidth,clip=true]{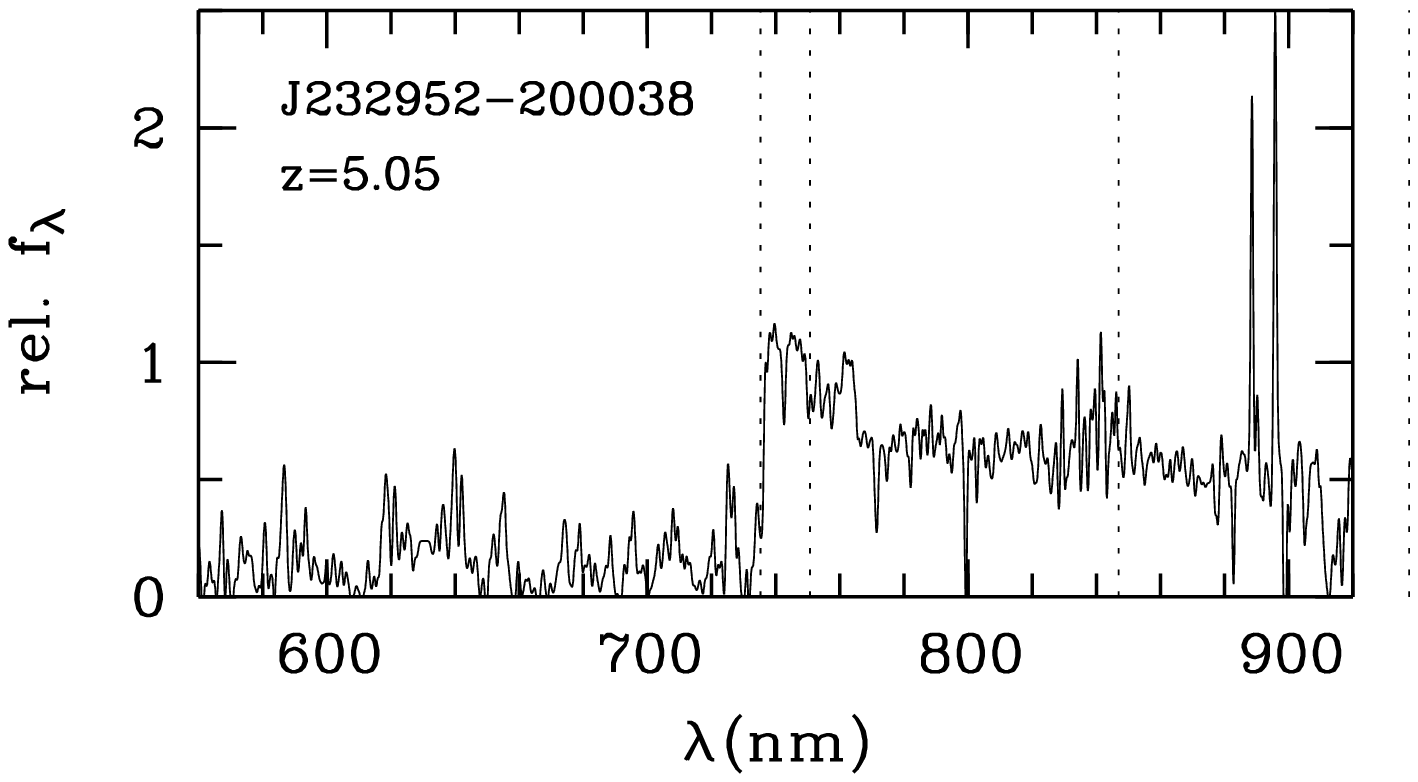}
\includegraphics[angle=270,width=0.32\textwidth,clip=true]{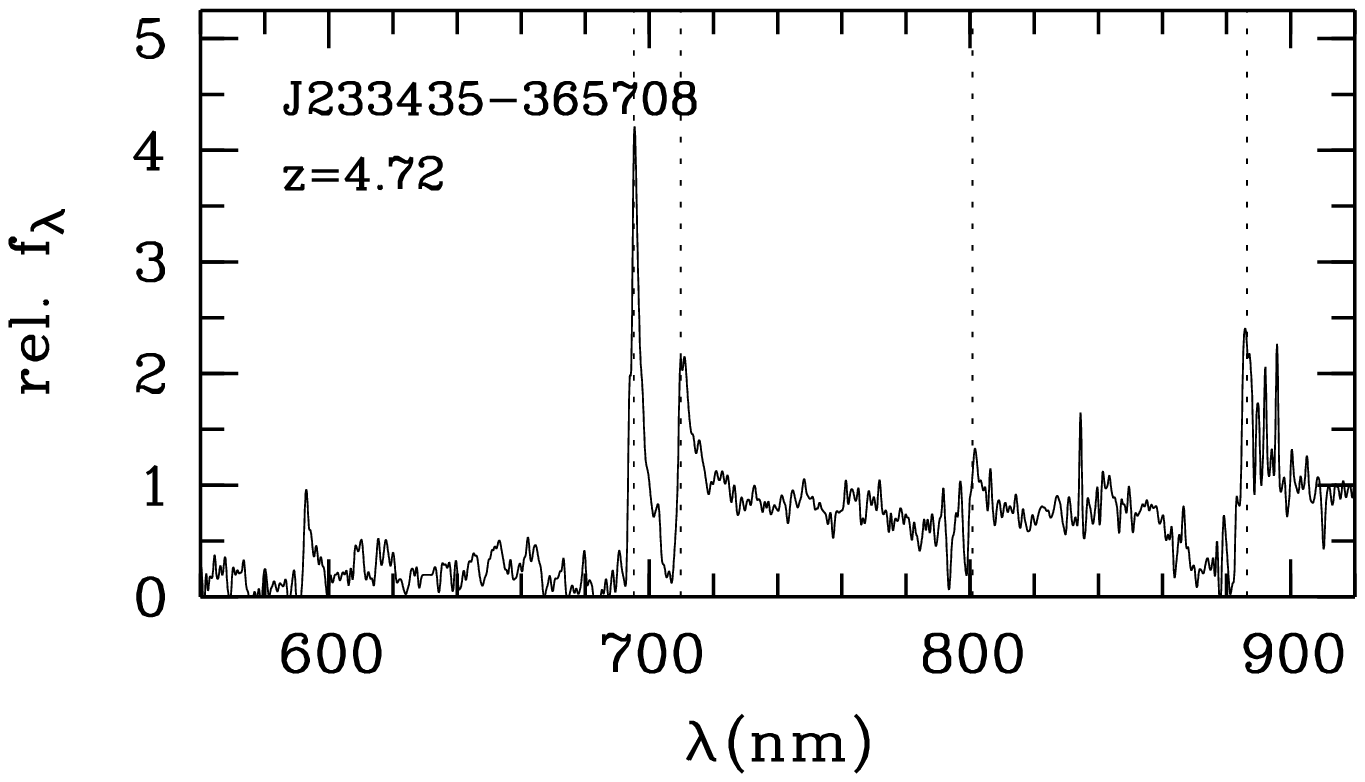}
\includegraphics[angle=270,width=0.32\textwidth,clip=true]{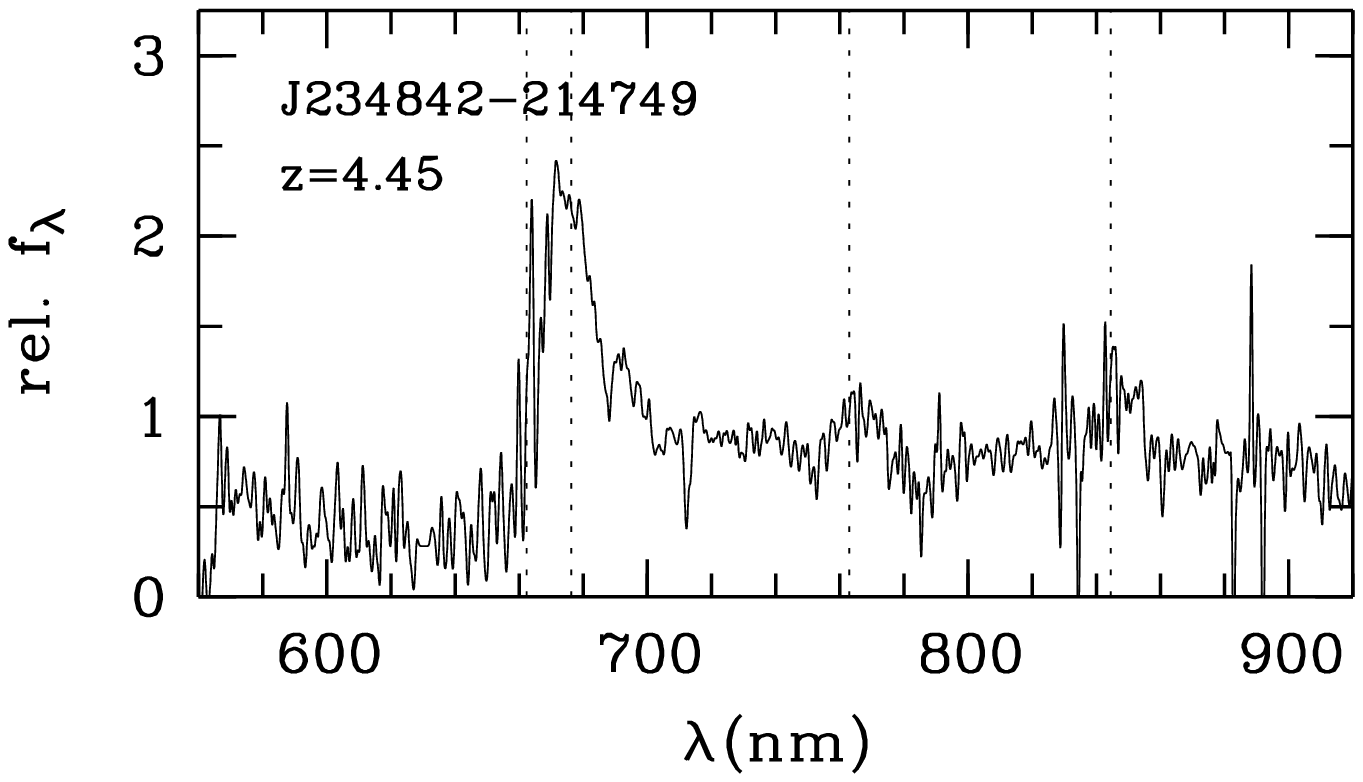}
\includegraphics[angle=270,width=0.32\textwidth,clip=true]{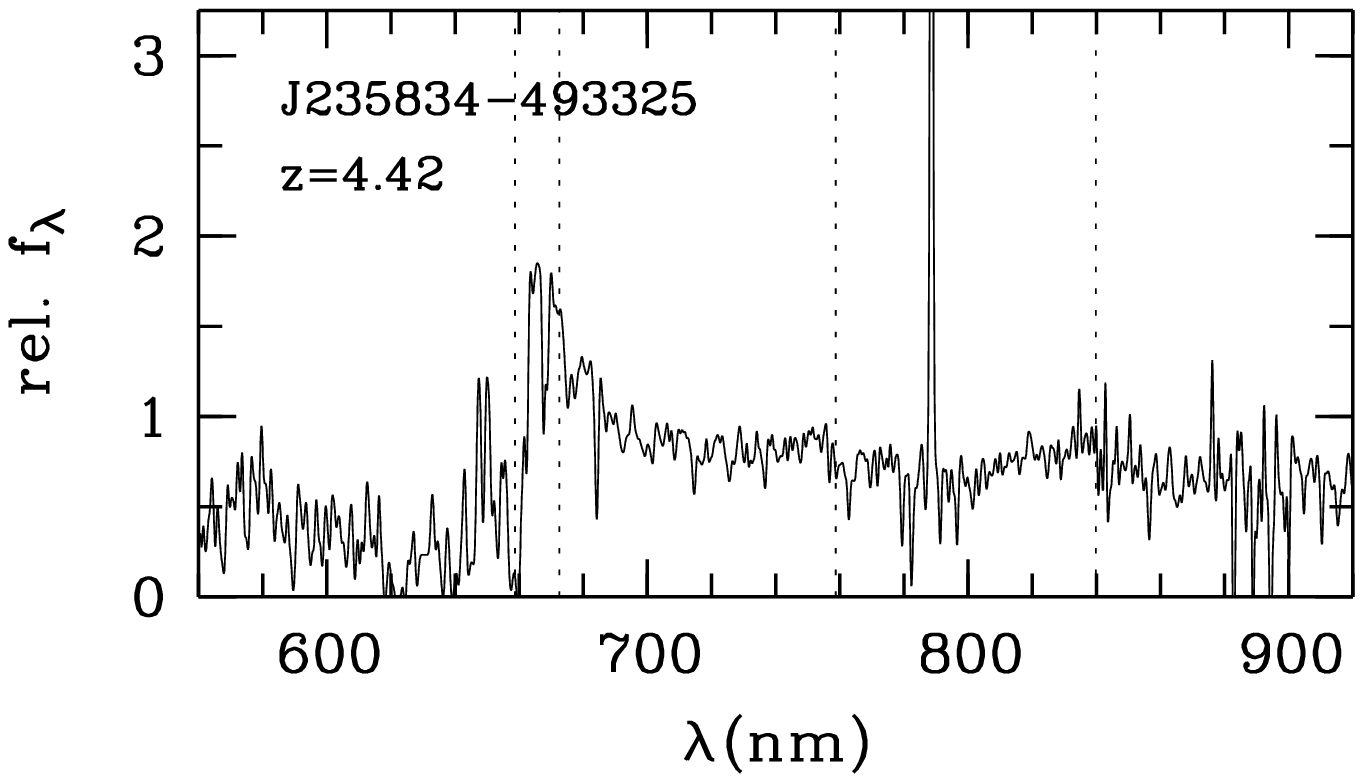}
\caption{Gallery of $3.8>z>5.5$ QSO spectra obtained in this work, ordered by RA, page 7.
\label{gallery7}}
\end{center}
\end{figure*}

\clearpage

\onecolumn
\section{SMSS~J164147.78-775029.8}\label{sec:J1641}
\counterwithin{figure}{section}

\begin{figure*}
\begin{center}
\includegraphics[angle=0,width=\textwidth,clip=true]{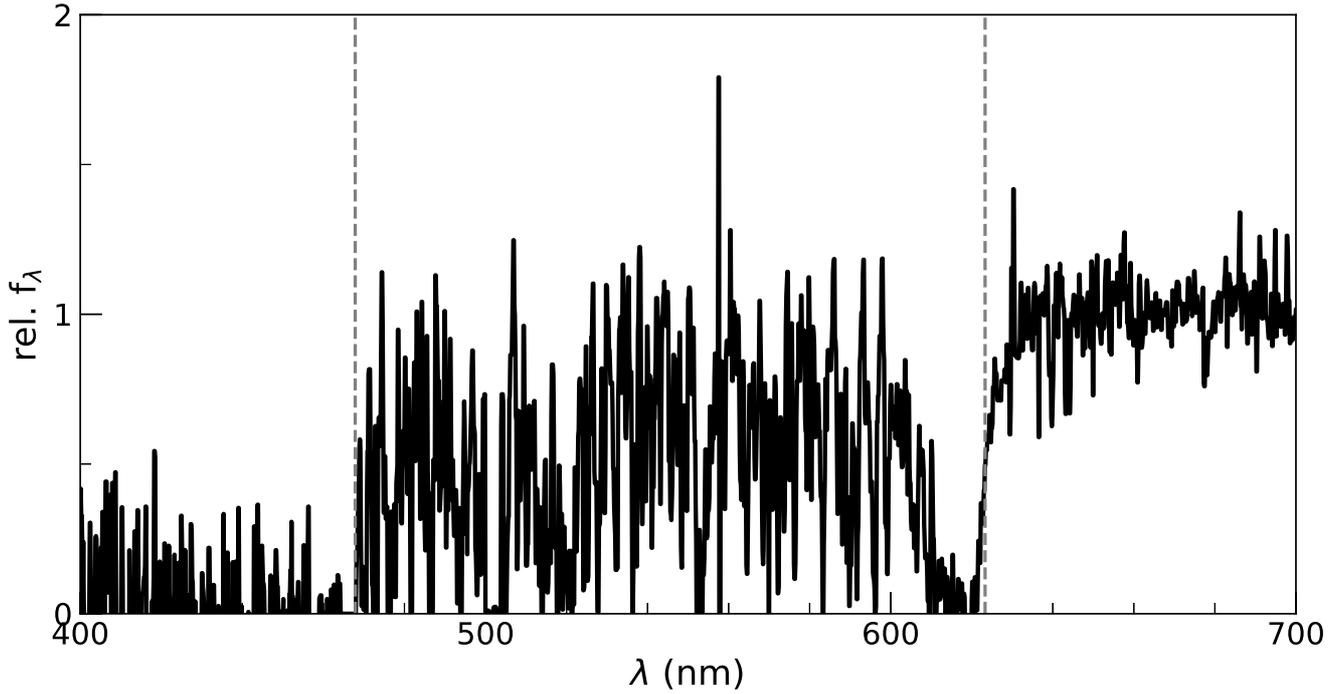}
\caption{Coadded observed-frame WiFeS spectrum of SMSS~J164147.78-775029.8 obtained with two different beam-splitters to better cover the wavelength range between Ly\,$\alpha$ and the Lyman limit (shown as the two dashed vertical lines for our adopted redshift of $z=4.13$.).
\label{fig:J1641}}
\end{center}
\end{figure*}

The lack of prominent emission lines in SMSS~J164147.78-775029.8 led us to acquire additional WiFeS spectroscopy with the RT480 beam-splitter on UT 2021-07-04, in an attempt to better examine the associated Ly\,$\alpha$ forest absorption and confirm the redshift. We present the weighted average spectrum in Figure~\ref{fig:J1641}, with a 3-pixel median smoothing applied to reduce the noise. The onset of absorption shortward of Ly\,$\alpha$ and the lack of transmitted flux shortward of the Lyman limit lend confidence to the inferred redshift of 4.13.

\bsp	
\label{lastpage}

\begin{thebibliography}{}

\bibitem[\protect\citeauthoryear{Baldwin}{1977}]{1977ApJ...214..679B} 
  Baldwin J.~A., 1977, ApJ, 214, 679. doi:10.1086/155294
\bibitem[\protect\citeauthoryear{Bessell et al.}{2011}]{2011PASP..123..789B} 
  Bessell M., Bloxham G., Schmidt B., Keller S., Tisserand P., Francis P., 2011, PASP, 123, 789. doi:10.1086/660849
\bibitem[\protect\citeauthoryear{Boutsia et al.}{2021}]{Boutsia21}
  Boutsia K., Grazian A., Fontanot F., Giallongo E., Menci N., Calderone G., Cristiani S., et al., 2021, ApJ, 912, 111. doi:10.3847/1538-4357/abedb5
\bibitem[\protect\citeauthoryear{Calderone et al.}{2019}]{Calderone19}
  Calderone G., Boutsia K., Cristiani S., Grazian A., Amorin R., D'Odorico V., Cupani G., et al., 2019, ApJ, 887, 268. doi:10.3847/1538-4357/ab510a 
\bibitem[\protect\citeauthoryear{Casagrande \& VandenBerg}{2018}]{CasaVDB18}
  Casagrande L., VandenBerg D.~A., 2018, MNRAS, 479, L102
\bibitem[\protect\citeauthoryear{Chambers et al.}{2016}]{2016arXiv161205560C} Chambers K.~C., Magnier E.~A., Metcalfe N., Flewelling H.~A., Huber M.~E., Waters C.~Z., Denneau L., et al., 2016, arXiv, arXiv:1612.05560
\bibitem[\protect\citeauthoryear{Childress et al.}{2014}]{Childress14}
  Childress, M. J., Vogt, F. P. A., Nielsen, J., \& Sharp, R. G. 2014, Ap\&SS, 349, 617
\bibitem[\protect\citeauthoryear{Cowie et al.}{1996}]{Cowie96}
  Cowie L.~L., Songaila A., Hu E.~M., Cohen J.~G., 1996, AJ, 112, 839. doi:10.1086/118058
\bibitem[\protect\citeauthoryear{Cross et al.}{2012}]{2012A&A...548A.119C} 
  Cross N.~J.~G., Collins R.~S., Mann R.~G., Read M.~A., Sutorius E.~T.~W., Blake R.~P., Holliman M., et al., 2012, A\&A, 548, A119. doi:10.1051/0004-6361/201219505
\bibitem[\protect\citeauthoryear{Dopita et al.}{2010}]{Dopita10} 
  Dopita, M., Rhee, J., Farage, C. et al. 2010, Ap\&SS, 327, 245
\bibitem[\protect\citeauthoryear{Edge et al.}{2013}]{Edge13}
  Edge A., Sutherland W., Kuijken K., Driver S., McMahon R., Eales S., Emerson J.~P., 2013, ESO Messenger, 154, 32
\bibitem[\protect\citeauthoryear{Fan et al.}{2001}]{Fan2001b}
  Fan X., Strauss M.~A., Schneider D.~P., Gunn J.~E., Lupton R.~H., Becker R.~H., Davis M., et al., 2001, AJ, 121, 54. doi:10.1086/318033
\bibitem[\protect\citeauthoryear{Ferrarese \& Merritt}{2000}]{Ferrarese00}
  Ferrarese L., Merritt D., 2000, ApJL, 539, L9. doi:10.1086/312838
\bibitem[\protect\citeauthoryear{Flesch}{2015}]{Flesch15}
  Flesch E.~W., 2015, PASA, 32, e010
\bibitem[\protect\citeauthoryear{Gaia Collaboration et al.}{2018}]{GaiaDR2}
  Gaia Collaboration, Brown A.~G.~A., Vallenari A., Prusti T., de Bruijne J.~H.~J., Babusiaux C., Bailer-Jones C.~A.~L., et al., 2018, A\&A, 616, A1. doi:10.1051/0004-6361/201833051
\bibitem[\protect\citeauthoryear{Gaia Collaboration et al.}{2021}]{Gaia_eDR3} 
  Gaia Collaboration, Brown A.~G.~A., Vallenari A., Prusti T., de Bruijne J.~H.~J., Babusiaux C., Biermann M., et al., 2021, A\&A, 649, A1. doi:10.1051/0004-6361/202039657
\bibitem[\protect\citeauthoryear{Grazian et al.}{2021}]{2021arXiv211013736G} 
  Grazian A., Giallongo E., Boutsia K., Calderone G., Cristiani S., Cupani G., Fontanot F., et al., 2021, ApJ (in press, arXiv:2110.13736)
\bibitem[\protect\citeauthoryear{Guarneri et al.}{2021}]{2021MNRAS.506.2471G} 
  Guarneri F., Calderone G., Cristiani S., Fontanot F., Boutsia K., Cupani G., Grazian A., et al., 2021, MNRAS, 506, 2471. doi:10.1093/mnras/stab1867
\bibitem[\protect\citeauthoryear{Hambly et al.}{2008}]{2008MNRAS.384..637H} 
  Hambly N.~C., Collins R.~S., Cross N.~J.~G., Mann R.~G., Read M.~A., Sutorius E.~T.~W., Bond I., et al., 2008, MNRAS, 384, 637. doi:10.1111/j.1365-2966.2007.12700.x
\bibitem[\protect\citeauthoryear{Harikane et al.}{2021}]{2021arXiv210801090H} 
  Harikane Y., Ono Y., Ouchi M., Liu C., Sawicki M., Shibuya T., Behroozi P.~S., et al., 2021, ApJS (in press, arXiv:2108.01090)
\bibitem[\protect\citeauthoryear{Hasinger, Miyaji, \& Schmidt}{2005}]{Hasinger05}
  Hasinger G., Miyaji T., Schmidt M., 2005, A\&A, 441, 417. doi:10.1051/0004-6361:20042134
\bibitem[\protect\citeauthoryear{Hinton et al.}{2016}]{Hinton16}
  Hinton S.~R., Davis T.~M., Lidman C., Glazebrook K., Lewis G.~F., 2016, Astronomy \& Computing, 15, 61
\bibitem[\protect\citeauthoryear{Hopkins, Richards, \& Hernquist}{2007}]{Hopkins07}
  Hopkins P.~F., Richards G.~T., Hernquist L., 2007, ApJ, 654, 731. doi:10.1086/509629
\bibitem[\protect\citeauthoryear{Irwin et al.}{2004}]{2004SPIE.5493..411I} 
  Irwin M.~J., Lewis J., Hodgkin S., Bunclark P., Evans D., McMahon R., Emerson J.~P., et al., 2004, SPIE, 5493, 411. doi:10.1117/12.551449
\bibitem[\protect\citeauthoryear{Jiang et al.}{2016}]{Jiang16}
  Jiang L. et al., 2016, ApJ, 833, 222
\bibitem[\protect\citeauthoryear{Jones et al.}{2001}]{SciPy}
  Jones E. et al., 2001, {\sc SciPy}: Open source scientific tools for {\sc Python}, \url{http://www.scipy.org/}
\bibitem[\protect\citeauthoryear{Karachentsev, Makarov, \& Kaisina}{2013}]{Karachentsev13}
  Karachentsev I.~D., Makarov D.~I., Kaisina E.~I., 2013, AJ, 145, 101. doi:10.1088/0004-6256/145/4/101
\bibitem[\protect\citeauthoryear{Kim et al.}{2020}]{Kim20}
  Kim Y., Im M., Jeon Y., Kim M., Pak S., Hyun M., Taak Y.~C., et al., 2020, ApJ, 904, 111. doi:10.3847/1538-4357/abc0ea (K20)
\bibitem[\protect\citeauthoryear{Kormendy \& Ho}{2013}]{KH13}
  Kormendy J., Ho L.~C., 2013, ARA\&A, 51, 511. doi:10.1146/annurev-astro-082708-101811
\bibitem[\protect\citeauthoryear{Lyke et al.}{2020}]{Lyke20}
  Lyke B.~W., Higley A.~N., McLane J.~N., Schurhammer D.~P., Myers A.~D., Ross A.~J., Dawson K., et al., 2020, ApJS, 250, 8. doi:10.3847/1538-4365/aba623
\bibitem[\protect\citeauthoryear{Marocco et al.}{2021}]{Marocco21}
  Marocco F., Eisenhardt P.~R.~M., Fowler J.~W., Kirkpatrick J.~D., Meisner A.~M., Schlafly E.~F., Stanford S.~A., et al., 2021, ApJS, 253, 8. doi:10.3847/1538-4365/abd805
\bibitem[\protect\citeauthoryear{Marshall et al.}{1983}]{1983ApJ...269...35M}
  Marshall H.~L., Tananbaum H., Avni Y., Zamorani G., 1983, ApJ, 269, 35. doi:10.1086/161016
\bibitem[\protect\citeauthoryear{McGreer et al.}{2013}]{2013ApJ...768..105M} 
  McGreer I.~D., Jiang L., Fan X., Richards G.~T., Strauss M.~A., Ross N.~P., White M., et al., 2013, ApJ, 768, 105. doi:10.1088/0004-637X/768/2/105
\bibitem[\protect\citeauthoryear{McGreer et al.}{2018}]{McGreer18}
  McGreer I.~D., Fan X., Jiang L., Cai Z., 2018, AJ, 155, 131. doi:10.3847/1538-3881/aaaab4
\bibitem[\protect\citeauthoryear{McMahon et al.}{2013}]{VHS}
  McMahon, R.~G., Banerji, M., Gonzalez, E. et al.\ 2013, The Messenger, 154, 35 
\bibitem[\protect\citeauthoryear{Mortlock et al.}{2012}]{Mortlock12}
  Mortlock D.~J., Patel M., Warren S.~J., Hewett P.~C., Venemans B.~P., McMahon R.~G., Simpson C., 2012, MNRAS, 419, 390. doi:10.1111/j.1365-2966.2011.19710.x
\bibitem[\protect\citeauthoryear{Nidever et al.}{2021}]{2021AJ....161..192N} 
  Nidever D.~L., Dey A., Fasbender K., Juneau S., Meisner A.~M., Wishart J., Scott A., et al., 2021, AJ, 161, 192. doi:10.3847/1538-3881/abd6e1
\bibitem[\protect\citeauthoryear{Niida et al.}{2020}]{Niida20}
  Niida M., Nagao T., Ikeda H., Akiyama M., Matsuoka Y., He W., Matsuoka K., et al., 2020, ApJ, 904, 89. doi:10.3847/1538-4357/abbe11 (N20)
\bibitem[\protect\citeauthoryear{Onken et al.}{2019}]{Onken19}
  Onken, C. A., Wolf, C., Shao, L., Luvaul, L. C. et al. 2019, PASA, 36, 33
\bibitem[\protect\citeauthoryear{Onken et al.}{2020}]{Onken20}
  Onken C.~A., Bian F., Fan X., Wang F., Wolf C., Yang J., 2020, MNRAS, 496, 2309. doi:10.1093/mnras/staa1635
\bibitem[\protect\citeauthoryear{Rakshit, Stalin, \& Kotilainen}{2020}]{2020ApJS..249...17R} 
  Rakshit S., Stalin C.~S., Kotilainen J., 2020, ApJS, 249, 17. doi:10.3847/1538-4365/ab99c5
\bibitem[\protect\citeauthoryear{Reed et al.}{2017}]{Reed17}
  Reed S.~L., McMahon R.~G., Martini P., Banerji M., Auger M., Hewett P.~C., Koposov S.~E., et al., 2017, MNRAS, 468, 4702. doi:10.1093/mnras/stx728
\bibitem[\protect\citeauthoryear{Richards et al.}{2006}]{Richards06}
  Richards G.~T., Strauss M.~A., Fan X., Hall P.~B., Jester S., Schneider D.~P., Vanden Berk D.~E., et al., 2006, AJ, 131, 2766. doi:10.1086/503559
\bibitem[\protect\citeauthoryear{Richards et al.}{2009}]{Richards09}
  Richards G.~T., Deo R.~P., Lacy M., Myers A.~D., Nichol R.~C., Zakamska N.~L., Brunner R.~J., et al., 2009, AJ, 137, 3884. doi:10.1088/0004-6256/137/4/3884
\bibitem[\protect\citeauthoryear{Riello et al.}{2021}]{Riello20} 
  Riello M., De Angeli F., Evans D.~W., Montegriffo P., Carrasco J.~M., Busso G., Palaversa L., et al., 2021, A\&A, 649, A3. doi:10.1051/0004-6361/202039587
\bibitem[\protect\citeauthoryear{Rodrigo, Solano, \& Bayo}{2012}]{2012ivoa.rept.1015R} 
  Rodrigo C., Solano E., Bayo A., 2012, ivoa.rept. doi:10.5479/ADS/bib/2012ivoa.rept.1015R
\bibitem[\protect\citeauthoryear{Rodrigo \& Solano}{2020}]{2020sea..confE.182R} 
  Rodrigo C., Solano E., 2020, sea..conf, 182
\bibitem[\protect\citeauthoryear{Ryan-Weber et al.}{2009}]{Ryan-Weber09}
  Ryan-Weber E.~V., Pettini M., Madau P., Zych B.~J., 2009, MNRAS, 395, 1476
\bibitem[\protect\citeauthoryear{Schindler et al.}{2019a}]{Schindler19}
  Schindler J.-T., Fan X., Huang Y.-H., Yue M., Yang J., Hall P.~B., Wenzl L., et al., 2019a, ApJS, 243, 5. doi:10.3847/1538-4365/ab20d0
\bibitem[\protect\citeauthoryear{Schindler et al.}{2019b}]{Schindler19b}
  Schindler J.-T., Fan X., McGreer I.~D., Yang J., Wang F., Green R., Fynbo J.~P.~U., et al., 2019b, ApJ, 871, 258. doi:10.3847/1538-4357/aaf86c
\bibitem[\protect\citeauthoryear{Schlafly \& Finkbeiner}{2011}]{SF11}
  Schlafly E.~F., Finkbeiner D.~P., 2011, ApJ, 737, 103. doi:10.1088/0004-637X/737/2/103
\bibitem[\protect\citeauthoryear{Schlegel, Finkbeiner, \& Davis}{1998}]{SFD98}
  Schlegel D.~J., Finkbeiner D.~P., Davis M., 1998, ApJ, 500, 525 
\bibitem[\protect\citeauthoryear{Selsing et al.}{2016}]{2016A&A...585A..87S} 
  Selsing J., Fynbo J.~P.~U., Christensen L., Krogager J.-K., 2016, A\&A, 585, A87. doi:10.1051/0004-6361/201527096 (S16)
\bibitem[\protect\citeauthoryear{Simcoe et al.}{2011}]{Simcoe11}
  Simcoe, R.~A., Cooksey, K.~L., Matejek, M. et al. 2011, \apj, 743, 21 
\bibitem[\protect\citeauthoryear{Skrutskie et al.}{2006}]{2MASS}
  Skrutskie, M.~F., Cutri, R. M., Stiening, R., Weinberg, M.~D., et al. 2006, AJ, 131, 1163 
\bibitem[\protect\citeauthoryear{Telfer et al.}{2002}]{2002ApJ...565..773T} 
  Telfer R.~C., Zheng W., Kriss G.~A., Davidsen A.~F., 2002, ApJ, 565, 773. doi:10.1086/324689
\bibitem[\protect\citeauthoryear{Vanden Berk et al.}{2001}]{2001AJ....122..549V} 
  Vanden Berk D.~E., Richards G.~T., Bauer A., Strauss M.~A., Schneider D.~P., Heckman T.~M., York D.~G., et al., 2001, AJ, 122, 549. doi:10.1086/321167
\bibitem[\protect\citeauthoryear{Wang et al.}{2016}]{Wang16}
  Wang, F., Wu, X.-B., Fan, X. et al. 2016, ApJ, 819, 24 (W16)
\bibitem[\protect\citeauthoryear{Wang \& Chen}{2019}]{WangChen19}
  Wang S., Chen X., 2019, ApJ, 877, 116. doi:10.3847/1538-4357/ab1c61
\bibitem[\protect\citeauthoryear{Wenzl et al.}{2021}]{2021AJ....162...72W} 
  Wenzl L., Schindler J.-T., Fan X., Andika I.~T., Ba{\~n}ados E., Decarli R., Jahnke K., et al., 2021, AJ, 162, 72. doi:10.3847/1538-3881/ac0254 (W21)
\bibitem[\protect\citeauthoryear{Wolf, Meisenheimer, \& R{\"o}ser}{2001}]{WMR01}
  Wolf C., Meisenheimer K., R{\"o}ser H.-J., 2001, A\&A, 365, 660. doi:10.1051/0004-6361:20000474
\bibitem[\protect\citeauthoryear{Wolf et al.}{2018a}]{Wolf18a}
  Wolf, C., Onken, C. A., Luvaul, L. C. et al. 2018a, PASA, 35, 10. doi: 10.4225/41/593620ad5b574
\bibitem[\protect\citeauthoryear{Wolf et al.}{2018b}]{Wolf18b}
  Wolf, C., Bian, F., Onken, C. A., Schmidt, B. P., Tisserand, P. et al. 2018b, PASA, 35, 24 
\bibitem[\protect\citeauthoryear{Wolf et al.}{2020}]{Wolf20}
  Wolf C., Hon W.~J., Bian F., Onken C.~A., Alonzi N., Bessell M.~A., Li Z., et al., 2020, MNRAS, 491, 1970. doi:10.1093/mnras/stz2955 (Paper~I)
\bibitem[\protect\citeauthoryear{Wright et al.}{2010}]{Wright10}
  Wright, E. L., Eisenhardt, P. R. M., Mainzer, A. K. et al. 2010, AJ, 140, 1868
\bibitem[\protect\citeauthoryear{Xie et al.}{2016}]{Xie16}
  Xie X., Shao Z., Shen S., Liu H., Li L., 2016, ApJ, 824, 38. doi:10.3847/0004-637X/824/1/38
\bibitem[\protect\citeauthoryear{Yang et al.}{2016}]{Yang16}
  Yang J. et al., 2016, ApJ, 829, 33 (Y16)
\bibitem[\protect\citeauthoryear{Yang et al.}{2017}]{Yang17}
  Yang J. et al., 2017, AJ, 153, 184
\bibitem[\protect\citeauthoryear{Yang et al.}{2019}]{Yang19}
  Yang J. et al., 2019, ApJ, 871, 199
\bibitem[\protect\citeauthoryear{York et al.}{2000}]{SDSS}
  York, D.~G., Adelman, J., Anderson, J.~E., Jr. et al.\ 2000, \aj, 120, 1579 

\end{thebibliography}
\end{document}